\documentclass[preprint]{aa}
\newcommand{\kms}{km~s$^{-1}$}
\usepackage{natbib}
\usepackage{graphicx}
\usepackage{multirow,bigdelim}
\usepackage[varg]{txfonts}
\usepackage{capt-of}
\usepackage{longtable}
\authorrunning{J.~K.~J{\o}rgensen et al.}

\begin{document}

\title{The ALMA-PILS survey: Isotopic composition of oxygen-containing
  complex organic molecules toward IRAS~16293--2422B}
\titlerunning{Isotopic composition of oxygen-containing
  complex organic molecules toward IRAS~16293--2422B}
\authorrunning{J.~K.~J{\o}rgensen et al.}

\author{J.~K. J{\o}rgensen\inst{1}, H.~S.~P. M{\"u}ller\inst{2},
  H. Calcutt\inst{1}, A. Coutens\inst{3}, M.~N. Drozdovskaya\inst{4},
  K.~I. \"{O}berg\inst{5} M.~V. Persson\inst{6}, V. Taquet\inst{7},
  E.~F. van Dishoeck\inst{8,9}, \and S.~F. Wampfler\inst{4}} 
\institute{Centre for Star and Planet Formation, Niels Bohr Institute \& Natural
  History Museum of Denmark, University of Copenhagen, {\O}ster
  Voldgade 5--7, DK-1350 Copenhagen {K}., Denmark 
\and I. Physikalisches Institut, Universit\"{a}t zu K\"{o}ln,
Z\"{u}lpicher Str. 77, 50937 K\"{o}ln, Germany
\and Laboratoire d'astrophysique de Bordeaux, Univ. Bordeaux, CNRS, B18N, all{\'e}e Geoffroy Saint-Hilaire, 33615 Pessac, France 
\and Center for Space and Habitability (CSH), University of Bern, Sidlerstrasse 5, CH-3012 Bern, Switzerland
\and Harvard-Smithsonian Center for Astrophysics, 60 Garden Street, Cambridge, MA 02138, USA
\and Department of Earth and Space Sciences, Chalmers University of Technology, Onsala Space Observatory, 439 92 Onsala, Sweden
\and INAF-Osservatorio Astrofisico di Arcetri, Largo E. Fermi 5, I-50125 Firenze, Italy
\and Leiden Observatory, Leiden University, PO Box 9513, NL-2300 RA Leiden, The Netherlands
\and Max-Planck Institut f\"{u}r Extraterrestrische Physik (MPE), Giessenbachstr. 1, 85748 Garching, Germany}

\abstract {One of the important questions of astrochemistry is how
  complex organic molecules, including potential prebiotic species,
  are formed in the envelopes around embedded protostars. The
  abundances of minor isotopologues of a molecule, in particular the
  D- and $^{13}$C-bearing variants, are sensitive to the densities,
  temperatures and time-scales characteristic of the environment in
  which they form, and can therefore provide important constraints on
  the formation routes and conditions of individual species.}
{The aim of this paper is to systematically survey the deuteration and
  the $^{13}$C content of a variety of oxygen-bearing complex organic
  molecules on Solar System scales toward the ``B component'' of
    the protostellar binary IRAS~16293--2422.}
{We use the data from an unbiased molecular line survey of the
  protostellar binary IRAS 16293$-$2422 between 329 and 363~GHz from
  the Atacama Large Millimeter/submillimeter Array (ALMA). The data
  probe scales of 60~AU (diameter) where most of the organic molecules
  are expected to have sublimated off dust grains and be present in
  the gas-phase. The deuterated and $^{13}$C-isotopic species of
  ketene, acetaldehyde and formic acid, as well as deuterated ethanol,
  are detected unambiguously for the first time in the interstellar
  medium.  These species are analysed together with the $^{13}$C
  isotopic species of ethanol, dimethyl ether and methyl formate along
  with mono-deuterated methanol, dimethyl ether and methyl formate. }
{The complex organic molecules can be divided into two groups with one
  group, the simpler species, showing a D/H ratio of $\approx$~2\% and
  the other, the more complex species, D/H ratios of 4--8\%. This
  division may reflect the formation time of each species in the ices
  before or during warm-up/infall of material through the protostellar
  envelope. No significant differences are seen in the deuteration of
  different functional groups for individual species, possibly a
  result of the short time-scale for infall through the innermost warm
  regions where exchange reactions between different species may be
  taking place. The species show differences in excitation
  temperatures between 125~K and 300~K. This likely reflects the
  binding energies/sublimation temperatures of the individual species,
  in good agreement to what has previously been found for high-mass
  sources. For dimethyl ether the $^{12}$C/$^{13}$C ratio is found to be lower by up to a
  factor of 2 compared to typical ISM values similar to what has
  previously been inferred for glycolaldehyde. Tentative
  identifications suggest
  that the same may apply for $^{13}$C
  isotopologues of methyl
  formate and ethanol. If confirmed, this may be a
  clue to their formation at the late prestellar/early protostellar
  phases with an enhancement of the available $^{13}$C relative to
  $^{12}$C related to small differences in binding energies for CO
  isotopologues or the impact of FUV irradiation by the central
  protostar.}
{The results point to the importance of ice surface chemistry for the
  formation of these complex organic molecules at different stages in
  the evolution of embedded protostars and demonstrate the use of
  accurate isotope measurements for understanding the history of
  individual species.}

\offprints{Jes K.\,J{\o}rgensen, \email{jeskj@nbi.ku.dk}}
\maketitle

\section{Introduction}\label{introduction}
The earliest stages of protostars are characterised by a rich
chemistry, which in some cases leads to abundant complex organic, and
even prebiotic, molecules present in the gas-phase close to the young
stars of both low- and high-mass. The general picture is that the
formation of icy grain mantles and chemistry in and on grains
facilitates the build-up of larger species through surface reactions
starting from CO
\citep[e.g.,][]{tielens82,hasegawa93,charnley97,watanabe02,fuchs09,oberg09,garrod13,fedoseev15,chuang17}. However,
there are still a number of important open questions concerning the
chemistry leading to the formation of these species, for example, what
specific reactions are dominant, what is the importance of, e.g., UV
irradiation, how important is the ice composition and the rate at
which the ices are warming up.

The isotopic composition of the gas is a particular characteristic of
the early stages of young stars, which may hold significant clues to
some of these networks. It has for some time been recognised that the
many species in the cold gas around embedded protostars are
significantly enhanced in deuterium relative to hydrogen
\citep[e.g.,][]{vandishoeck95,ceccarelli01d2co} compared to the cosmic
D/H ratio of $1.5-2.0\times 10^{-5}$ \citep[e.g,.,][]{linsky03,prodanovic10}. In particular,
the deuterated isotopologues of species such as formaldehyde (H$_2$CO)
and methanol (CH$_3$OH) are found to be significantly enhanced up to
levels of $\sim$~10\% or above compared to the regular isotopologues
in single-dish observations \citep[e.g.,][]{parise06deuterium} and
some of these sources even show detections of multiply deuterated
species (e.g., CD$_2$HOH and CD$_3$OH; \citealt{parise02} and
\citealt{parise04}) indicating very high degrees of deuteration
\citep[see also overview in][]{jorgensen16}.

These enhancements are thought to be a consequence of the exothermic reaction
\begin{equation}
{\rm H}_3^+ + {\rm HD} \quad\leftrightarrows\quad {\rm H}_2{\rm D}^+ + {\rm H}_2+ \Delta E
\end{equation}
(where $\Delta E = 232$~K), which followed by dissociative
recombination of H$_2$D$^+$ with electrons leads to an enhanced atomic
D/H ratio that can be transferred to grains when the neutral species
freeze-out \citep{tielens83}. Other isotopic systems such as
  $^{12}$C and $^{13}$C-containing isotopologues may also show
  variations compared to the local ISM due to the fractionation
  processes in the cold phases although the variations of those
    are expected to be at much smaller levels than those of
    deuterium vs. hydrogen. Isotope selective photodissociation or ion-molecule
  reactions in the gas through the reaction:
\begin{equation}
{\rm ^{13}C}^+ + {\rm CO}\quad \leftrightarrows \quad {\rm C}^+ +
{\rm^{13}CO}+\Delta E
\end{equation}
(where $\Delta E = 35$~K; \citealt{langer84,furuya11}) may enhance the
$^{13}$CO abundance in the gas-phase relative to $^{12}$CO. Isotope
selective photodissociation may decrease the amount of $^{13}$CO
relative to $^{12}$CO, but at the same time enhancing the amount of
$^{13}$C. Once on the grains, the enhanced $^{13}$CO or $^{13}$C can
then be incorporated into the complex organic molecules. Other
mechanisms, such as segregation between $^{12}$CO and $^{13}$CO in
the ices due to slight differences in their binding energies
\citep[e.g.,][]{smith15} may also affect the amount of $^{12}$C
relative to $^{13}$C that is available for incorporation into larger
molecules.

Since all of these fractionation processes are very sensitive to the
gas physics (density and temperature) it has therefore been suggested
that measurements of the relative abundances of different
isotopologues of individual molecules may hold strong clues to the
formation pathways constrained by chemical models or alternatively to
the physical conditions and early evolution of young stars through
their pre- or protostellar stages
\citep[e.g.][]{charnley04,cazaux11,taquet13,taquet14}. With the
Atacama Large Millimeter/submillimeter Array (ALMA) it is becoming
possible to push forward in providing accurate measurements of the
isotopic composition of complex organics.

As an example of such a study, \cite{belloche16} presented a
systematic survey of the deuteration of molecules toward the
Sagittarius B2 high-mass star forming region in the Central Molecular
Zone close to the Galactic Center. Utilising ALMA's high sensitivity,
\citeauthor{belloche16} presented new detections as well as strict
upper limits for the deuterated variants of a number of complex
organics. Typically the abundances of the deuterated species were
found to be lower than what is seen in nearby molecular clouds and
also predicted by astrochemical models. \cite{belloche16} speculated
that this may reflect a lower overall deuterium abundance or the
higher temperatures in the clouds toward the Galactic Center.

One of the prime targets for similar studies of solar-type stars is
the Class~0 protostar IRAS~16293--2422. This source has long been
recognised as having large abundances of deuterated species compared
to high-mass protostellar sources \citep{vandishoeck95} and was also
the first solar-type protostar for which abundant complex organic
molecules were found \citep{cazaux03}. Combining these two properties,
it has been a natural target for deuteration studies.  This is the
source toward which the above mentioned multi-deuterated species, as
well as some deuterated complex organics, e.g., methyl formate
(CH$_3$OCHO; \citealt{demyk10}) and dimethyl ether (CH$_3$OCH$_3$;
\citealt{richard13}), were first detected. Recent observations of the
THz ground state transitions of H$_2$D$^+$ \citep{bruenken14} and
D$_2$H$^+$ \citep{harju17} toward the cold envelope around
IRAS~16293--2422 suggest very similar amount of both cations in this
source. Due to sensitivity limitations, however, the single-dish
observations typically targeted the brightest, low excitation,
transitions that in some cases may be optically thick both for the
main isotopologues as well as for the deuterated
variants. Furthermore, single-dish observations naturally encompass
the entire protostellar envelope and are thus intrinsically biased
toward emission on larger scales where most of the material is located
and molecules may still be largely frozen-out on dust grains due to
the lower temperatures.

Using ALMA we have recently completed a large unbiased survey, the
\emph{Protostellar Interferometric Line Survey (PILS)}, of a key
frequency window around 345~GHz \citep{jorgensen16}. These data are up
to two orders of magnitude more sensitive than previous single-dish
studies and thus provide excellent opportunities for systematic
studies of complex organics and their isotopologues. In particular,
the interferometric observations make it possible to zoom in on the
inner 30--60~AU around the central protostars, i.e., Solar System
scales, where the temperatures are high and the ices have sublimated
and the complex organic molecules therefore are all present in the
gas-phase. The initial results presented the first detections of the
deuterated isotopologues of isocyanic acid and formamide
\citep{coutens16} as well as the deuterated and $^{13}$C-isotopologues
of glycolaldehyde \citep{jorgensen16}. These species all show
significant deuterium enhancements -- although not on the levels
hinted by the previous single-dish observations -- but also
indications of differences that may exactly reflect their
chemistries. Specifically, glycolaldehyde is more enhanced in
deuterium than the other species by a factor of a few. The
$^{13}$C-isotopologues of glycolaldehyde further indicate an
enhancement of the $^{13}$C isotopologues above that of the local ISM,
again possibly reflecting formation of this species in the coldest
part of the envelope.

In the present work, we present an extensive investigation into the
deuteration and the $^{13}$C content of oxygen-bearing complex organic
molecules. These include methanol, its next heavier homologue ethanol
along with its isomer dimethyl ether, the dehydrogenated relatives of
ethanol, acetaldehyde, and ketene, as well as methyl formate, an
isomer of glycolaldehyde, and formic acid. The paper is laid out as
follows: Sect.~\ref{analysis} summarises the main points about the
ALMA data and spectroscopy while Sect.~\ref{results} describes the
data analysis, including line identification and column
density/abundance determinations of the complex organics and their
isotopologues. Sect.~\ref{discussion} discusses the implications in
relation to the formation of the species.

\section{Data analysis}\label{analysis}
In the following we present the results concerning identification of
the isotopologues of oxygen-bearing complex organic
molecules. Sect.~\ref{observations} summarises the key aspects of the
observations and Sect.~\ref{lab-spec} describes the information about
laboratory spectroscopy. Further information about the history of
detections and issues with the fits for individual species are given
in Appendix~\ref{indspecies}.

\subsection{The ALMA PILS survey}\label{observations}
In this paper we utilise data from the ALMA-PILS survey targeting the
protostellar binary IRAS~16293--2422 (IRAS16923 hereafter) using the
Atacama Large Millimeter/submillimeter Array.  An overview of the data
and reduction as well as first results from the survey are presented
in \cite{jorgensen16}; here we highlight a few specifics. The PILS
survey provides a full survey of IRAS16293 covering the frequency
range in ALMA's Band~7 from 329.1 to 362.9~GHz with 0.2~\kms\ spectral
and $\approx$~0.5\arcsec\ angular resolution (60~AU diameter)
performed in ALMA's Cycle 2 (project-id: 2013.1.00278.S). Also,
additional targeted observations were performed in ALMA's Bands~3
($\approx$~100~GHz) and 6 ($\approx 230$~GHz) taken as part of a
Cycle~1 program (project-id: 2012.1.00712.S). The Band~7 part of the
survey is one-to--two orders of magnitude more sensitive than previous
surveys, reaching a sensitivity of
$\approx$5~mJy~beam$^{-1}$~km~s$^{-1}$ across the entire frequency
range. In this paper, we focus on the ``B'' component of the
  source, which, with its narrow lines ($\sim$~1~km~s$^{-1}$;
  e.g., \citealt{iras16293sma}),
  is an ideal target for searches for rare species.

\subsection{Laboratory spectroscopic information}
\label{lab-spec}
The spectroscopic data were taken from the Cologne Database of
Molecular Spectroscopy, CDMS \citep{cdms1,cdms2} and the Jet
Propulsion Laboratory (JPL) \citep[][]{jpl} catalog. These databases
compile, and in some cases extend, spectroscopic information and
partition functions based on reports in literature. Some care must be
taken when comparing across these spectroscopic entries. In
particular, for some entries, the partition functions only contain the
ground vibrational state while the excited vibrational states also may
be populated at the temperatures of up to a few hundred K
characterising the gas on small scales toward IRAS16293. Calculated
column densities considering only the ground vibrational state will
therefore underestimate the true values.  We therefore carefully
examined the original literature for each species as described in the
following, e.g., evaluating the vibrational correction factors.
Particular attention was paid to make sure that the partition
functions were in fact derived using the same degeneracies and thus
line strengths as listed in the line catalogs.

For the analysis of methanol we used the $^{18}$O isotopologue using
the catalogue entry from the CDMS. The entry is based on
\citet{18O-MeOH_rot_2007} who combined new far-infrared (FIR) data and
rotational data with microwave accuracy summarised in
\citet{18O-MeOH_rot_1998}. The partition function should be converged
at 300~K. For CH$_2$DOH the entry from the JPL catalogue was used:
this entry is based on \citet{CH2DOH_rot_2012}. The partition function
in that entry takes only the ground vibrational state into
account. The vibrational correction factor is estimated to be 1.457 at
300~K based on torsional data from \citet{CH2DOH_FIR_2009}. For
CH$_3$OD transition frequencies from \citet{CH3OD_rot_2003} with
published line strengths from \citet{CH3OD_rot_1988} and with
partition function values scaled with those from CH$_3^{18}$OH are
used.

Ethanol, C$_2$H$_5$OH, possesses two large amplitude motions, the
torsions of the CH$_3$ and the OH groups. The methyl torsion is
usually neglected in the ground vibrational state. The OH torsion
leads to two distinguishable conformers, the lower lying \textit{anti}
conformer and the doubly degenerate \textit{gauche}
conformer. Tunneling between the two \textit{gauche} minima leads to
two distinguishable states, the symmetric \textit{gauche}$^+$ and the
antisymmetric \textit{gauche}$^-$.  \citet{EtOH_rot_2008} present an
extensive analysis of the ground state rotational spectrum of ethanol
and determined the energy differences with great accuracy. The results
of the analysis are available in the JPL catalogue. \citet{muller16},
however, found that the predicted intensities do not match those of
ethanol emissions in an ALMA survey of Sagittarius (Sgr) B2(N2) at
3~mm. Intensity discrepancies were also detected in laboratory spectra
(F. Lewen, unpublished) and in our ALMA data. A new catalogue entry
was created for the CDMS as a consequence. The line and parameter
files are those provided in the JPL catalogue. The entry is based on
\citet{EtOH_rot_2008} with additional data in the range of our survey
from \citet{anti-EtOH_rot_1995} and \citet{gauche-EtOH_rot_1996} for
\textit{anti} and \textit{gauche} ethanol, respectively.

Information on the $^{13}$C and D containing isotopomers of ethanol
were taken from the CDMS.  The entries were based on
\citet{13C-EtOH_rot_2012} and \citet{D-EtOH_rot_2015}, respectively.
The entries deal with the \textit{anti} conformer only in each case
which can be treated separately up to moderately high quantum numbers
\citep{anti-EtOH_rot_1995}. Transitions are modeled well for
$J + 2K_a \le 32$ in the case of the $^{13}$C isotopomers, slightly
higher in $J$ or $K_a$ for mono-deuterated ethanol. The column
densities of ethanol with $^{13}$C or D have to be multiplied by
$\sim$2.69 in order to account for the presence of the \textit{gauche}
conformer. This factor was evaluated from the ground state partition
functions of the main isotopic species involving the \textit{anti}
conformer only and both conformers, respectively.  The correction
factors are expected to be very similar for the $^{13}$C isotopomers,
but may be slightly larger for the deuterated isotopomers. The
vibrational factor at 300~K is 2.824 \citep{EtOH_EtSH_IR_1975}.

The dimethyl ether, CH$_3$OCH$_3$, data were taken from the CDMS. They
are based on \citet{DME_rot_2009} with additional data in the range of
our survey from \citet{DME_rot_1998}. Data of mono-deuterated dimethyl
ether were taken from \citet{richard13}, those of $^{13}$CH$_3$OCH$_3$
from \citet{13C-DME_lab_det_2013}. The partition function of the main
isotopologue includes contributions from vibrational states up to
$\varv _7 = 1$ (\citealt{DME_FIR_Ra_1977} \& C.~P. Endres,
unpublished). The difference at 125~K is only a factor of 1.185.

For acetaldehyde, CH$_3$CHO, the JPL catalogue entry based on
\citet{CH3CHO_catalog_1996} is used.  Experimental transition
frequencies in the range of our survey are from
\citet{CH3CHO_rot_1993}.  \citet{CH3CDO_rot_2010} provides data on
CH$_3$CDO while \citet{13C-CH3CHO_rot_2015} published data on the
isotopomers with one $^{13}$C. The partition function of the main
isotopologue includes contributions from the first and second excited
methyl torsional mode. Their contributions are 0.243 at 125~K for the
first and 0.049 for the second.

The ketene entries are taken from the CDMS catalogue. The main,
$^{13}$C and $^{18}$O data are based on \citet{ketene_plus_rot_2003}
with additional data in the range of our survey from
\citet{ketene_plus_rot_1990}. The entries with one and two D are based
on \citet{ketene-D1_2_etc_rot_2005}.  Vibrational contributions to the
partition function are very small at 125~K.

We took the methyl formate (CH$_3$OCHO) data from the JPL
catalogue. The entry is based on \citet{CH3OCHO_analysis_2009} with
data in the range of our survey from \citet{A-CH3OCHO_rot_1984},
\citet{CH3OCHO_rot_1999}, and \citet{CH3OCHO_rot_2008}. The CH$_3$OCDO
data were based on \citet{CH3OCDO_rot_1995} and \cite{margules09}, and
those of CH$_2$DOCHO on \citealt{CH2DOCHO_rot_2009} \citep[available
in][]{coudert13}. We consulted the CDMS catalogue for
CH$_3$O$^{13}$CHO data. The entry is based on
\citet{CH3OC-13-HO_rot_2010} with additional data in the range of our
survey from \citet{CH3OC-13-HO_rot_2006}, \citet{CH3OC-13-HO_rot_2008}
and \citet{CH3OCHO_rot_2008}. The vibrational factors for entries
containing only $v_t=0$ and those containing both $v_t=0$ and $v_t=1$
are listed in the CDMS.

Formic acid, HCOOH, occurs in two conformers, \textit{trans} and
\textit{cis}, with the latter being almost 2000~K higher in
energy. Predictions for the main species were taken from the CDMS
which were based on \citet{winnewisser02}.  Predictions for
H$^{13}$COOH, DCOOH and HCOOD were taken from the JPL catalogue.  They
were based on \citet{lattanzi08} with (additional) data in the range
of our survey from
\citet{HCOOH_13C_cis_rot_2002,HC-13-OOH_rot_nu7_nu9_2006} for
H$^{13}$COOH, from \citet{DCOOH_etc_rot_1996} for DCOOH and summarised
in \cite{HCOOD_nu7_nu9_1999} for HCOOD. Formic acid has two relatively
low-lying vibrational modes, $\nu _7$ and $\nu _9$.  Vibrational
corrections to the ground state partition function at 300~K are 1.103,
1.105, 1.107 and 1.177 for HCOOH, H$^{13}$COOH, DCOOH and HCOOD,
respectively
\citep{HCOOH_nu7_nu9_2002,HC-13-OOH_rot_nu7_nu9_2006,DCOOH_nu7_nu9_2003,HCOOD_nu7_nu9_1999}.

\subsection{Species identification, modeling and uncertainties}\label{modeling}
For the identification and modeling of individual species we adopted
the same procedure as described in \cite{jorgensen16}: since the
density of lines is high enough to approach the confusion limit at
different points in the spectra it is not practical to identify all
clean transitions of a specific species and, e.g., create rotation
diagrams based on their (measured) intensities. In fact, for a
reliable identification of a given species it is important to
demonstrate that there are no anti-coincidences, i.e., the inferred
column density does not predict bright lines that are not
observed. Also, inherent to the rotation diagram method is that the
line emission is optically thin, which may not be the case for the
main isotopologues of the complex organics on the scales that we
consider. Rotation diagrams may therefore lead to underestimated
column densities, while fainter optically thin transitions that are
not considered in the fits, and for the derived column density
predicted not to be observable, in fact are present at a significant
level. Nevertheless, carefully constructed rotation diagrams may be
useful to check the consistency of derived temperatures and column
densities.

We calculate synthetic spectra for each species and compare those to
the data for the most part similar to what is done in many other
surveys of regions with dense spectra \citep[e.g.][as well as previous
PILS
papers]{comito05,zernickel12,crockett14,neill14,belloche13,belloche16,cernicharo16}. Rather
than fitting the individual lines, synthetic spectra are calculated
assuming that the emission from the molecules are in Local
Thermodynamic Equilibrium (LTE), taking into account optical depth and
line overlaps. With this approach, the parameters going into the
modeling are the column density of the molecule, its excitation
temperature, the LSR velocity, the width of the lines and the extent
of the emission on the sky. These synthetic spectra can then be
directly compared to the observed spectrum over the full spectral
range and constrain those parameters. In this paper, the LTE
assumption for the complex organic molecules is justified by the high
density, $n_{{\rm H}_2}\gtrsim 10^{10}$~cm$^{-3}$, in the environment
of IRAS16293B on the 30--60~AU scales considered here, as demonstrated
by the non-LTE calculations for methanol in \cite{jorgensen16}. As
noted in previous PILS papers \citep{coutens16,jorgensen16,lykke17},
all features toward IRAS16293B are well reproduced with a source size
of 0.5$''$, a line width (FWHM) of 1.0~\kms\ and a LSR velocity of
2.5--2.7~\kms. This likely reflects the fact that the material on the
scales of the ALMA resolution is relatively homogeneously distributed
and dominated by one physical component.

With those parameters fixed, the remaining quantities to be
  derived for each species are excitation temperatures and column
  densities for each species. The uncertainties for the typical fits
  can be estimated comparing synthetic diagrams to the observed
  spectra. Fig.~\ref{a-ch3od_texerror} and \ref{a-ch3od_coldenserror}
  in Sect.~\ref{methanolfits} of the Appendix compare the best fit for
  CH$_3$OD to models where the derived excitation temperatures and
  column densities are varied. As seen, models with either excitation
  temperatures or column densities higher or lower by 20\% start
  providing fits that are visibly worse and not reproducing the
  observed spectra. These percentages can therefore be considered
  conservative uncertainties on each fit. Naturally, these two
  parameters cannot be considered as completely independent, but
  varying the excitation temperature within the $\pm$20\% results in
  changes of the derived column densities of the species of
  $<$~10\%. 

  For rarer isotopologues with weaker lines than shown in
  Fig.~\ref{a-ch3od_coldenserror} the excitation temperature typically
  cannot be derived independently but that from the main isotopologue
  is adopted, in which case the column density is the only free
  parameter and the uncertainty remains the same as long as lines with
  S/N larger than 5 (i.e., statistical uncertainty is $< 20$\% for the
  intensity of each line) are considered. Likewise, the statistical
  uncertainty on the line fits for species with more, and stronger,
  transitions may be smaller. However, those are also the species
  where optical depth become more important and where the range of
  lines start to be more sensitive to the exact structure of the
  emitting region -- including, whether, e.g., a gradient in
  temperature should be considered: close to the central protostars
  the temperatures are strongly varying and therefore the single
  excitation temperature derived for each species likely represents
  the temperature in the gas where the bulk of the emission from that
  species originate.

 For different
  isotopologues of the same molecule, it is a reasonable first
  assumption that the excitation temperature is unchanged as long as
  only optically thin lines are considered. As found in previous PILS
  papers \citep[e.g.,][]{coutens16,lykke17} there are some differences
  with respect to the optimal excitation temperatures for different
  groups of species. While glycolaldehyde, ethylene glycol
  \citep{jorgensen12,jorgensen16} and formamide \citep{coutens16} are
  well-modeled with an excitation temperature of around 300~K, lines
  of other species such as acetaldehyde and ethylene oxide are best
  reproduced with a lower excitation temperature of around 125~K. It
  seems that other species follow this trend. Within the uncertainties
  quoted above (25~K for an excitation temperature of 125~K and 60~K
  for a temperature of 300~K) and the small dependence of column
  density on those we therefore broadly group the species into 
  ``hot'' (300~K) or ``cold'' (125~K) and use those excitation
  temperatures. While, in principle, small variations may be present
  on a species-to-species level within these groups, those variations
  will more likely be statistical at this level rather than
  representing the physical structure. Again, any differences at this
  level will only cause minor variations in the derived column
  densities. For relative abundances expressed as ratios of column
  densities between different species, error propagation translate the
  20\% uncertainties into an uncertainty of $\approx
  30\%$~$(= \sqrt{2}\,\times\, 20\%)$ -- or, e.g., 0.6 or 1.5
  percentage points for D/H ratios of 2\% or 5\%, respectively. But,
  the error could be lower if only calibration uncertainty is an issue
  and other parameters such as $T_{\rm ex}$ can be assumed to be the
  same. This should certainly be valid for the comparison between the
  $^{12}$C- and $^{13}$C-isotopologues and likely also some of the
  deuterated species.

\begin{figure}
\resizebox{\hsize}{!}{\includegraphics{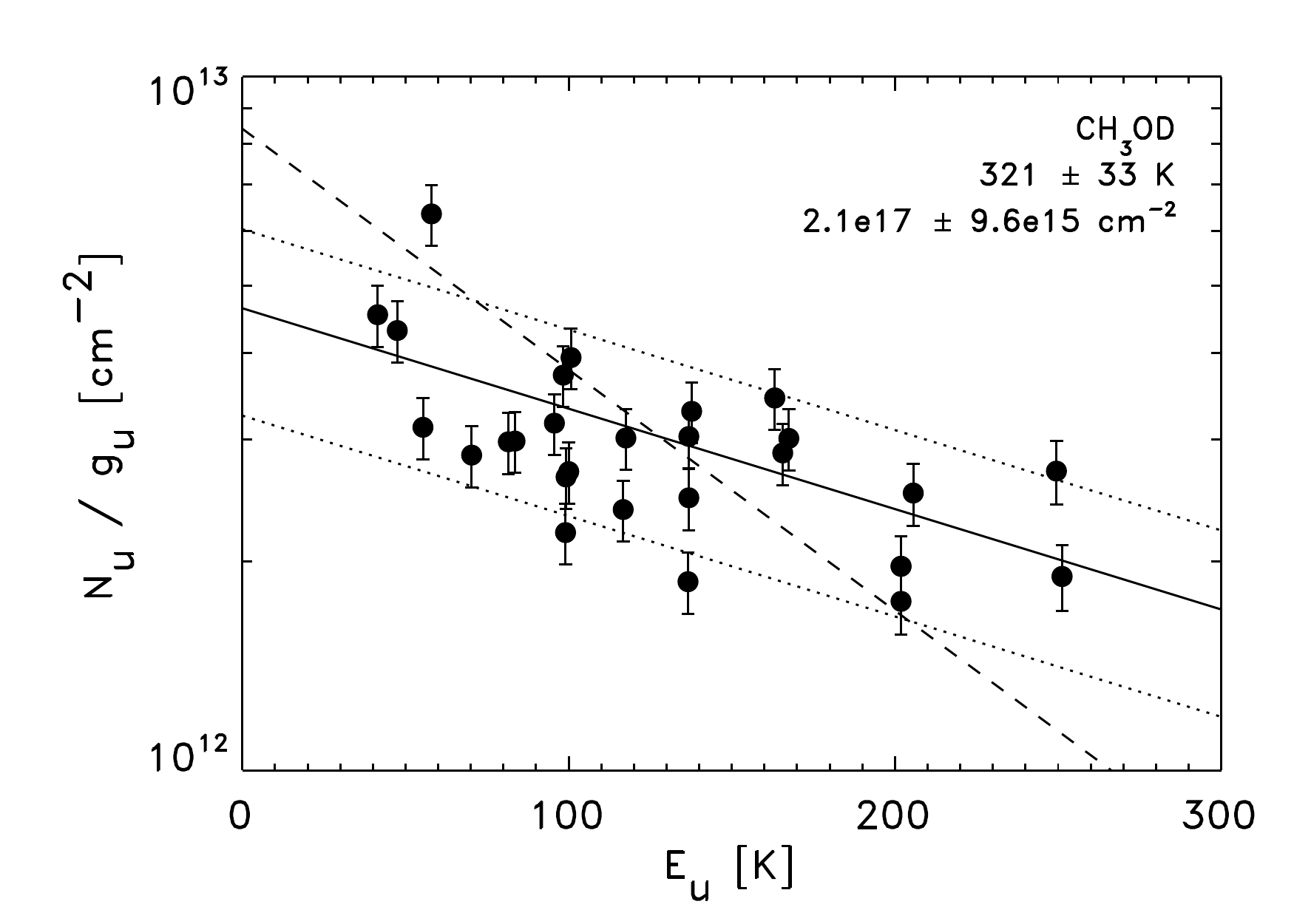}}
\resizebox{\hsize}{!}{\includegraphics{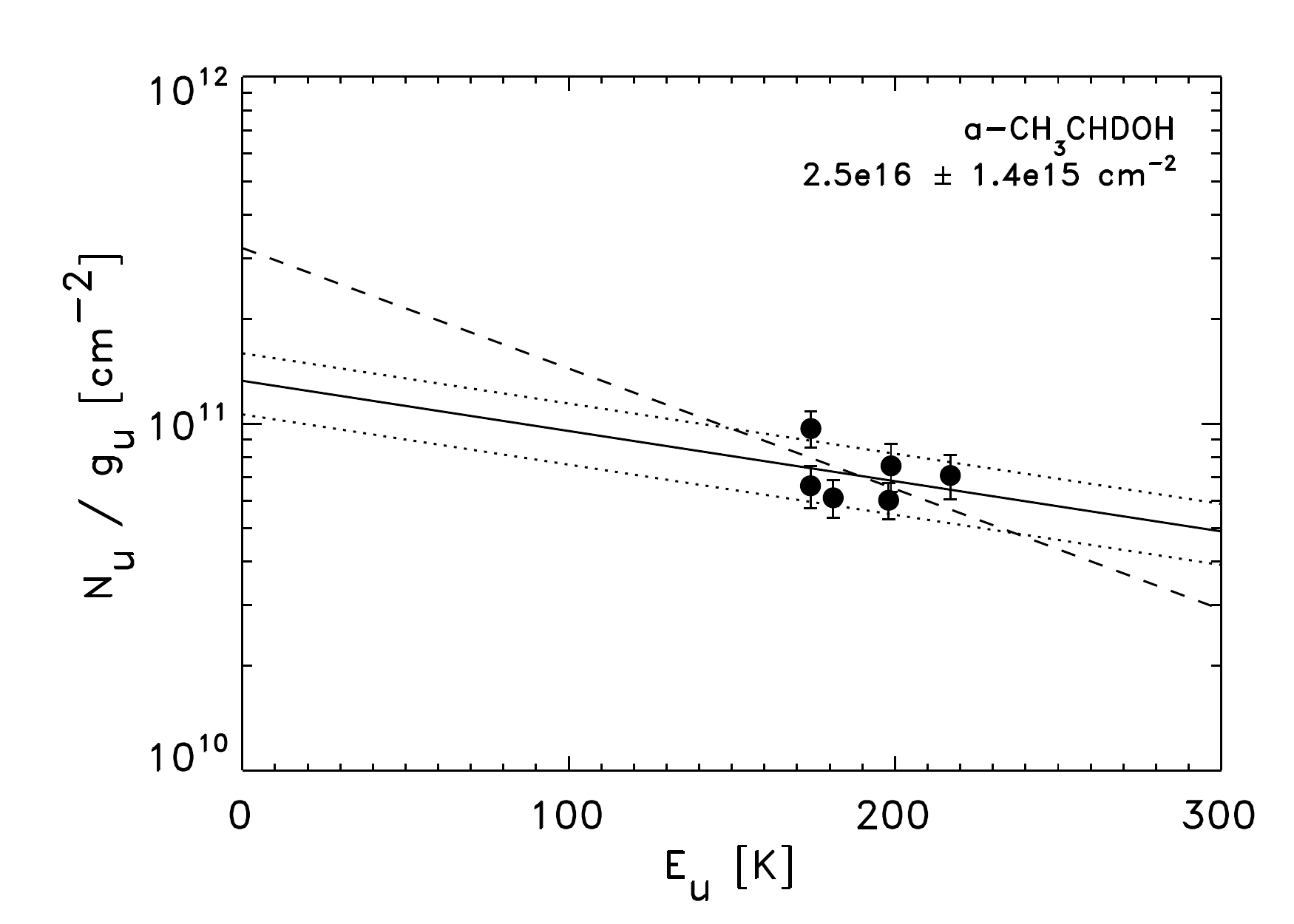}}
\caption{Rotation diagrams for well-isolated lines of CH$_3$OD
  (upper panel) and CH$_3$CHDOH (lower panel). For the former, both
  rotation temperature and column density can be constrained, while
  the limited range of CH$_3$OD transitions only allow us to constrain
  the column density for that species. In both panels, the best fit is
  shown with the solid line with our conservative estimates on the
  uncertainties represented by the dotted lines (the best fit values
  given in the upper right corner of each panels). The dashed lines in
  both panels indicate the best fits assuming a fixed rotation
  temperature of 125~K: this temperature clearly does not match the
  data for CH$_3$OD over the range of upper energies for the observed
  lines.}\label{rot_diagrams}
\end{figure}

  As a comparison to our method, rotation diagrams for CH$_3$OD
    and a-CH$_3$CHDOH are shown in Fig.~\ref{rot_diagrams}. CH$_3$OD
    is an example of a species with a significant number of completely
    well-isolated lines spanning a range of energy levels. The errors
    on the derived excitation temperatures and column densities are
    $\approx$10\% and 5\%, respectively (taking into account the
    statistical noise on the derived line strengths and calibration
    uncertainties) with the numbers otherwise in agreement with those
    derived above. For a-CH$_3$CHDOH the isolated lines represent a
    narrow range in energy levels, so for this species only the column
    density can be properly constrained. The derived uncertainty on
    the column density is also about 6\%. This value reflects that the
    fitted, most isolated, transitions have S/N ratios of 8--16, i.e.,
    that the statistical uncertainties on their line fluxes are about
    6--12\% (9\% on the average). When fitting just the column density
    by matching the intensities, the derived error goes down with the
    square root of the number of lines -- in this case six. I.e., the
    derived uncertainty is a factor of about 2.5 below the average
    statistical uncertainties of the fluxes of the individual
    lines. Other separate checks that these uncertainties are
    reasonable come from statistical fits to species with few
    well-isolated lines where either rotation diagrams
    \citep{jorgensen16} or full $\chi^2$-analyses \citep{persson16}
    give similar uncertainties.

Uncertainties arising from vibrational contributions to the partition
functions are small (about 5\% or less) for the $^{13}$C isotopologues
of molecules with low excitation temperatures ($\sim$125~K) or of
those for which the vibrational states are well-characterised. For
deuterated species or for molecules with large correction factors,
i.p., ethanol and methyl formate, this uncertainty may be up to
10--25\%. Generally, the assumption that the vibrational factor of an
isotopologue of a heavier molecule is the same as that of its main
isotopic species will lead to an underestimate of its column density
and thus lead to an estimated D/H ratio that is somewhat too small or
a $^{12}$C/$^{13}$C ratio somewhat too large.

\section{Results}\label{results}
Many of these detections mark the first reports of these species in
the ISM, including the detections of the deuterated isotopologues of
ethanol, acetaldehyde, ketene, formic acid. Other species have
previously been seen toward Sgr B2(N2) (e.g., the
$^{13}$C-isotopologues of ethanol by \citealt{muller16}) or Orion KL
(e.g., CH$_2$DOCHO by \citealt{coudert13}) and a few have been
identified from single-dish observations of IRAS~16293 (i.e., methyl
formate, CH$_3$OCDO, tentatively by \citealt{demyk10} and dimethyl
ether, CH$_2$DOCH$_3$, by \citealt{richard13}). However, for those
species the systematic study presented here allows for the first time
a direct comparison between these species with a high number of
identified lines originating in a region that is spatially resolved on
Solar System scales and with what is thought to be under relatively
homogeneous physical conditions in terms of temperatures and
densities.

\subsection{Emission morphology}\label{morphology}
The bulk of the analysis in this paper concerns the identification of
lines toward single positions, but one of the main strengths of the
ALMA data is to provide fully sampled images. Therefore a few words
about the emission morphology are included here.

As discussed in \cite{jorgensen12,jorgensen16} the emission around
IRAS16293B is complex: toward the location of the protostar itself the
emission from the dust continuum as well as that of many lines is
optically thick. This causes many of the lines in the spectra to
appear in absorption hampering interpretation. For identification and
modeling of species, it is instead more useful to consider positions
offset by 0.5--1.0$''$ from the continuum peak where the line emission
from many of the complex organic molecules appears brighter and where
absorption is less of an issue.

\begin{figure}
\resizebox{\hsize}{!}{\includegraphics{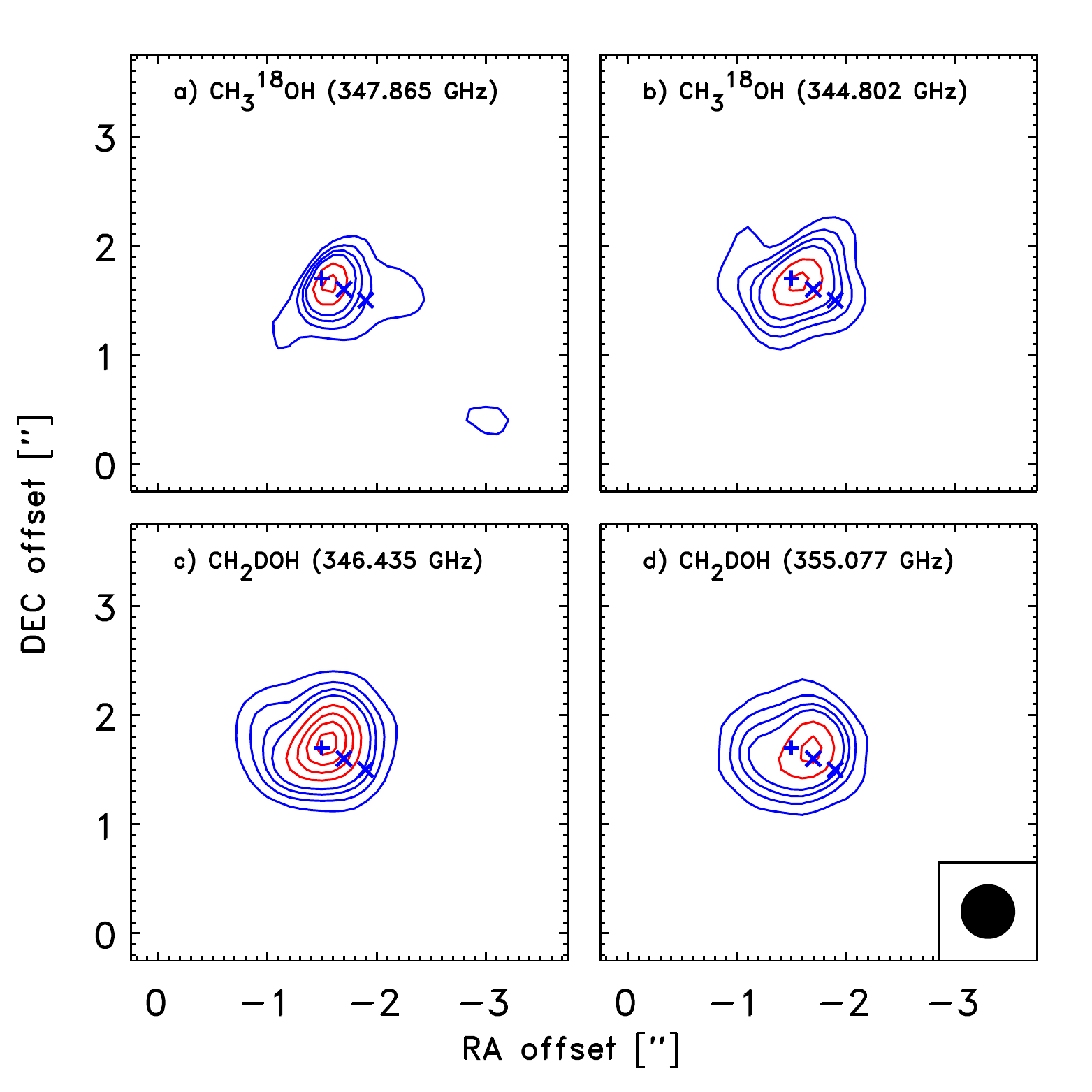}}
\caption{Representative maps for lines of CH$_3^{18}$OH
  $8_{3,6}-9_{2,8}$ ($\nu$=347.865~GHz; $E_{\rm up}$=143~K) and
  $17_{1,17}-16_{2,15}$ ($\nu$=344.802~GHz; $E_{\rm up}$=345~K) in the
  upper left and right panels (a and b), respectively and CH$_2$DOH
  $7_{5,2/3}-8_{4,5/4}$ ($\nu$=346.43471~GHz; $E_{\rm up}$=176~K) and
  $17_{2,15}-16_{3,14}$ ($\nu$=355.077~GHz; $E_{\rm up}$=364~K) in the
  lower left and right panels (c and d), respectively. In all panels,
  the integrations were performed over $\pm 0.5$~km~s$^{-1}$ around
  the systemic velocity and the contours shown at 3$\sigma$,
  6$\sigma$, 9$\sigma$ and 12$\sigma$ (blue contours) and from there
  upwards in steps of 6$\sigma$ (red contours).}\label{examplemaps}
\end{figure}

Fig.~\ref{examplemaps} compares the integrated emission around
IRAS16293B for two transitions of CH$_3^{18}$OH and CH$_2$DOH
representing different excitation levels. This highlights that the
emission morphologies are remarkably similar for all species and lines
of different excitation levels, what is also seen for other complex
organic species identified from the PILS data
\citep{jorgensen16,coutens16,lykke17}. This reenforces the point above
that the bulk of the emission is dominated by relatively compact
material with relatively homogeneous physical conditions.

\begin{figure*}[!htb]\centering
\resizebox{0.9\hsize}{!}{\includegraphics{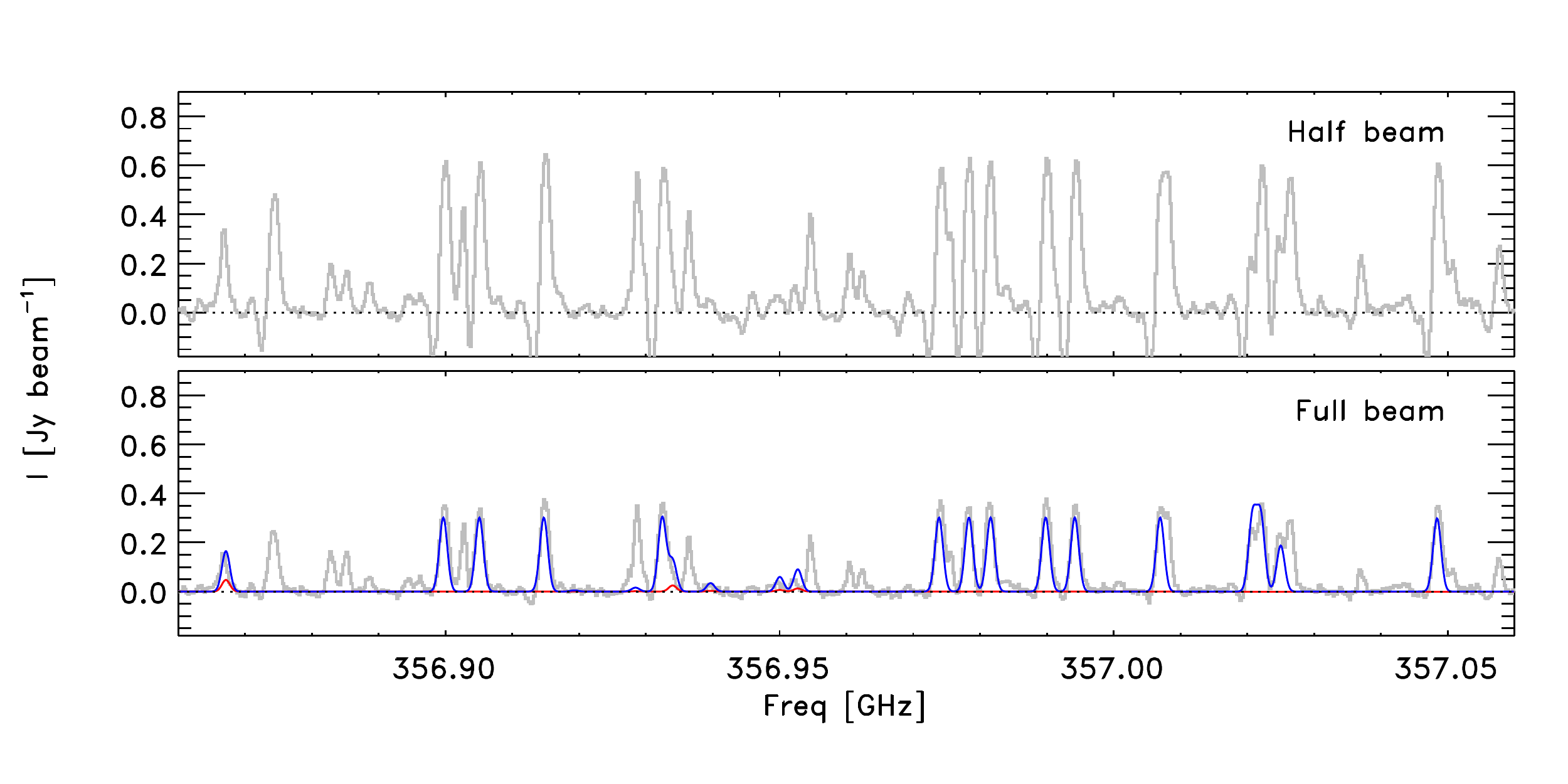}}
\caption{Spectrum of optically thick transitions of CH$_2$DOH at
  356.95~GHz toward the half-beam (upper) and full-beam (lower) offset
  positions from IRAS16293B. In the lower panel, synthetic spectra for
  CH$_2$DOH are overlaid: those are both calculated for the same
  column density (Table~\ref{isotopologuetable}) but with temperatures
  of 50~K (blue) and 300~K (red).}\label{ch2doh_exspectrum}
\end{figure*}

Also marked in Fig.~\ref{examplemaps} are the continuum peak and the
two offset positions at one half and one full beam (0.5\arcsec\ and
1\arcsec): the emission peaks close to the half-beam offset position
with its strength reduced by a factor of 2 at the full beam offset
position. This is consistent with the emission peak being marginally
resolved in that direction. Fig.~\ref{ch2doh_exspectrum} compares
spectra toward the two offset positions over a frequency range where
prominent (low excitation) lines of CH$_2$DOH are seen. The figure
also shows a synthetic spectrum calculated at low temperatures
indicating the location of the CH$_2$DOH transitions. It is clearly
seen that many of these lines that show absorption at the half-beam
offset position are in emission at the full beam offset
position. Indeed, the synthetic spectrum predicts that these specific
transitions will be optically thick ($\tau > 1$).

In fitting the data, we therefore iteratively selected transitions
that have $\tau < 0.1$. Focusing only on the optically thin
transitions it is found that there is about a factor of 2 difference
between the derived column densities (assuming the same filling
factor) toward the full and half beam offset positions. Given that the
emission is only marginally resolved, this likely reflects a
combination of the actual drop in column density of the species as
well as the fact that the emission does fill a smaller fraction of the
beam at the full beam offset position (Fig.~\ref{examplemaps}). In the
following work we therefore predominantly focus on the full beam
position, but note \emph{(i)} that we check for consistency of the
fits toward the half-beam position and \emph{(ii)} that for comparison
to other results from the half-beam positions one can multiply the
quoted column densities by a factor of 2.

\subsection{Derived column densities}\label{columndensities}
Figures~\ref{first_spectra}--\ref{last_spectra} show the observed
spectra for each of the species, compared to the predictions by the
synthetic spectra and the individual fits are discussed in the
appendix. The data show detections of each of the (singly)
  deuterated versions as well as the $^{13}$C isotopologues of the
  targeted oxygen-bearing complex organic molecules
  (the $^{13}$C isotopologues for ethanol only tentatively).

Table~\ref{isotopologuetable} lists the derived column densities for
each of the species and, for the rarer isotopologues, the abundances
relative to the main isotopologue. The column densities of the
deuterated species range from 2\% up to 18\% relative to the main
isotopologues. A significant part of this variation is due to the
composition of the functional groups. For the same D/H ratio one
should expect a molecule where the deuterium is substituted into one
of the hydrogen atoms in the methyl CH$_3$ group to be three times
more abundant than a molecule where the substitution is into a
functional group such as OH with only one hydrogen atom. If this
effect is taken into account, surprisingly small variations are seen
in the D/H ratios for the individual isotopologues. A distinction is
apparent with the simpler molecules are consistent with a D/H ratio of
$\approx$2\%, and the more complex species methyl formate,
acetaldehyde, ethanol and dimethyl ether show higher ratios ranging
from about 4 to 8\%, similar to the values of glycolaldehyde. For
  dimethyl ether a lower $^{12}$C/$^{13}$C ratio is found, about half that of the
standard ratio for the local ISM \citep[68;][]{milam05}, similar to
what has previously been inferred for glycolaldehyde. The marginal
cases of $^{13}$C isotopologues of methyl formate and ethanol are also
consistent with lower $^{12}$C/$^{13}$C ratios but additional
observations are needed to confirm and solidify the numbers for these two
species.

\section{Discussion}\label{discussion}
\begin{figure}[!htb]
\resizebox{\hsize}{!}{\includegraphics{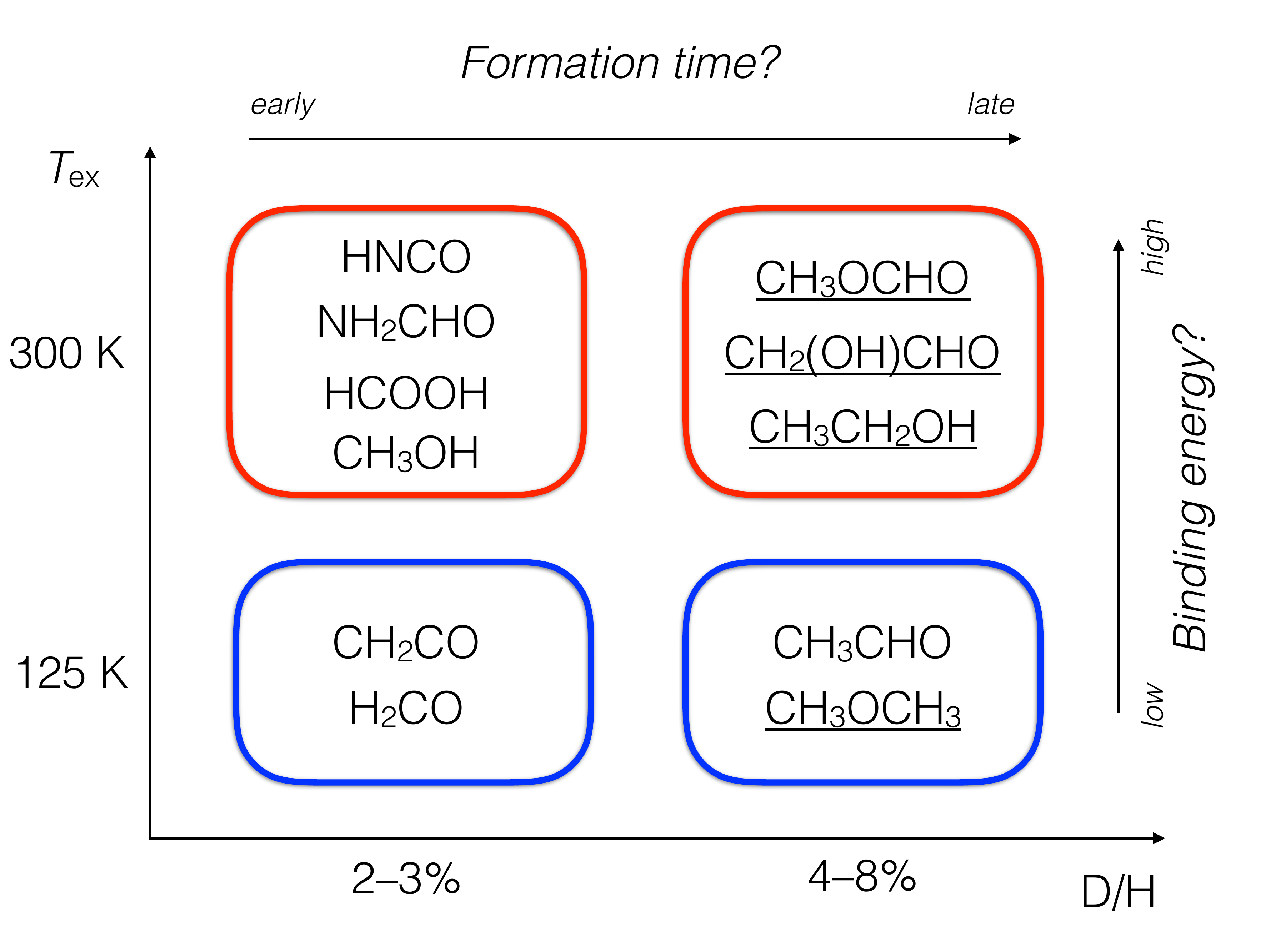}}
\caption{Schematic representation of the results from this paper: the
  discussed molecules are grouped according to their D/H ratios and
  excitation temperatures. The species underlined can be shown to have
  lower $^{12}$C/$^{13}$C ratios by a factor 2 than typical values for
  the local ISM (although only tentatively for
    methyl formate and ethanol), while the remainder are either consistent with a
  standard ISM ratio or have too optically thick lines of the main
  isotopologue for a reliable independent
  estimate.}\label{schematic_results}
\end{figure}

The presented analysis provides important clues concerning the origin
of various complex organic molecules in different ways. As summarised
in Fig.~\ref{schematic_results}, the species fall in a number of
different groups depending on their excitation temperatures and the
deuteration levels. In this section we try to reconcile these
observations to discuss the implications for formation of the
different complex organics.
\begin{table*}[!hbt!]\begin{center}
    \caption{Derived column densities for the isotopologues of the
      complex organics and abundances relative to the main
      isotopologue at the position offset by 0.5$''$ (60~AU) from
      IRAS16293B. As discussed in the text the uncertainties on the
      individual column densities are about 20\%, while a conservative
      estimate of the
      uncertainties on the ratios between species are 30\% from error
      propagation. Square brackets indicate tentative fits to marginal
    detections.}\label{isotopologuetable}
\begin{tabular}{lcccc}\hline\hline
Species                 & $N$         & $N / N_{\rm main}$$^{a}$ & $N_{\rm corr} / N_{\rm main}$$^{a}$\\
                        & [cm$^{-2}$] & & \\ \hline \\[-1.5ex]
\multicolumn{4}{c}{Methanol}\\
CH$_3$OH$^{b}$          & $1.0\times 10^{19}$ & $\ldots$ & $\ldots$  \\ 
CH$_2$DOH               & $7.1\times 10^{17}$ & 0.071  & 0.024 \\
CH$_3$OD                & $1.8\times 10^{17}$ & 0.018  & 0.018 \\[1.0ex]
\multicolumn{4}{c}{Ethanol}\\[0.5ex]
CH$_3$CH$_2$OH          & $2.3\times 10^{17}$ & $\ldots$  \\
a-a-CH$_2$DCH$_2$OH     & $2.7\times 10^{16}$ & 0.12 & 0.059  \\
a-s-CH$_2$DCH$_2$OH     & $1.3\times 10^{16}$ & 0.057 & 0.057  \\
a-CH$_3$CHDOH           & $2.3\times 10^{16}$ & 0.10 & 0.050  \\
a-CH$_3$CH$_2$OD        & [$1.1\times 10^{16}$] & [0.050] & [0.050]  \\[0.5ex]
a-CH$_3^{13}$CH$_2$OH \phantom{xx}\rdelim\}{2}{0mm}[] & & & \\[-1.0ex]
a-$^{13}$CH$_3$CH$_2$OH & \raisebox{1.5ex}{[$9.1\times 10^{15}$]} & \multicolumn{2}{c}{\raisebox{1.5ex}{[0.040 (1/25)]}}\\[1.0ex]
\multicolumn{4}{c}{Methyl formate}\\[0.5ex]
CH$_3$OCHO              & $2.6\times 10^{17}$   & $\ldots$ & $\ldots$     \\
CH$_3$OCDO              & $1.5\times 10^{16}$   & 0.057    & 0.057 \\
CH$_2$DOCHO             & $4.8\times 10^{16}$   & 0.18     & 0.061 \\
CH$_3$O$^{13}$CHO     & [$6.3\times 10^{15}$]   &  \multicolumn{2}{c}{[0.024 (1/41)]} \\[1.0ex]
\multicolumn{4}{c}{Ketene$^c$}\\[0.5ex]
CH$_2$CO                & $4.8\times 10^{16}$   & $\ldots$     & $\ldots$ \\
$^{13}$CH$_2$CO \phantom{x}\rdelim\}{2}{0mm}[]        &   & \\[-1.0ex]
CH$_2$$^{13}$CO         & \raisebox{1.5ex}{$7.1\times 10^{14}$}   & \multicolumn{2}{c}{\raisebox{1.5ex}{0.015 (1/68)$^c$}} \\
CHDCO                   & $2.0\times 10^{15}$   & 0.042        & 0.021 \\[1.0ex]
\multicolumn{4}{c}{Dimethyl ether}\\[0.5ex]
CH$_3$OCH$_3$           & $2.4\times 10^{17}$   & $\ldots$     & $\ldots$ \\
$^{13}$CH$_3$OCH$_3$    & $1.4\times 10^{16}$   & \multicolumn{2}{c}{0.029 (1/34)$^d$} \\
asym-CH$_2$DOCH$_3$     & $4.1\times 10^{16}$   & 0.17         & 0.043  \\
sym-CH$_2$DOCH$_3$      & $1.2\times 10^{16}$   & 0.050        & 0.025 \\[1.0ex]
\multicolumn{4}{c}{Acetaldehyde}\\[0.5ex]
CH$_3$CHO               & $1.2\times 10^{17}$   & $\ldots$      & $\ldots$ \\
CH$_3$CDO               & $9.6\times 10^{15}$   & 0.080         & 0.08 \\
$^{13}$CH$_3$CHO \phantom{x}\rdelim\}{2}{0mm}[] & & \\[-1.0ex]
CH$_3$$^{13}$CHO        & \raisebox{1.5ex}{$1.8\times 10^{15}$}   & \multicolumn{2}{c}{\raisebox{1.5ex}{0.015 (1/67)}} \\[1.0ex]
\multicolumn{4}{c}{Formic acid$^e$}\\[0.5ex]
t-HCOOH                 & $5.6\times 10^{16}$   & $\ldots$      & $\ldots$ \\
t-H$^{13}$COOH          & $8.3\times 10^{14}$   & \multicolumn{2}{c}{0.015 (1/68)$^c$} \\[0.5ex]
t-DCOOH \phantom{x}\rdelim\}{2}{0mm}[]                &    & & \\[-1.0ex]
t-HCOOD                 & \raisebox{1.5ex}{$1.1\times 10^{15}$}   & \raisebox{1.5ex}{0.020} & \raisebox{1.5ex}{ 0.020} \\ \hline
\end{tabular}
\end{center}

$^{a}$Column density of isotopologue relative to that of the main
species given respectively without ($N / N_{\rm main}$) and with the
statistical corrections ($N_{\rm corr} / N_{\rm main}$). $^{b}$Derived
from the $^{18}$O-isotopologue assuming a $^{16}$O/$^{18}$O abundance
ratio of 560 \citep{wilson94}. $^c$Many
lines of the main isotopologues of ketene re optically thick. The most reliable
column density comes from the $^{13}$C isotopologues but are
consistent with a standard $^{12}$C/$^{13}$C ratio for the few
relatively thin transitions of the main isotopologues. $^d$For the $^{13}$C
isotopologue of dimethyl ether the ratios with respect to the main isotopologue
have been corrected by the statistical factor corresponding to the two
indistinguishable carbon-atoms. $^{e}$Due to optical depth issues and
fainter lines the
fits to formic acid rely on simultaneous fits of the main isotopologue
and $^{13}$C isotopologue combined and the two deuterated variants,
respectively (see discussion in Sect.~\ref{formicacid_discussion}).
\end{table*}

\subsection{Excitation temperatures}
An important aspect of the presented analysis is the variations in the
excitation temperatures between the different species with one group
consistent with temperatures of approximately 100--150~K and one with
temperatures of 250--300~K. The former group includes CH$_2$CO,
CH$_3$CHO, CH$_3$OCH$_3$ as well as H$_2$CO \citep{persson18} and
c-C$_2$H$_4$O \citep{lykke17}, while the latter group includes
CH$_3$OH, C$_2$H$_5$OH, CH$_3$OCHO as well as NH$_2$CHO and HNCO
\citep{coutens16} and CH$_2$OHCHO and (CH$_2$OH)$_2$
\citep{jorgensen16}. These divisions are mostly in agreement with a
similar division into ``cold'' ($T_{\rm ex} \lesssim 100$~K) and
``hot'' ($T_{\rm ex} \gtrsim 100$~K) molecules by \cite{bisschop07}
from a survey of high-mass hot cores and a follow-up study by
\cite{isokoski13}. Also, the molecules with low excitation
temperatures are typically those identified in the large single-dish
survey of complex organics toward IRAS~16293--2422 by
\cite{jaber14}. Table~\ref{coldvshot} compares these detections: in
general it seems that the hot and cold molecules from the high-mass
surveys and the interferometric data agree well with a few
exceptions. Also, all the cold molecules from the interferometric data
are detected in the IRAS16293 single-dish survey again illustrating
that that survey may be biased toward the colder gas on larger
scales. Of the warm molecules in the interferometric data, CH$_3$OCHO,
CH$_3$OH and NH$_2$CHO are seen in the single-dish survey: those are
all relatively abundant and thus may contribute significantly enough
to be picked out in the larger beam. There are no species detected
from the single-dish observations that are not detected in the
interferometric data.

As pointed out previously \citep[e.g.,][]{garrod08}, a straightforward
explanation for the distinction between hot and cold molecules may lie
in the binding energies of the different species (also listed in
Table~\ref{coldvshot}).  A very clear distinction is seen between
species that have high excitation temperatures in IRAS16293
interferometric data, which have binding energies of 5000--7000~K
(ethylene glycol as high as 10200~K), and those that have low
excitation temperatures, which have binding energies in the
2000--4000~K range.  In the context of the three phase chemical model
by \cite{garrod13} this would imply that the bulk of the colder
species simply sublimate at lower temperatures and vice versa for the
warm species: the lower binding energies would correspond to
sublimation temperatures of $\approx$~50~K and the higher ones to
$\approx$~100~K. One should note though that these binding energies
are for pure ices. If the species are mixed in ice with H$_2$O or
CH$_3$OH the binding energies would increase to the values of those,
$\approx 5500$~K.

The difference between these (lower) sublimation temperatures of
50--100~K and the higher inferred excitation temperatures (100--300~K)
may be a result of the rapid infall (and short distances) in the dense
innermost region of the envelope around IRAS16293B. For a
typical infall speed of $\sim$~1~km~s$^{-1}$ the time it would take
material to move $\sim$50~AU (typical of these temperature regimes in
the IRAS16293) is $\sim$~200~years. Thus, while the separation of the
molecules into ``hot'' and ``cold'' species likely reflects where
those species originally sublimate, this short time-scale would imply
that the derived values for the excitation temperatures reflect the
current location of the bulk of the species closer to the
protostar. As argued by \cite{schoeier02}, this short time-scale may
also inhibit any ``second generation'' formation of complex organics,
at least based on those species sublimating only at the higher
temperatures. The ambiguity for dimethyl ether may reflect that
this species toward the high mass sources can be formed by ``second
generation'' chemistry through gas-phase reactions
\citep[e.g.,][]{charnley92} as is suggested by the multiple
components observed toward the multi-line study of this species toward
the high-mass hot core G327.3$-$0.6 \citep{bisschop13}.
\begin{table*}\centering
\caption{Division of molecules into ``hot'' and ``cold'' from the
  interferometric observations of IRAS16293 (this study) and high-mass hot cores
  \citep{bisschop07}, an indication of whether or not the species are detected in
  the single-dish study of IRAS16293 \citep{jaber14} and the binding energies of
  each species \citep[from the tabulation by][]{garrod13}.}\label{coldvshot}
\begin{tabular}{llllll} \hline\hline
Species        & & \multicolumn{2}{c}{IRAS16293} & High-mass hot cores & Binding energy$^a$ \\
               & & Interferometric & Single-dish &                     & [K]            \\ \hline
Formaldehyde    & H$_2$CO        & Cold            & +           & Cold / Hot          & 2050           \\
Ketene          & CH$_2$CO       & Cold            & +           & Cold                & 2200           \\ 
Acetaldehyde    & CH$_3$CHO      & Cold            & +           & Cold                & 2775           \\ 
Dimethyl ether  & CH$_3$OCH$_3$  & Cold            & +           & Cold / Hot $^b$     & 3675           \\ 
Methyl formate  & CH$_3$OCHO     & Hot             & +           & Hot                 & 5200           \\ 
Methanol        & CH$_3$OH       & Hot             & +           & Hot                 & 5530           \\
Formamide       & NH$_2$CHO      & Hot             & +           & Hot                 & 5560           \\ 
Formic acid     & HCOOH          & Hot             & -           & Cold                & 5570           \\ 
Ethanol         & C$_2$H$_5$OH   & Hot             & -           & Hot                 & 6260           \\ 
Glycolaldehyde  & CH$_2$OHCHO    & Hot             & -           & -                   & 6680           \\ 
Ethylene glycol & (CH$_2$OH)$_2$ & Hot             & -           & -                   & 10200          \\ 
\hline
\end{tabular}

Notes: $^{a}$The listed binding energies for pure ices
\citep{garrod13}. $^b$Although, dimethyl ether is classified as hot in
\cite{bisschop07}, temperatures of 80--130~K are found toward SgrB2(N) \citep{belloche13} and G327.3$-$0.6 \citep{bisschop13}, i.e., relatively ``cold'' compared to other species.
\end{table*}

\subsection{Deuteration}
\subsubsection{Observed ratios}
The analysis presents new accurate measurements for the D/H ratios of
complex organics representative for the warm inner envelope
$T > 100$~K of IRAS~16293B where most of the molecules have sublimated
off the dust grains. For the species considered in this paper, and for
glycolaldehyde, isocyanic acid and formamide
\citep{jorgensen16,coutens16}, no differences are seen between the
fractionation in different functional groups. It is therefore possible
to assign one value per species (after taking the statistics for
species with multiple H atoms into account). This is summarised in
Table~\ref{dratios} and compared to the single-dish estimates from the
literature.
\begin{table*}
  \caption{Inferred abundances and deuterium fractionation (from
    interferometric and single-dish data, respectively) for different
    organic molecules. For the deuterium fractionation, the
    numbers refer to the inherent D/H ratio (i.e., taking into account
    the statistical ratios for the functional groups with multiple
    H-atoms) for the singly deuterated species. As noted in the text
      the uncertainties on the relative abundances, e.g., D/H ratios,
      from the ALMA data are about 30\%; 0.6 or 1.5
      percentage points for ratios of 2 or 5\%, respectively.}\label{dratios}
\begin{tabular}{llc@{\hskip 0.4in}|@{\hskip 0.4in}cccc} \hline\hline
               &                 &  & \multicolumn{4}{c}{[D/H] ratios} \\
Species        &                 & [$X$/CH$_3$OH] & Single-dish             & Interferometric       & \multicolumn{2}{c}{Model \citep{taquet14}} \\
               &                 & & ($\sim$~1000~AU) & ($\sim$~50~AU) & $1.1\times 10^5$~yr & $2.0\times 10^5$~yr \\ \hline
Formaldehyde   & H$_2$CO         & 19\%   & 7.5\%$^f$               & 3\%$^{a}$             & 2.1\%           & 0.21\%      \\
Methanol       & CH$_3$OH        & ---    & 1.8--5.9\%$^{f,g}$      & 2\%$^{b}$             & 3.5--1.8\%      & 0.28--0.17\% \\
Ethanol        & CH$_3$CH$_2$OH  & 2.3\%  & $\ldots$                & 5\%$^{b}$             & 0.63--9.7\%$^m$ & 0.090\%--0.12\%$^m$  \\
Dimethyl ether & CH$_3$OCH$_3$   & 2.4\%  & 3\%$^j$                 & 4\%$^{b}$             & 3.7\%           & 0.10\%    \\
Glycolaldehyde & CH$_2$OHCHO     & 0.34\% & $\ldots$                & 5\%$^{c}$             & 6.5--13\%       & 0.22--0.38\% \\
Methyl formate & CH$_3$OCHO      & 2.6\%  & 6\%$^{h}$               & 6\%$^{b}$             & 7--9\%          & 0.25--0.22\% \\
Acetaldehyde   & CH$_3$CHO       & 1.2\%  & $\ldots$                & 8\%$^{b}$             & 9.2\%$^{l}$     & 0.068\%$^{l}$  \\
Ketene         & CH$_2$CO        & 0.48\% & $\ldots$                & 2\%$^{b}$             & 0.25\%          & 0.015\%  \\
Formic acid    & HCOOH           & 0.56\% & $\ldots$                & 2\%$^{b}$             & 2.3--1.0\%$^{k}$& 0.58--0.66\%$^{k}$  \\
Isocyanic acid & HNCO            & 0.27\% & $\ldots$                & 1\%$^{d}$             & $\ldots$        & $\ldots$  \\
Formamide      & NH$_2$CHO       & 0.10\% & $\ldots$                & 2\%$^{d}$             & $\ldots$        & $\ldots$  \\
Water          & H$_2$O          & $\ldots$ & 0.25\%$^i$     & 0.046\%$^{e}$         & 1.2\%           & 0.33\%    \\ \hline
\end{tabular}

Notes: $^a$\cite{persson18}. $^b$This paper. $^c$\cite{jorgensen16}.
$^d$\cite{coutens16}. $^e$\cite{persson13} referring to IRAS16293A. $^f$\cite{parise06deuterium}.
$^g$\citeauthor{parise06deuterium} infers a different deuterium fractionation for
CH$_3$OD (lower value) and CH$_2$DOH (higher value). In the models of
\cite{taquet14}, the D/H ratio for CH$_3$OD is higher than that of CH$_2$DOH
when corrected for the statistics (last two columns). We note that according to
\cite{belloche16} issues were identified with the spectroscopic predictions for
CH$_2$DOH causing the column densities for CH$_2$DOH to be over-predicted by a
factor $2.1\pm 0.4$. We have corrected for this factor in the number quoted here.
$^h$From \cite{demyk10} inferred for CH$_3$OCDO: in the
paper a D/H ratio of 15\% is quoted but according to
the actual numbers for the column densities given in the paper, the fraction
is in fact 6\%. $^{i}$From \cite{coutens12} for cold water in the outer envelope
based on detailed radiative transfer modeling of the water abundance
profiles. $^{j}$From \cite{richard13}. $^{k}$\cite{taquet14} predict different D/H ratios for DCOOH and HCOOD of 2.3\% and 1\%, respectively, for the early model and 0.58\% and 0.66\%, respectively, for the late model. $^l$For CH$_3$CDO detected in PILS. For CH$_2$DCHO \cite{taquet14} predicts D/H ratios of 0.90\% and 0.040\% for the early and late model, respectively. $^m$\cite{taquet14} predicts D/H ratios of 0.63\%, 4.6\% and 9.7\% for CH$_2$DCH$_2$OH, CH$_3$CHDOH and CH$_3$CH$_2$OD, respectively in the early model and 0.090\%, 0.095\% and 0.12\% in the late model.
\end{table*}

The three most important take-aways from this comparison are
\emph{(i)} the consistency between D/H ratios on the $\sim 50$~AU
scales probed by the ALMA observations and those on $\sim 1000$~AU
scales probed by the single-dish data (except for H$_2$O), \emph{(ii)}
the small, but significant variations between the interferometric
measurements for the different species and \emph{(iii)} the lack of
variations in the deuteration for the different functional groups.

The strength of this study, besides providing new detections for a number of
deuterated isotopologues, is the systematic census.  Previous (single-dish)
studies have typically focused on brighter lines from one or two species and, as
noted above, care must be taken with estimates of column densities for the main
isotopologues, in particular, due to optical depth issues. Also, the larger
number of lines for many of the observed isotopologues means that the accuracy
of the absolute column density determinations are at the 10--20\% level and
relative abundances between isotopologues of individual species likely better
than that. Differences of factors of 1.5--2.0 in the D/H ratios in
Table~\ref{dratios} are therefore significant.  Table~\ref{dratios} still shows
that there is a good agreement between the single-dish and interferometric D/H
ratios with the exception of water, for which the interferometric value is a
factor of 5 lower than that based on single-dish measurements.

It appears that there are some systematic variations between the
organics: methanol, ketene, formic acid all have D/H ratios of
approximately 2\% similar to formamide and isocyanic acid
\citep{coutens16} and formaldehyde \citep{persson18}. Dimethyl ether
has a D/H ratio of $\approx$~4\%, ethanol, methyl formate and
glycolaldehyde $\approx$~5--6\% and acetaldehyde the highest ratio of
$\approx$~8\%. Water shows much lower
D/H ratios than the organics, likely reflecting its formation in the
earlier prestellar stages where the degree of deuteration is lower
\citep[e.g.,][]{taquet14,furuya16}.

Compared to the study of the Sgr~B2(N2) by \cite{belloche16} the
detections of the deuterated isotopologues of ethanol and methyl
formate are new and all ratios for the low-mass IRAS16293 are higher
than those for the high-mass source. For methanol, our CH$_2$DOH
measurement is a factor of 50 above the estimate for CH$_2$DOH of
\citeauthor{belloche16} and our CH$_3$OD estimate more than a factor
of 100 above the upper limit for Sgr~B2(N2). The derived column
density ratios for the different deuterated ethanol variants relative
to the main isotopologue is a factor of 2--4 above the upper limits
toward Sgr~B2(N2). The column density ratio for one variant of
deuterated methyl formate, CH$_2$DOCHO, is about a factor of 7 higher
than the upper limit of \cite{belloche16}.

It is interesting to note that there is a lack of variation in the D/H
ratios for the different functional groups, e.g., the
[CH$_2$DOH]/[CH$_3$OD] ratio is very close to a factor of 3 as expected from
statistics. This is different from what is found in prestellar cores
with much higher ratios $\gtrsim 10$ \citep[e.g.,][]{bizzocchi14} or
high-mass star-forming regions such as Orion KL with lower ratios
$\sim 1$ \citep[e.g.,][]{jacq93,peng12,neill13}. Such differences from
the statistical ratios led \cite{faure15} to propose that exchange
reactions between the -OH groups of H$_2$O and CH$_3$OH during the
warm-up phase could significantly alter the relative abundance
ratios. This in turn would predict an inverse scaling between the
[CH$_2$DOH]/[CH$_3$OD] ratio on the one hand and the [HDO]/[H$_2$O]
ratio on the other. Also, they noted that the previous high
[CH$_2$DOH]/[CH$_3$OD] ratios reported based on single-dish
observations for IRAS16293 compared to those in Orion KL could reflect
the presence of heterogeneous ices with respect to D/H ratios in
IRAS16293. However, as noted above revised methanol values from the
single-dish data are in agreement with that inferred from our analysis
here. Also, the revised measurements of the [HDO]/[H$_2$O] ratio
toward IRAS16293 from interferometric measurements \citep{persson13}
representing the fully sublimated ice mantles, is lower than what was
adopted in \cite{faure15}. \citeauthor{faure15} quote a timescale of
$\sim 10^3$~years for equilibrium in their calculations, which may be
too long given the short dynamical timescale governing the infall of
material in the inner envelope of IRAS16293 as noted above. It is
possible that exchange reactions could play a role in the upper layers
of the ice mantles reflecting the higher D/H ratios of water measured
in the ambient cloud \citep{coutens12} or that CH$_3$OH and H$_2$O
simply are not co-located \citep{cuppen09}.

\subsubsection{Comparison with models}
A qualitative comparison can be made to the models by
\cite{garrod13}. Those models followed the formation of complex
organic molecules through radical-radical recombination reactions
taking place during the warm-up of the core surrounding of a high-mass
protostar. When looking at the species in those models, it is
interesting to note that the species with D/H ratios of 2\%
(formaldehyde, methanol and ketene) are all abundant in the ices from
the very beginning of the warm-up, ethanol and dimethyl ether
production in those models set in at temperatures around 20~K, while
methyl formate and glycolaldehyde only appears in the ices at
temperatures $\gtrsim 25$~K. The same models also predict significant
isocyanic acid ice present at low temperatures, again consistent with
its low D/H ratio of $\approx$1\% \citep{coutens16} and some formamide
ice is formed through reactions between NH$_2$ and H$_2$CO already at
low temperatures as also seen in recent laboratory experiments
(\citealt{fedoseev15,fedoseev16}; see also the discussion in
\citealt{coutens16}). Formic acid, HCOOH, is the only species that
seems to be an exception of this trend. In the models of
\cite{garrod13}, formic acid predominantly forms in the gas phase when
formaldehyde evaporates off dust grains and then freezes out on the
dust grains immediately. However, \cite{taquet14} point to recent
laboratory and theoretical studies \citep{goumans08,ioppolo11} and
argue that formic acid more likely is formed in the cold ices through
reactions between CO and OH. In either case, it is not unreasonable to
expect the $\approx$~2\% D/H ratio for formic acid observed, either
because it is inherited from the formaldehyde or because it is set in
the early phases in the ices together with most of the other species.

A direct quantitative comparison can be made to the results of
\cite{taquet14} who presented an extensive model for the formation and
deuteration of interstellar ices during the collapse of a protostellar
core including also the sublimation of these species close to the
protostar.  In those models, species are predominantly formed on ice
surfaces during the warm-up or collapse of material in the pre- and
protostellar core and thereby typically inherit the D/H ratios at that
specific time or layer in the ices. This means that the species
forming later or in the outer parts of the ices would be characterised
by higher D/H ratios than those formed earlier/deeper in the
ices. Also, in those models, the D/H ratios for individual functional
groups should be correlated across the different species, reflecting
the formation pathways based on the radical recombinations.
\begin{figure*}
\resizebox{\hsize}{!}{\includegraphics{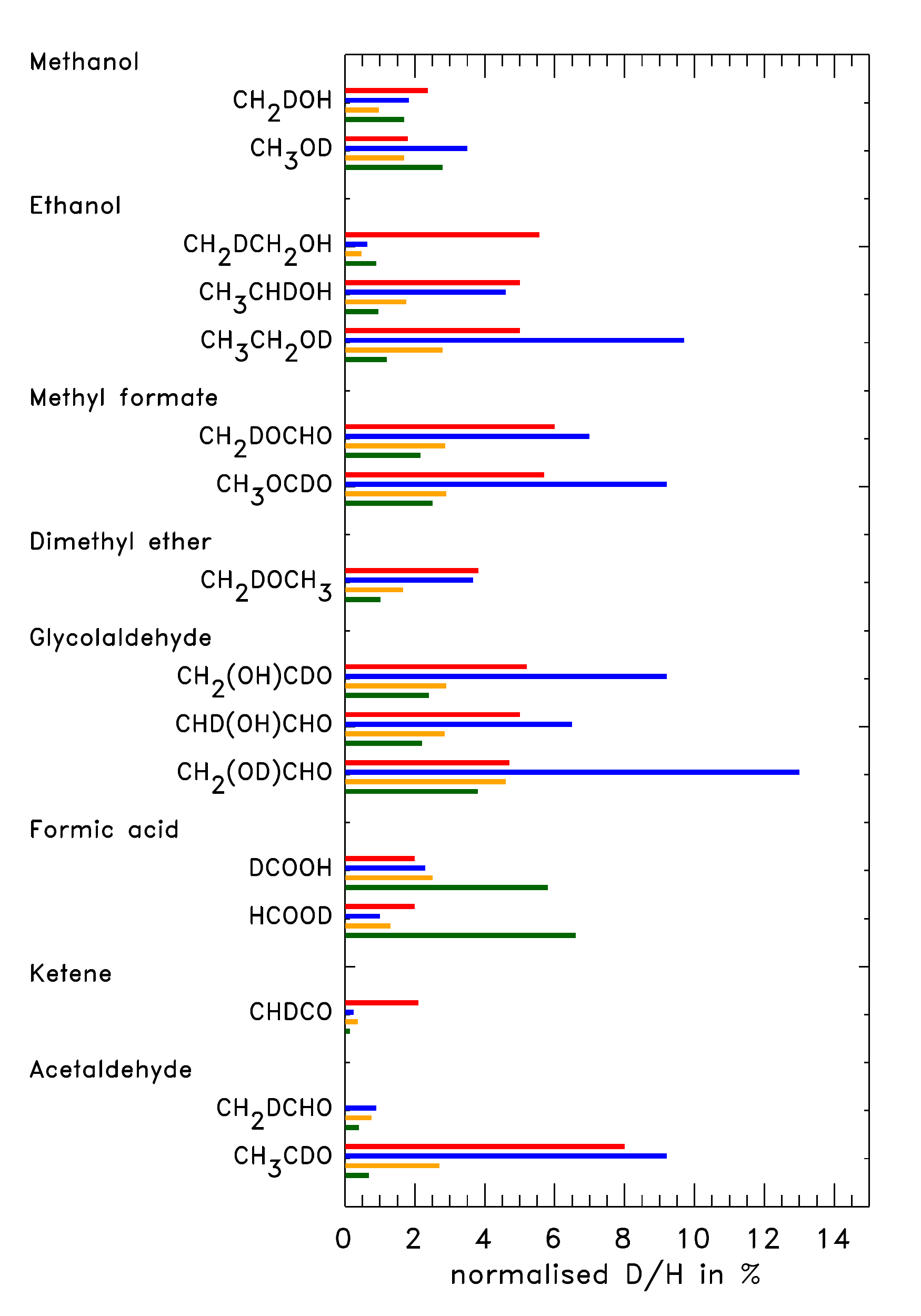}\includegraphics{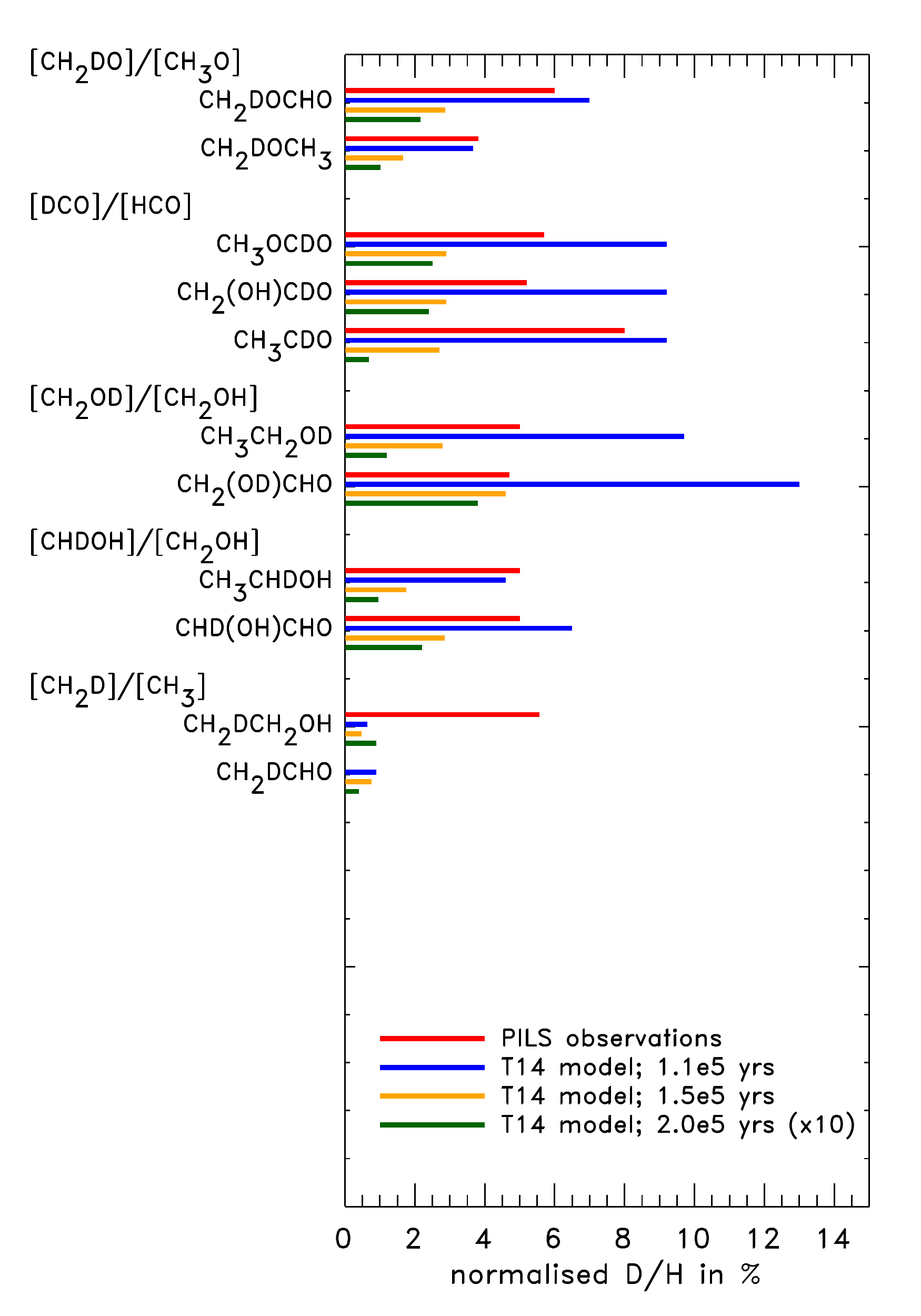}}
\caption{Comparison between the D/H ratios (per H-atom) for the complex organics
  sorted by species (left) and the leading radical in the formation (right). For
  each species, the PILS observations are represented by the first, red, bar and
  the predictions in the models by \cite{taquet14} after $1.1\times 10^5$,
  $1.5\times 10^5$ and $2.0\times 10^5$~years by the second (blue), third
  (orange) and fourth (green) bars (see also legend in the part of the right
  panel). For the model at $2.0\times 10^5$~years, the D/H ratios have been
  multiplied by a factor of 10 to enable comparisons.}\label{T14_comp}
\end{figure*}

The models of \cite{taquet14} predict lower D/H values on smaller
(interferometric) than larger (single-dish) scales, but as noted above
this is not supported by the current observations, except for
water. Still, care must be taken that although the single-dish
observations are weighted toward the gas on larger scales, the strong
enhancement of the species close to the two components of IRAS16293
will weigh heavily there. Fig.~\ref{T14_comp} compares the measured
D/H ratios to the predictions on interferometric scales at three
separate times after collapse -- 1.1$\times 10^5$~years (``early''),
1.5$\times 10^5$~years (``intermediate'') and 2.0$\times 10^5$~years
(``late'') -- in the models by \cite{taquet14}. The D/H ratios
(corrected for statistics) are sorted by species (left) and
radical/functional group determining the D/H ratio (right). Generally
it is seen that while the ``late'' model significantly under-predicts
the observed D/H ratios for each species (predictions of order 0.1\%
vs. the observed few \%) the ``early'' and ``intermediate'' models
provide a much better agreement, although, in particular, the
``early'' model slightly overestimates the observed values (with a few
exceptions, i.p., methanol). The ``early'' model predicts large
variations between the D/H ratios for different functional groups for
individual molecules (in particular, ethanol and glycolaldehyde),
which is not observed. These variations become smaller in the
``intermediate'' and ``late'' models. Taken together these
comparisons, in particular, the slightly over-prediction of the D/H
ratios in the ``early'' model compared to the observations and more
homogeneous functional group D/H ratios at late times, could be taken
as an indication that the biggest challenge in fact is to determine
the initial conditions and/or time in the models that applies best in
the specific case.

A slightly different paradigm to those presented by \cite{garrod13}
and \cite{taquet14} has been offered by \cite{drozdovskaya15} who
modeled the 2D structure of the complex organic chemistry of a static
protostellar envelope with a significant outflow cavity. The
introduction of such an outflow cavity would significantly alter the
chemistry by allowing dust to be heated and FUV photons escaping to
larger scales leading to enhanced
photodissociation. \citeauthor{drozdovskaya15} demonstrated that this
would lead to a time ordering of the appearance of complex organic
molecules (their Fig.~11) with, in particular, ethanol and dimethyl
ether forming during the protostellar stages where the FUV photons
from the young protostar enhance the radicals resulting in their
formation, compared to the other species being formed during the
prestellar core stage. Such a time-ordering would thus imply a
three-step formation of the complex organics: formaldehyde, methanol,
ketene and formic acid would be formed early in the prestellar core
where the D/H ratio is low, acetaldehyde, methyl formate and
glycolaldehyde at the late prestellar/early protostellar stage where
the D/H ratio reaches its maximum and dimethyl ether and ethanol in
the protostellar envelope when the FUV field becomes important.

In either scenario the distinct formation time of groups of species
may also reflect in the resulting $^{12}$C/$^{13}$C ratios: as
discussed in \cite{jorgensen16} and noted in the introduction, a lower
$^{12}$C/$^{13}$C ratio than the ISM value may reflect
fractionation or differences in sublimation temperatures of, i.p.,
$^{12}$CO and $^{13}$CO. A possible difference in the sublimation
temperatures for $^{12}$CO and $^{13}$CO may play a role at the later
stages when the protostellar heating is starting to increase the
temperature to the point when CO starts sublimating. \cite{smith15}
demonstrated that small differences in the binding energies between
$^{12}$CO and $^{13}$CO (the latter more tightly bound) of up to 10~K
could lead to an enhanced $^{12}$CO:$^{13}$CO gas-phase abundance
ratio by up to a factor of 2 observed at infrared wavelengths toward some
sources. This would naturally have the opposite effect of the
$^{12}$C/$^{13}$C of the organics derived from the CO ``left behind in
the ices'' that would be deficient in $^{12}$C relative to
$^{13}$C. These effects would mainly play a role in the region where
the temperature is just around what is required for CO to sublimate
and thus mainly affecting species formed at the onset of warming-up by
the protostar itself. Fractionation due to the \emph{ambient} UV field
in contrast is unlikely to introduce any significant differences as it predominantly occurs during the
early cloud stages: an anomalous $^{12}$C/$^{13}$C ratio for
  specific species therefore
would have to be carried through to  their formation time of
without affecting the other
species. However, if in fact the formation of some of the complex
organics require the FUV irradiation provided by the \emph{central
  protostar} such as in the scenario described by
\cite{drozdovskaya15}, this could perhaps trigger an anomalous
$^{12}$C/$^{13}$C ratio for some species.

Clearly, more comparisons like these are needed to move forward and
test the different scenarios.  This could, for example, be through
similar comprehensive comparisons for more sources but also to other
predictions in the models (e.g., the absolute abundances for the
individual organics). Such comparisons would, in particular, make it
possible to address the question of whether the source physical evolution
during the early stages is the critical parameter in determining the
D/H ratio. Also, observations at higher angular resolution should be
able to resolve the emission regions of the groups of species with
different excitation temperatures and thus test that spatial
segregation. In this context, it would naturally be interesting to see
whether the relatively high excitation temperatures inferred for
IRAS16293B and some high-mass hot cores relative to the Galactic
Center apply in general, whether they reflect the distribution of the
material around those sources or whether, e.g., additional production
of some species at high temperatures come into play.

\section{Conclusions}\label{conclusions}
We have presented a systematic survey of the isotopic content of
oxygen-bearing (and some nitrogen-bearing from \citealt{coutens16})
complex organic molecules toward the low-mass protostar
IRAS16293B. Using data from the ALMA \emph{Protostellar
  Interferometric Line Survey (PILS)} program we provide an inventory
of those species and their relative abundances in the warm gas close
to the central protostar where ices are completely sublimated. The
main conclusions are:

\begin{itemize}
\item For the first time we identify the deuterated and
  $^{13}$C-isotopic species of ketene, acetaldehyde and formic acid as
  well as deuterated ethanol in the interstellar medium and present
  observations of $^{13}$C isotopic species of dimethyl ether,
    methyl formate and ethanol (the latter two only tentatively
    detected) along with mono-deuterated methanol, dimethyl ether and
  methyl formate. Systematic derivations of excitation temperatures and column densities
    result in small uncertainties in
    their relative abundances.
\item Some differences are found in excitation temperatures for
  different species, with the lines for one group (formaldehyde,
  dimethyl ether, acetaldehyde and formic acid, together with
  formaldehyde, \citealt{persson18}), best fit with an excitation
  temperature of approximately 125~K with the remaining species,
  together with formamide and glycolaldehyde
  \citep{coutens16,jorgensen16}, better fit with a temperature of
  300~K. This possibly reflects different binding energies of the
  individual species to the grains with the former group having
  binding energies in the 2000--4000~K range and the latter in the
  5000--7000~K range. The division is very similar to what has
  previously been found for high-mass sources \citep{bisschop07}.
\item The D/H ratios of the complex organic molecules can be divided
  into two groups with some of the simpler species (methanol, ketene,
  formic acid) as well as formaldehyde, formamide and isocyanic acid)
  showing D/H ratio of $\approx$~2\% and the more complex species
  dimethyl ether, ethanol, methyl formate, glycolaldehyde and
  acetaldehyde showing higher ratios ranging from about 4 to 8\%.
    Conservative estimates of the errors on these ratios are 0.6 percentage points
    for the lower value and 1.5 percentage points for the higher
    values. This distinction may reflect the formation time of each
  species in the ices before or during warm-up/infall of material
  through the protostellar envelope.
\item No significant differences are seen in the deuteration of different
  functional groups for individual species, possibly a result of the
  short time-scale for infall through the innermost regions where
  exchange reactions between different species may be taking place.
\item The $^{12}$C/$^{13}$C ratio of dimethyl ether is found to
    be lower than
    that of the local ISM, similar to the case of
    glycolaldehyde \citep{jorgensen16}. Marginal detections of the
    $^{13}$C isotopologues of methyl formate and ethanol are also
    consistent with a lower ratio. Low $^{12}$C/$^{13}$C ratios may
    reflect the formation histories of individual species; that they form in the ices at a
  late point where more $^{13}$C is available due to a slightly
  lower binding energy of $^{12}$CO compared to $^{13}$CO ice -- or
  fractionation triggered by FUV irradiation from the central
  protostar.
\end{itemize}

This study is an important illustration of how the isotopic
composition can be used to determine the formation pathways for
different molecular species.  Efforts on analysing the
nitrogen-containing organics toward IRAS16293B
\citep{ligterink17,ligterink18,coutens18,calcutt18}, as well as
  establishing comparable inventories of both oxygen- and
  nitrogen-bearing species toward IRAS16293A, are ongoing. Together
these studies will serve as an important constraint on astrochemical
models attempting to account for the fractionation processes in
protostellar environment. It would be worthwhile revisiting the
abundances of species on larger scales for IRAS16293 and other
sources, given the issues with opacity effects for many of the main
isotopologues. As noted above, additional efforts are needed to
  investigate any deviations of the $^{12}$C:$^{13}$C ratios for the
  organics that are only marginally detected or those suffering from
  optical depth issues. Finally, extending this survey to other
embedded protostars, possibly in other regions, could shed light on
the importance of environment versus evolutionary histories of the
sources for the formation of complex molecules.

\begin{acknowledgements}
  The authors wish to thank the referee for good,
    constructive comments that significantly improved the paper. The
  authors are grateful to the various spectroscopy groups without
  whose systematic efforts, studies like this would not be
  possible. In particular, thanks go to Adam Walters and Frank Lewen
  for their contributions in Cologne and Laurent Margul\`es and Roman
  Motiyenko for their work in Lille. This paper makes use of the
  following ALMA data: ADS/JAO.ALMA\#2013.1.00278.S. ALMA is a
  partnership of ESO (representing its member states), NSF (USA) and
  NINS (Japan), together with NRC (Canada), NSC and ASIAA (Taiwan),
  and KASI (Republic of Korea), in cooperation with the Republic of
  Chile. The Joint ALMA Observatory is operated by ESO, AUI/NRAO and
  NAOJ. The group of JKJ is supported by the European Research Council
  (ERC) under the European Union's Horizon 2020 research and
  innovation programme through ERC Consolidator Grant ``S4F'' (grant
  agreement No~646908). Research at Centre for Star and Planet
  Formation is funded by the Danish National Research
  Foundation. A.C. postdoctoral grant is funded by the ERC Starting
  Grant 3DICE (grant agreement 336474). MND acknowledges the financial
  support of the Center for Space and Habitability (CSH) Fellowship
  and the IAU Gruber Foundation Fellowship. The group of EvD
  acknowledges ERC Advanced Grant ``CHEMPLAN'' (grant agreement
  No~291141).
\end{acknowledgements}

\begin{appendix}
\section{Discussion of individual species}\label{indspecies}
\subsection{History of detections}
Methanol, CH$_3$OH, \citep{ball70}, acetaldehyde, CH$_3$CHO, also
known as ethanal, \citep{det-MeCHO_1973,MeCHO_1974}, dimethyl ether,
CH$_3$OCH$_3$, \citep{det-DME_1974}, ethanol, C$_2$H$_5$OH,
\citep{a-EtOH_det_1975}, methyl formate, CH$_3$OCHO,
\citep{det-MeFo_1975a,MeFo_1975b} and also ketene, CH$_2$CO, also
known as ethenone, \citep{det-ketene_1977}, were all among the early
molecules detected by means of radio astronomy in the 1970s. All
detections were made toward high-mass star-forming regions, mostly in
the prolific Galactic Centre source Sagittarius~B2. Methanol and
ethanal were detected also in Sagittarius~A; dimethyl ether was
detected in the Orion Molecular Cloud. Unambiguous detections of minor
isotopic species have also been reported for several of these
molecules: The $^{13}$C \citep{det-13C-MeOH_1979} and $^{18}$O
\citep{CH3O-18-H_det_1989} species of methanol were detected, and in
the case of deuterated isotopologues even CD$_3$OH
\citep{parise06deuterium}. In the case of methyl formate both $^{13}$C
isotopologues were detected, along with CH$_3$OCDO
\citep[tentatively;][]{demyk10}, CH$_2$DOCHO \citep{coudert13} and
both $^{18}$O isotopologues \citep{18O-MeFo_det_2012}.  Detections of
deuterated and $^{13}$C dimethyl ether were reported by
\cite{richard13} and \cite{13C-DME_lab_det_2013}, respectively, while the
detection of the $^{13}$C isotopomers of ethanol toward Sagittarius~B2
were reported by \cite{muller16}. Formic acid was detected first
toward Sgr~B2 \citep{HCOOH_det-1_1971,HCOOH_det-2_1975}.  It was also
detected in a cold cloud \citep{HCOOH_dark-cloud_1990} and toward
IRAS~16293 \citep{cazaux03}. To our knowledge, there are no secure
detections of minor isotopic species of HCOOH.

\subsection{Methanol, CH$_3$OH}
As mentioned above one of the biggest problems about fitting CH$_2$DOH
is that many of its transitions become optically thick at low
temperatures. For the first we therefore only utilise lines that are
predicted to have $\tau < 0.1$ and further indicate additional lines
that are strongly optically thick at either 50 or 300~K separately.
Those lines are typically overproduced in the synthetic spectra here,
likely because their emission is quenched by the colder material at
larger scales. Three transitions (one seen at 349.62~GHz and two at
346.82~GHz) are still significantly overproduced by the models. This
likely reflects issues with the catalog entries for these high
excitation transitions. The fits to CH$_3^{18}$OH at the half beam
offset position were discussed in \cite{jorgensen16}. For completeness
(Fig.~\ref{18methanol_spectra}) shows the fits at the one beam offset
position where the derived column density is taken to be a factor 2
lower than at the half beam offset position. The CH$_3$OD lines are
well-reproduced for both the a- and e-type transitions, although these
lines span a more limited set of upper energy levels than for many of
the other species. For all methanol transitions, an excitation
temperature of 300~K works well. For CH$_3^{18}$OH and CH$_3$OD
  36 and 29 lines can be assigned, respectively.

\subsection{Ethanol, CH$_3$CH$_2$OH}
For the main isotopologue of ethanol similar issues as for CH$_2$DOH
arise with lines that become optically thick at larger scales. Those
are indicated in the figures. Still, more than hundred lines of
ethanol are predicted to be optically thin above $5\sigma$. The
  lines of the different deuterated variants vary somewhat in
  strength: for the two variants of CH$_2$DCH$_2$OH and CH$_3$CHDOH
  $>$60 and 37 transitions are identified, respectively providing
  reliable estimates of their column densities. For CH$_3$CH$_2$OD
  about 5--10 features can be assigned: these are blended to varying
  degree making the assignment more ambiguous and the derived column
  density tentative. As for methanol, the lines of ethanol are
well-produced with an excitation temperature of 300~K but it could be
10--20\% higher or lower without significantly altering the fits. For
the two $^{13}$C isotopologues a few plausible assignments are seen in
the spectra. Individually they are not sufficient to claim a secure
detection of any of the $^{13}$C isotopologues, however, if they are
considered together about a handful of lines can be modeled with a
column density corresponding to a $^{12}$C/$^{13}$C ratio lower by
about a factor 2 than the canonical ISM value.

\subsection{Methyl formate, CH$_3$OCHO}
The lines of methyl formate are well-fit with an excitation
temperature of 300~K. For the
  main isotopologue and CH$_2$DOCHO several hundreds of lines are
  present above 5$\sigma$ (about half of those of the main
  isotopologue are optically thin), while only about 10--15 lines are
  seen of the rarer CH$_3$OCDO isotopologue. The implied D/H ratio
  derived from this variant is close to that of the CH$_2$DOCHO
  isotopologue corrected for statistics, however, lending further
  credibility to the assignment. The lines of the $^{13}$C
isotopologue are relatively faint with most of the lines at the
5--10$\sigma$ levels blended to varying degrees. The best fit column
density imply a $^{12}$C/$^{13}$C
abundance ratio of $\approx$~30 but with large uncertainties.

\subsection{Ketene, CH$_2$CO}
The analysis of ketene is based on the $^{13}$C and singly deuterated
isotopologue and we present tentative detections/upper limits for the
$^{18}$O and doubly deuterated isotopologue. For the main isotopologue
of ketene only one out of the fourteen transitions predicted in
  the observed range is optically thin. The column density is
therefore derived based on the $^{13}$C isotopologues and assuming a
standard $^{12}$C/$^{13}$C ratio. These species are best fit with an
excitation temperature of 125~K.  Should it turn out that the
$^{12}$C/$^{13}$C ratio is lower for ketene, similar to the cases of
glycolaldehyde and methyl formate, the D/H ratio inferred here would
be an underestimate by that factor. We also show the fits for
brightest predicted lines for the $^{18}$O and doubly deuterated
isotopologues: for the $^{18}$O isotopologue we assume a standard
$^{16}$O:$^{18}$O ratio of 560 \citep{wilson94} from the abundance of
the main isotopologue inferred from the $^{13}$C species above. 
  The data are consistent with this fit, implying that the column
  density of the main isotopologue is not strongly underestimated, but
  the transitions are not sufficiently strong to consider a
  (tentative) detection. The doubly deuterated variant is not seen
with an upper limit relative to the main isotopologue of 0.2\%
($<10$\% relative to the singly deuterated version corrected for
statistics).

\subsection{Dimethyl ether, CH$_3$OCH$_3$}
Like for ketene, but in contrast to its isomer ethanol, dimethyl ether
is best fit with low excitation temperature of 125~K. The deuterated
version is naturally highly abundant due to the larger number of
indistinguishable H-atoms (six). For both the main and deuterated
  isotopologue about 100 transitions can be assigned. About 30 lines
  are present at the 5$\sigma$ level for the $^{13}$C isotopologue:
  those lines are also well-fit with this excitation temperature. The
ratio of the column densities between the $^{12}$C and $^{13}$C
isotopologues is a factor 17, which, corrected for the statistics
($^{13}$CH$_3$OCH$_3$ and CH$_3$O$^{13}$CH$_3$ again being
indistinguishable), implies a $^{12}$C/$^{13}$C ratio of 34, about
half that of the canonical ISM value and close to those measured for
glycolaldehyde, methyl formate and ethanol.

\subsection{Acetaldehyde, CH$_3$CHO}
Acetaldehyde is the third species best fit with a low excitation
temperature (see also \citealt{lykke17}). Like for the other
  species a number of transitions of the main isotopologue are
  optically thick, but still more than 100 optically thin transitions
  are present at the 5$\sigma$ level allowing a good estimate of its
  column density. For the $^{13}$C isotopologues a sufficient number
  of lines (about 60 and 80 transitions for CH$_3^{13}$CHO and
  $^{13}$CH$_3$CHO, respectively) are present to allow for a direct
  measurement of the $^{12}$C/$^{13}$C isotope ratio, which is found
  to be consistent with that of the local ISM. Spectroscopic data
only exists for the deuterated isotopologue with the D substituted 
on the -CHO group. This isotopologue is clearly detected.

\subsection{Formic acid, HCOOH}\label{formicacid_discussion}
For a reliable estimate of the column density of formic acid we rely
on lines of both the main and $^{13}$C isotopologues: the stronger
lines of the main isotopologue of \emph{(trans-)}formic acid are
strongly optically thick (about half of the 30 lines above
5$\sigma$). We therefore utilise the 17 lines of the $^{13}$C
isotopologue to derive a first estimate of the column density and from
that simulate the spectrum of the main isotopologue scaled by the
standard ISM ratio. Utilising this column density and focusing only on
the fainter lines of the main isotopologue, a number of assignments
can be thereby be made confirming the derived column density. In
Fig.~\ref{thcooh_spectra0} one line is seen at 338.202~GHz, which is
clearly over-predicted by the model. That line is an example of one of
the transitions being strongly optically thick ($\tau \gtrsim 1$). The
line at 338.192~GHz in that specific panel is centered on is optically
thin ($\tau \sim 0.01$), though.  Lines of the two deuterated
isotopologues are detected as well, but a number show severe
blending. We consequently derive a common column density for both
species based on the about 30 lines present above 5$\sigma$. It seems
that such a column density does reproduce the lines of both species
equally well, but the problems with the fits does make this number
more uncertain. The deuterated isotopologues are fit assuming the same
column density for both variants.

The lines of formic acid are in general well-produced with a high
excitation temperature of 300~K. In addition to the
\emph{trans-}formic acid we also checked for the presence of the high
energy \emph{cis} conformer that has recently been seen with
relatively high abundances relative to the \emph{trans} conformer
toward the Orion Bar \citep{cuadrado16}, likely a result of the strong
UV field there leading to photo-switching between the two
conformers. In our data the \emph{cis} conformer remains undetected
down to an abundance of $\sim 0.1$\% with respect to the \emph{trans}
conformer, which would be consistent with the expected ratio at
equilibrium at the temperature of IRAS16293B. It does mark a
difference though from the recent tentative detection of one line of
the \emph{cis} conformer toward the dark cloud B5 \citep{taquet17}.

\section{Spectra}
On the following pages the observed and modeled spectra are shown for
the lines predicted to be the brightest and optically thin according
to the synthetic spectra for each invidual species.

\clearpage
\subsection{Methanol}\label{methanolfits}
\noindent\begin{minipage}{\textwidth}
\resizebox{0.88\textwidth}{!}{\includegraphics{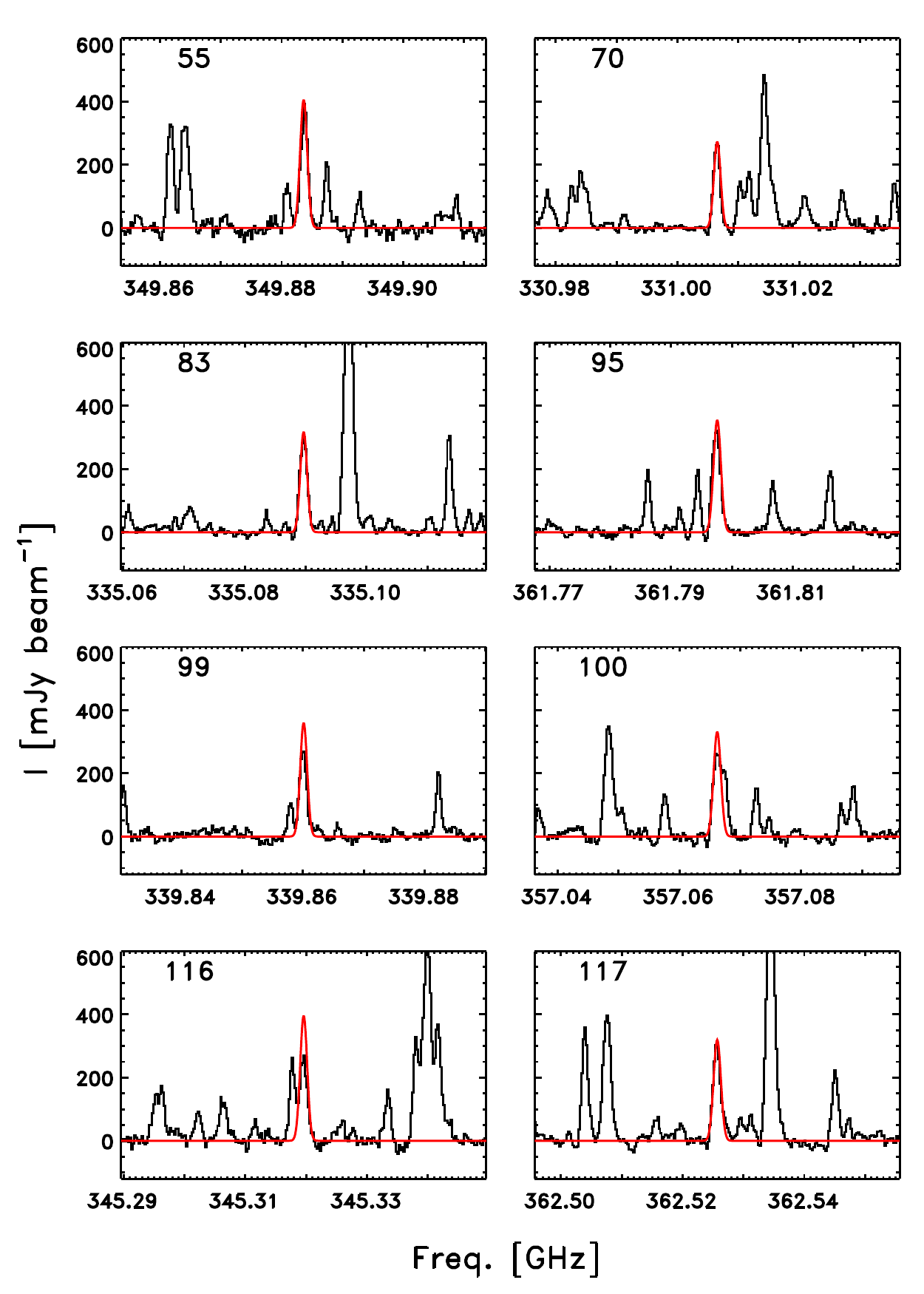}\includegraphics{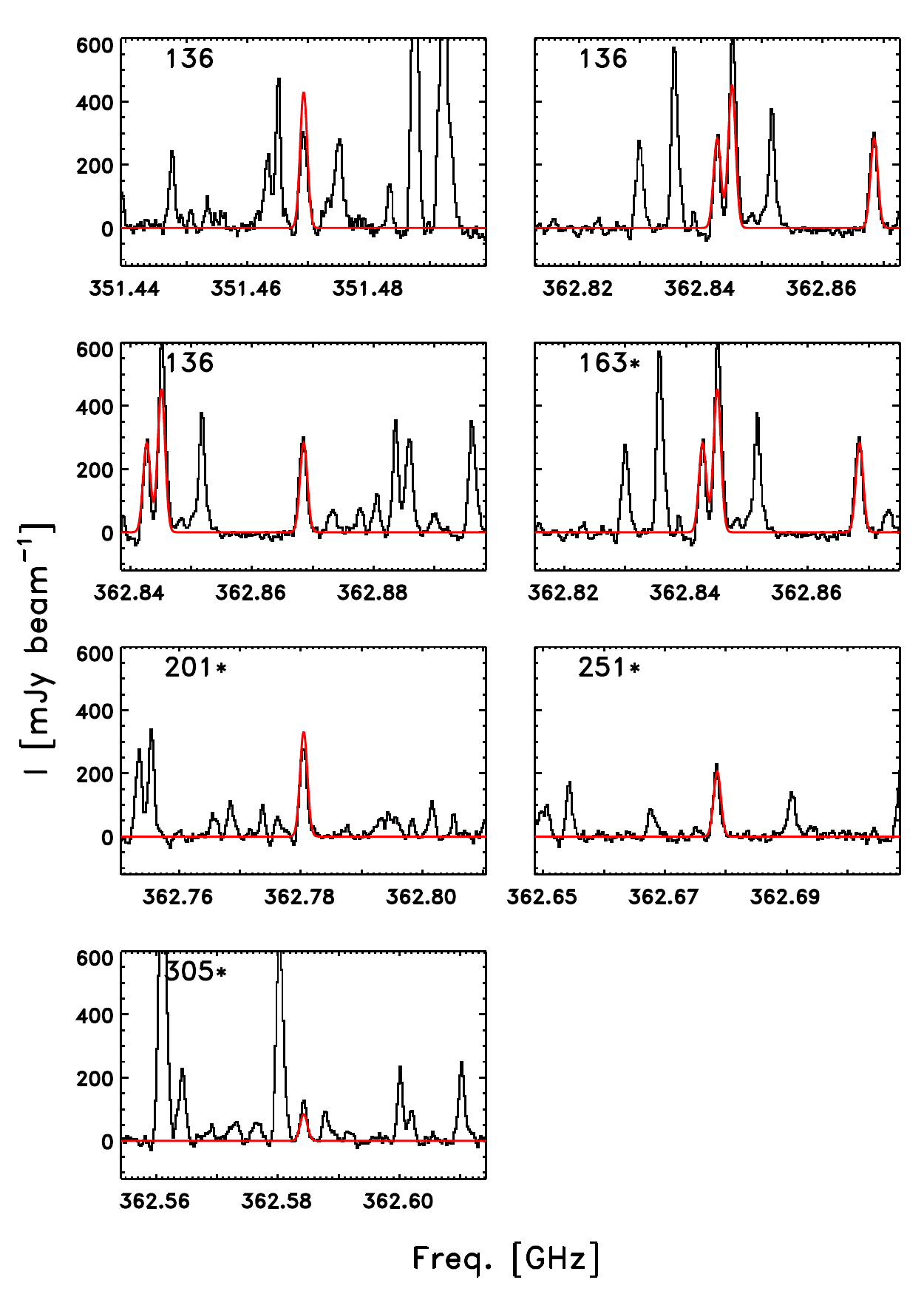}}\\
\resizebox{0.88\textwidth}{!}{\includegraphics{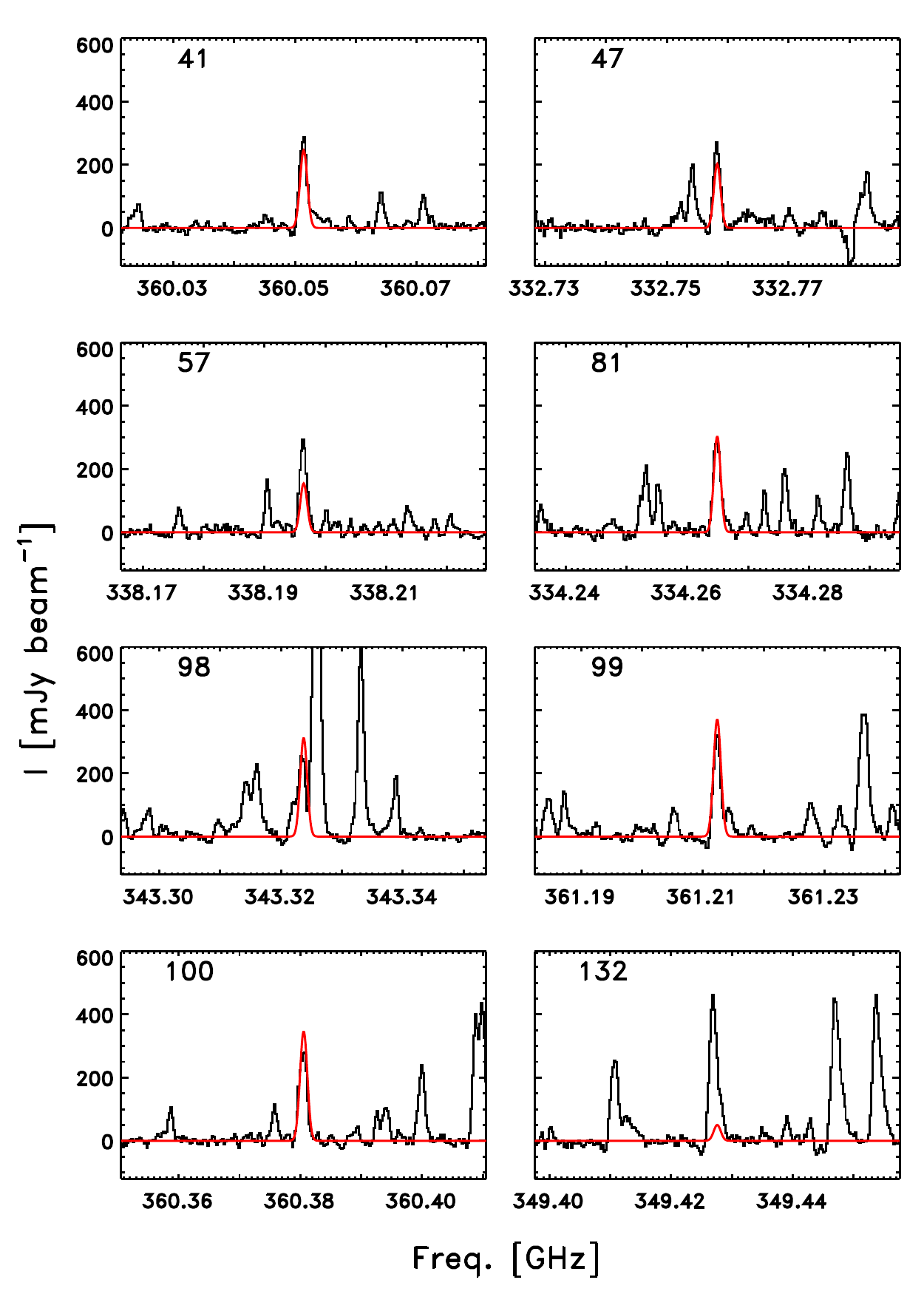}\includegraphics{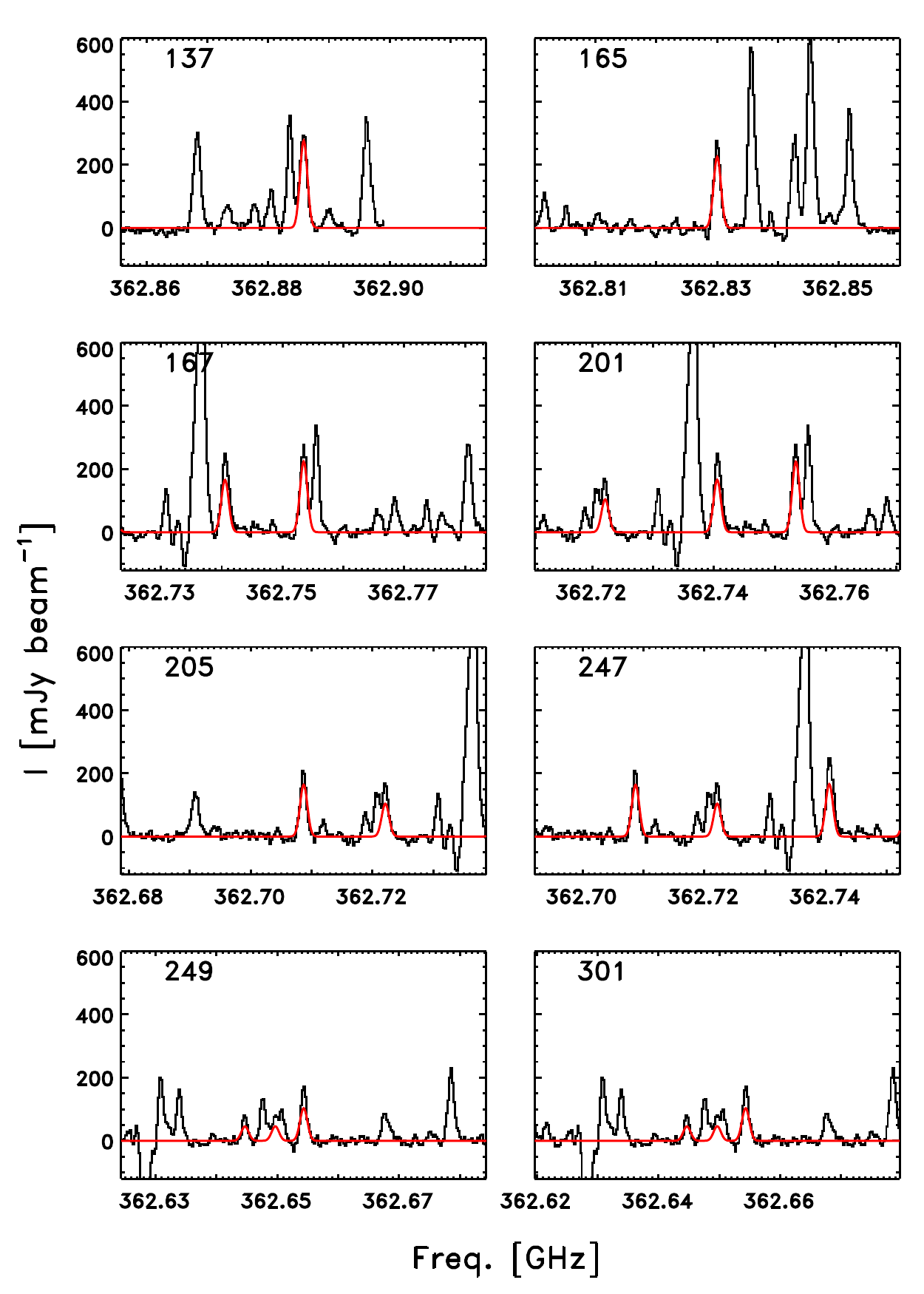}}
\captionof{figure}{The 15+16
  brightest lines of CH$_3$OD (a-type in the upper panels and e-type
  in the lower panels) toward the full-beam offset position. The lines
  are sorted according to $E_{\rm up}$ given in K in the upper left
  corner of each panel. The windows are centered on these specific
  lines.}\label{a-ch3od_spectra}\label{first_spectra}
\end{minipage}

\clearpage
\noindent\begin{minipage}{\textwidth}
\resizebox{0.88\textwidth}{!}{\includegraphics{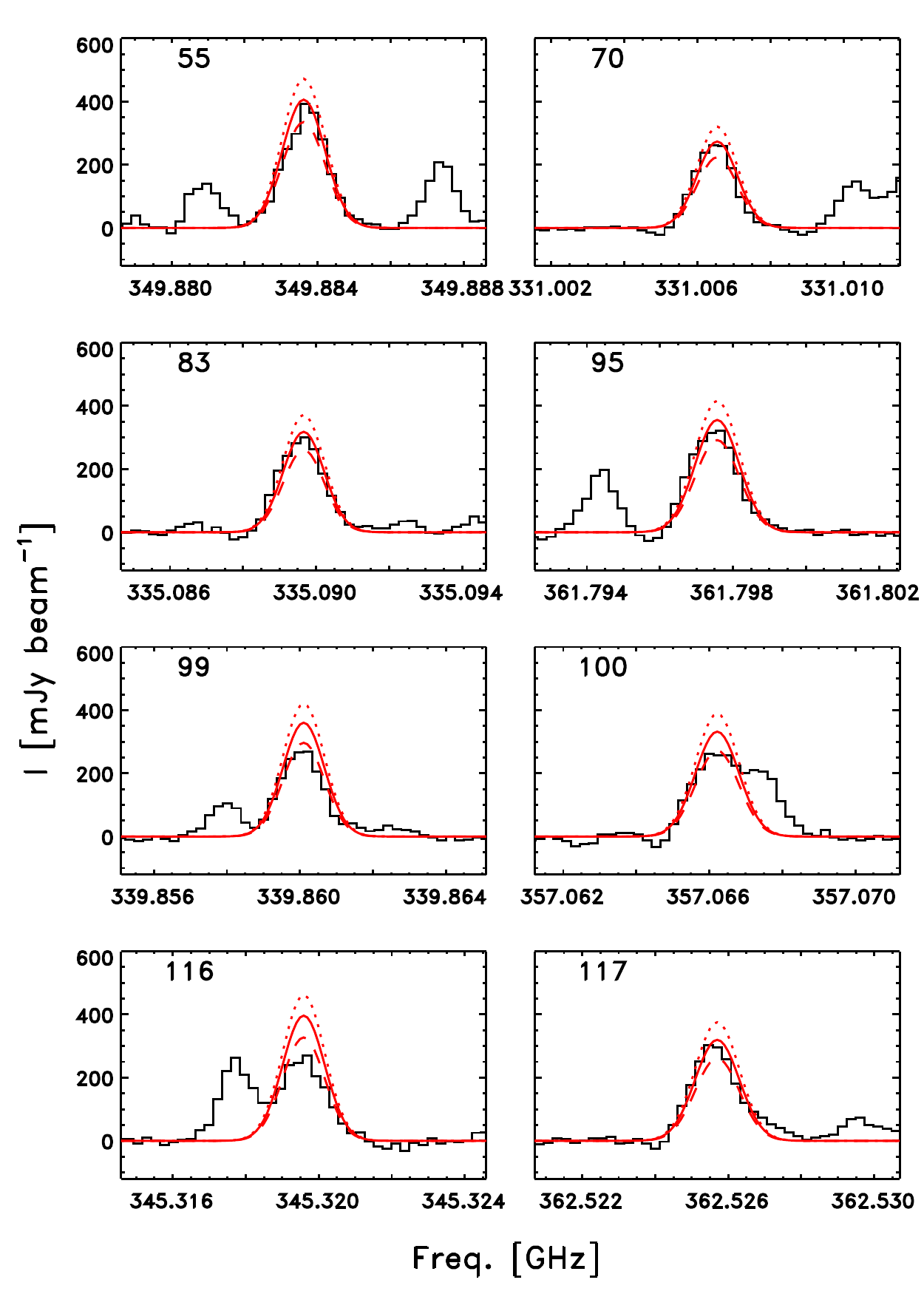}\includegraphics{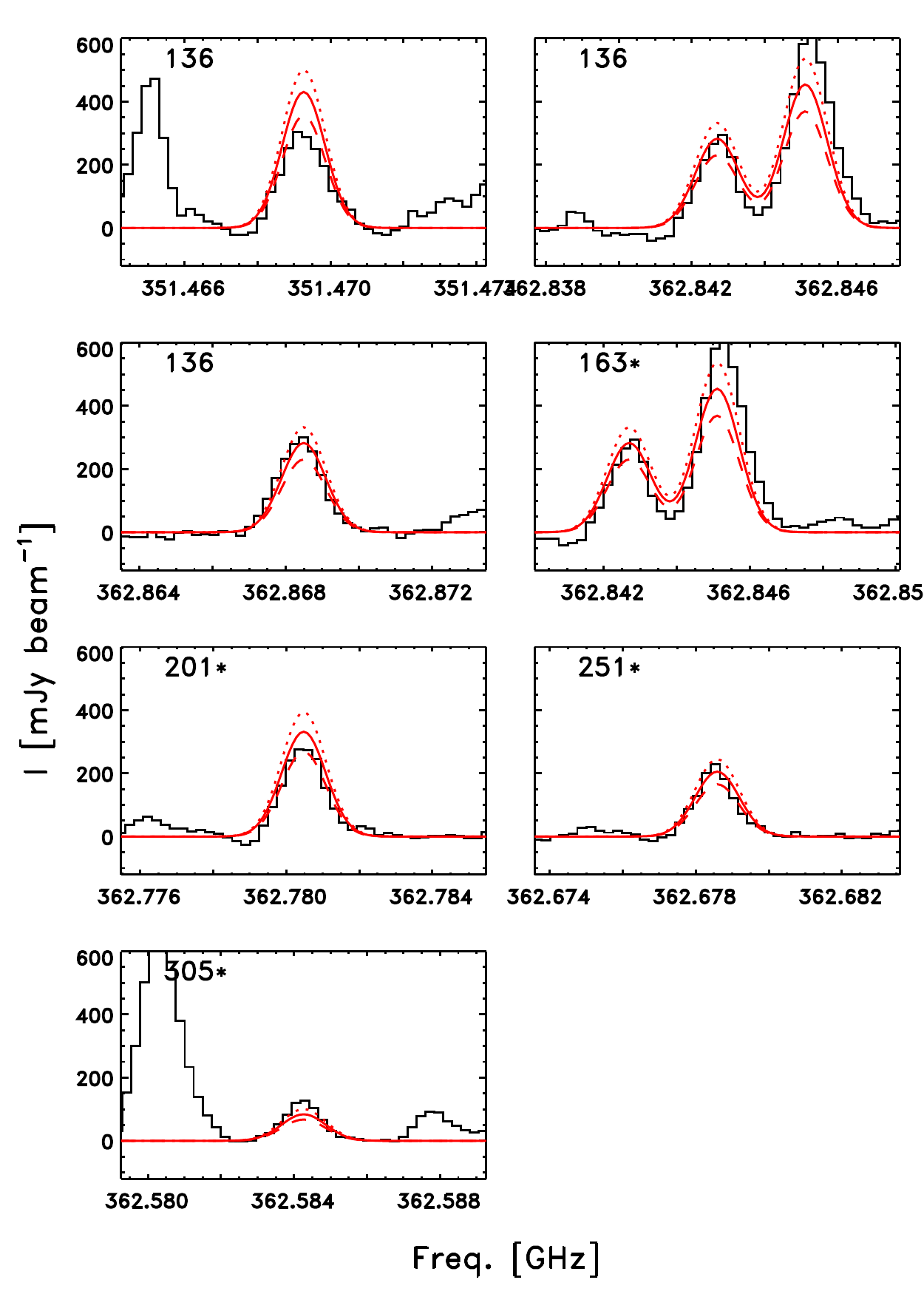}}\\
\resizebox{0.88\textwidth}{!}{\includegraphics{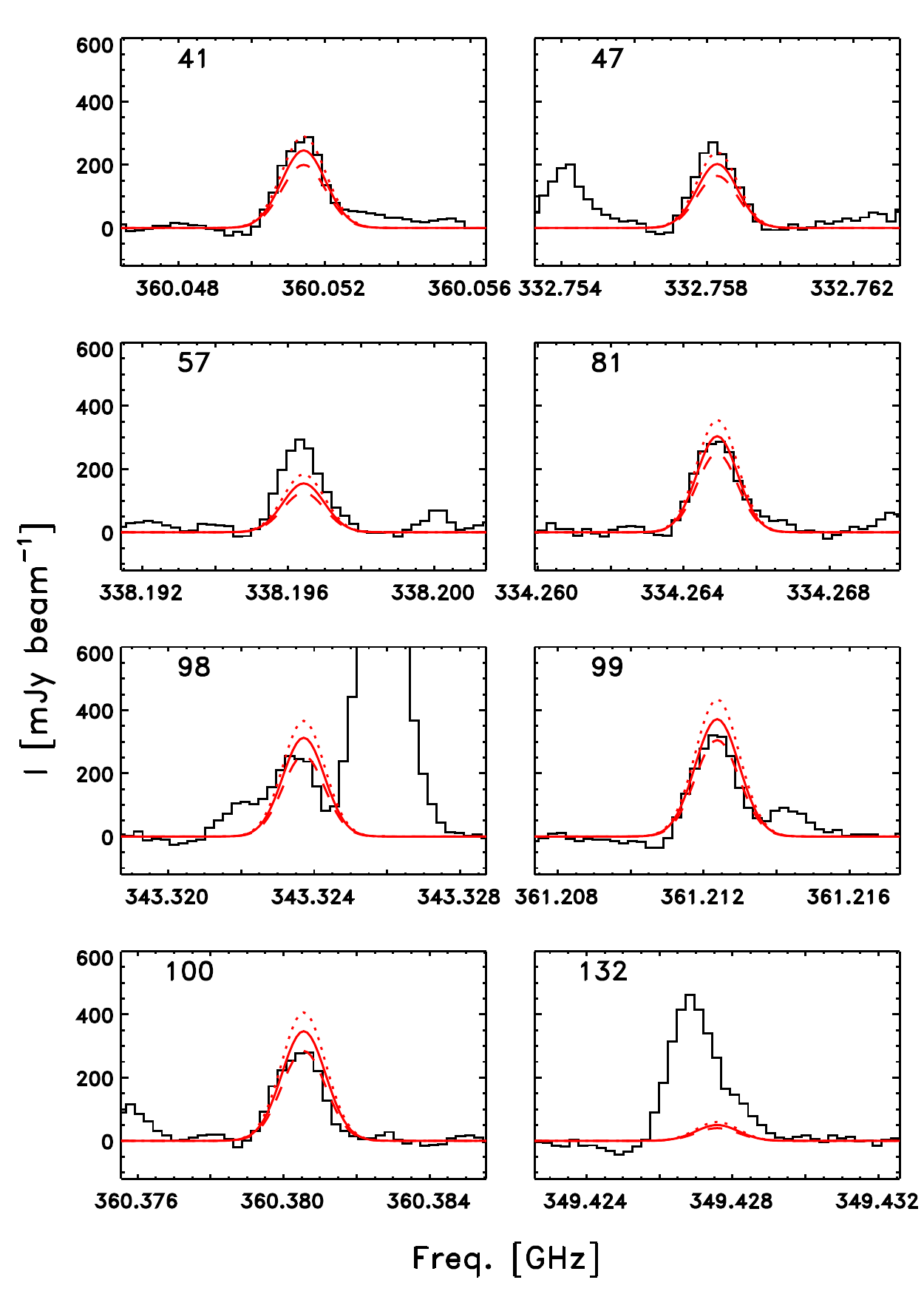}\includegraphics{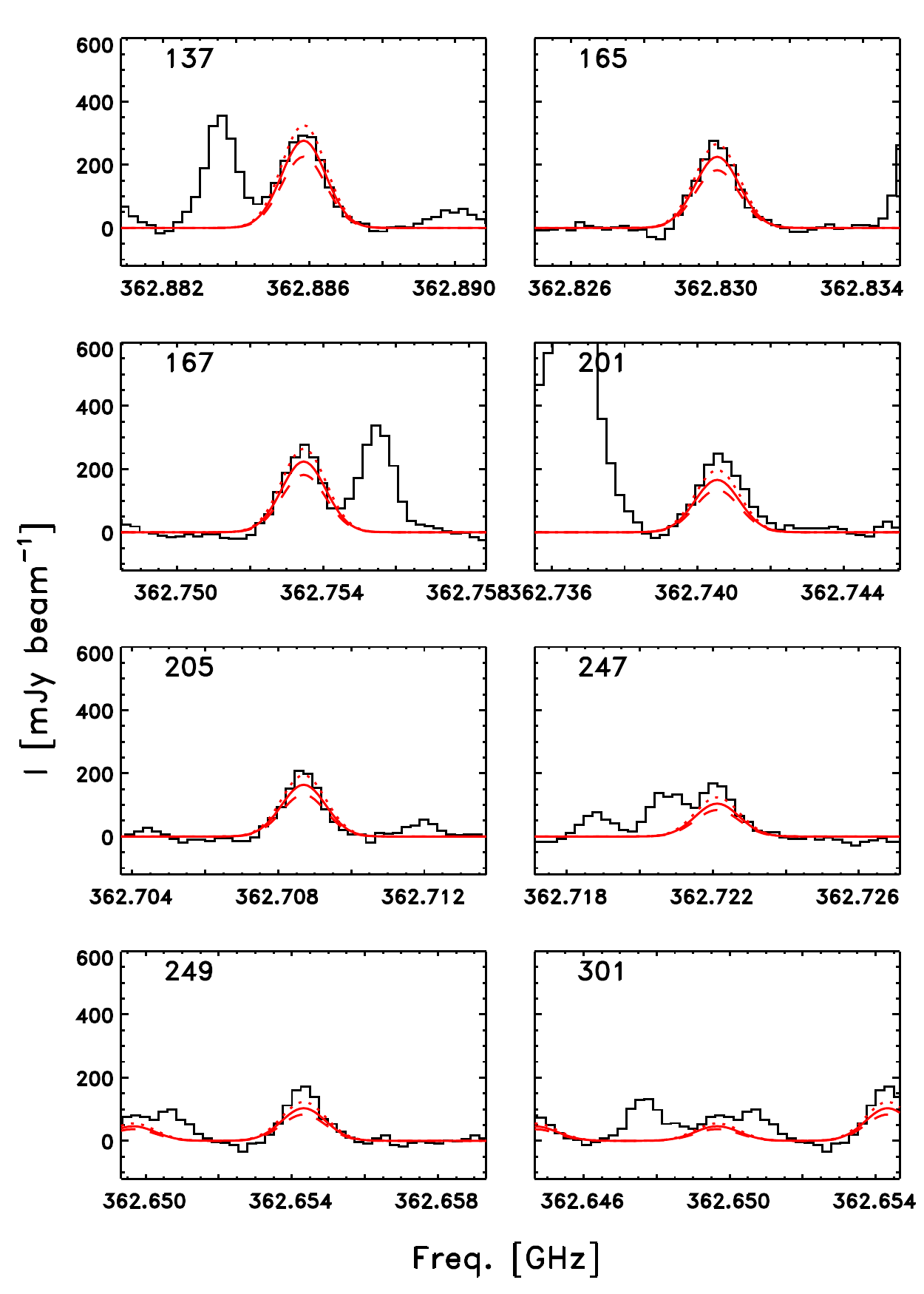}}
\captionof{figure}{As in Fig.~\ref{first_spectra}, zoomed-in to
  highlight the fits. Overplotted are the best fit models using 
  excitation temperatures 20\% above (dotted) and 20\% (dashed) below the
  best fit value.}\label{a-ch3od_texerror}
\end{minipage}

\clearpage
\noindent\begin{minipage}{\textwidth}
\resizebox{0.88\textwidth}{!}{\includegraphics{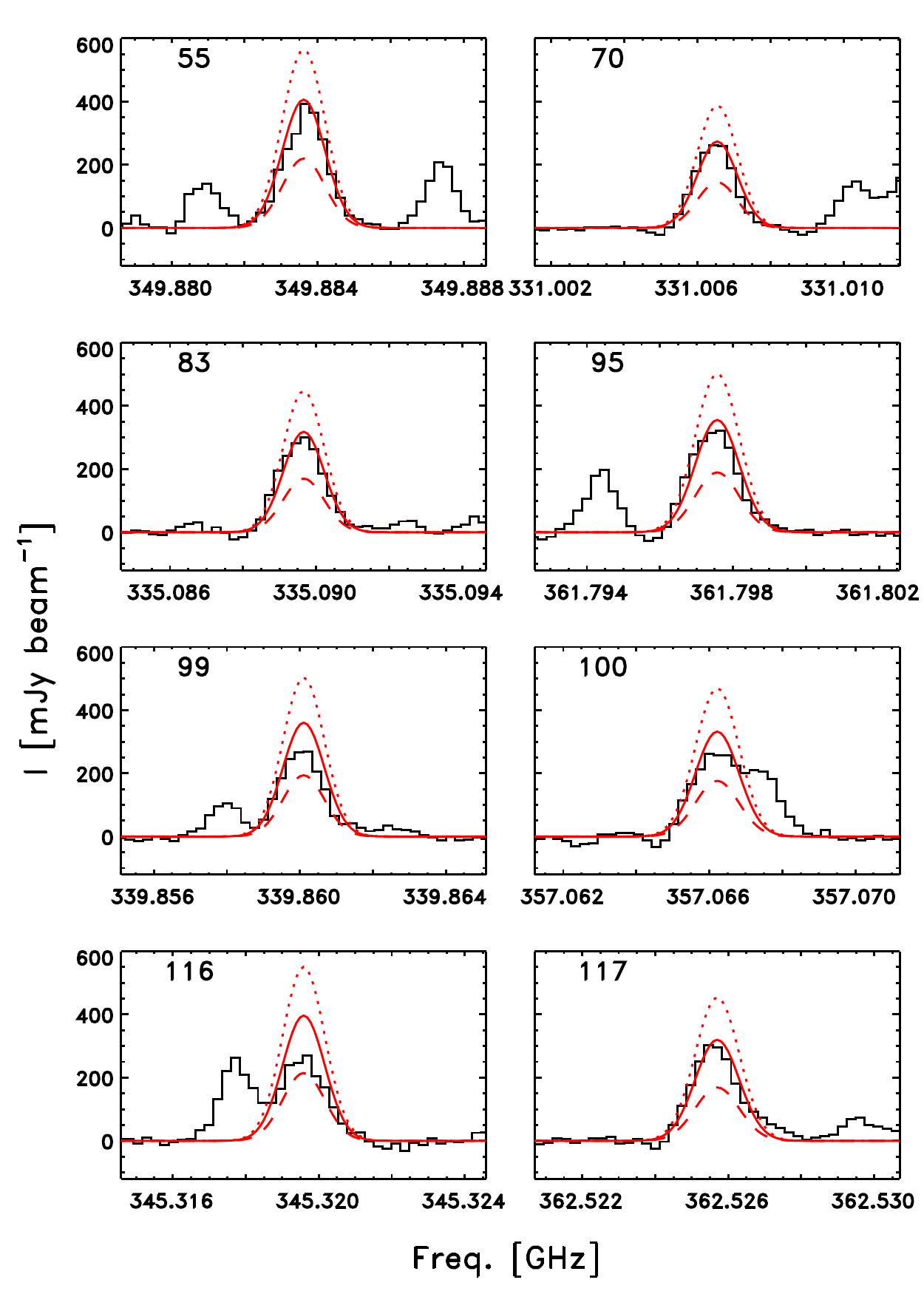}\includegraphics{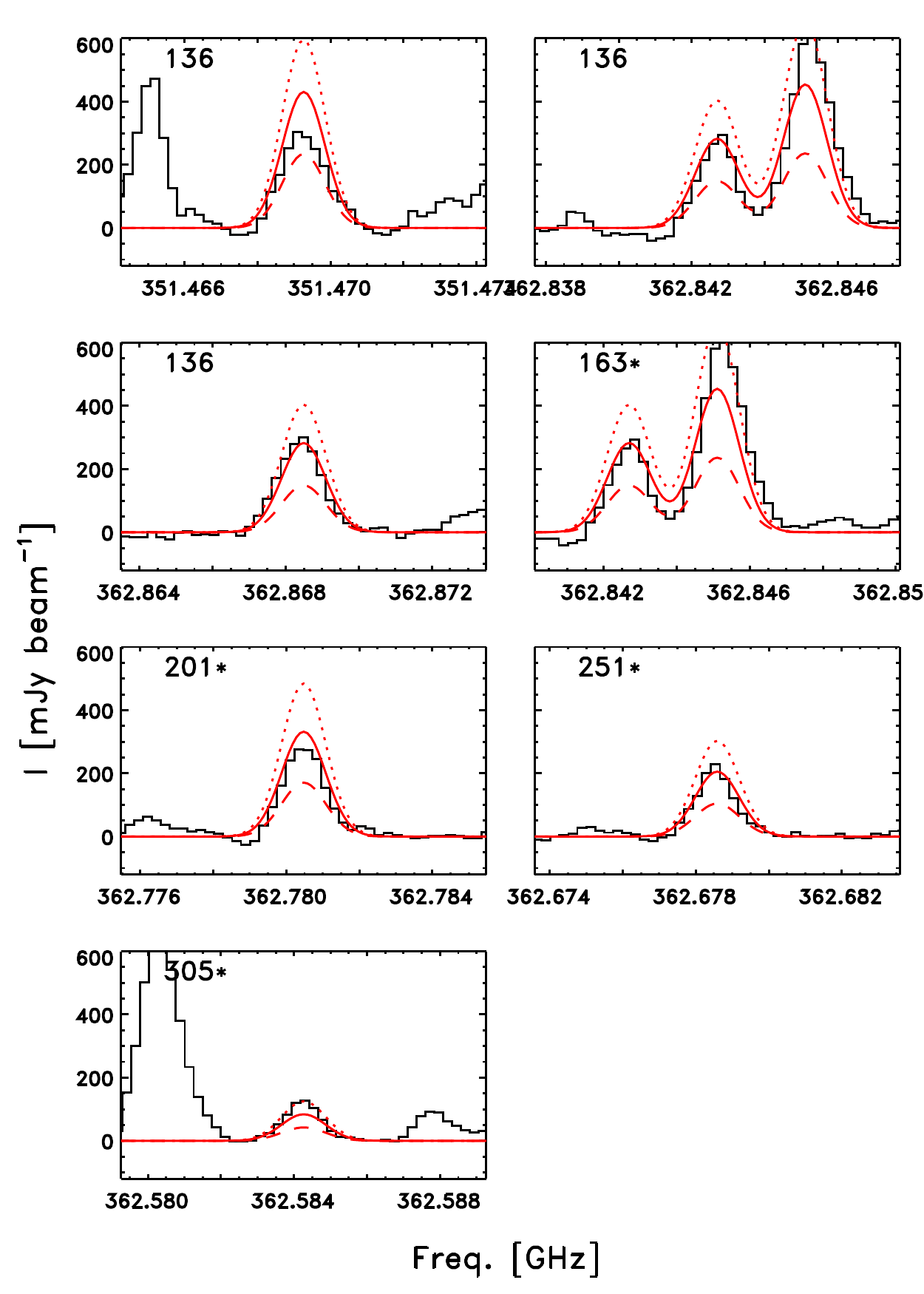}}\\
\resizebox{0.88\textwidth}{!}{\includegraphics{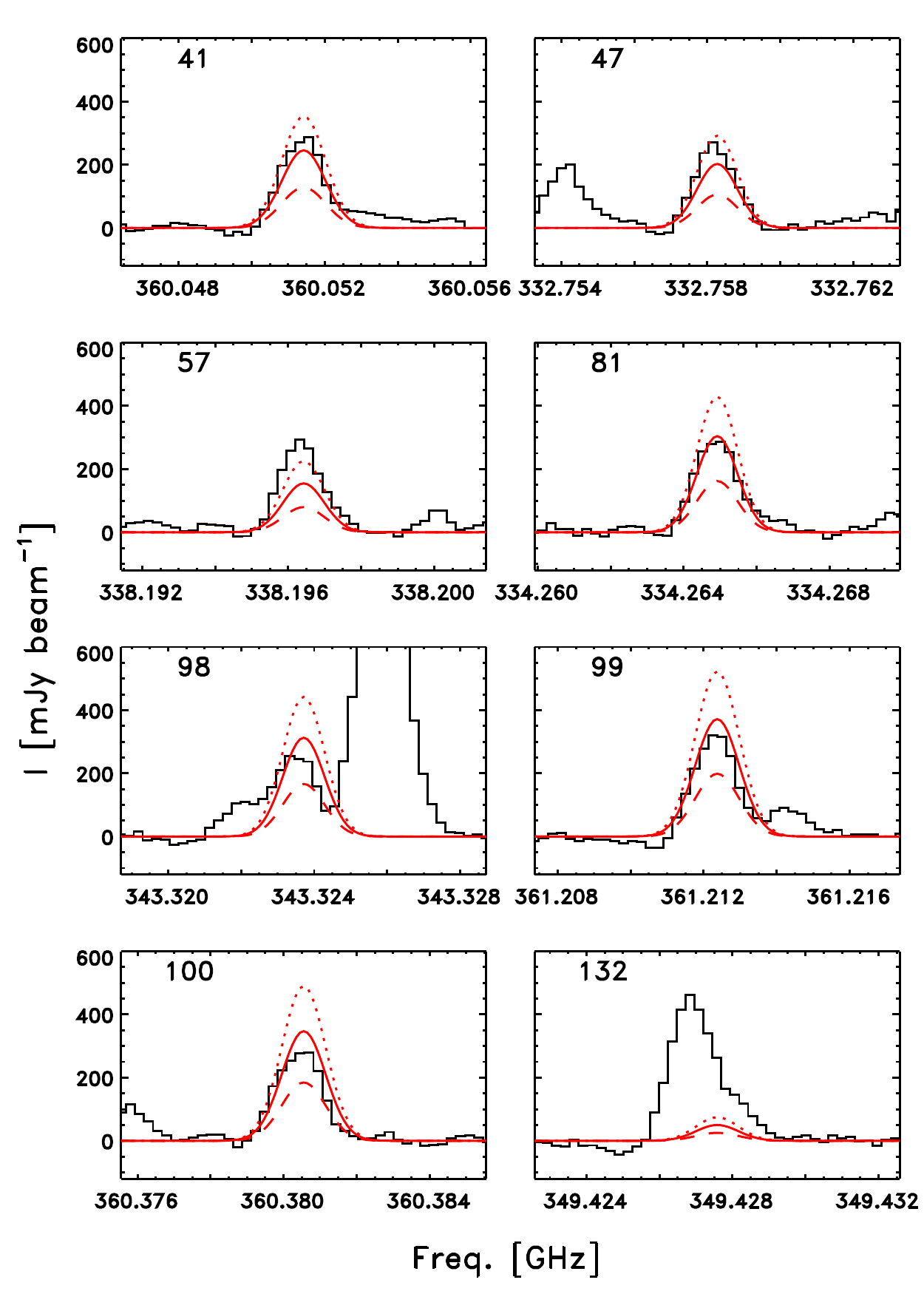}\includegraphics{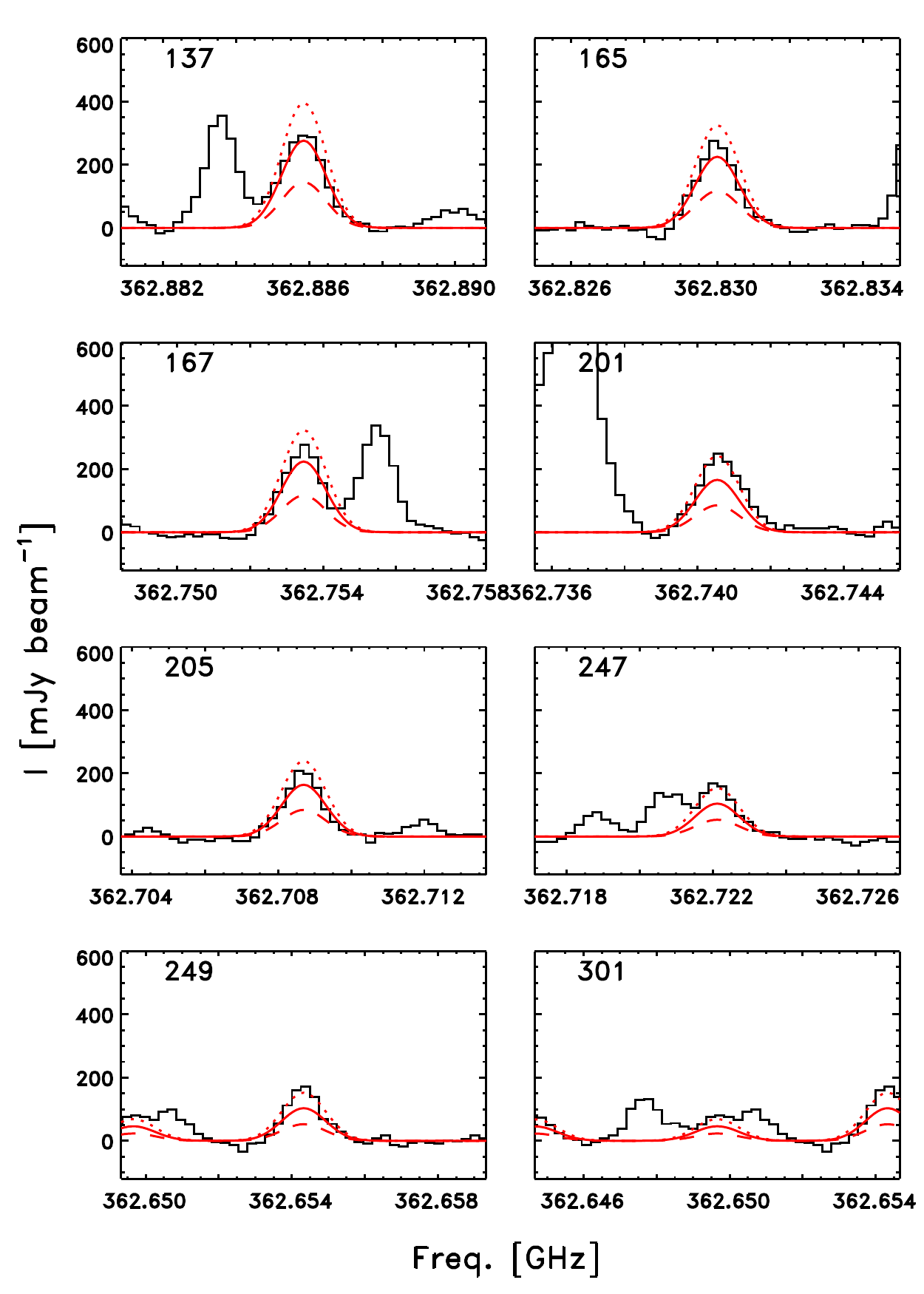}}
\captionof{figure}{As in Fig.~\ref{first_spectra}, zoomed-in to
  highlight the fits. Overplotted are the best fit models using
  column densities 20\% above (dotted) and 20\% (dashed) below the best fit
  value.}\label{a-ch3od_coldenserror}
\end{minipage}

\clearpage
\begin{minipage}{\textwidth}
\resizebox{0.88\textwidth}{!}{\includegraphics{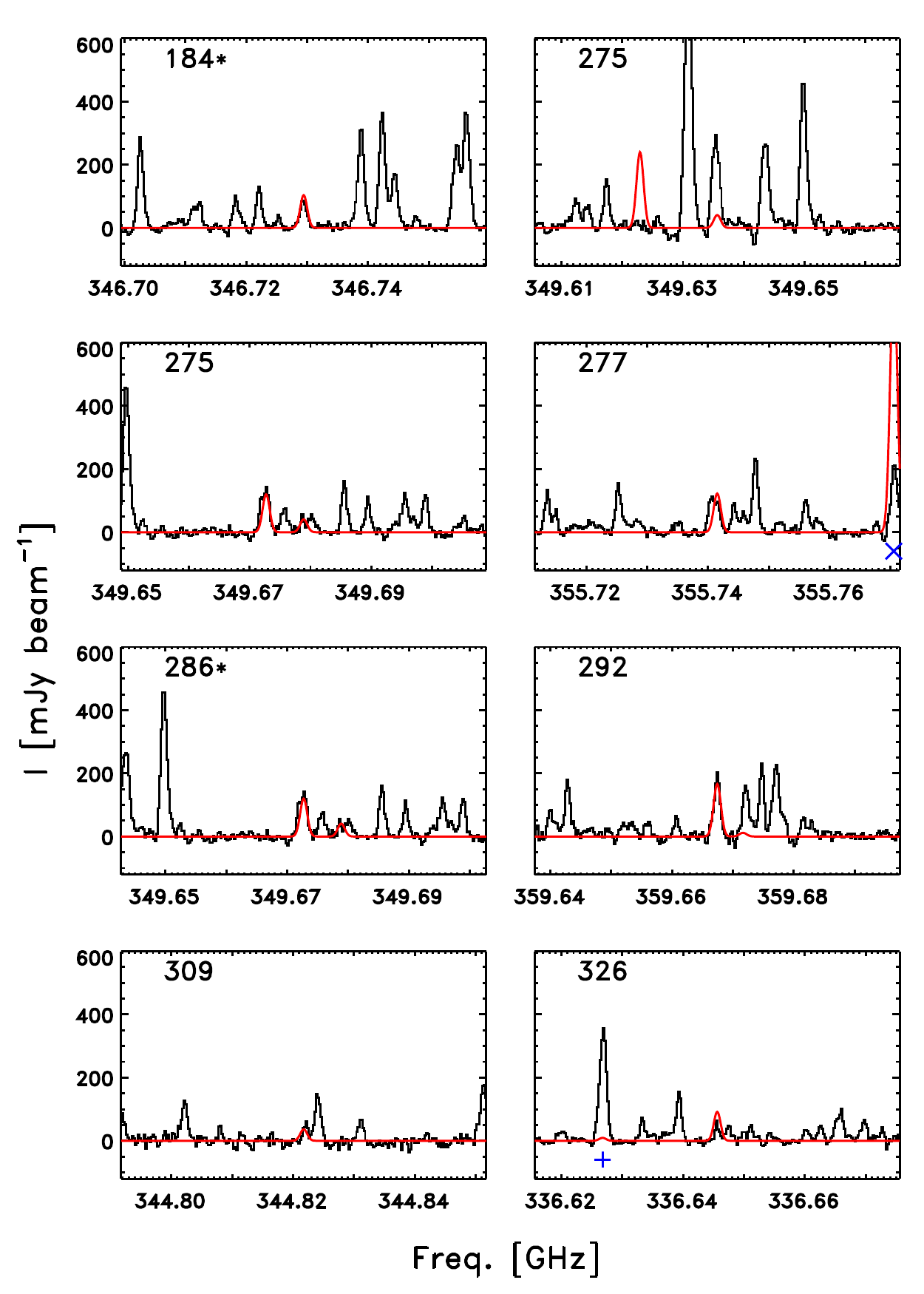}\includegraphics{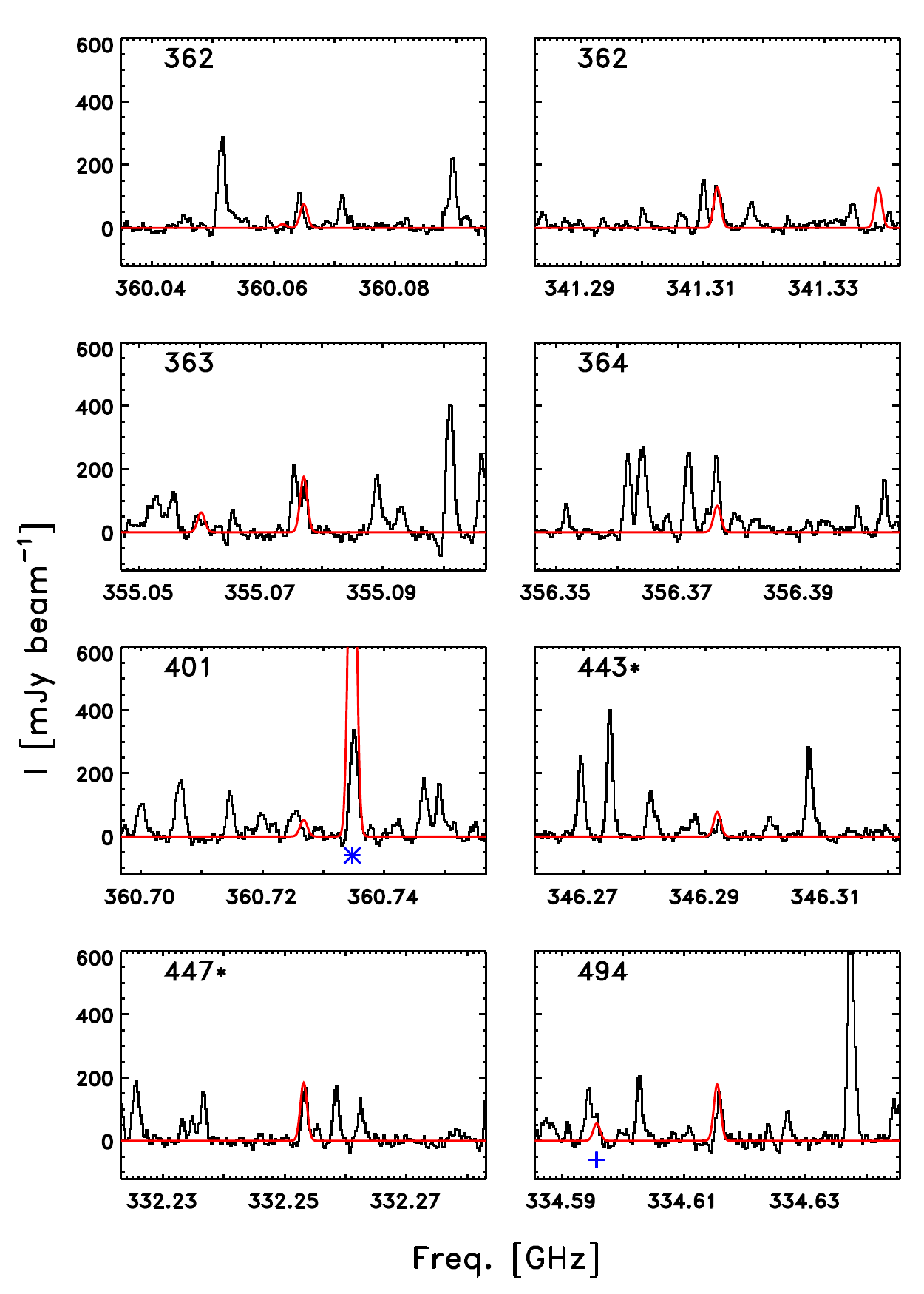}}
\resizebox{0.44\textwidth}{!}{\includegraphics{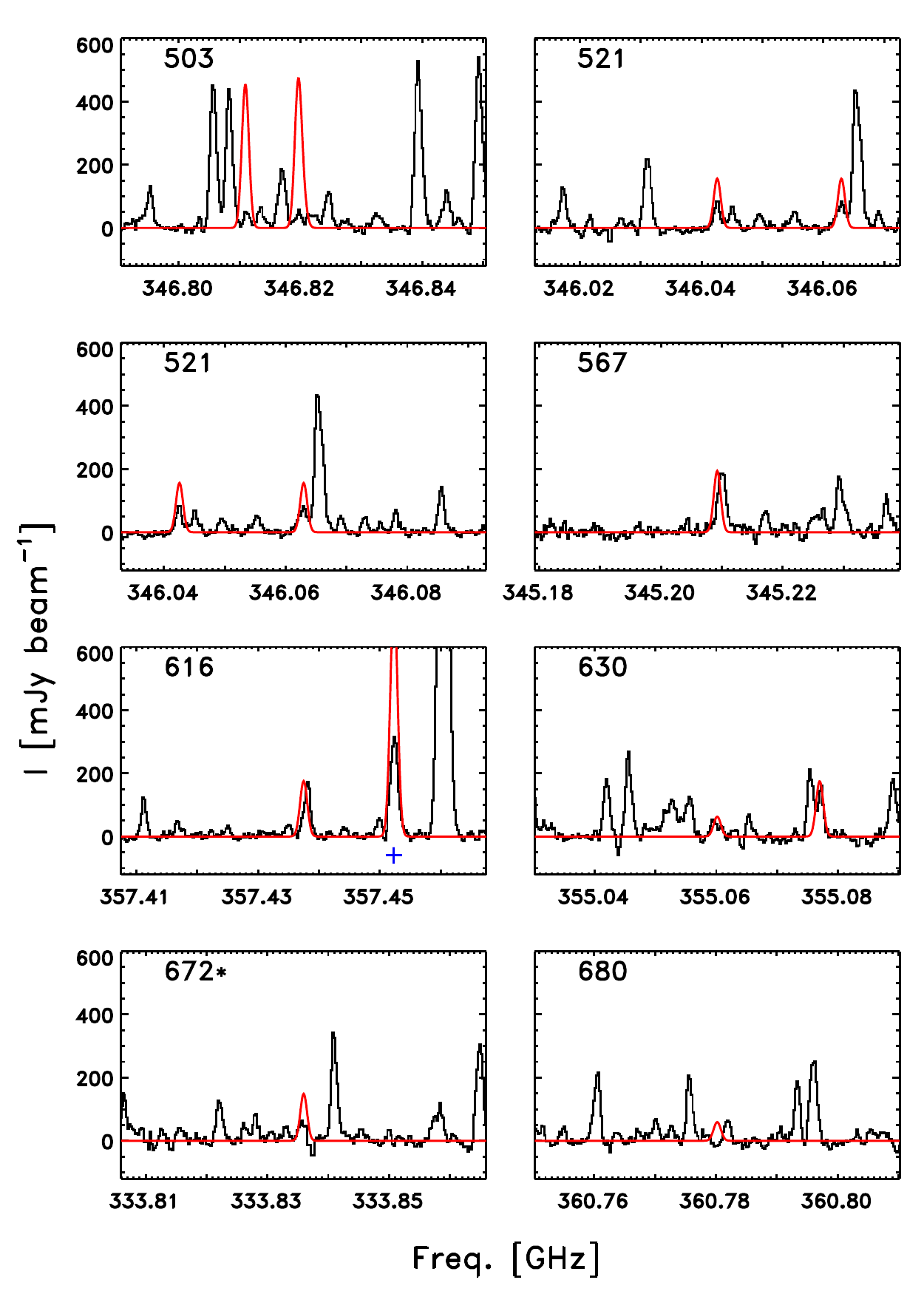}}
\captionof{figure}{The 24 brightest optically thin lines of CH$_2$DOH
  as expected from the synthetic spectrum toward the full-beam offset
  position. In the cases, where other lines fall in the band that are
  strongly optically thick at 50~K or 300~K they have been marked with
  a plus-sign or cross, respectively. The discrepancies for the lines
  at 349.62~GHz and 346.82~GHz likely reflect issues with the catalog
  entries for these high excitation
  transitions. }\label{ch2doh_spectra}
\end{minipage}

\clearpage

\begin{minipage}{\textwidth}
\resizebox{0.88\textwidth}{!}{\includegraphics{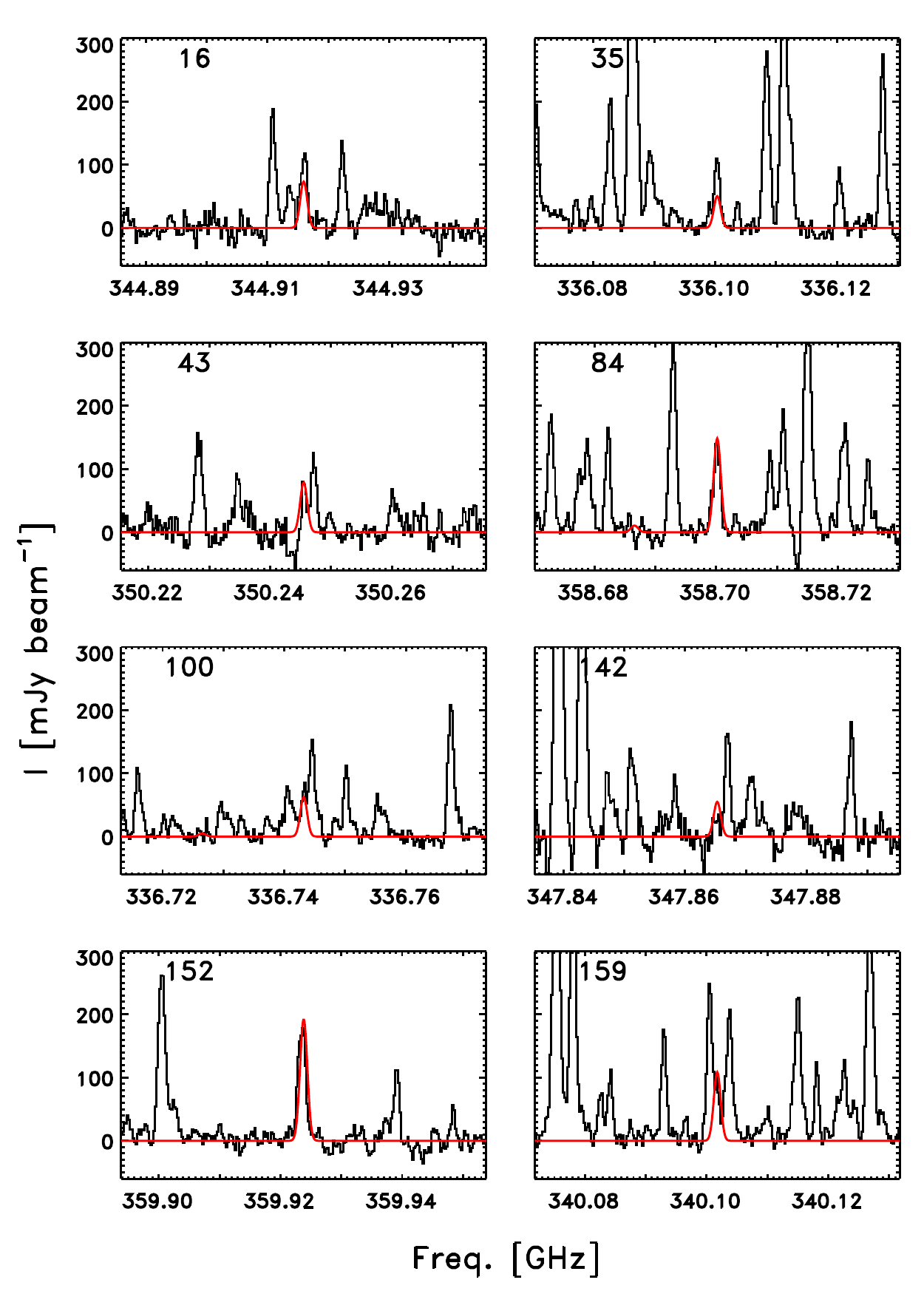}\includegraphics{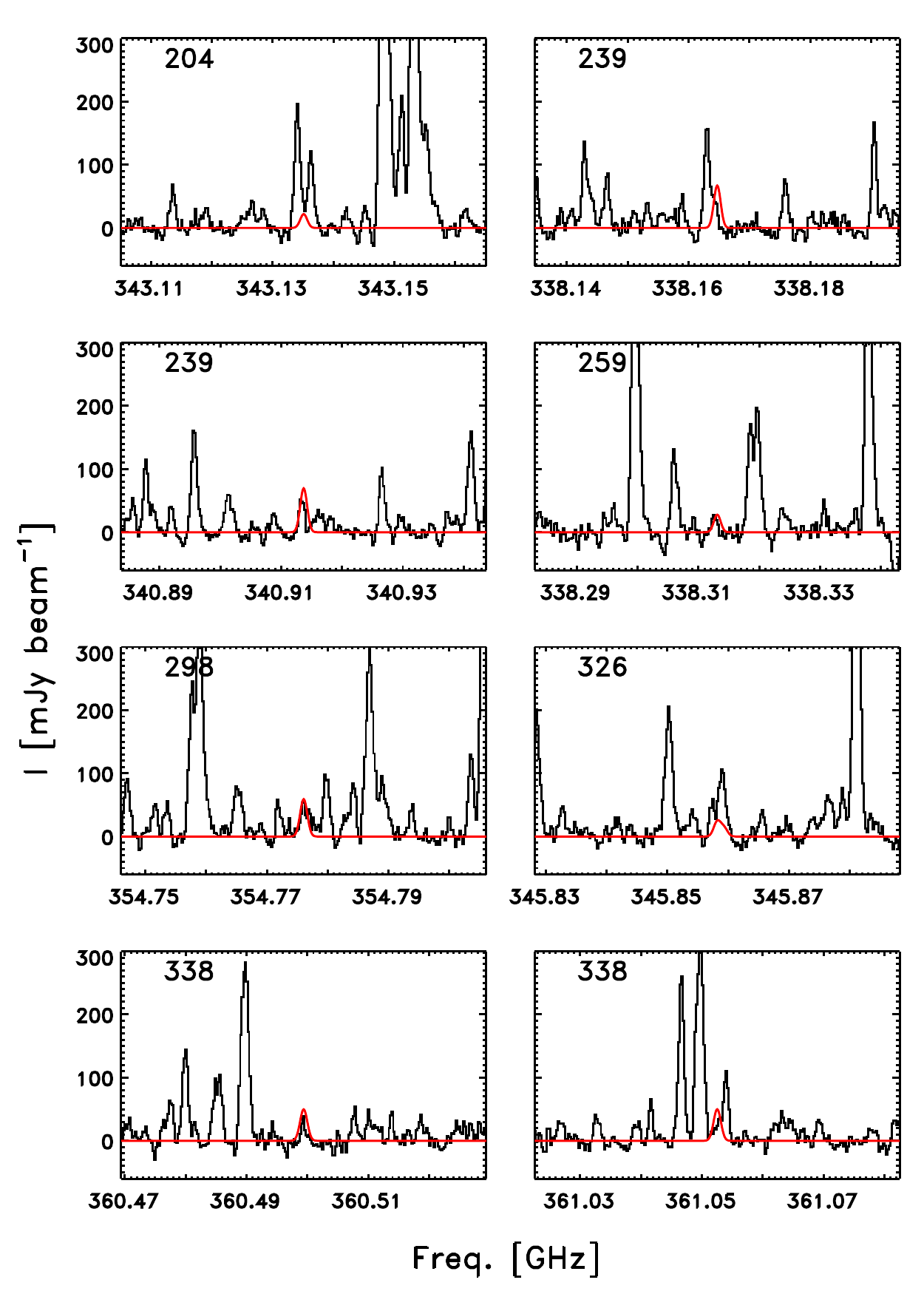}}
\resizebox{0.44\textwidth}{!}{\includegraphics{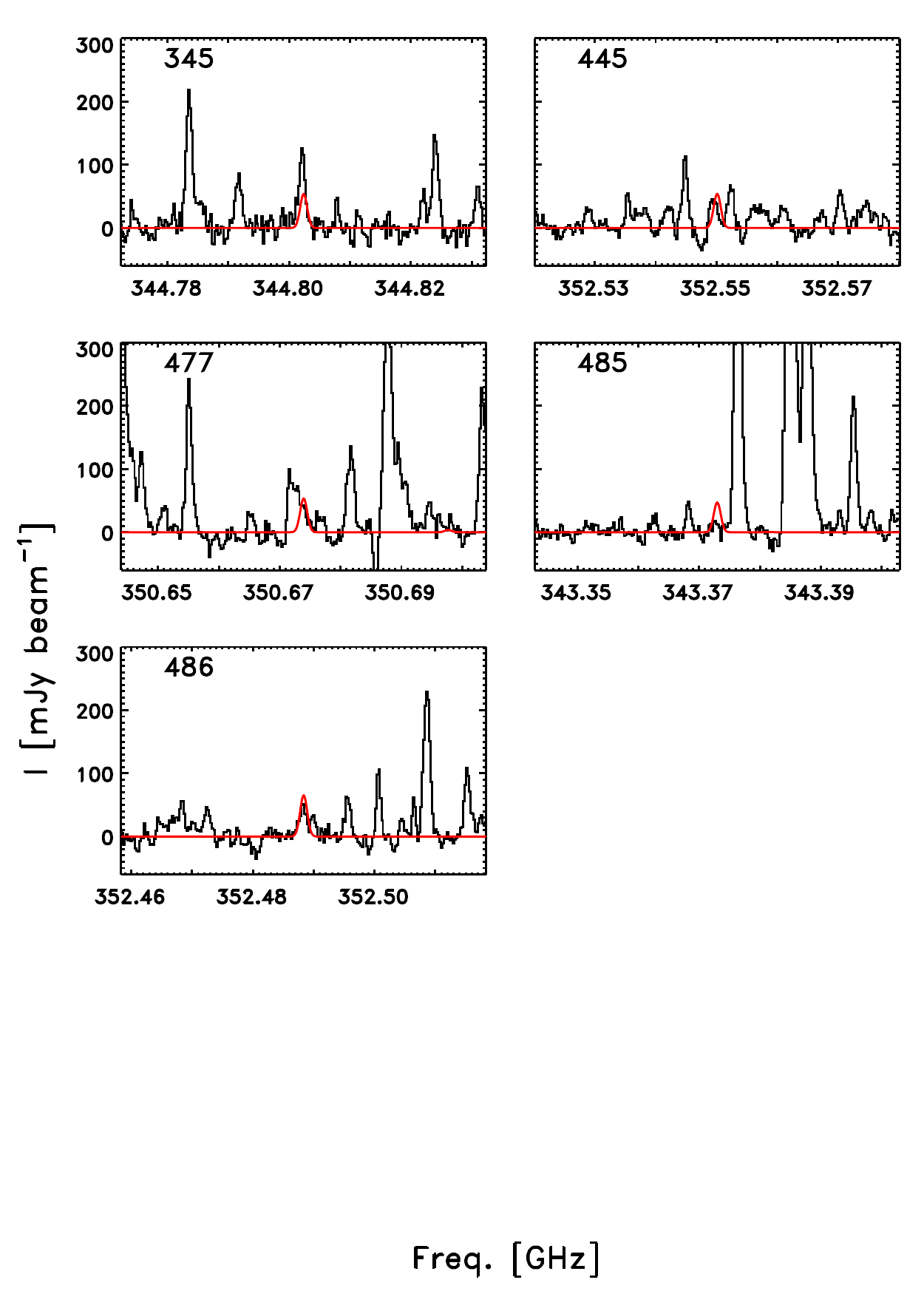}}
\captionof{figure}{As in Fig.~\ref{first_spectra} for the 21
  brightest lines of CH$_3^{18}$OH toward the full-beam offset
  position (for the fit to the half-beam offset position; see
  \citealt{jorgensen16}).}\label{18methanol_spectra}
\end{minipage}
\clearpage

\subsection{Ethanol}
\begin{minipage}{\textwidth}
\resizebox{0.88\textwidth}{!}{\includegraphics{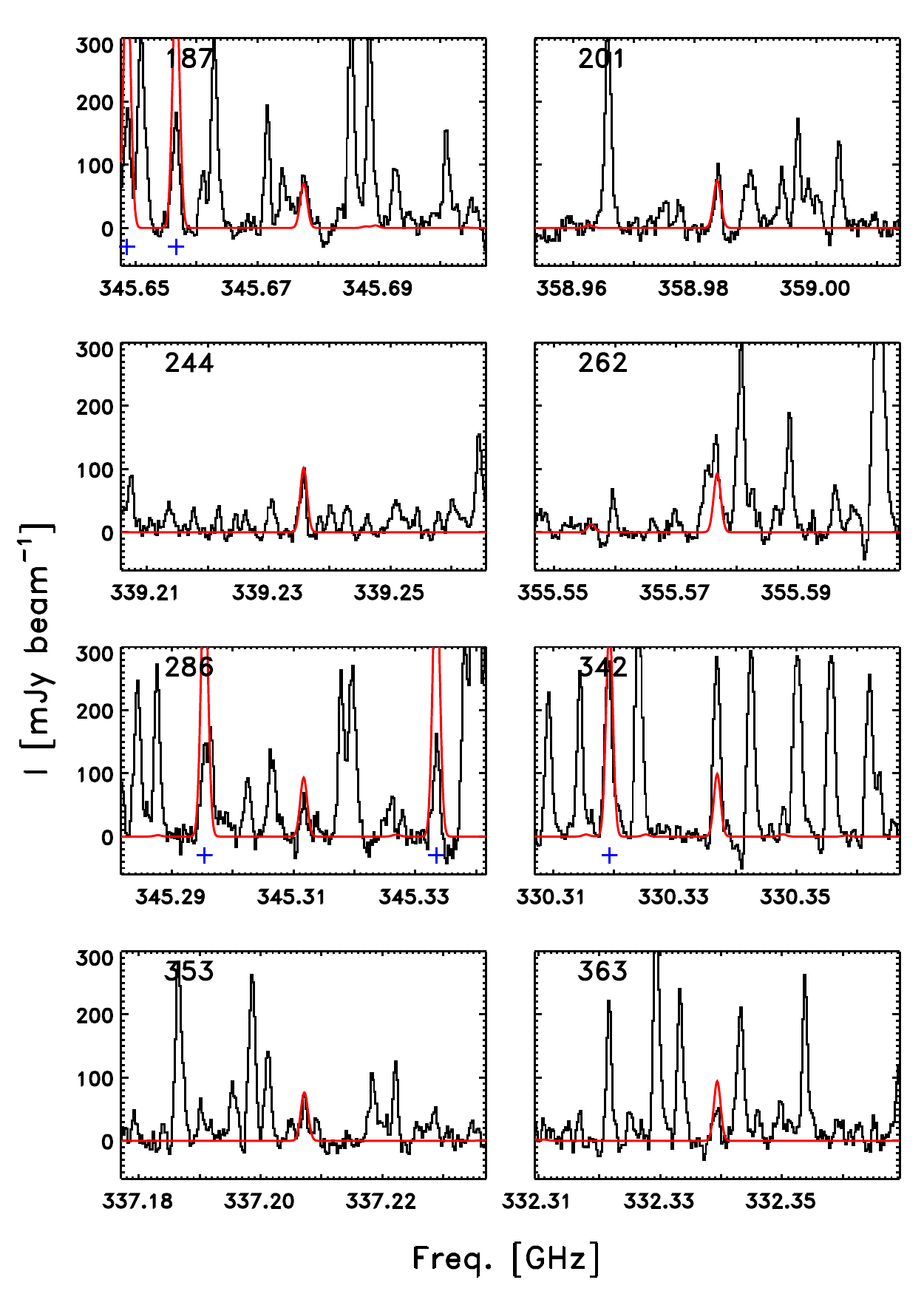}\includegraphics{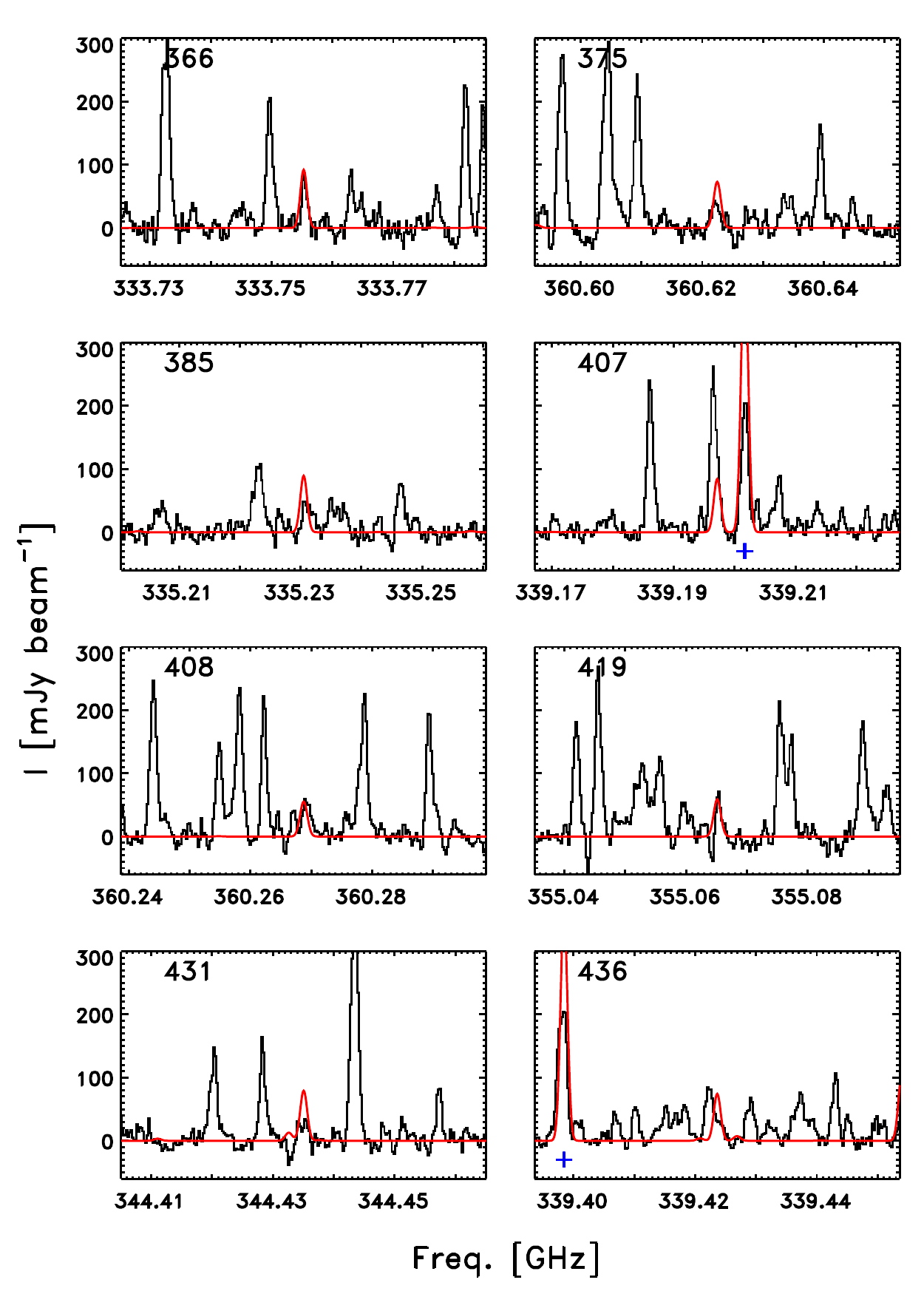}}
\resizebox{0.88\textwidth}{!}{\includegraphics{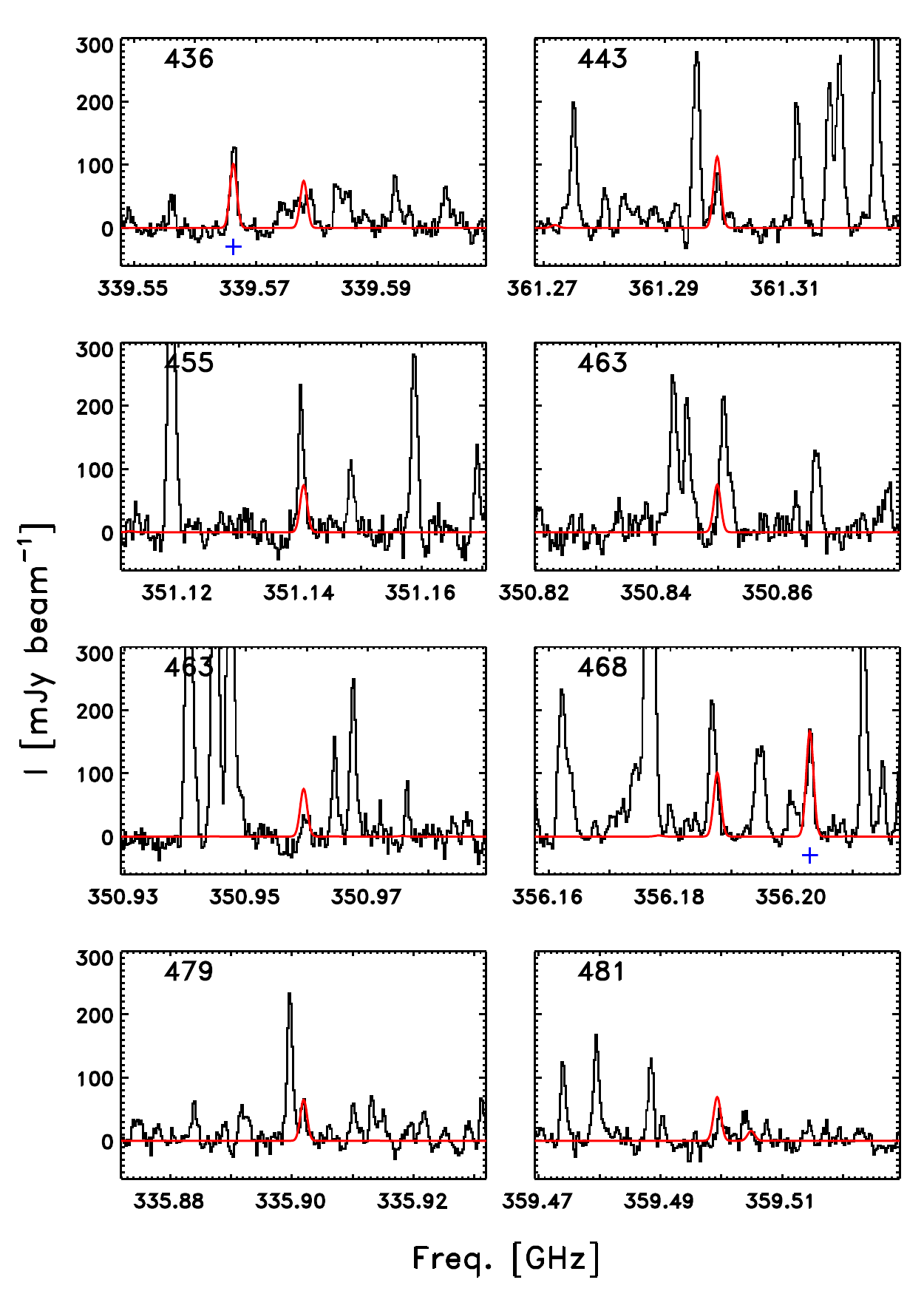}\includegraphics{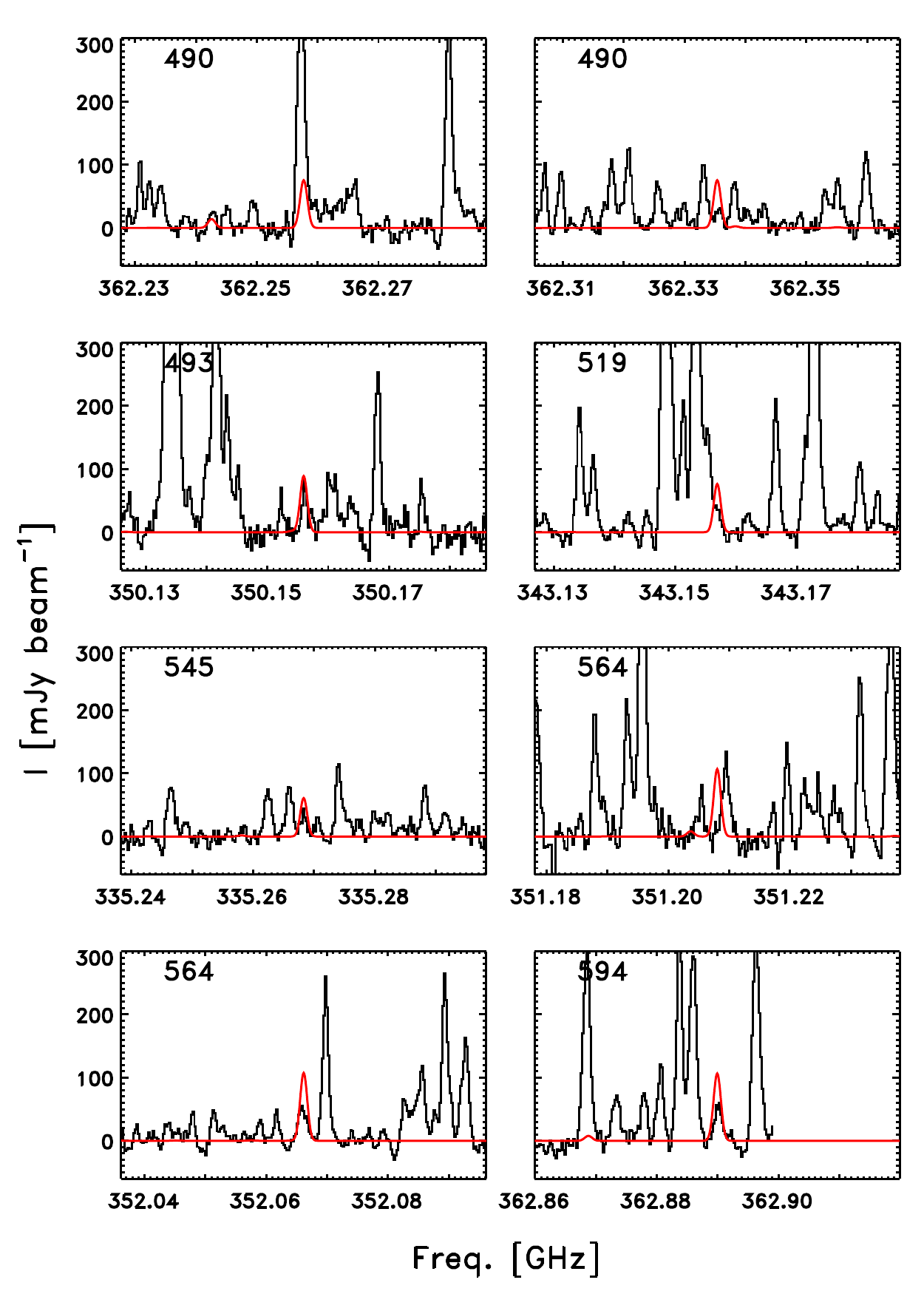}}
\captionof{figure}{As in Fig.~\ref{first_spectra} for the 32 brightest lines of ethanol as
  expected from the synthetic spectrum toward the full-beam offset position.}\label{ethanol_spectra}
\end{minipage}

\clearpage

\begin{minipage}{\textwidth}
\resizebox{0.88\textwidth}{!}{\includegraphics{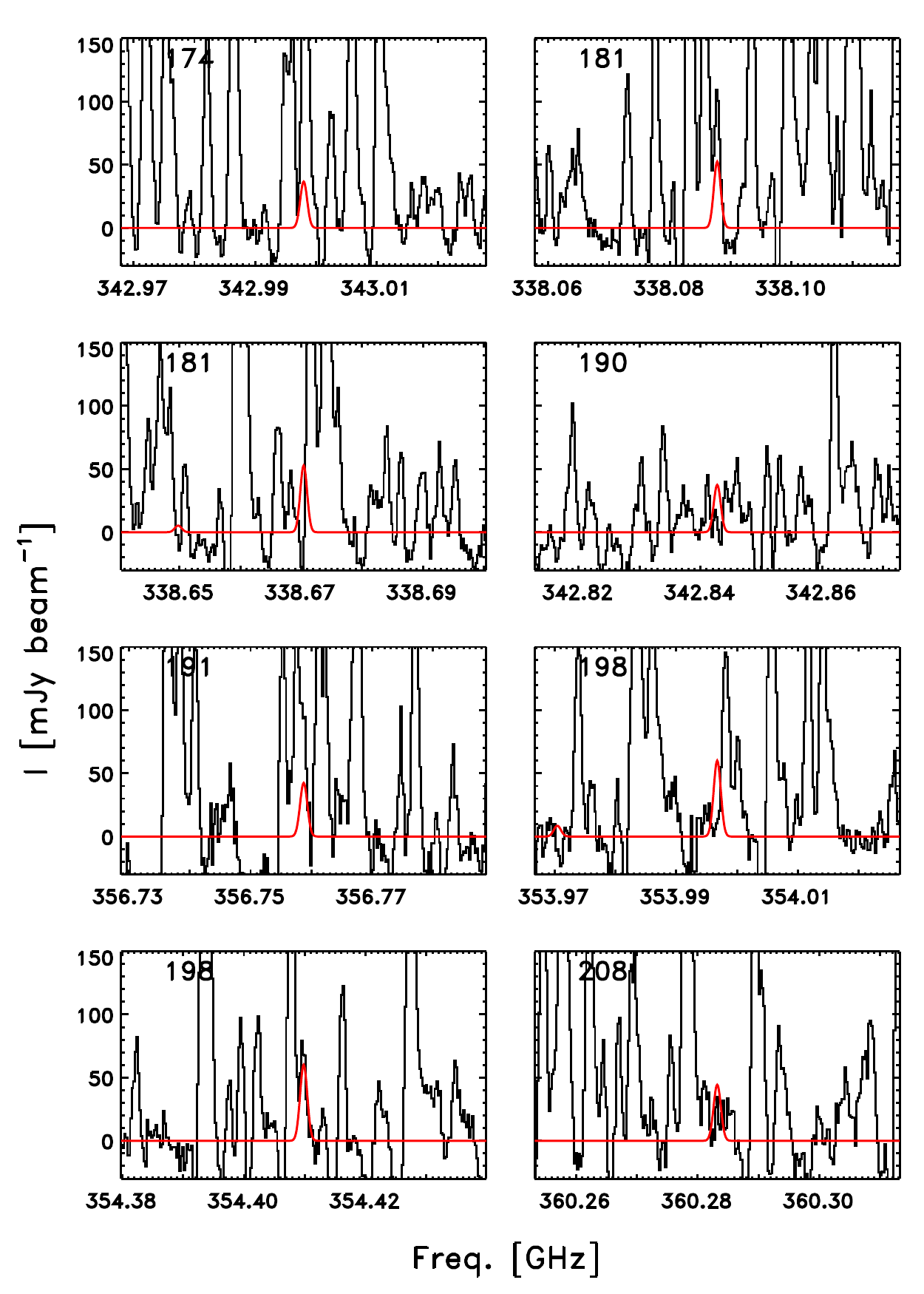}\includegraphics{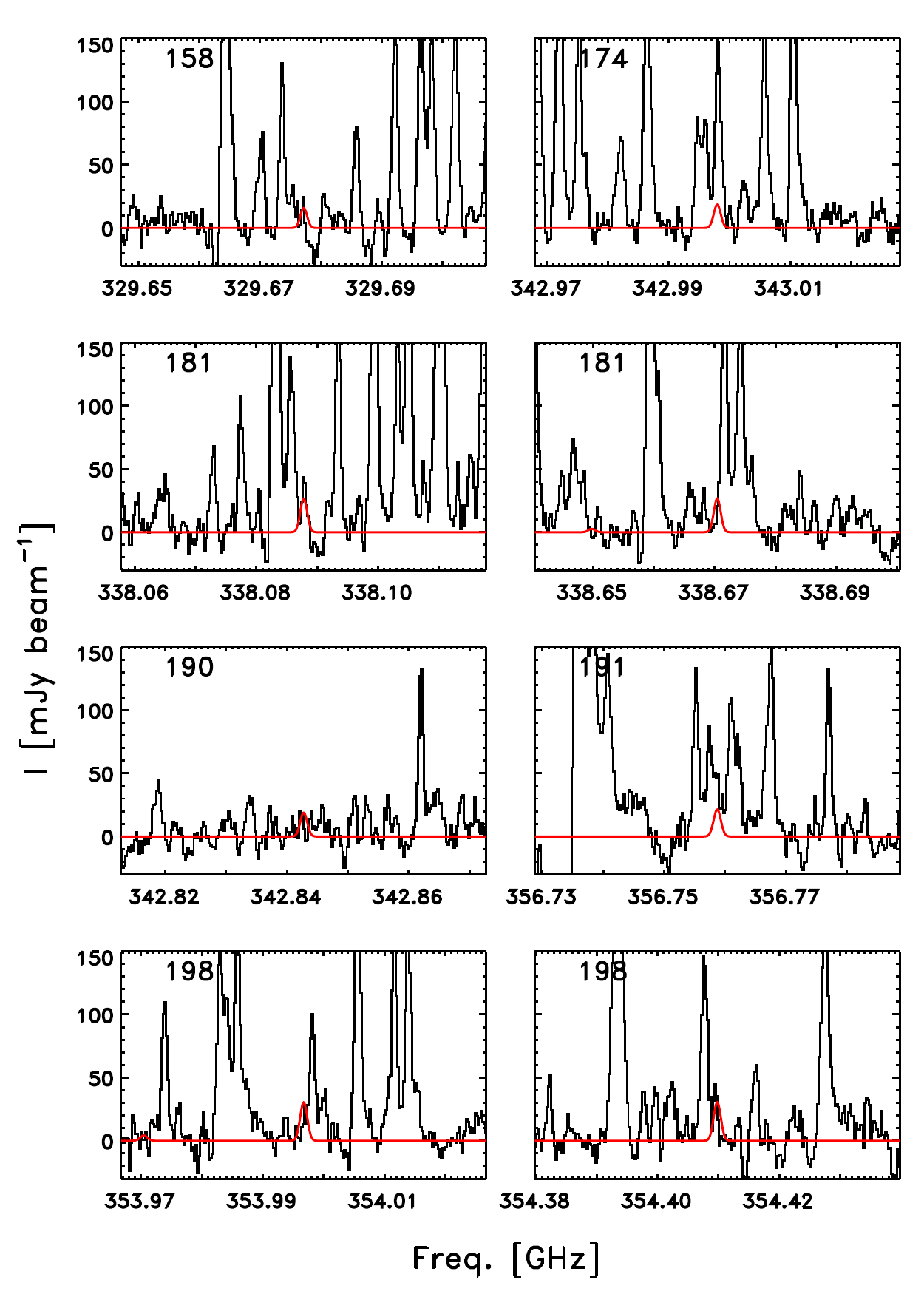}}
\resizebox{0.88\textwidth}{!}{\includegraphics{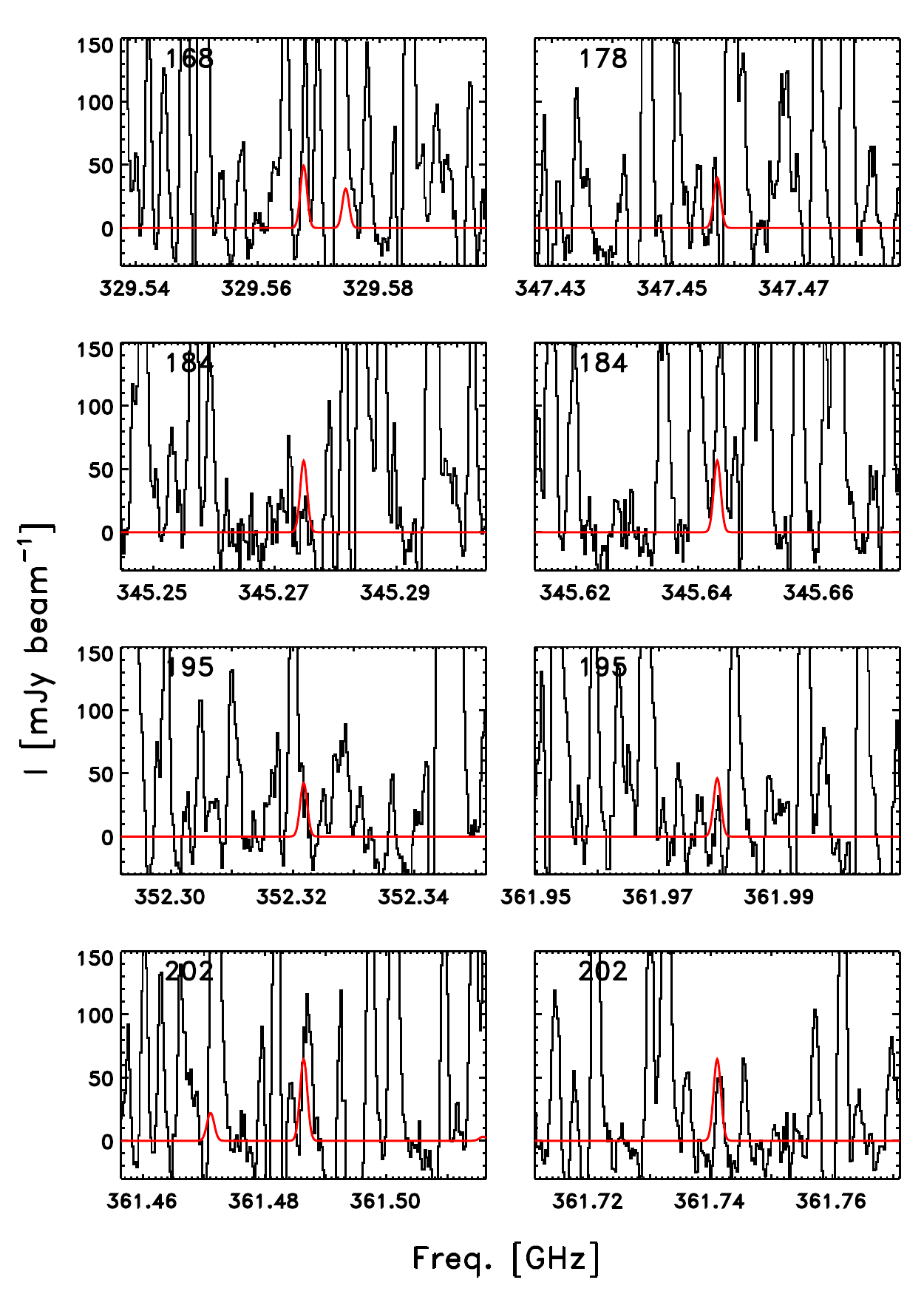}\includegraphics{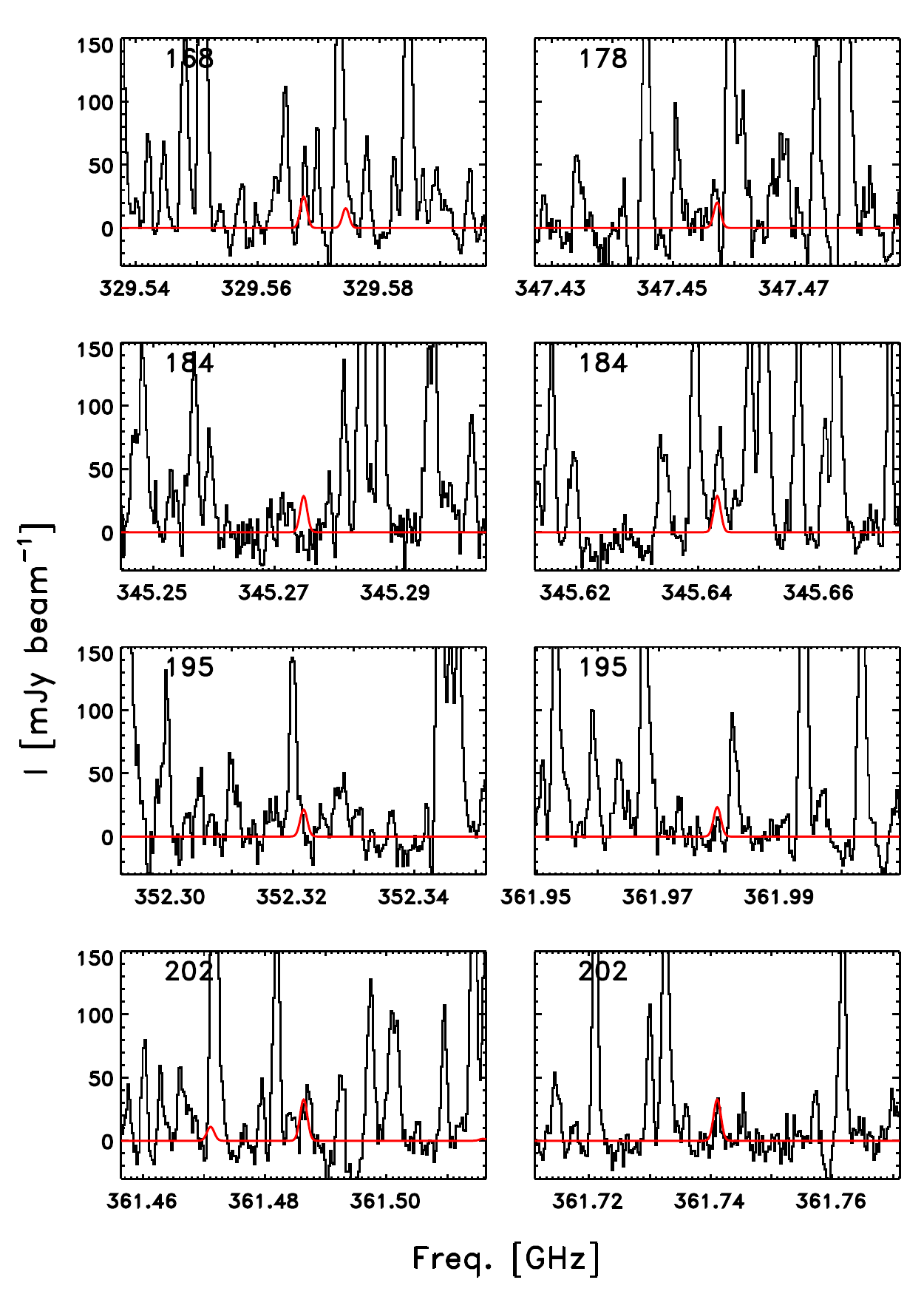}}
\captionof{figure}{As in Fig.~\ref{first_spectra} for the 16
  brightest lines of a-$^{13}$CH$_3$CH$_2$OH (upper panels) and
  a-CH$_3^{13}$CH$_2$OH (lower panels) toward the half-beam offset position
  (left set of panels) and full-beam offset position (right set of
  panels).}\label{ethanol13_spectra1}
\end{minipage}

\clearpage

\begin{minipage}{\textwidth}
\resizebox{0.88\textwidth}{!}{\includegraphics{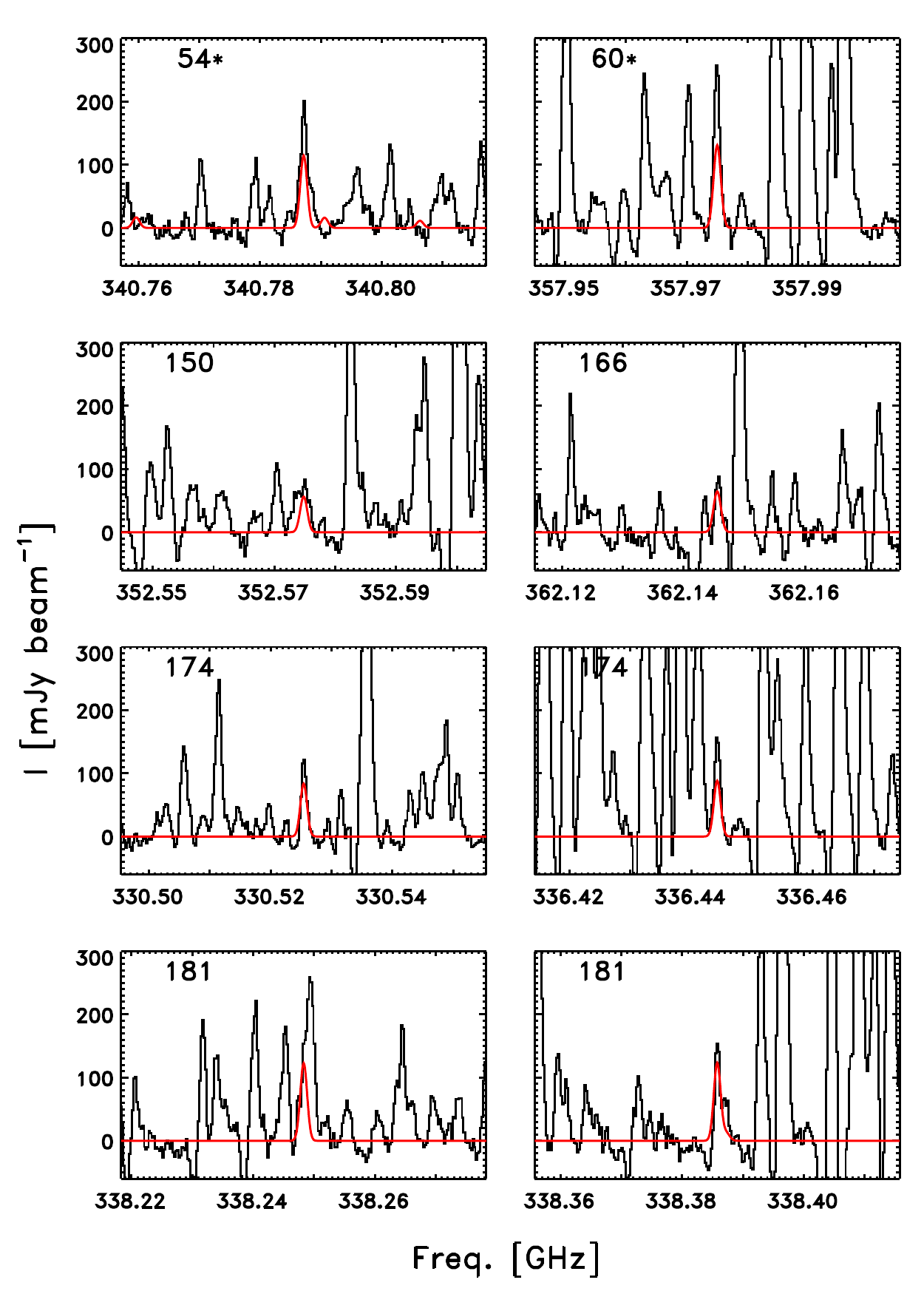}\includegraphics{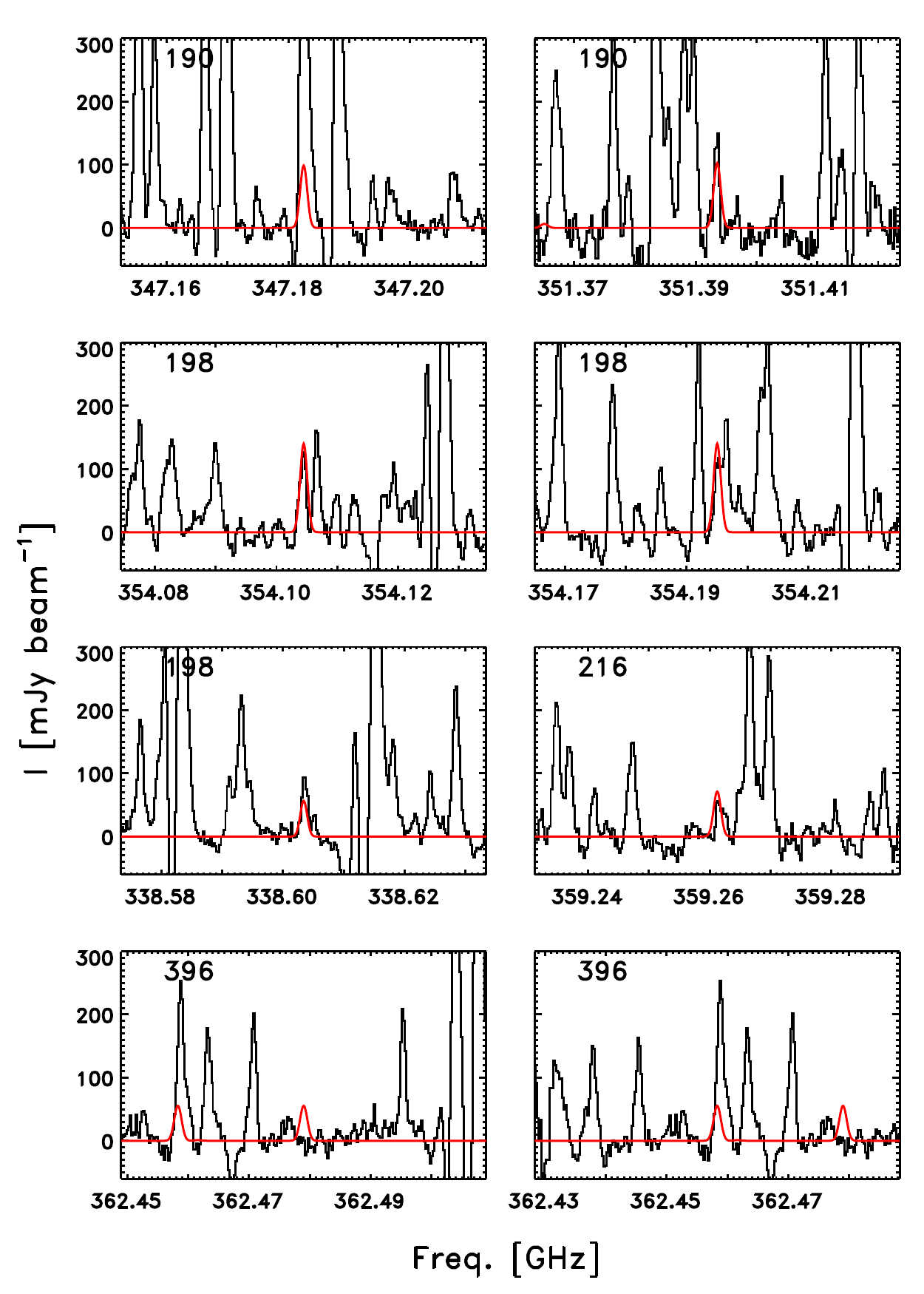}}
\resizebox{0.88\textwidth}{!}{\includegraphics{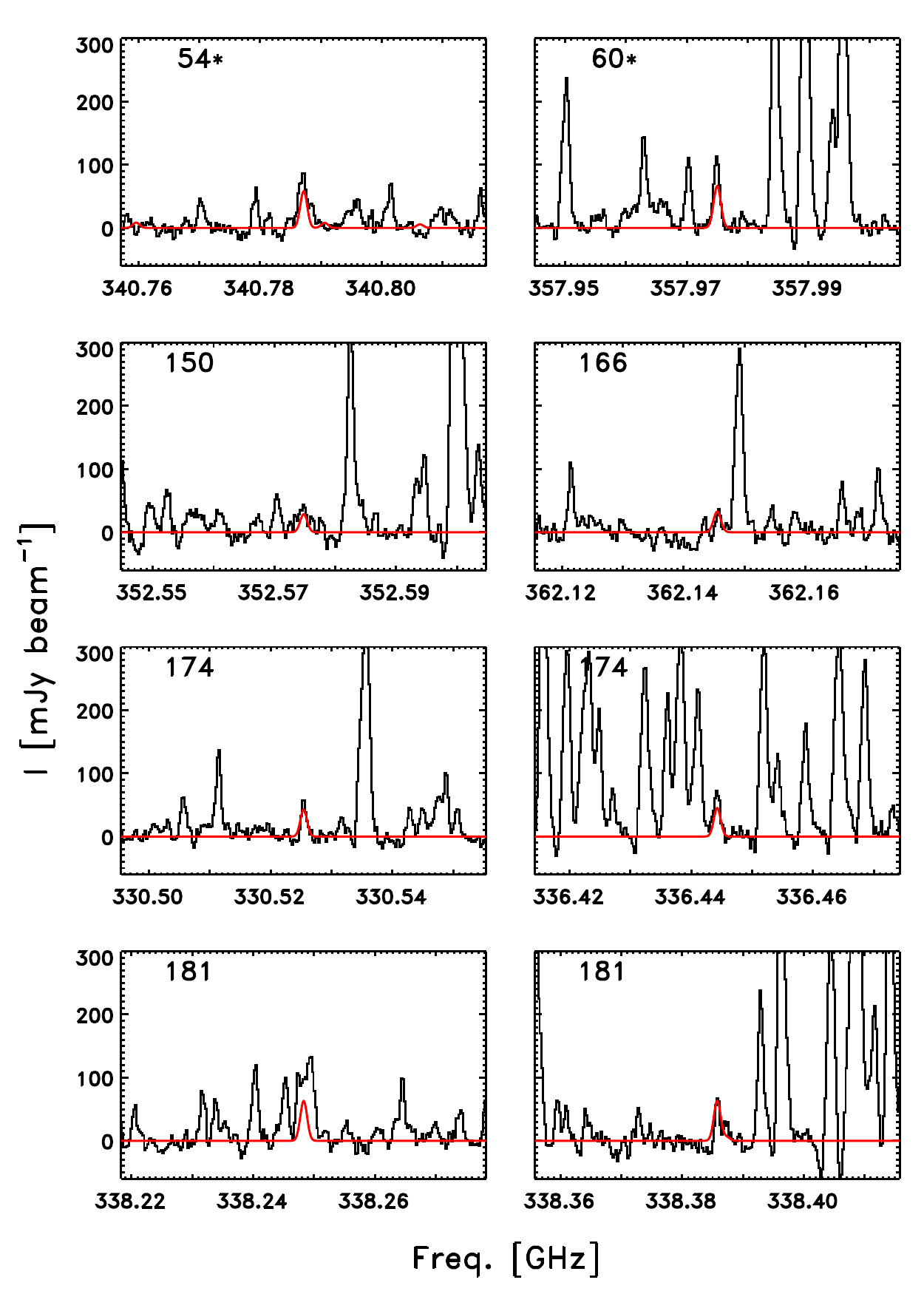}\includegraphics{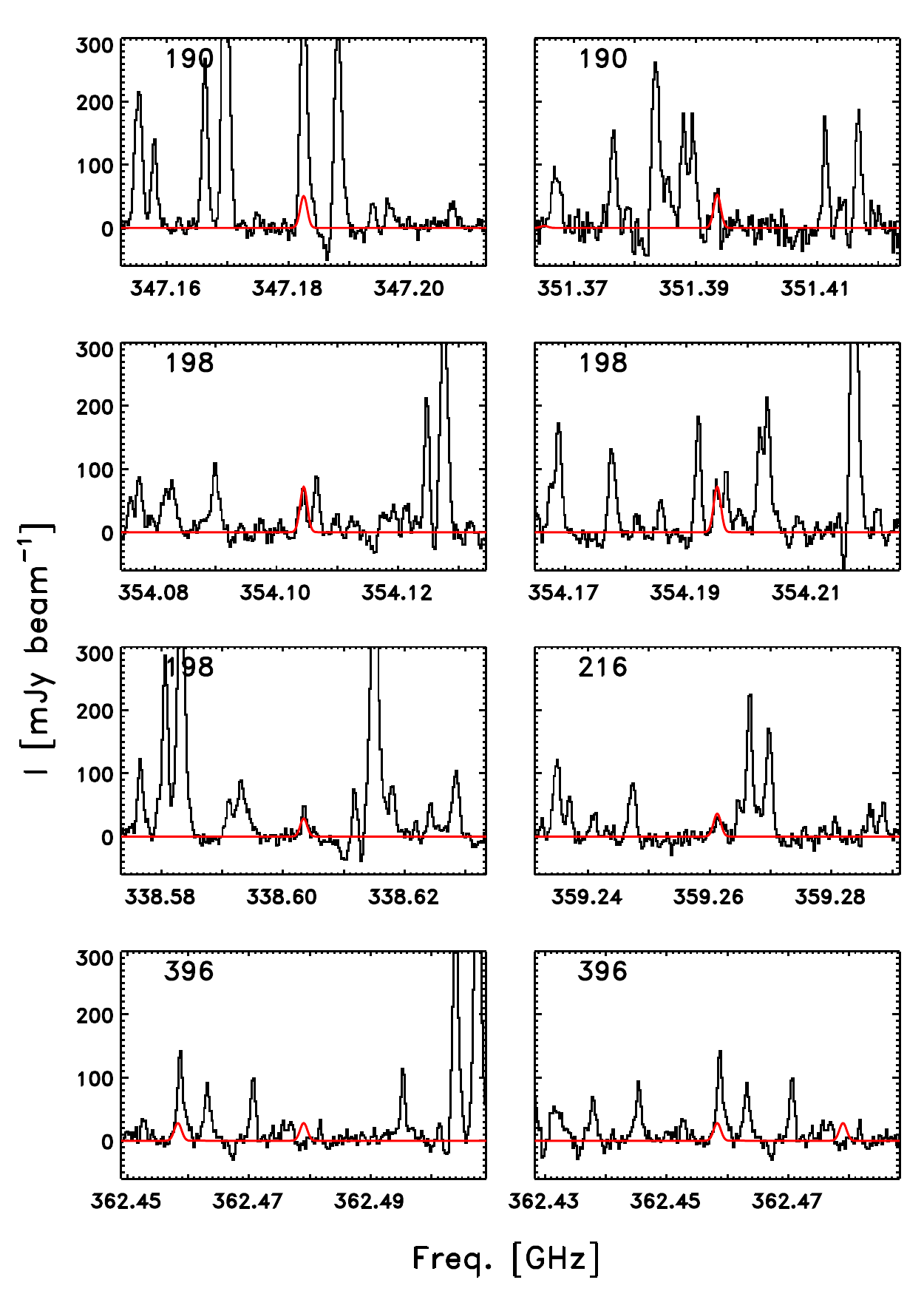}}
\captionof{figure}{As in Fig.~\ref{first_spectra} for the 16 brightest lines of
  a-CH$_3$CHDOH toward the half-beam (upper) and full-beam (lower)
  offset positions. The asterisks next to the upper energy level in
  some panel indicate features for which the multiple 
  transitions (from the same species) overlap.}\label{dethanol_spectra1}
\end{minipage}

\clearpage

\begin{minipage}{\textwidth}
\resizebox{0.88\textwidth}{!}{\includegraphics{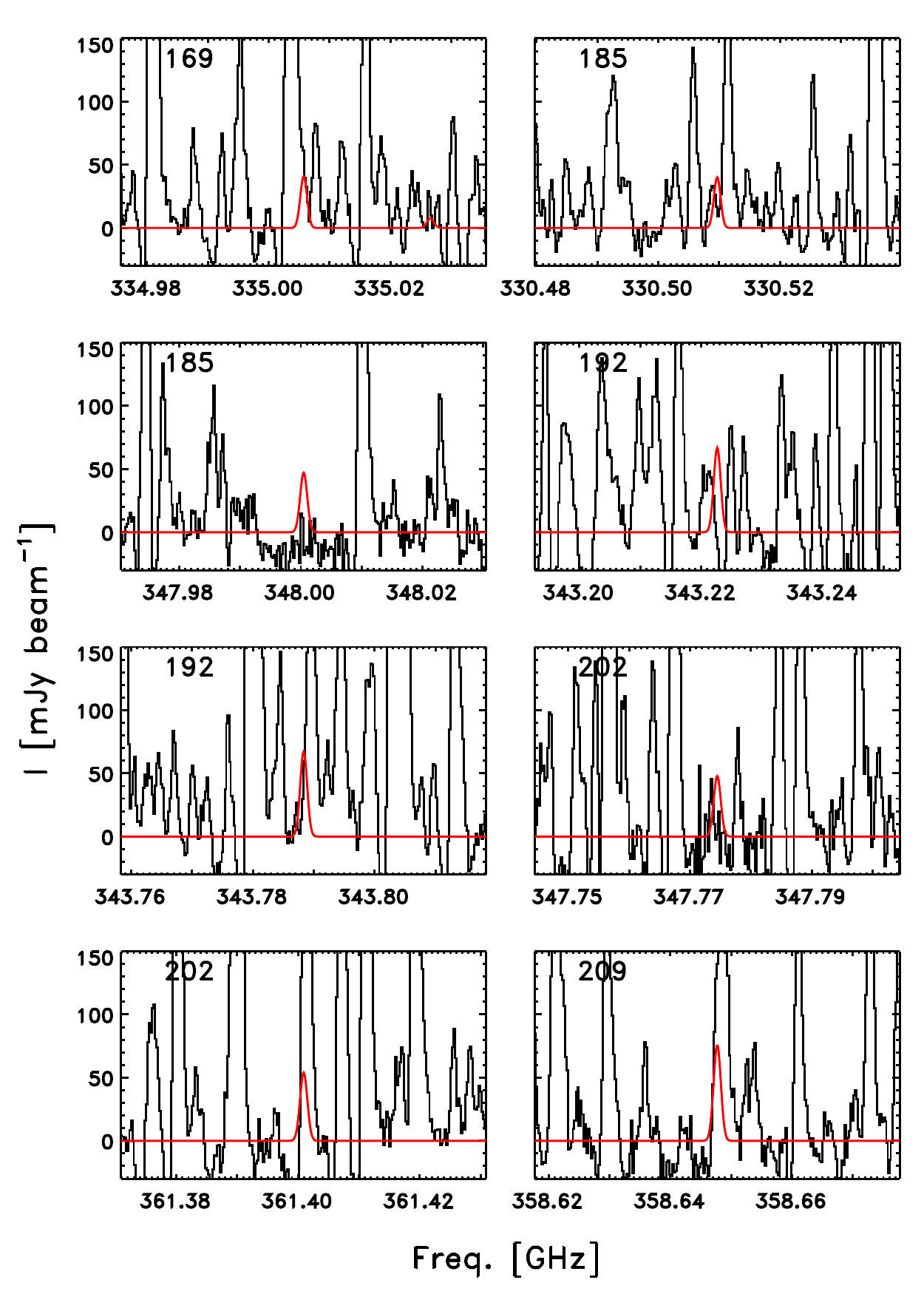}\includegraphics{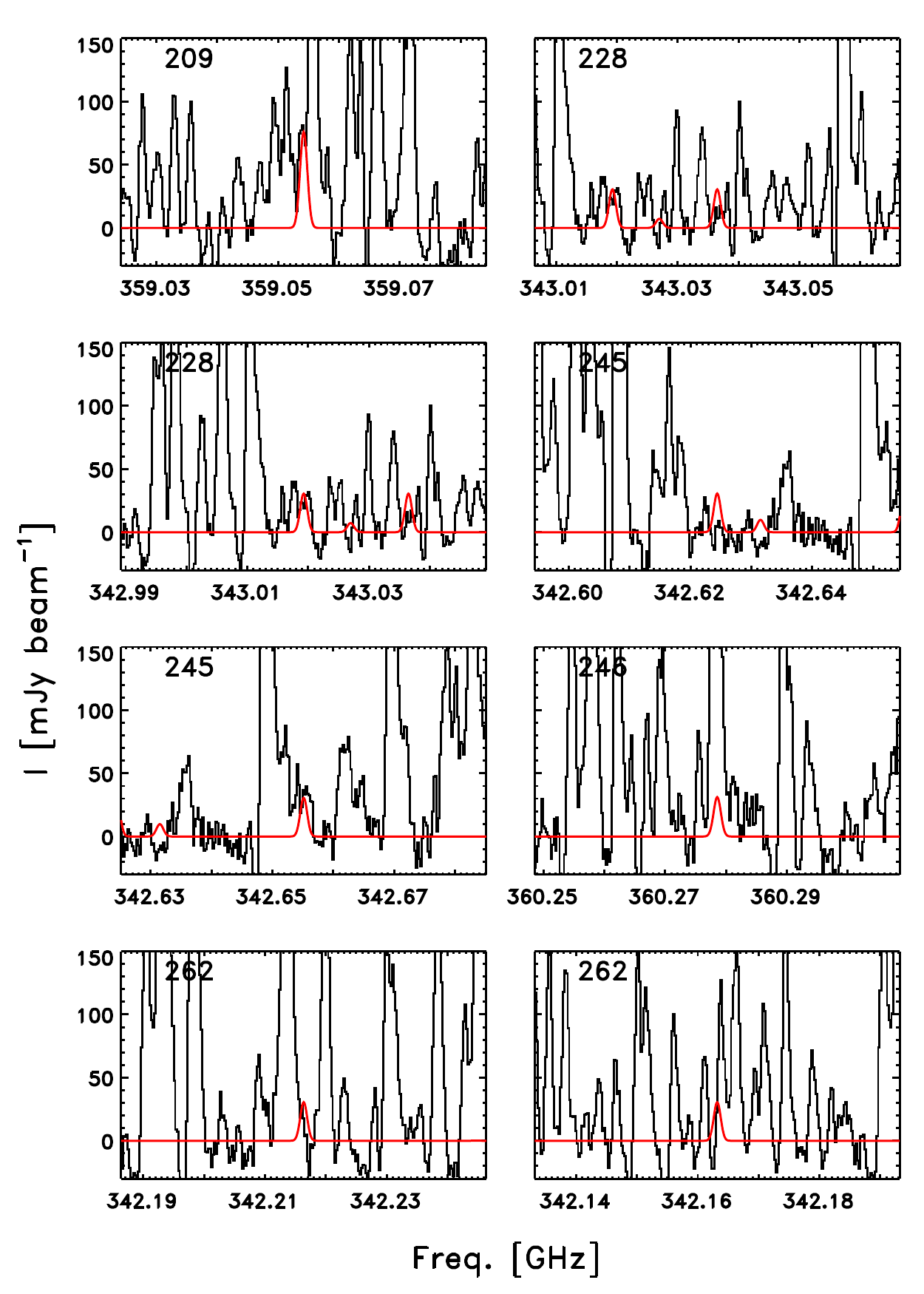}}
\resizebox{0.88\textwidth}{!}{\includegraphics{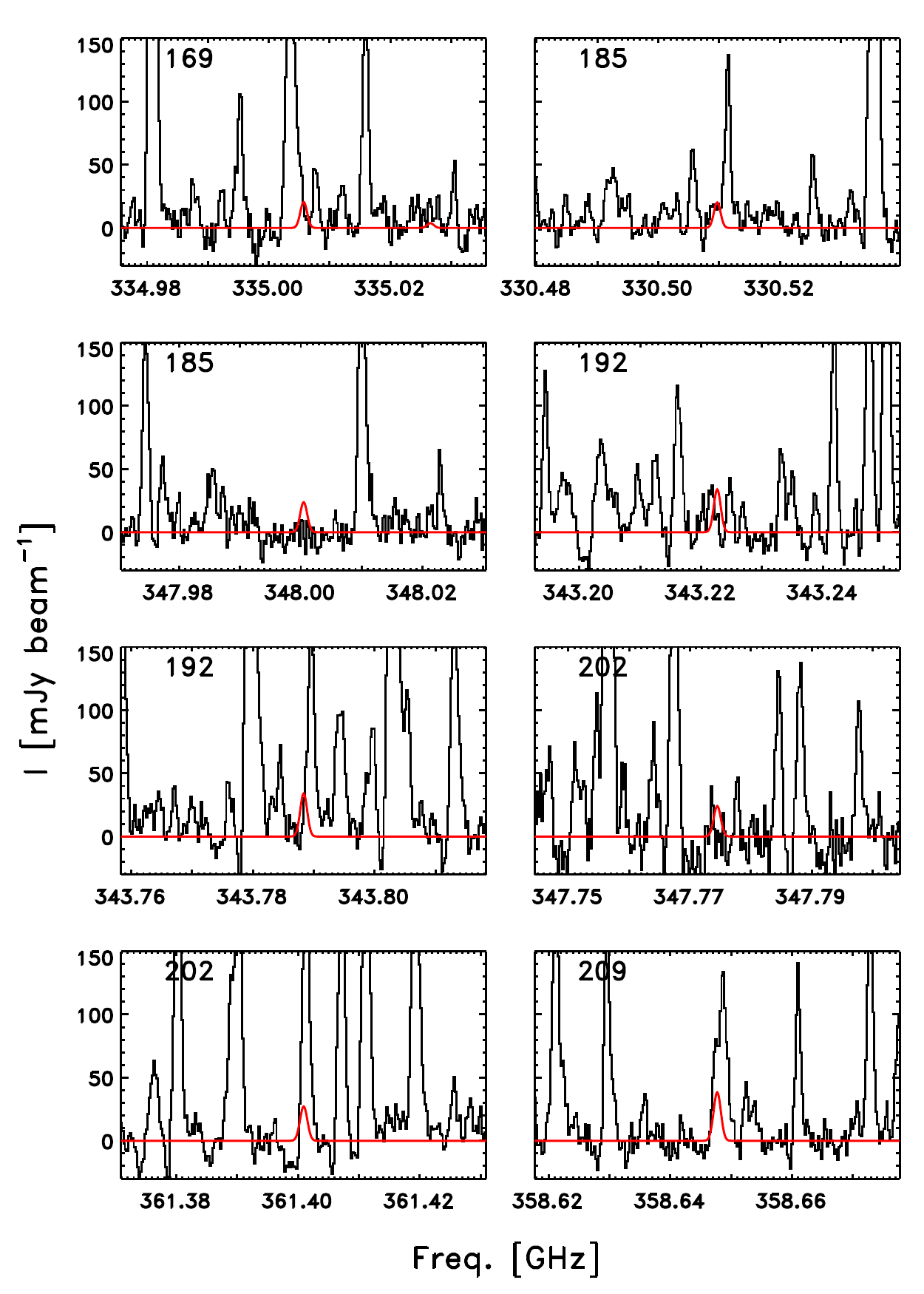}\includegraphics{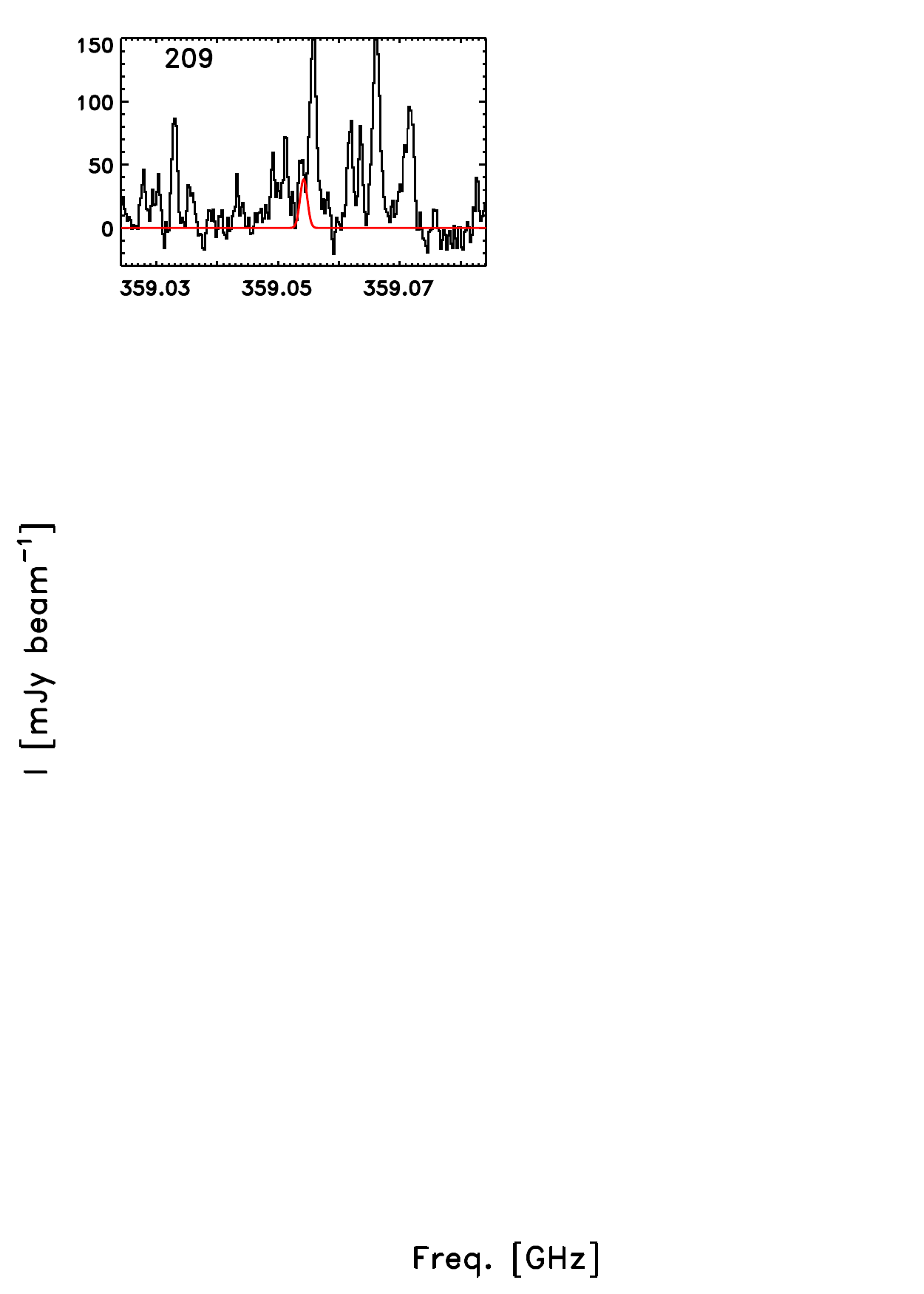}}
\captionof{figure}{As in Fig.~\ref{first_spectra} for
  a-CH$_3$CH$_2$OD: the upper panels show the spectra toward the
  half-beam position and the lower panels the full-beam (lower) offset
  position.}\label{dethanol_spectra2}
\end{minipage}
\clearpage

\begin{minipage}{\textwidth}
\resizebox{0.88\textwidth}{!}{\includegraphics{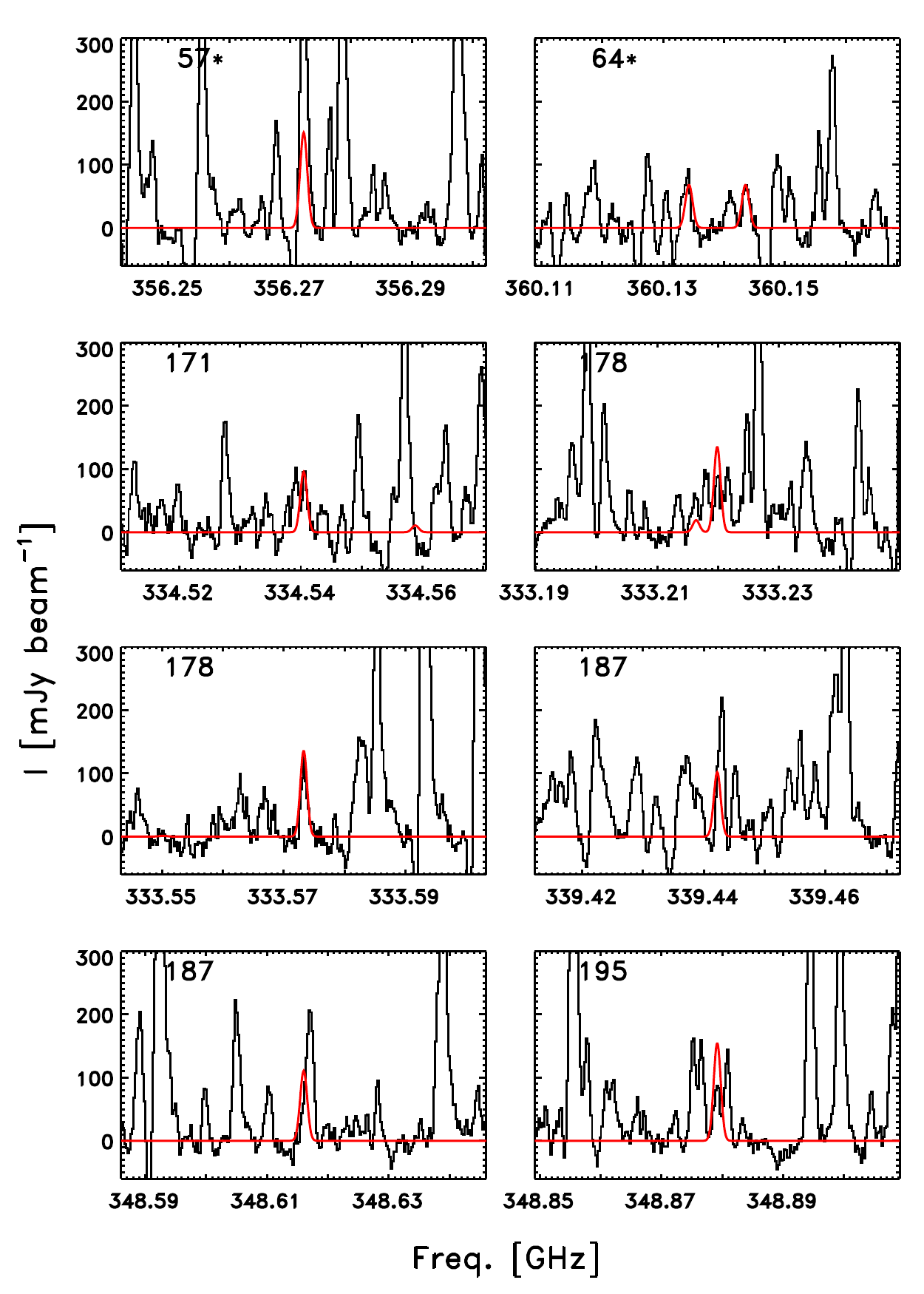}\includegraphics{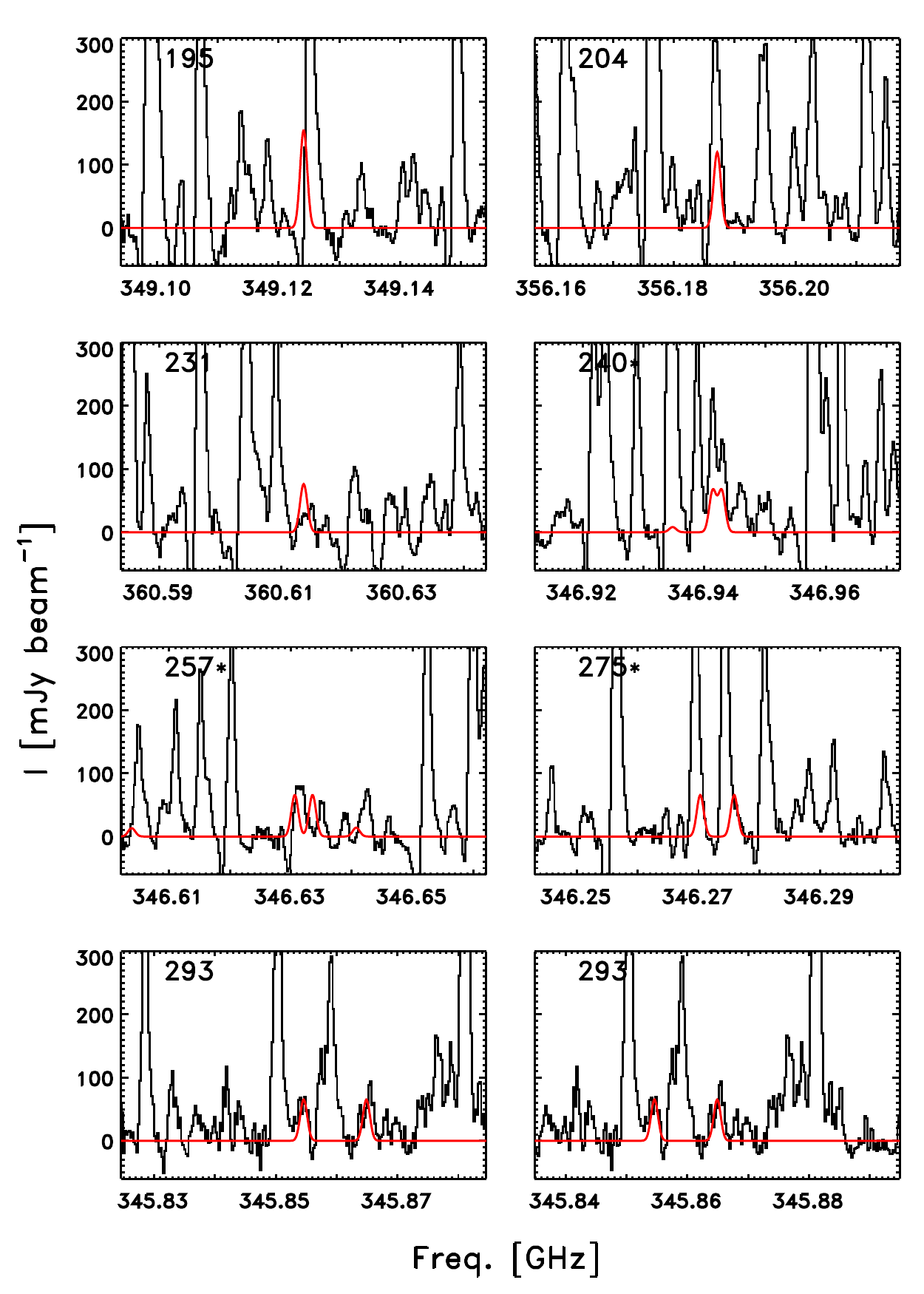}}
\resizebox{0.88\textwidth}{!}{\includegraphics{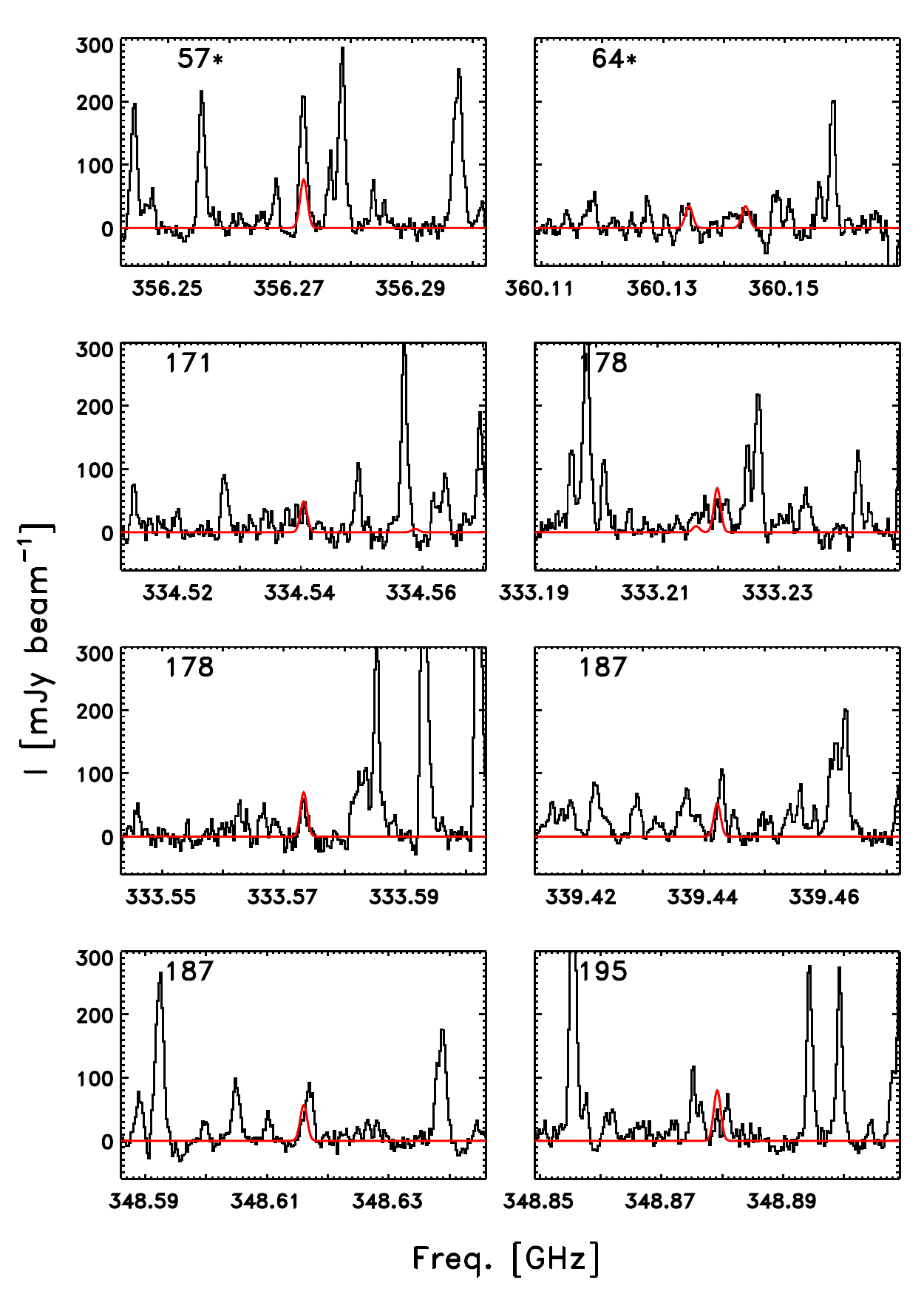}\includegraphics{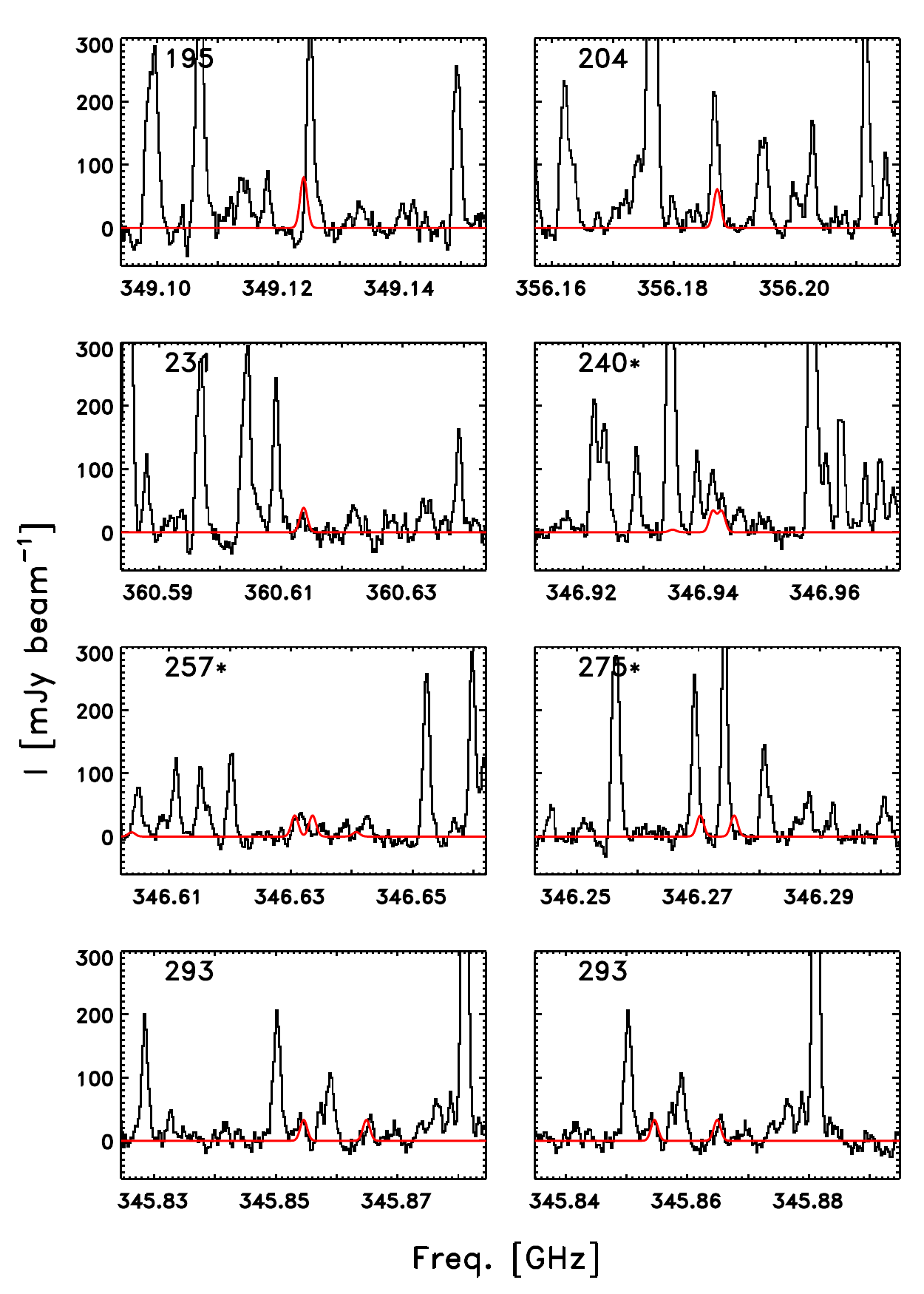}}
\captionof{figure}{As in Fig.~\ref{dethanol_spectra1} and Fig.~\ref{dethanol_spectra2} for aa-CH$_2$DCH$_2$OH.}\label{dethanol_spectra3}
\end{minipage}

\clearpage

\begin{minipage}{\textwidth}
\resizebox{0.88\textwidth}{!}{\includegraphics{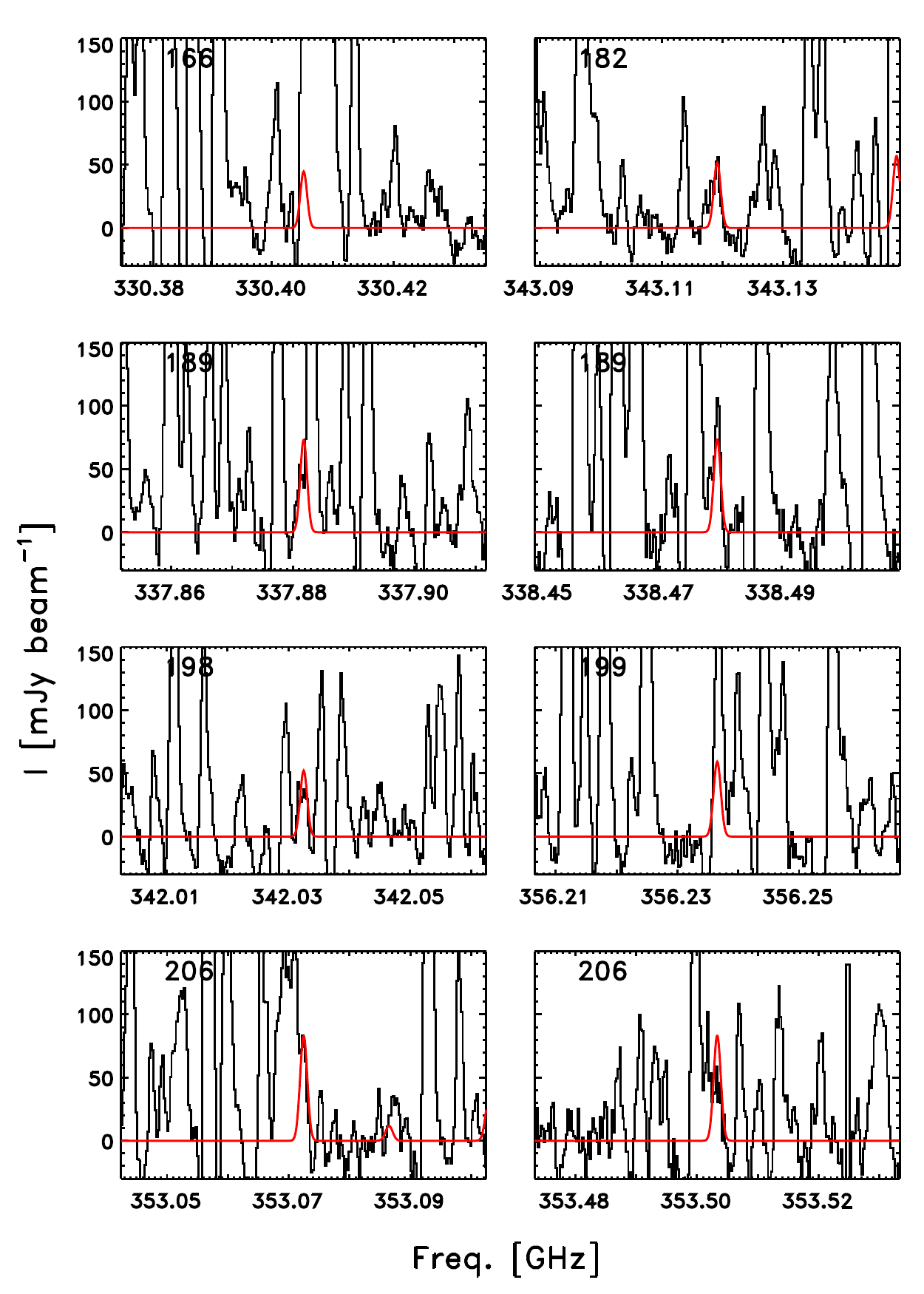}\includegraphics{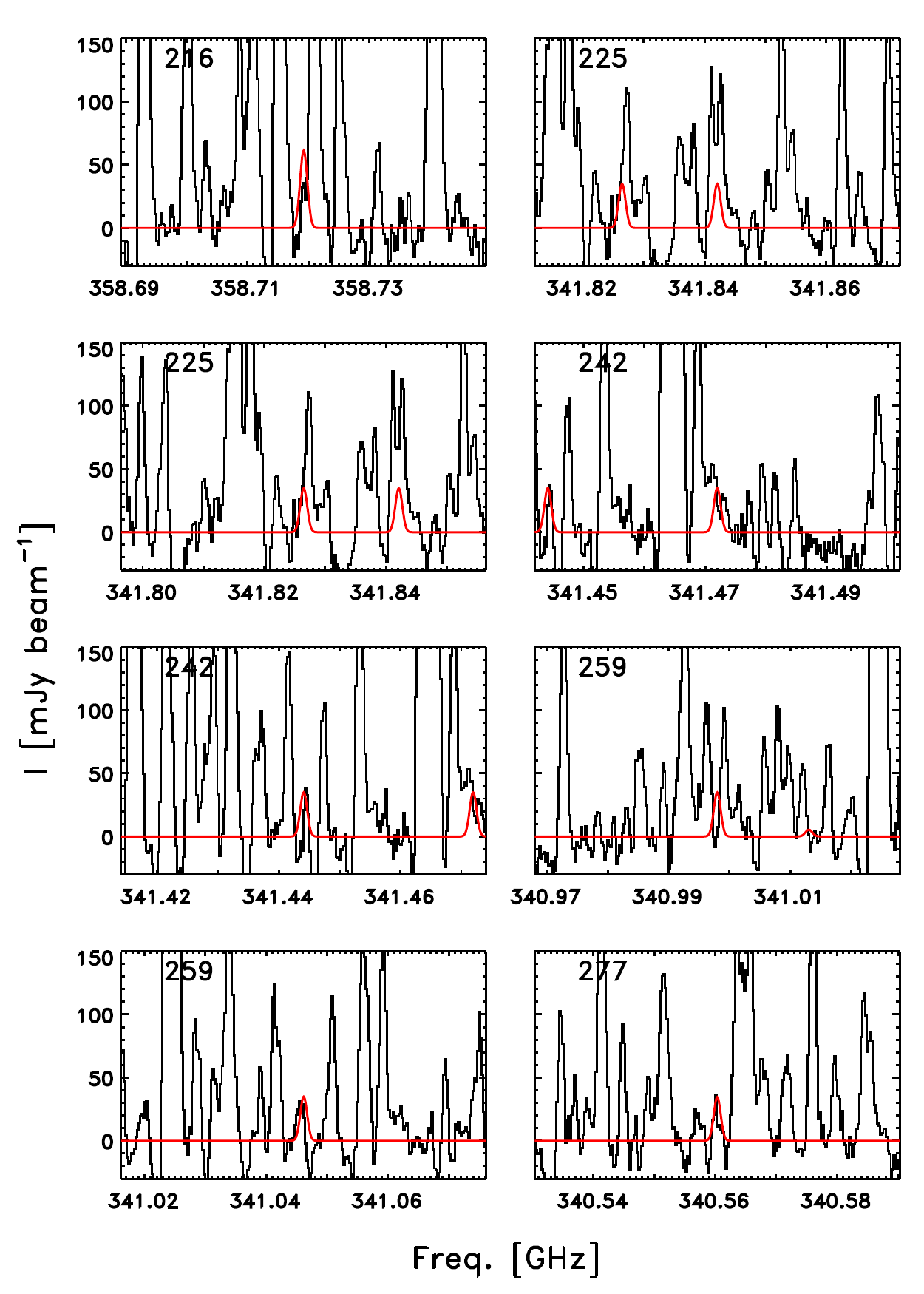}}
\resizebox{0.88\textwidth}{!}{\includegraphics{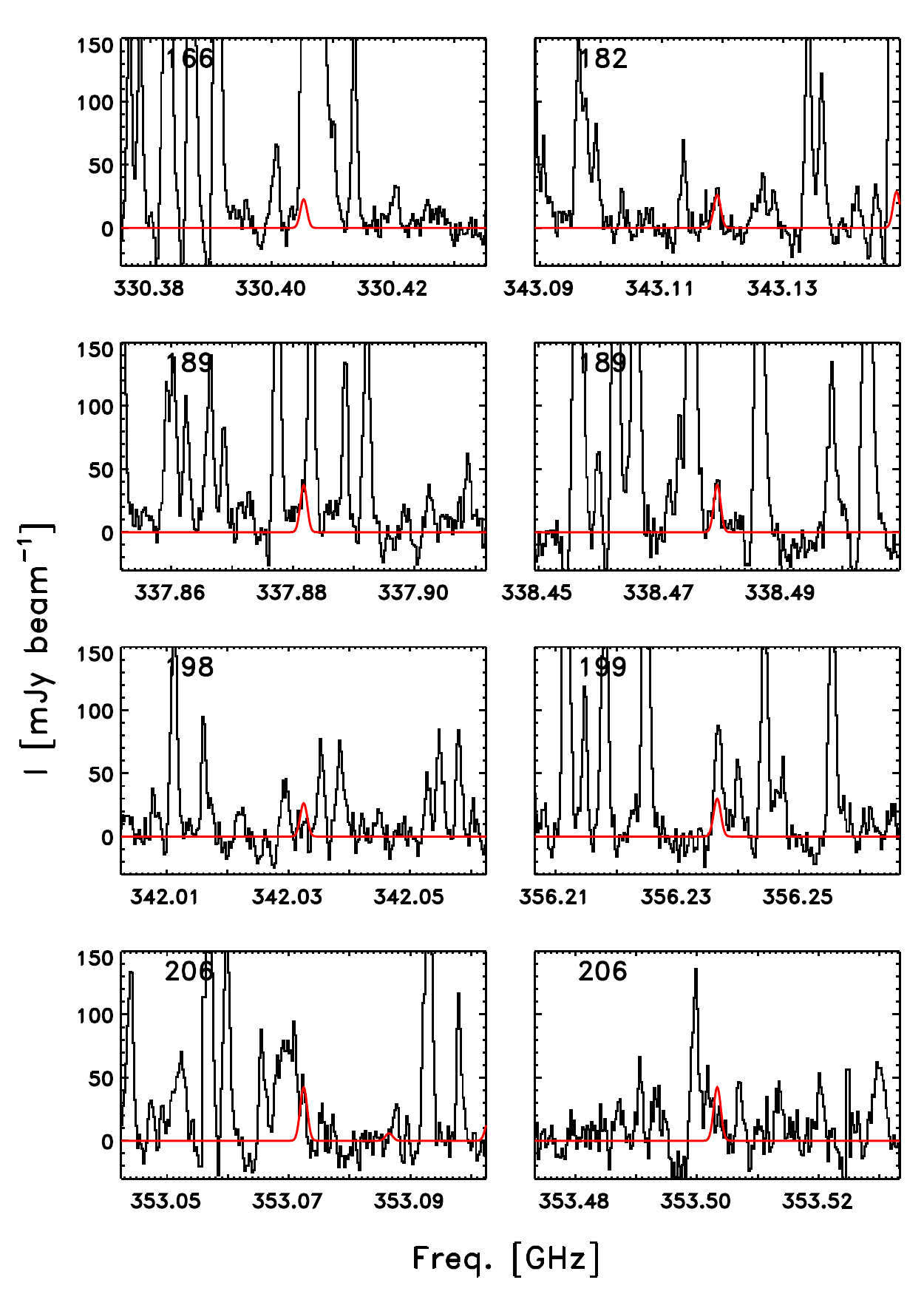}\includegraphics{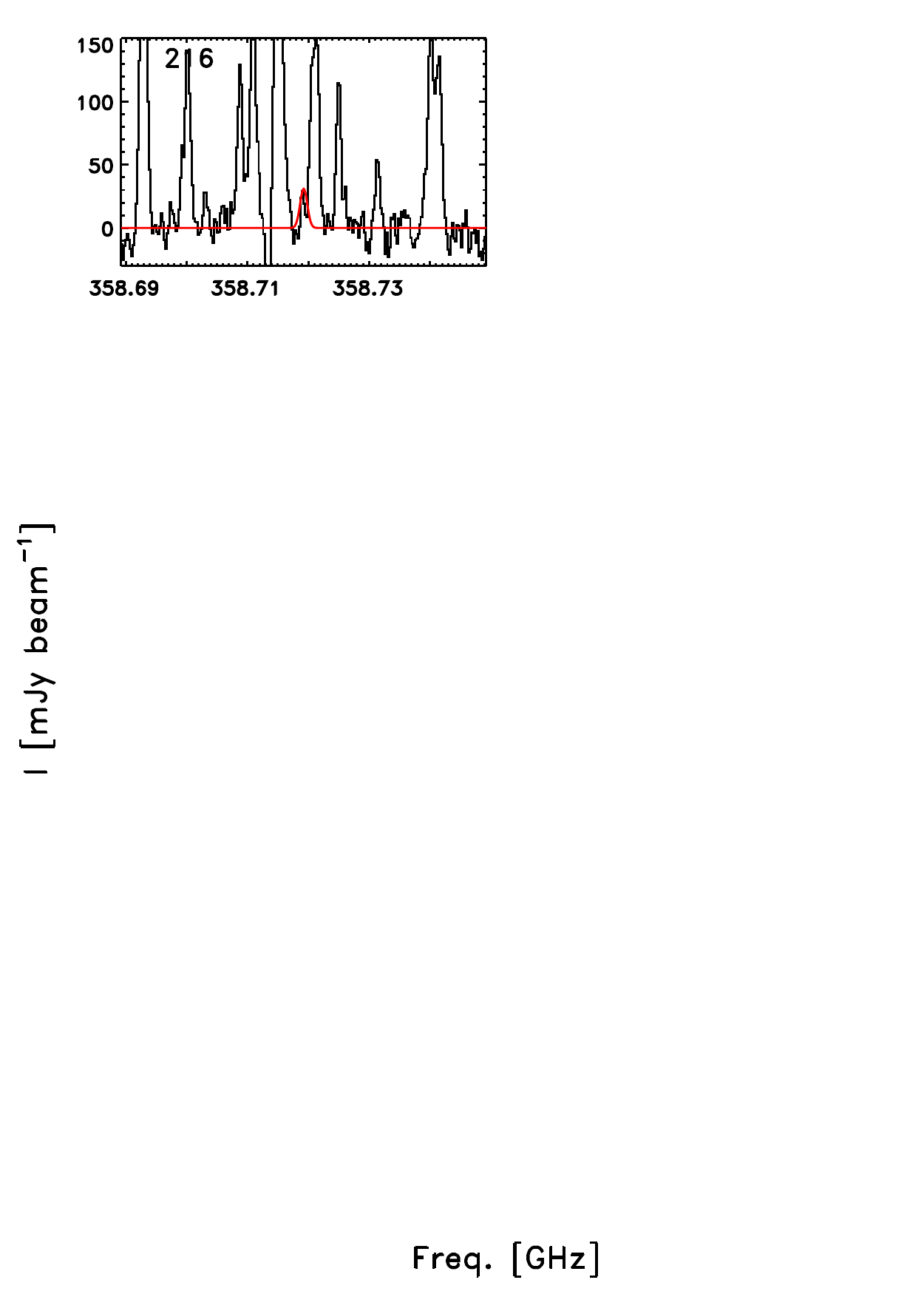}}
\captionof{figure}{As in Fig.~\ref{dethanol_spectra1}--\ref{dethanol_spectra3} for as-CH$_2$DCH$_2$OH.}\label{dethanol_spectra4}
\end{minipage}
\clearpage

\subsection{Methyl formate}
\begin{minipage}{\textwidth}
\resizebox{0.88\textwidth}{!}{\includegraphics{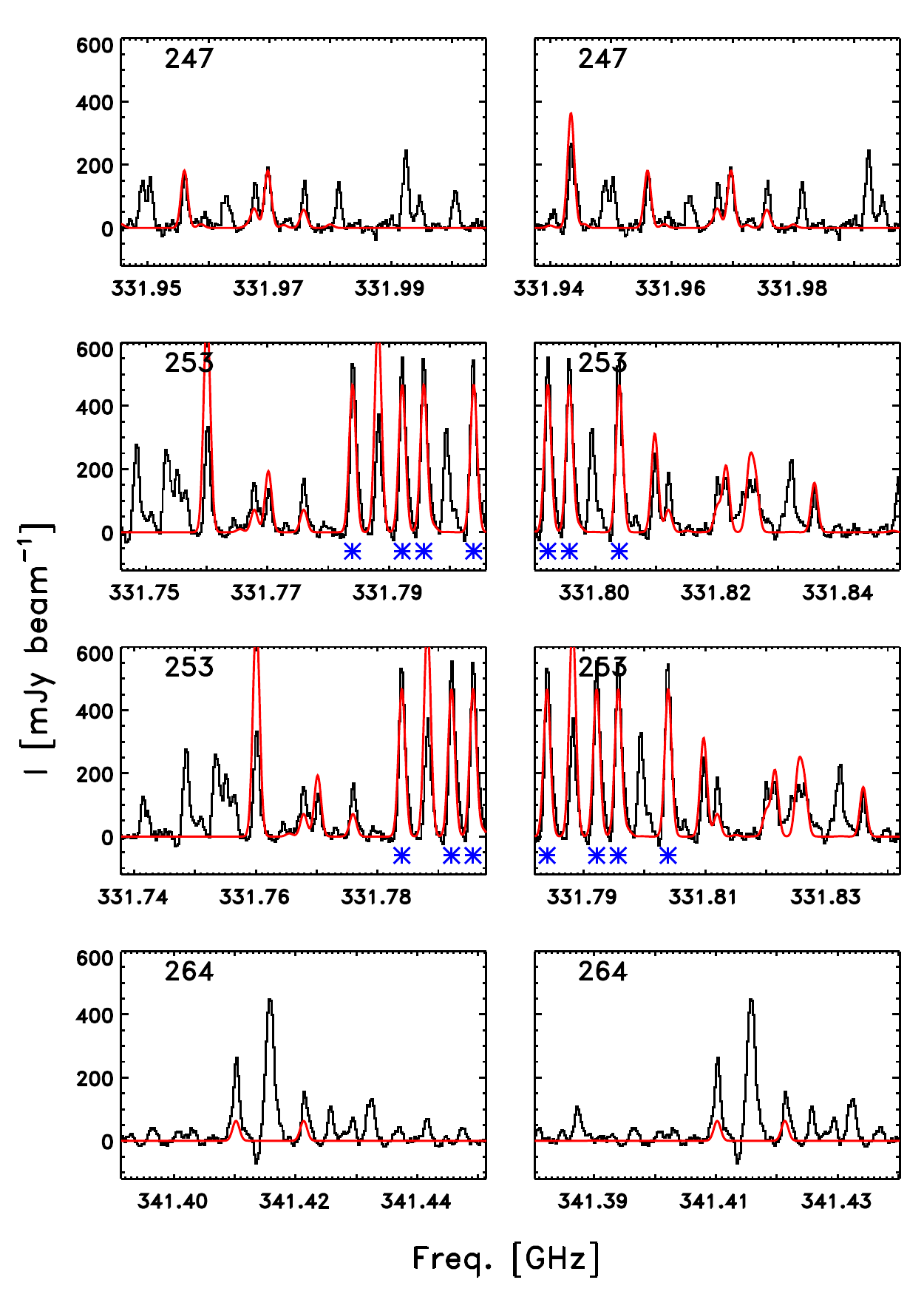}\includegraphics{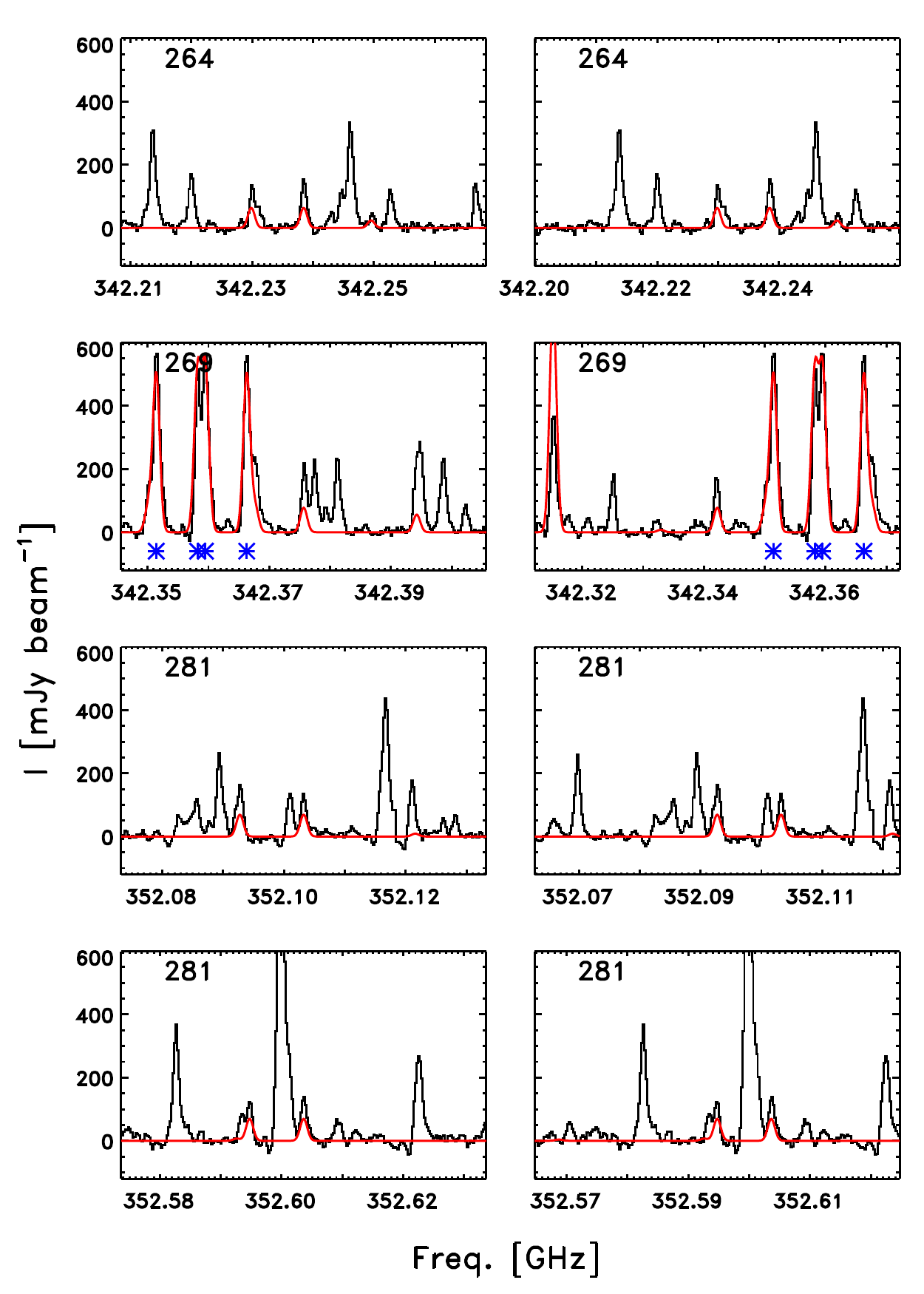}}
\resizebox{0.44\textwidth}{!}{\includegraphics{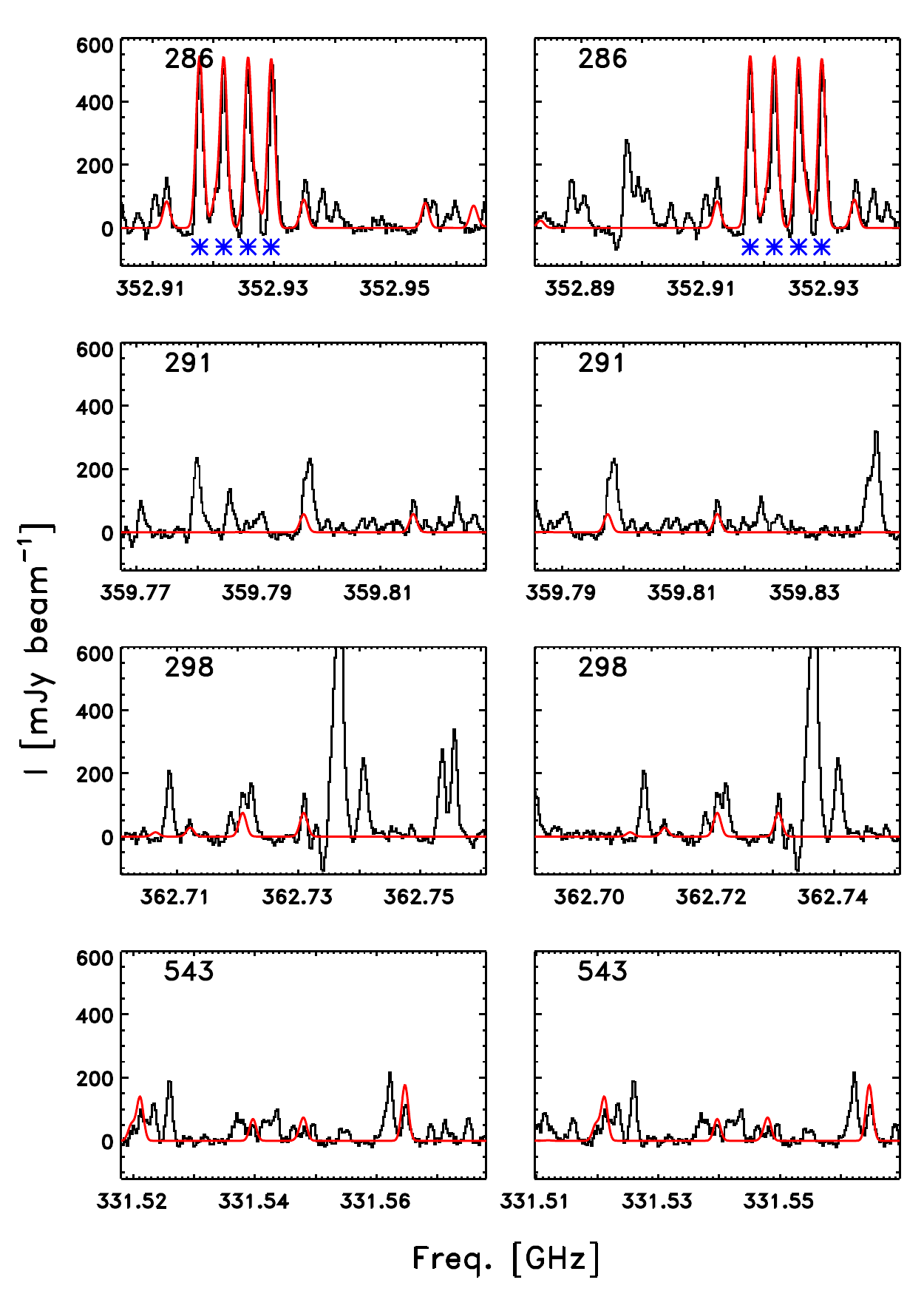}}
\captionof{figure}{As in Fig.~\ref{first_spectra} for the 24 brightest lines of CH$_3$OCHO as expected from the synthetic spectrum.}\label{methylformate_spectra}
\end{minipage}
\clearpage

\noindent\begin{minipage}{\textwidth}
\resizebox{0.88\textwidth}{!}{\includegraphics{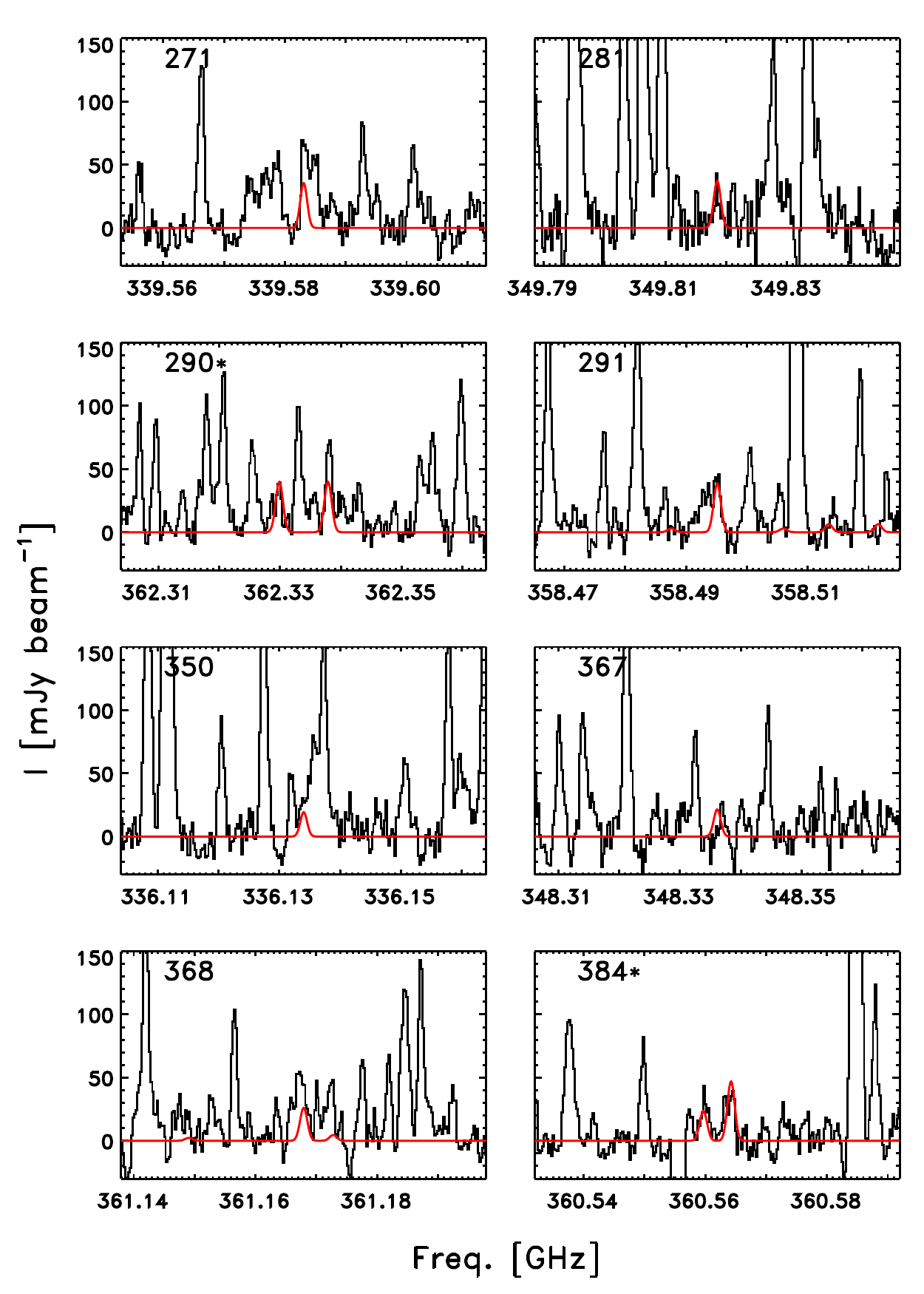}\includegraphics{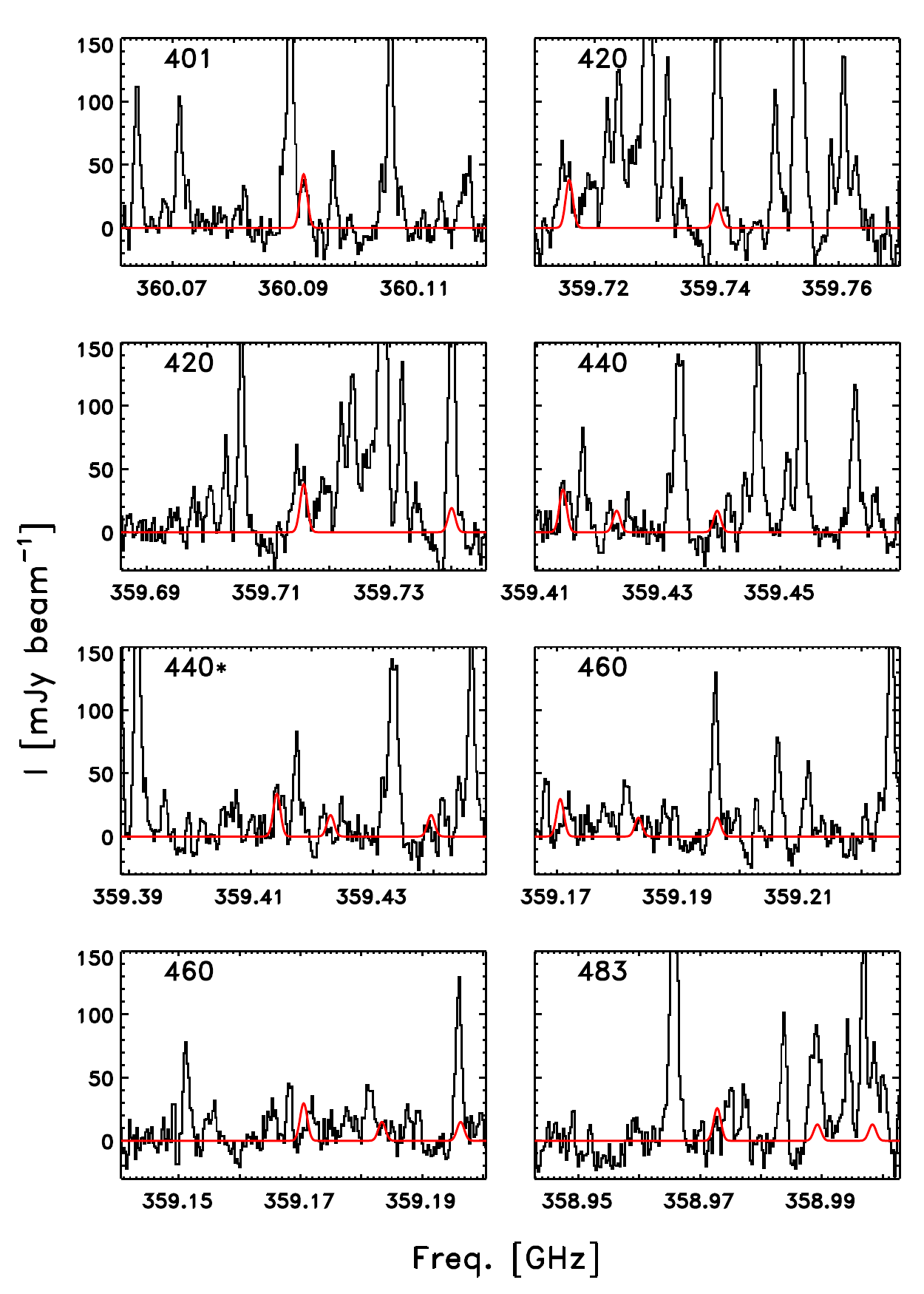}}\\
\resizebox{0.88\textwidth}{!}{\includegraphics{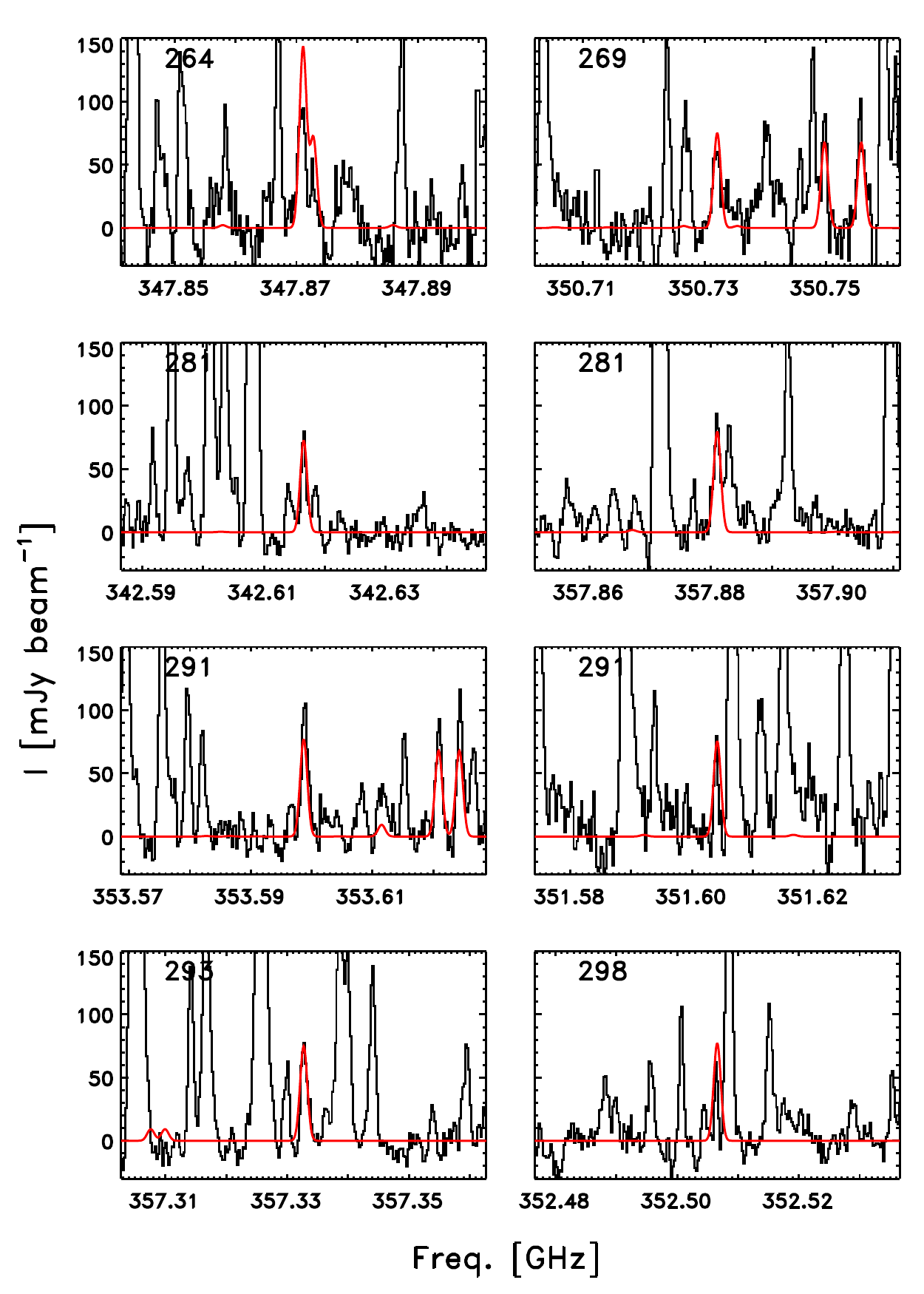}\includegraphics{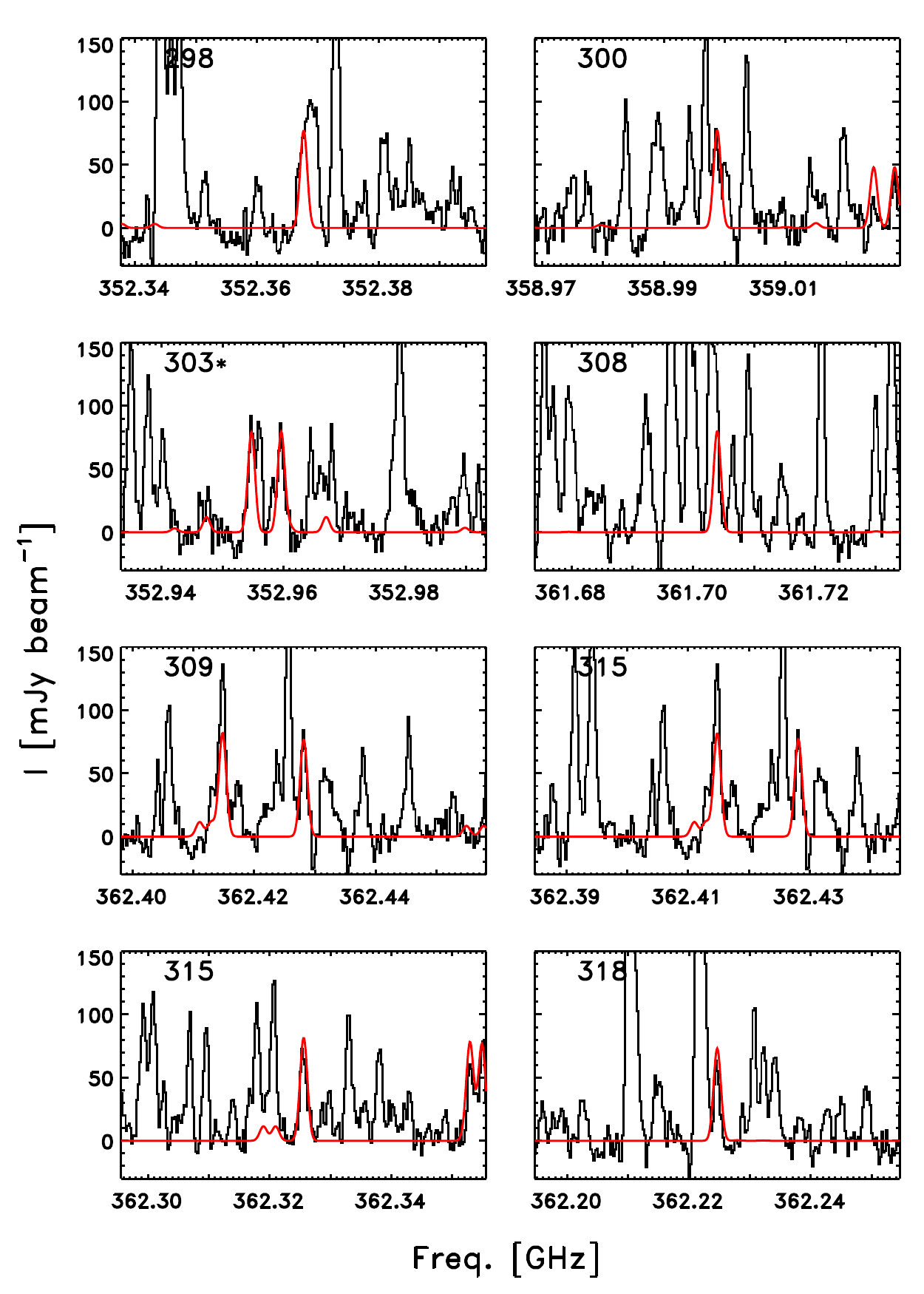}}
\captionof{figure}{As in Fig.~\ref{first_spectra} for the 16+16
  brightest lines of deuterated methyl formate: DCOOCH$_3$ in the
  upper panels and CH$_2$DOCHO in the lower panels.}\label{d-mfe_spectra}
\end{minipage}
\clearpage

\noindent\begin{minipage}{\textwidth}
\resizebox{0.88\textwidth}{!}{\includegraphics{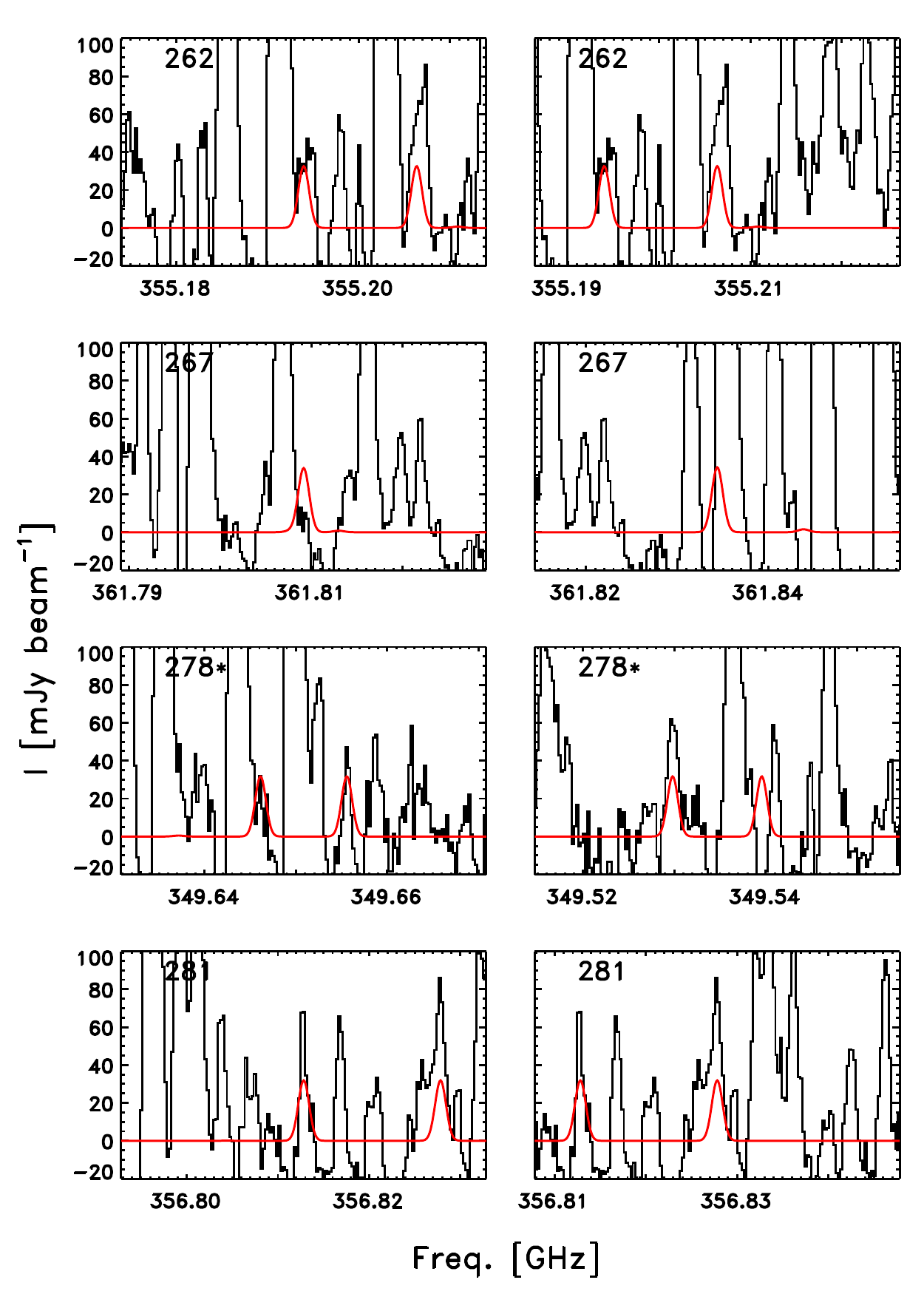}\includegraphics{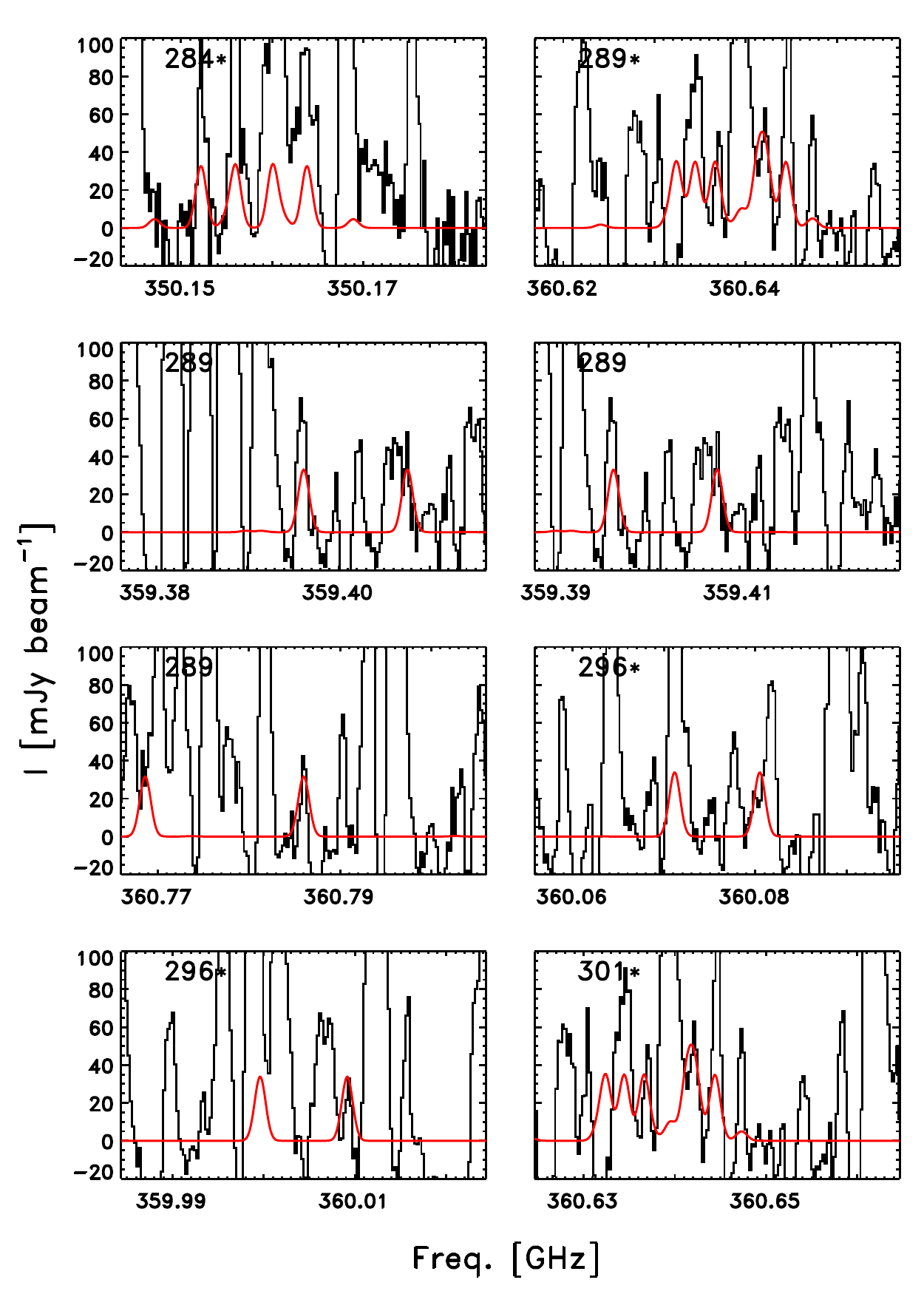}}
\resizebox{0.88\textwidth}{!}{\includegraphics{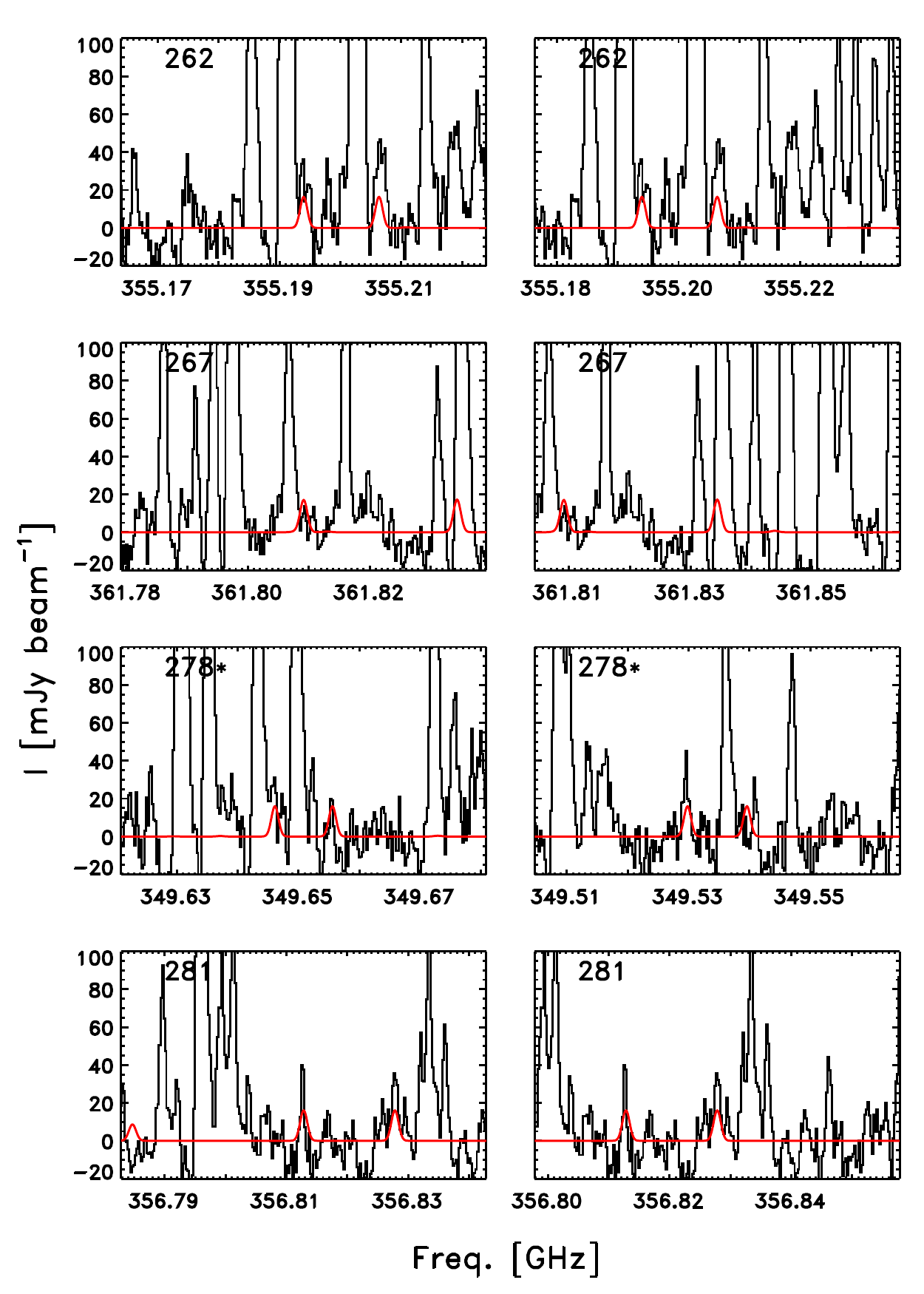}\includegraphics{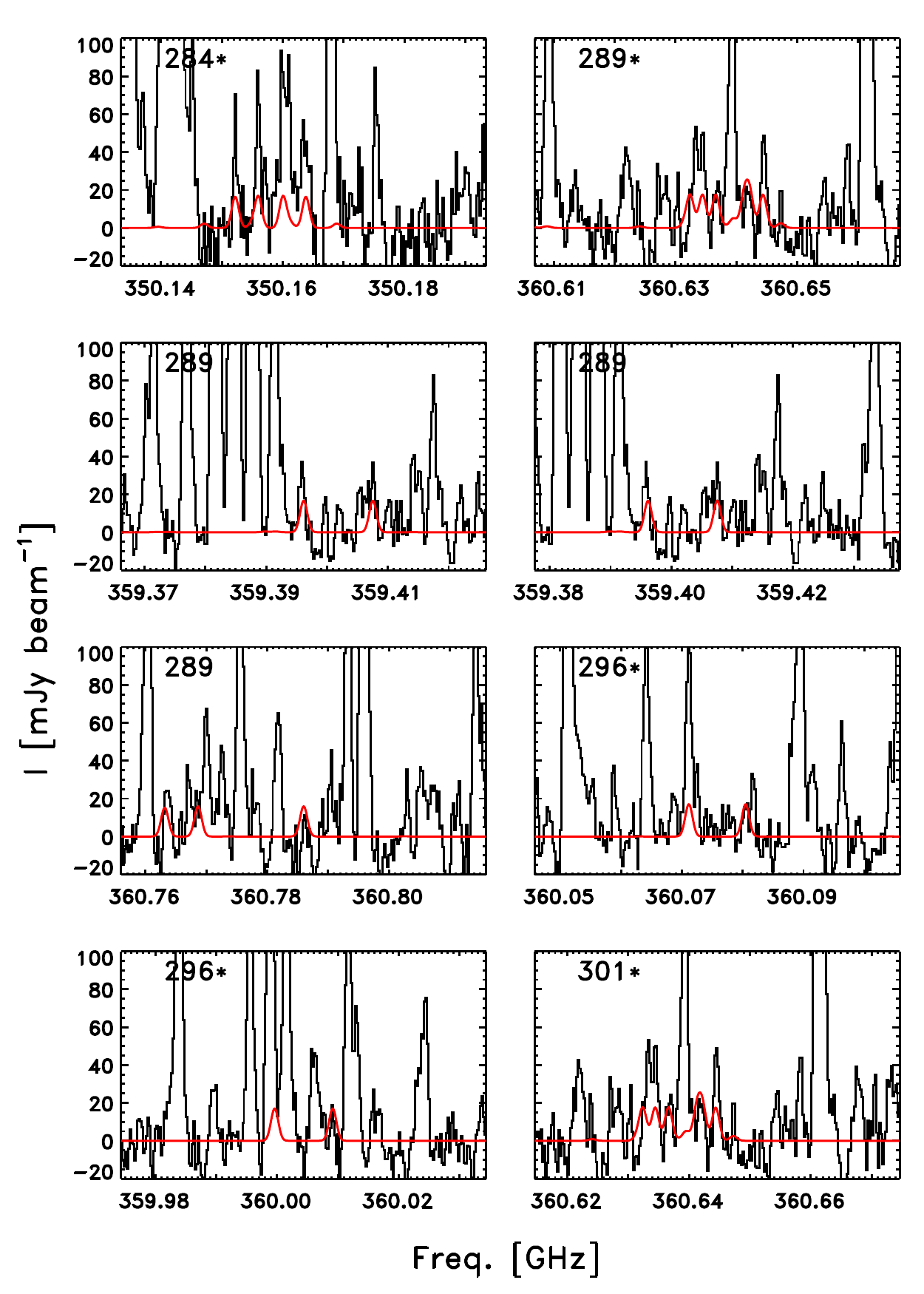}}
\captionof{figure}{As in Fig.~\ref{first_spectra} for the 16
  brightest lines of $^{13}$C methyl formate, CH$_3$O$^{13}$CHO 
    toward the half-beam (upper) and full-beam (lower) offset positions.}\label{mfe13_spectra}
\end{minipage}
\clearpage

\subsection{Ketene}

\noindent\begin{minipage}{\textwidth}
\resizebox{0.88\textwidth}{!}{\includegraphics{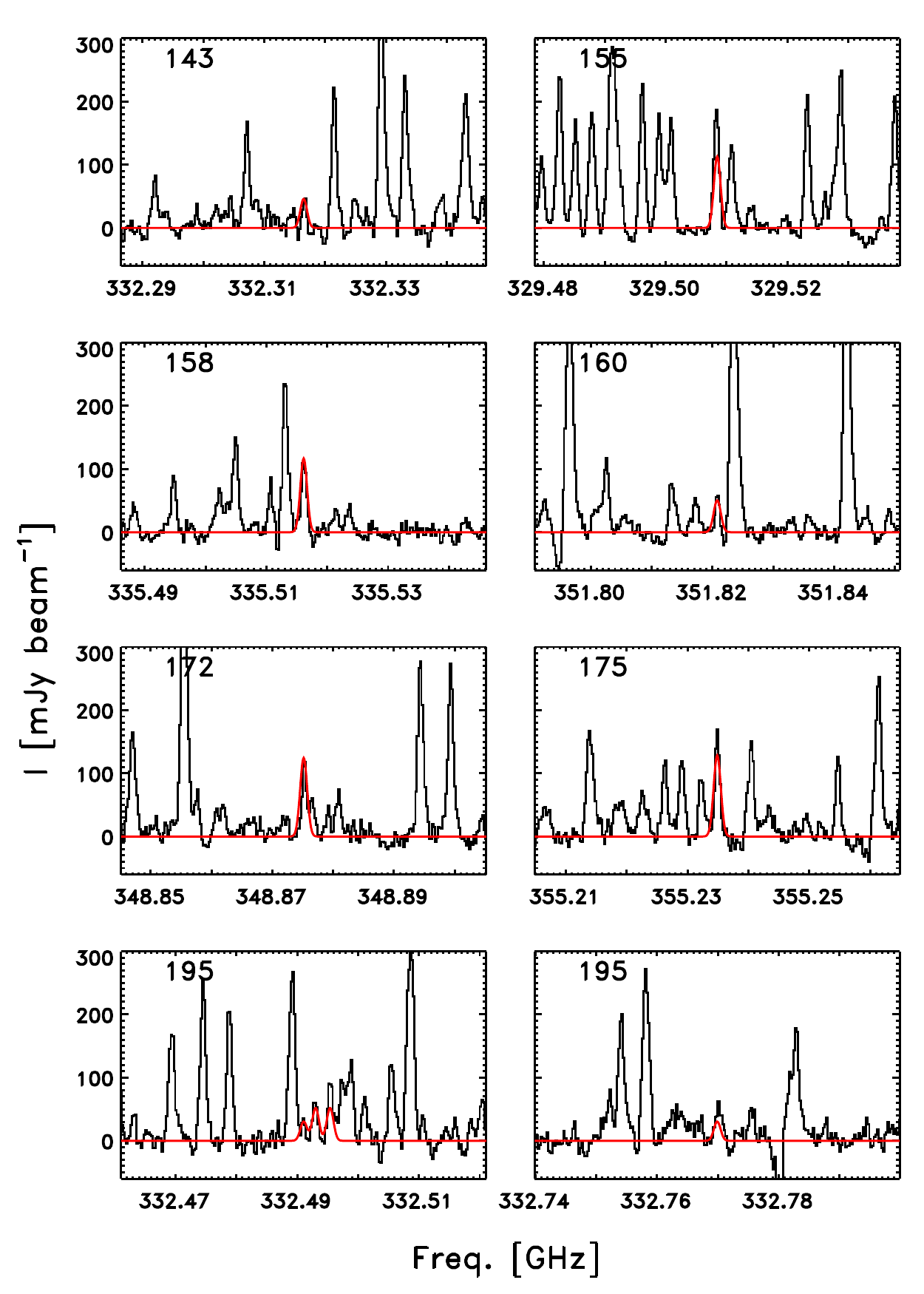}\includegraphics{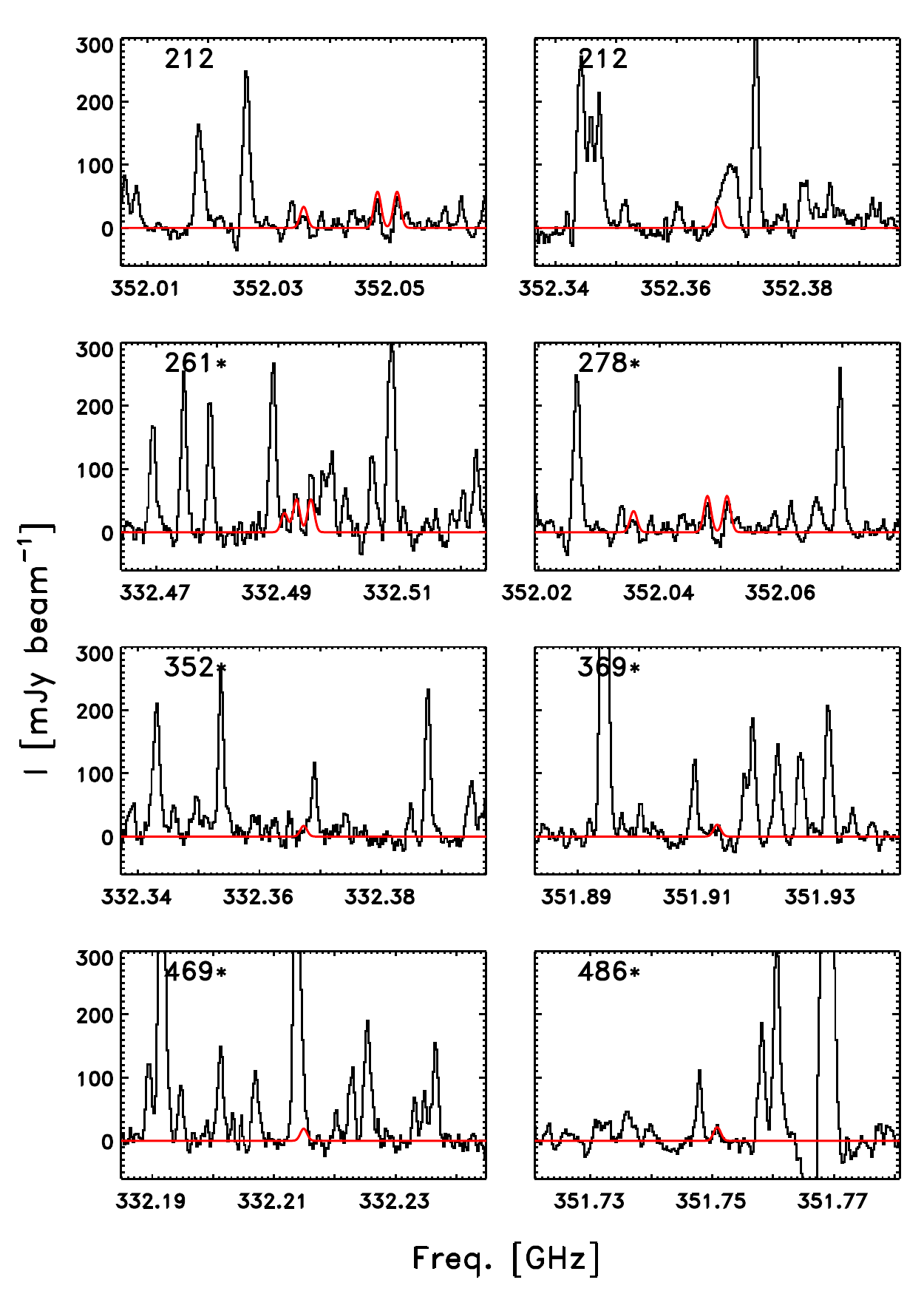}}\\
\resizebox{0.88\textwidth}{!}{\includegraphics{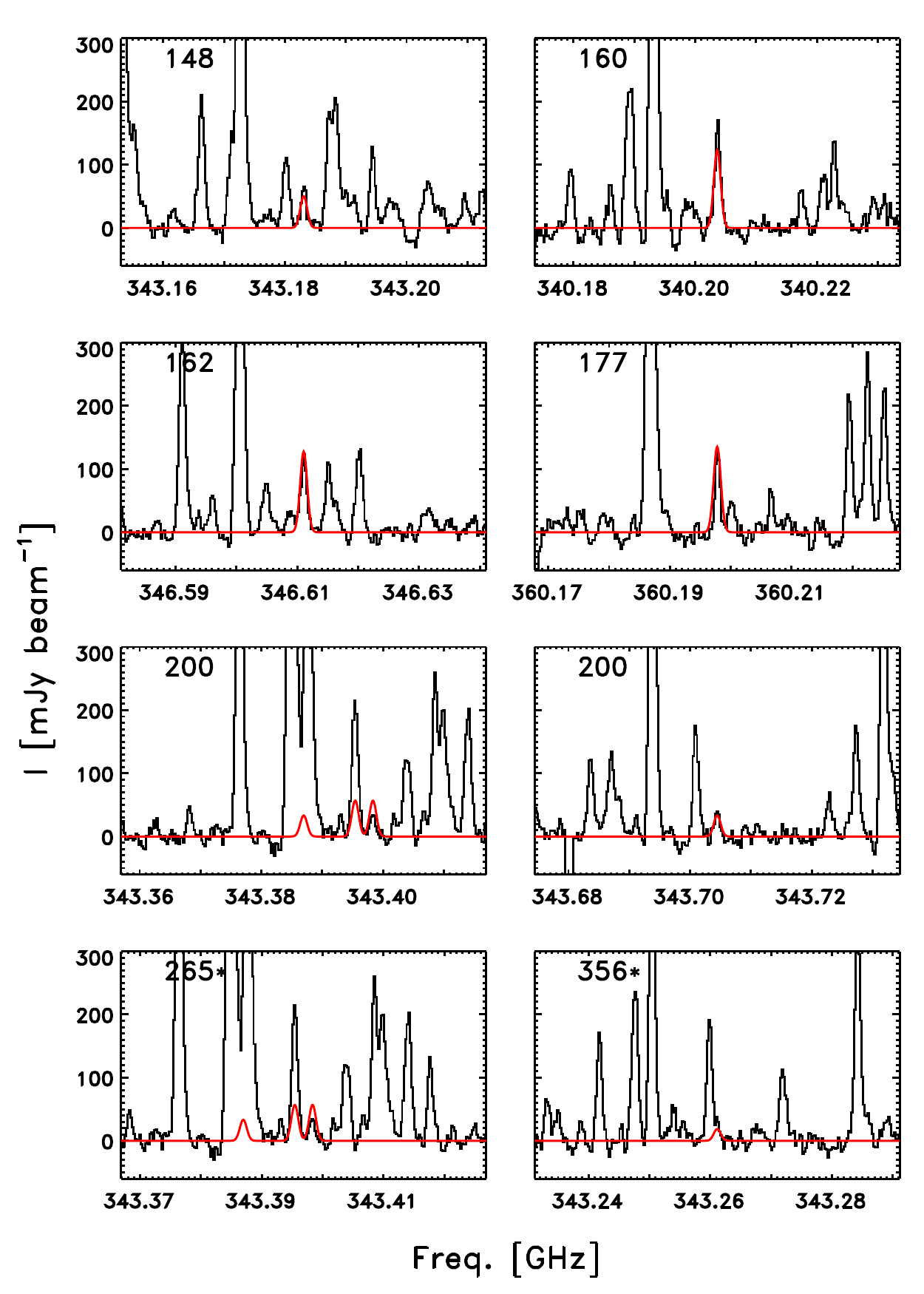}\includegraphics{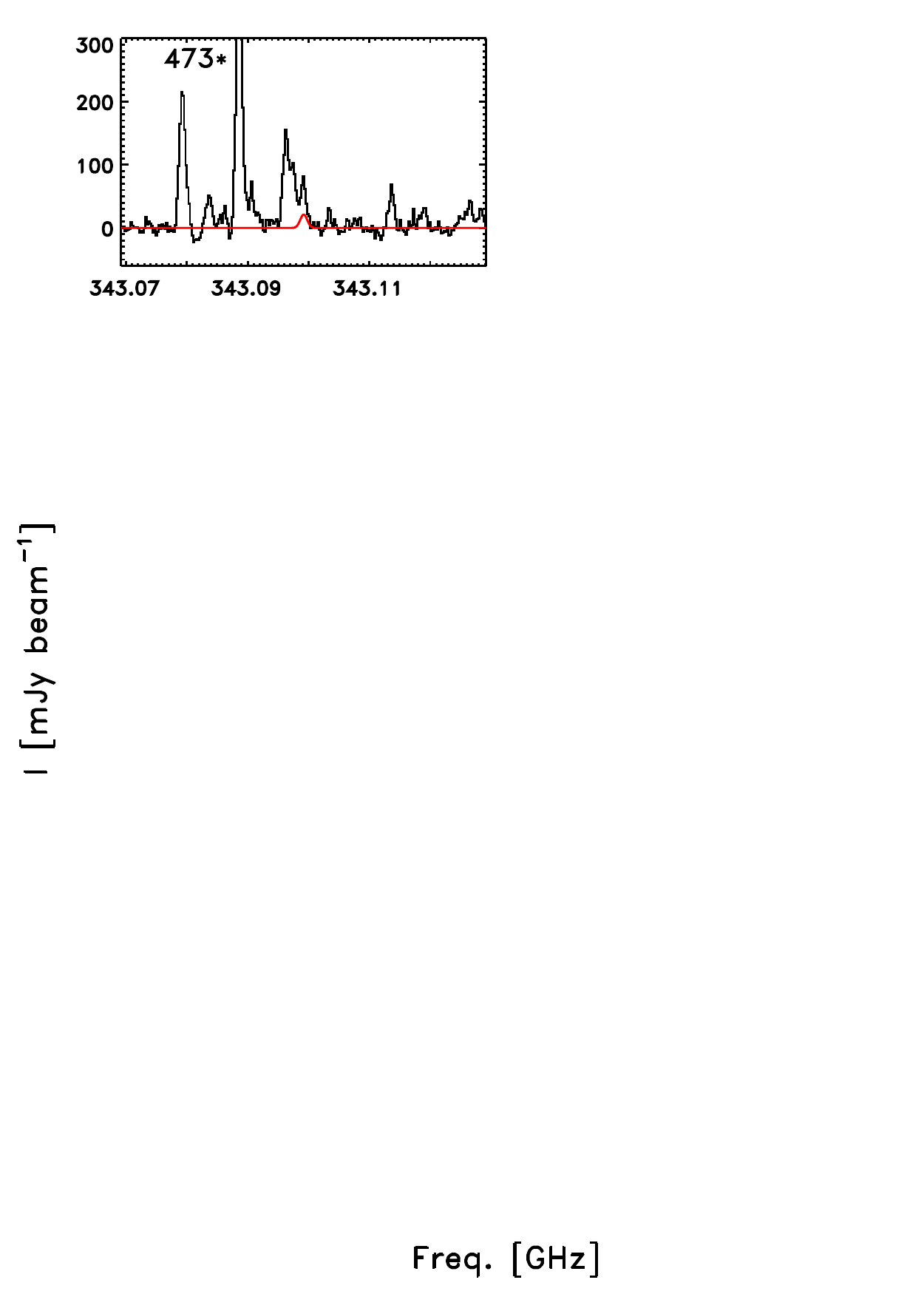}}
\captionof{figure}{As in Fig.~\ref{first_spectra} for the
  brightest lines of the $^{13}$C-isotopologues of ketene with
  $^{13}$CH$_2$CO shown in the upper panels and CH$_2$$^{13}$CO in the
  lower panels.}\label{ketene13_spectra}
\end{minipage}

\clearpage

\noindent\begin{minipage}{\textwidth}
\resizebox{0.88\textwidth}{!}{\includegraphics{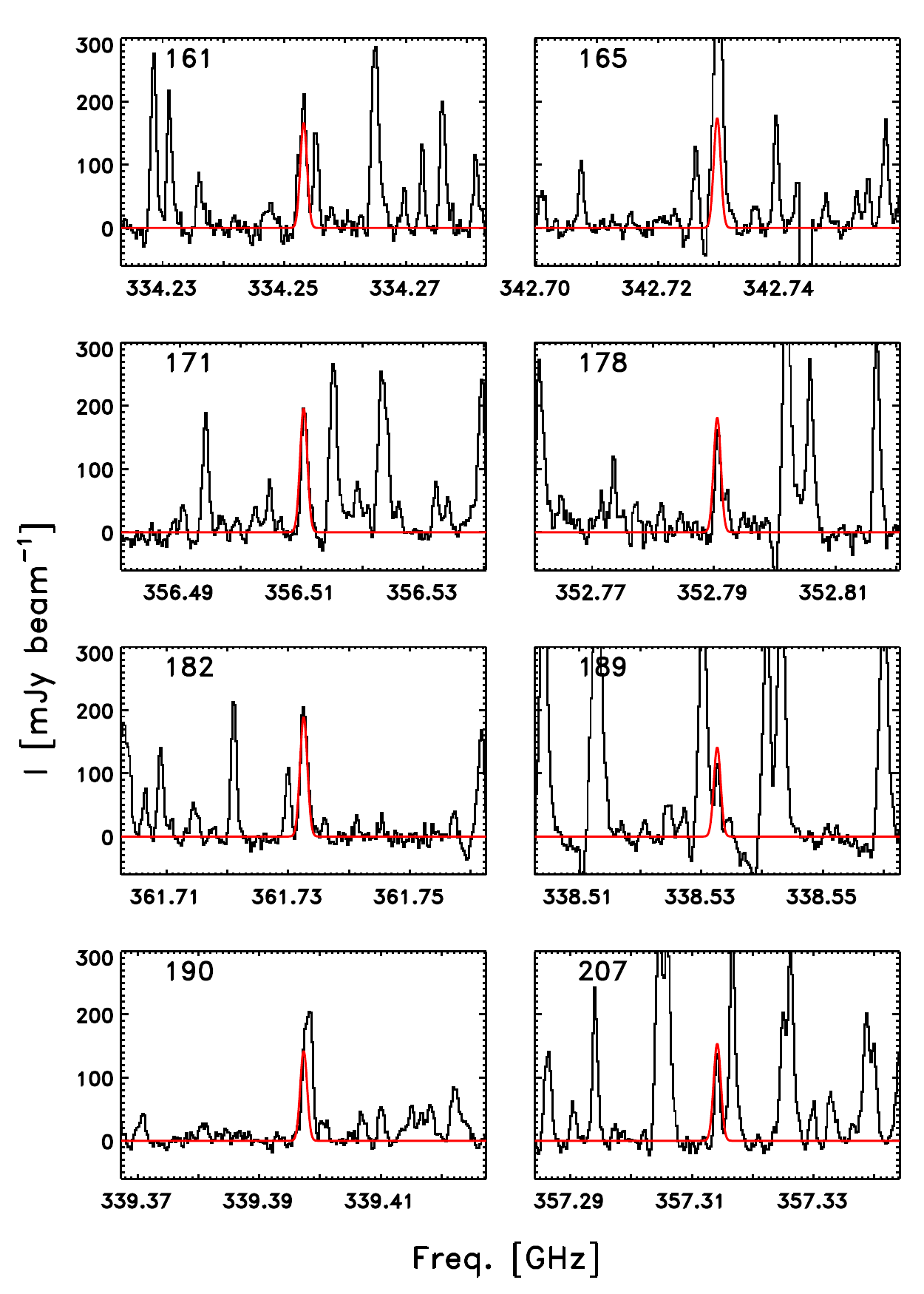}\includegraphics{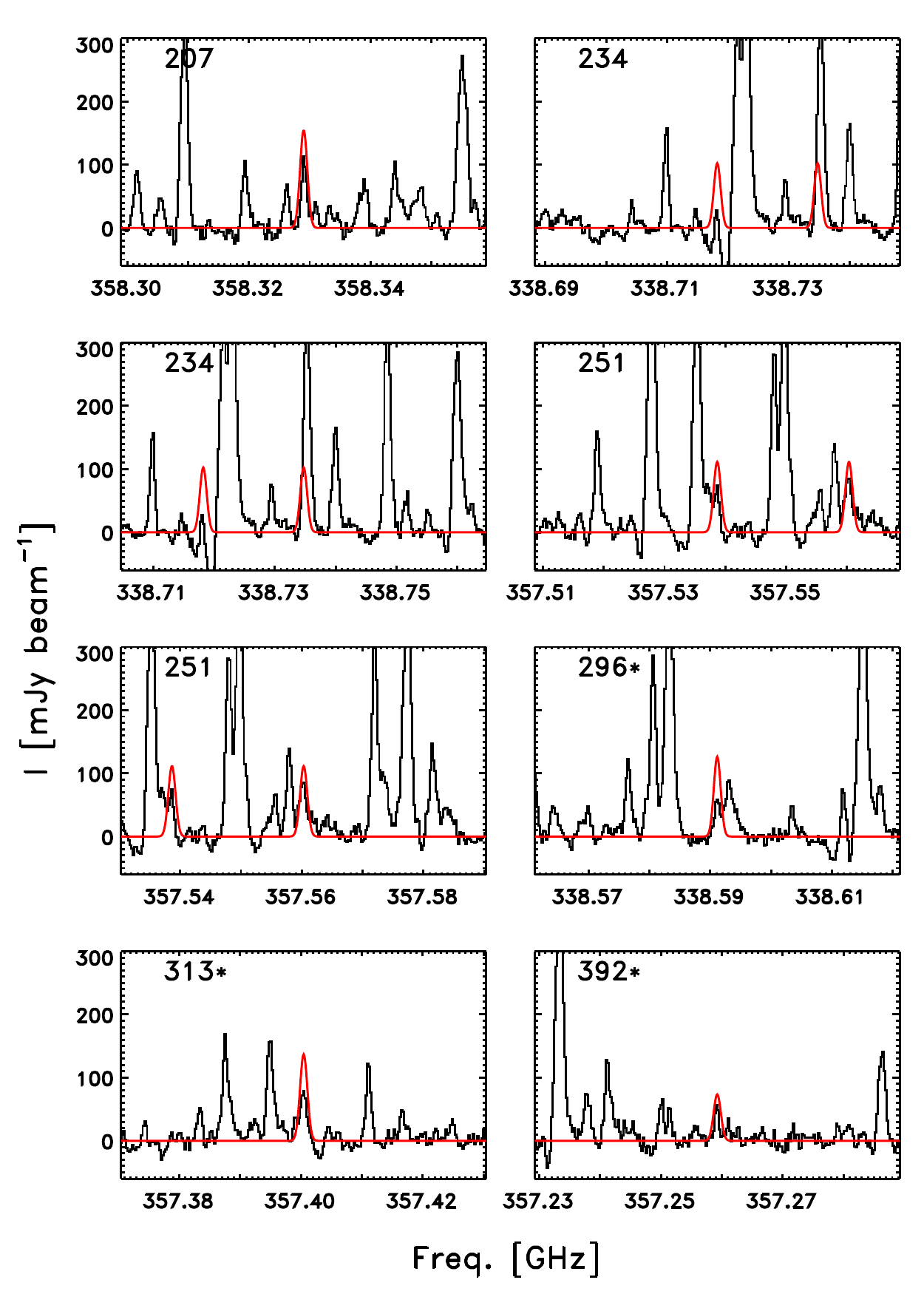}}\\
\resizebox{0.88\textwidth}{!}{\includegraphics{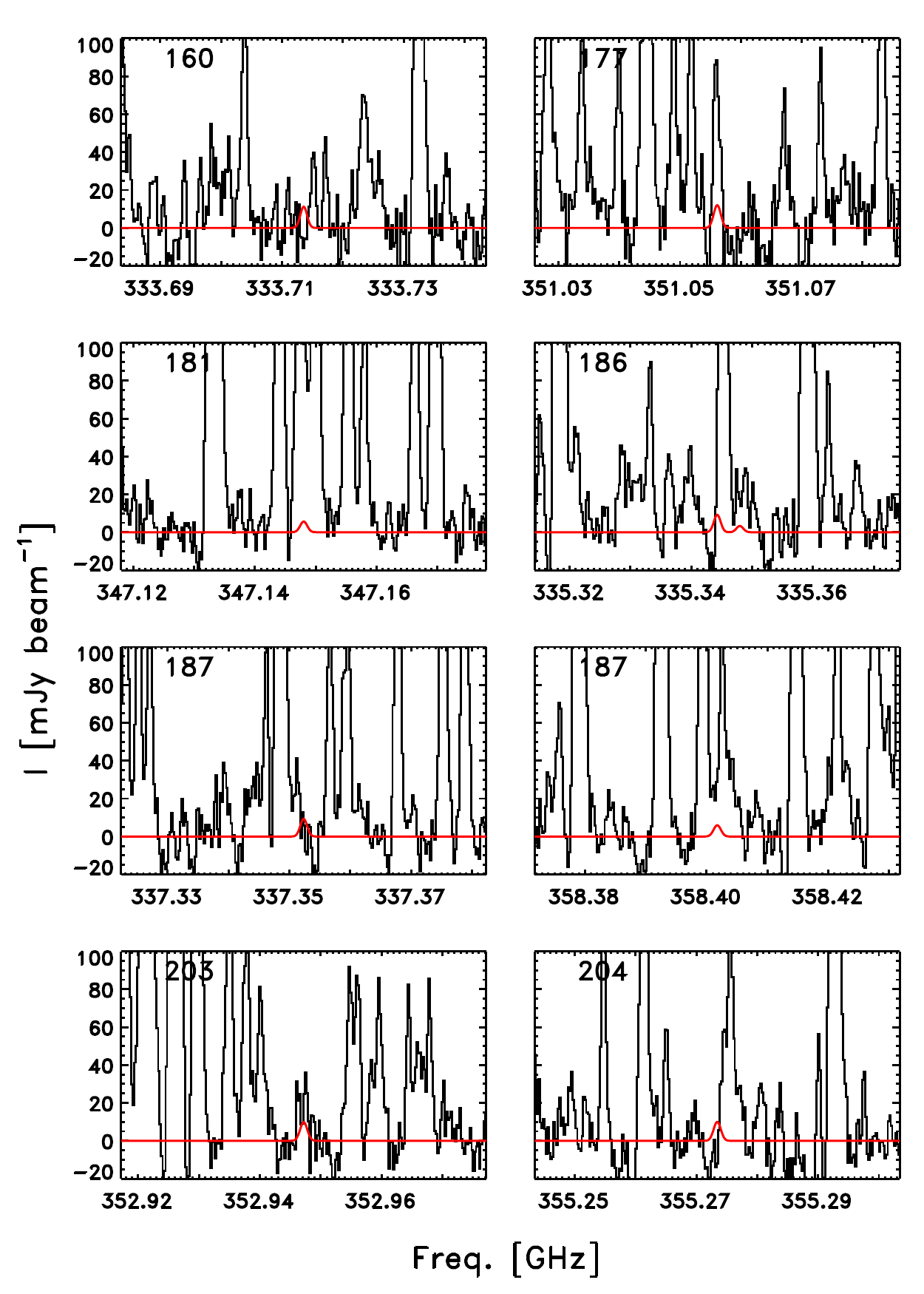}\includegraphics{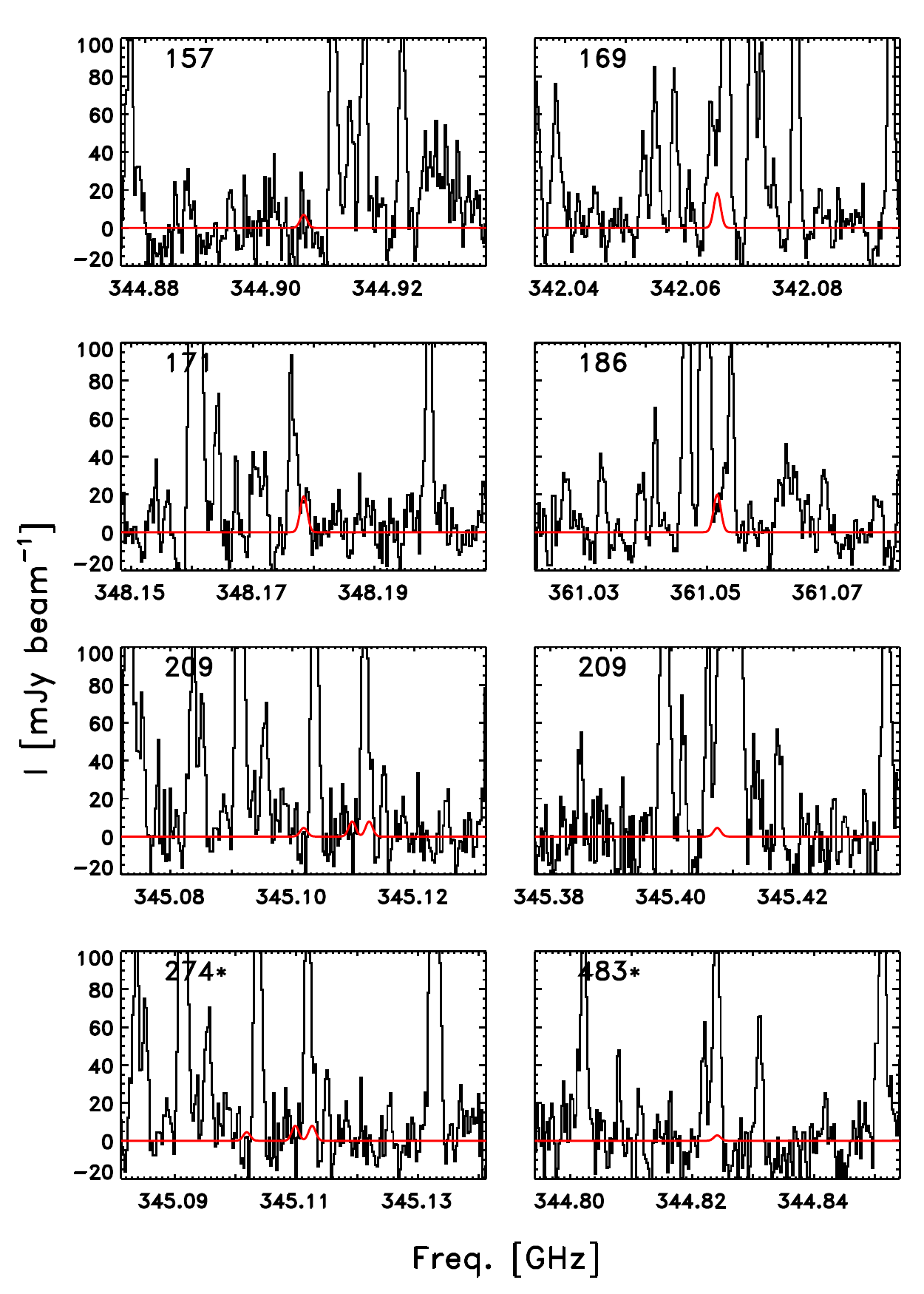}}
\captionof{figure}{As in Fig.~\ref{first_spectra} for the 16 brightest
  lines of the singly deuterated-isotopologues of ketene CHDCO (upper
  panels). The lower panels show the 8 brightest lines of doubly
  deuterated ketene, CD$_2$CO (left) and the $^{18}$O isotopologue
  CH$_2$C$^{18}$O (right). For those we assume the same D/H ratio as
  for the singly deuterated species  represents an upper
  limit/tentative identification. }\label{dketene_spectra}
\end{minipage}
\clearpage

\subsection{Dimethyl ether}
\noindent\begin{minipage}{\textwidth}
\resizebox{0.88\textwidth}{!}{\includegraphics{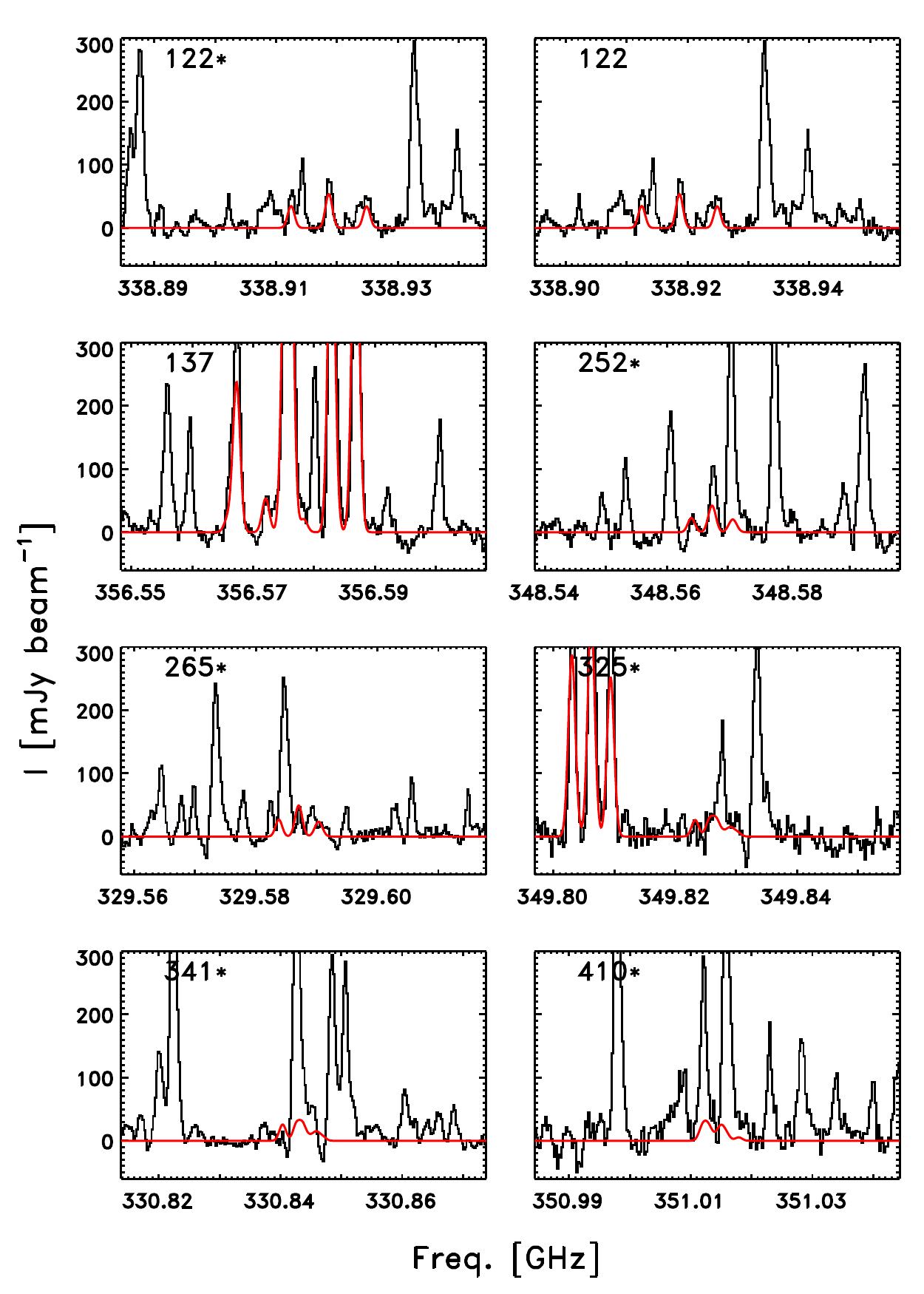}\includegraphics{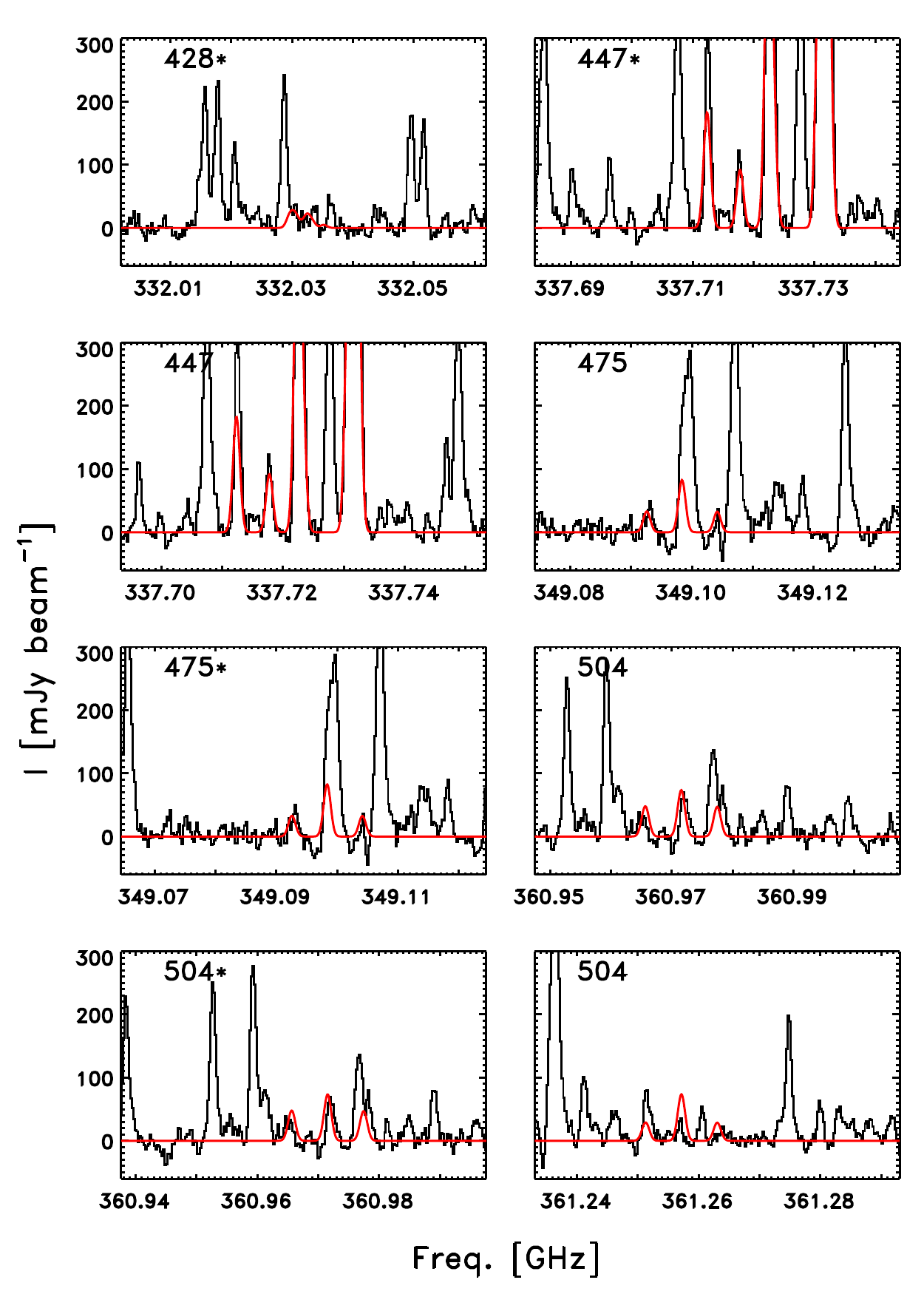}}
\resizebox{0.88\textwidth}{!}{\includegraphics{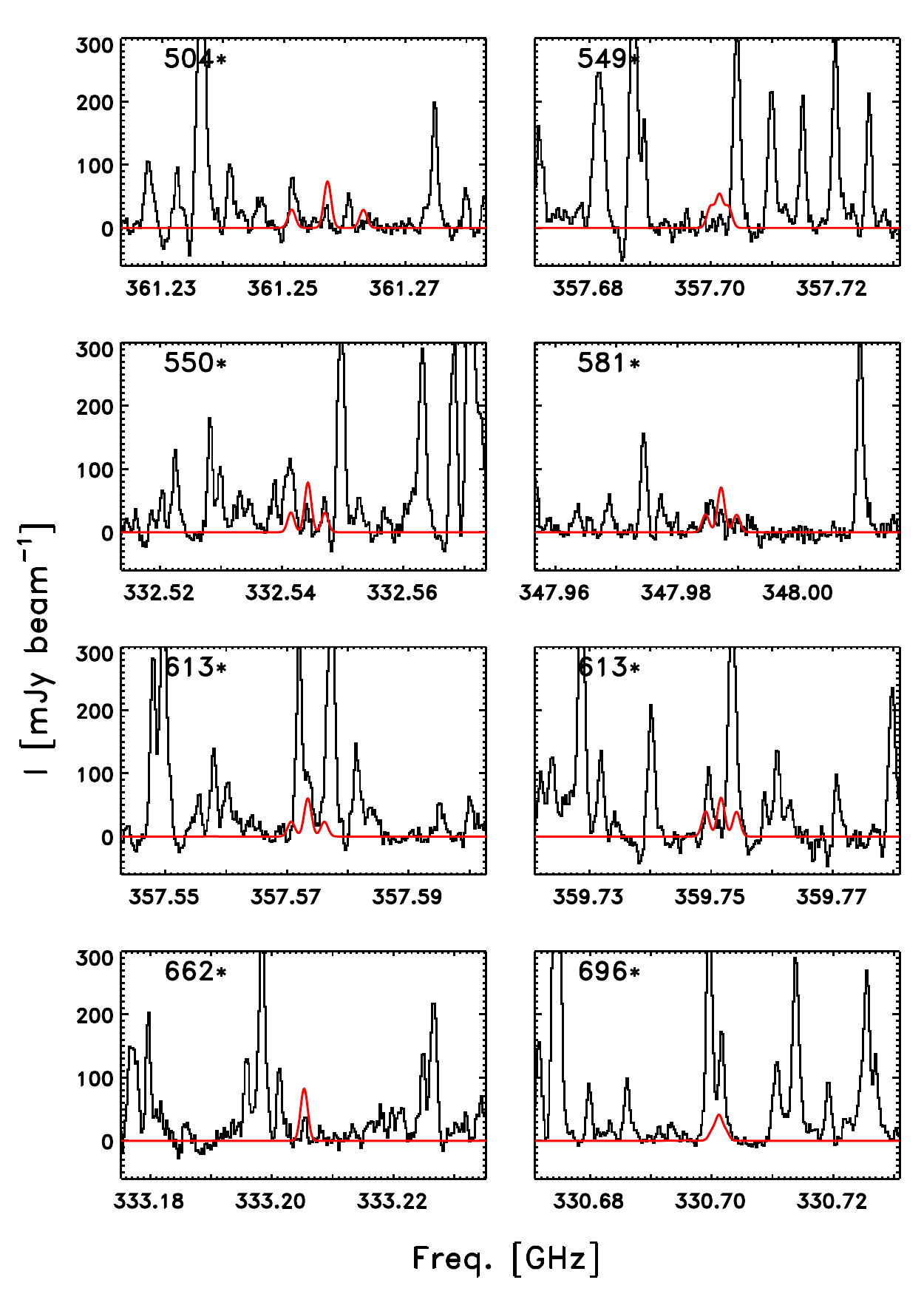}\includegraphics{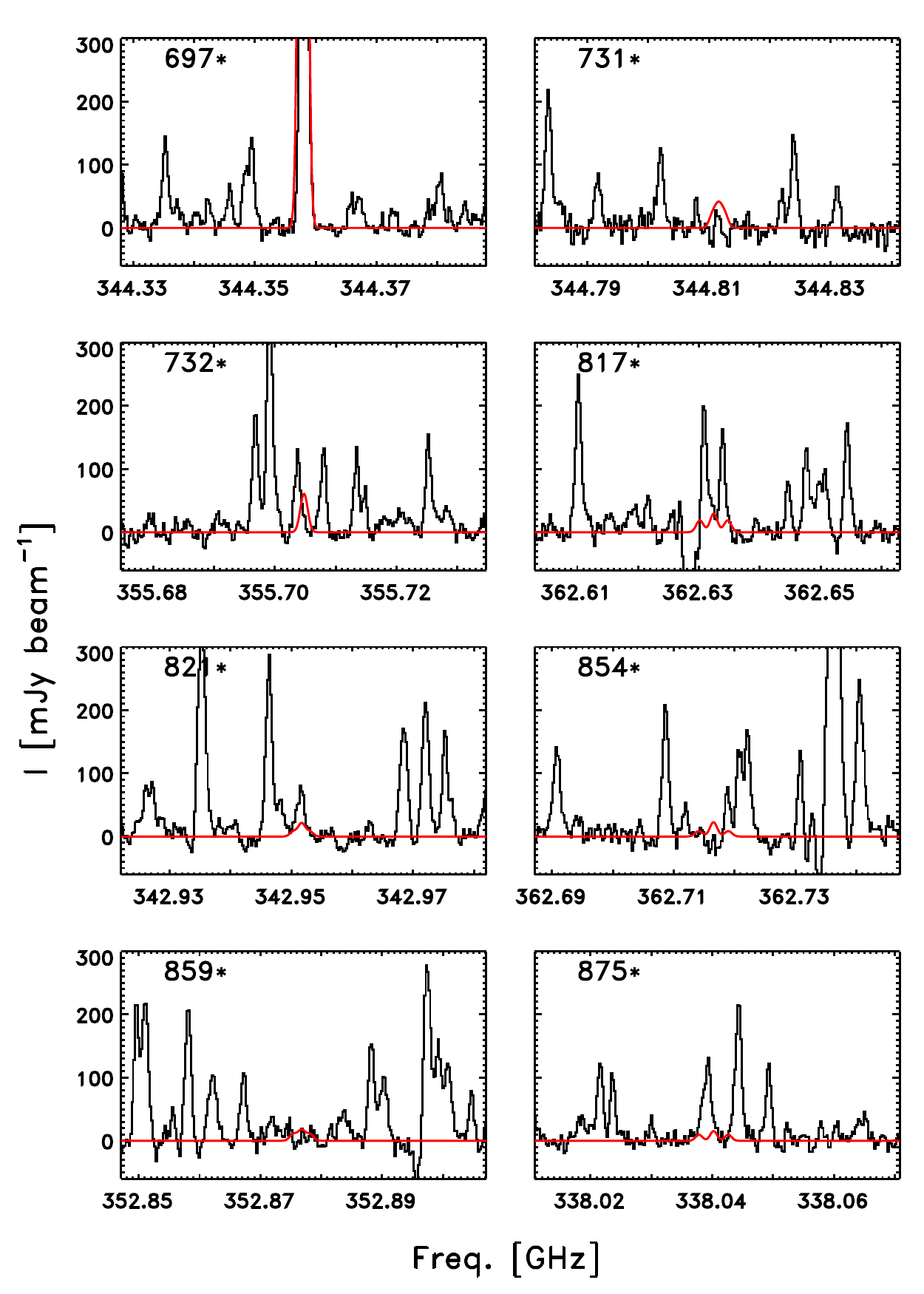}}
\captionof{figure}{As in Fig.~\ref{first_spectra} for the 32
  brightest lines of the dimethyl ether, CH$_3$OCH$_3$. Like in
  Fig.~\ref{dethanol_spectra1} the asterisks next to some of the
  upper energy values indicate features where multiple transitions of
  the species contribute to the observed line.}  \label{dme_spectra}
\end{minipage}
\clearpage

\noindent\begin{minipage}{\textwidth}
\resizebox{0.88\textwidth}{!}{\includegraphics{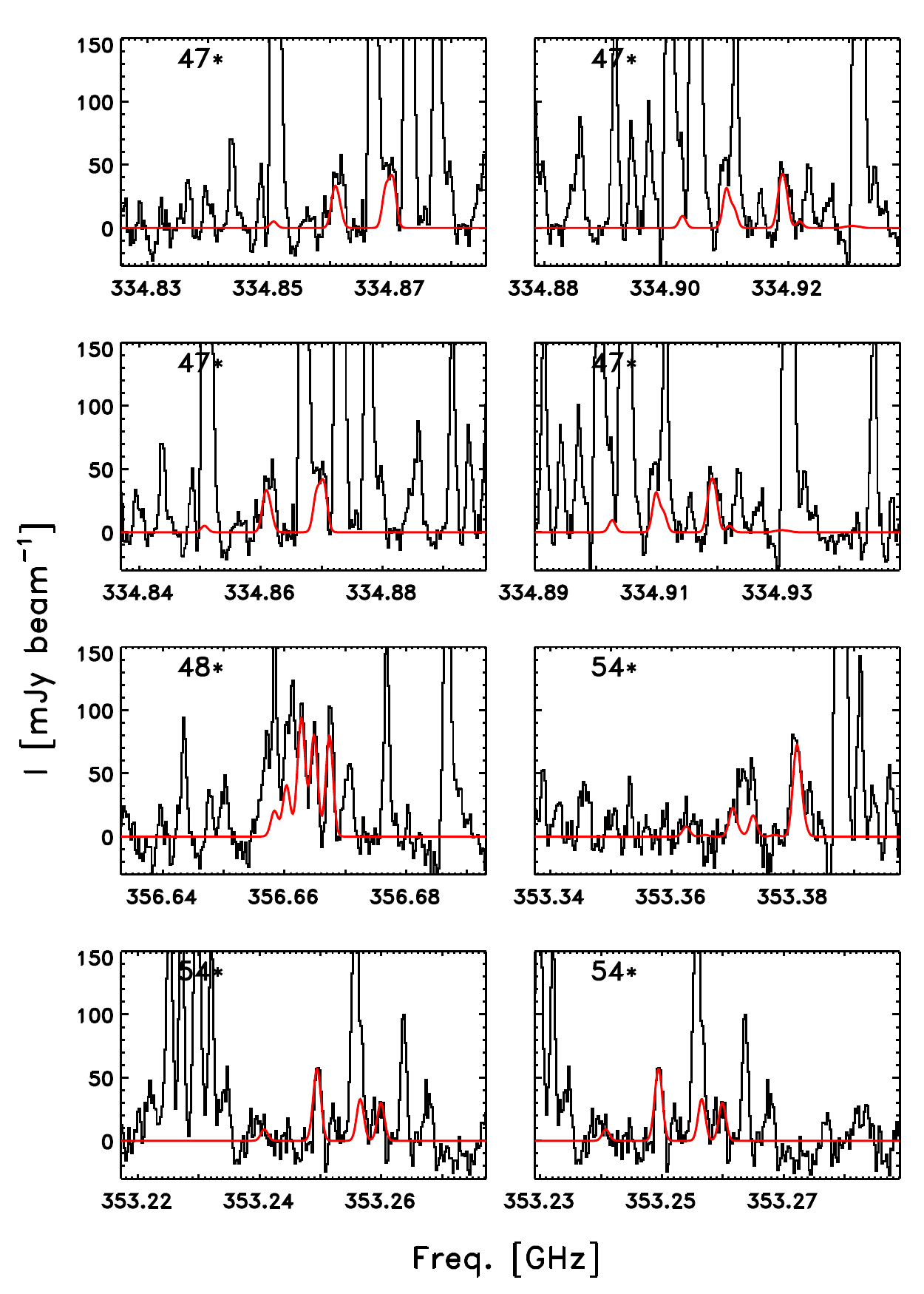}\includegraphics{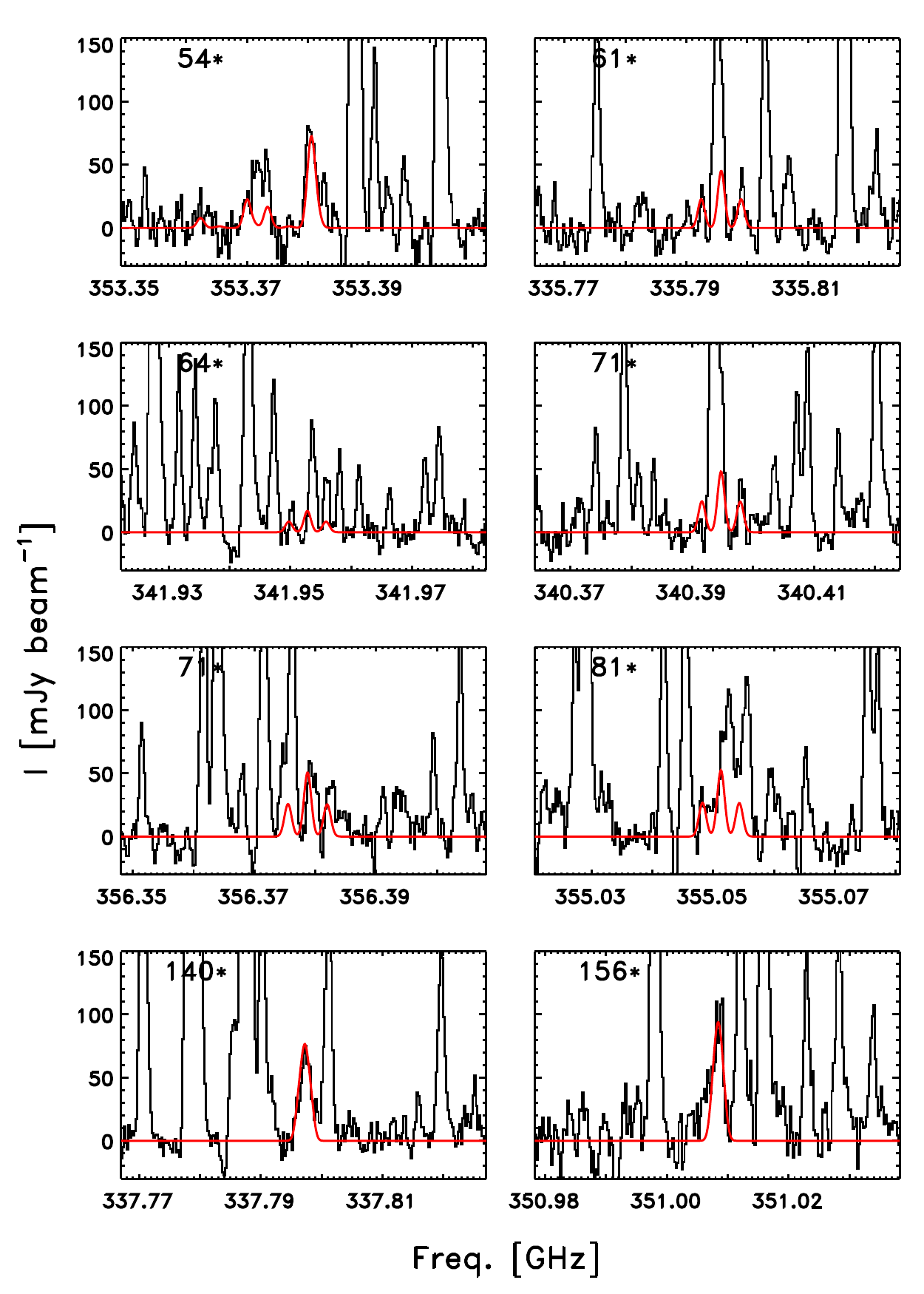}}\\
\resizebox{0.44\textwidth}{!}{\includegraphics{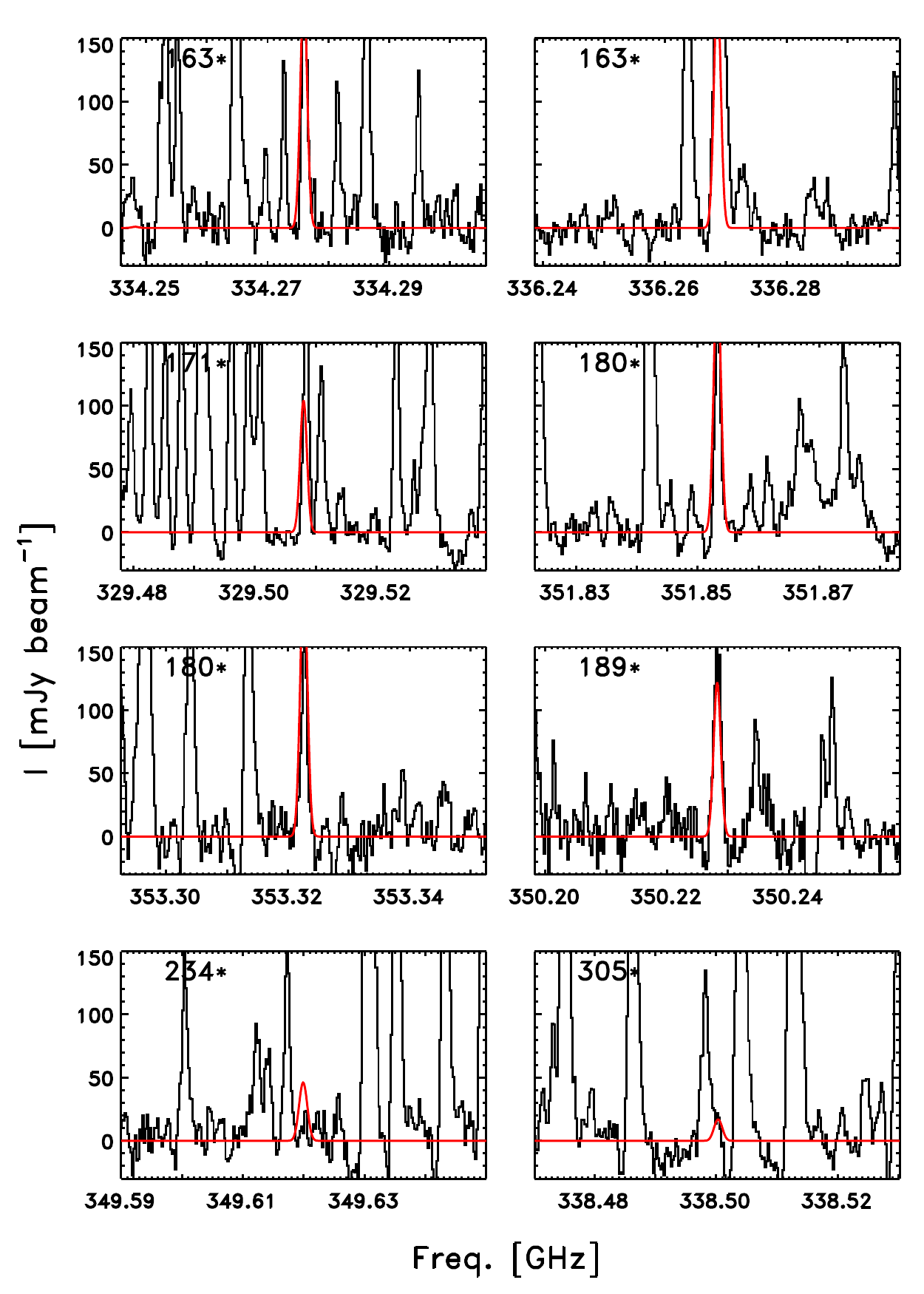}}
\captionof{figure}{As in Fig.~\ref{first_spectra} for the 24
  brightest lines of the $^{13}$C isotopologue of dimethyl ether,
  $^{13}$CH$_3$OCH$_3$ (see also Fig.~\ref{dme_spectra}).}\label{13dme_spectra}
\end{minipage}
\clearpage

\noindent\begin{minipage}{\textwidth}
\resizebox{0.88\textwidth}{!}{\includegraphics{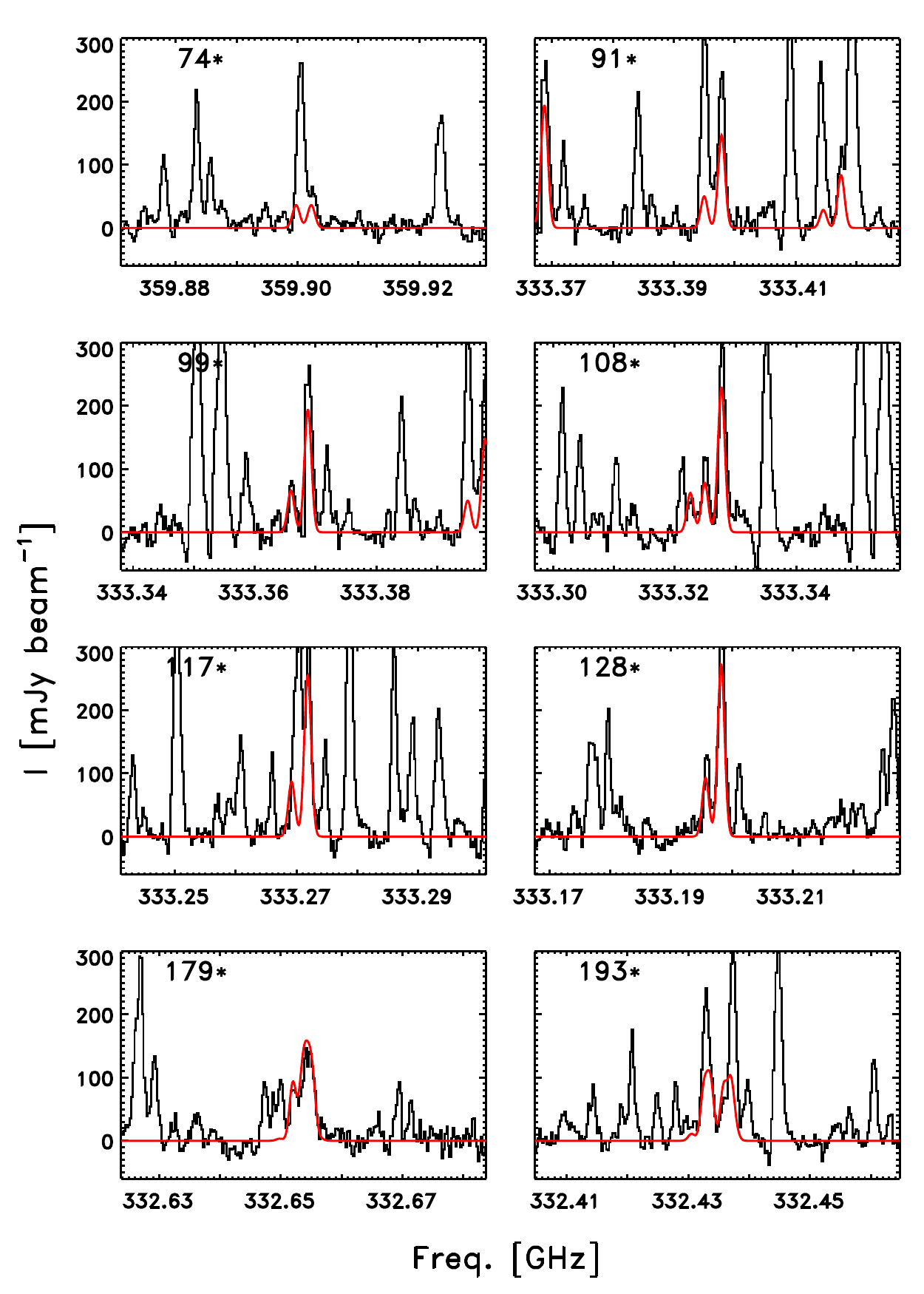}\includegraphics{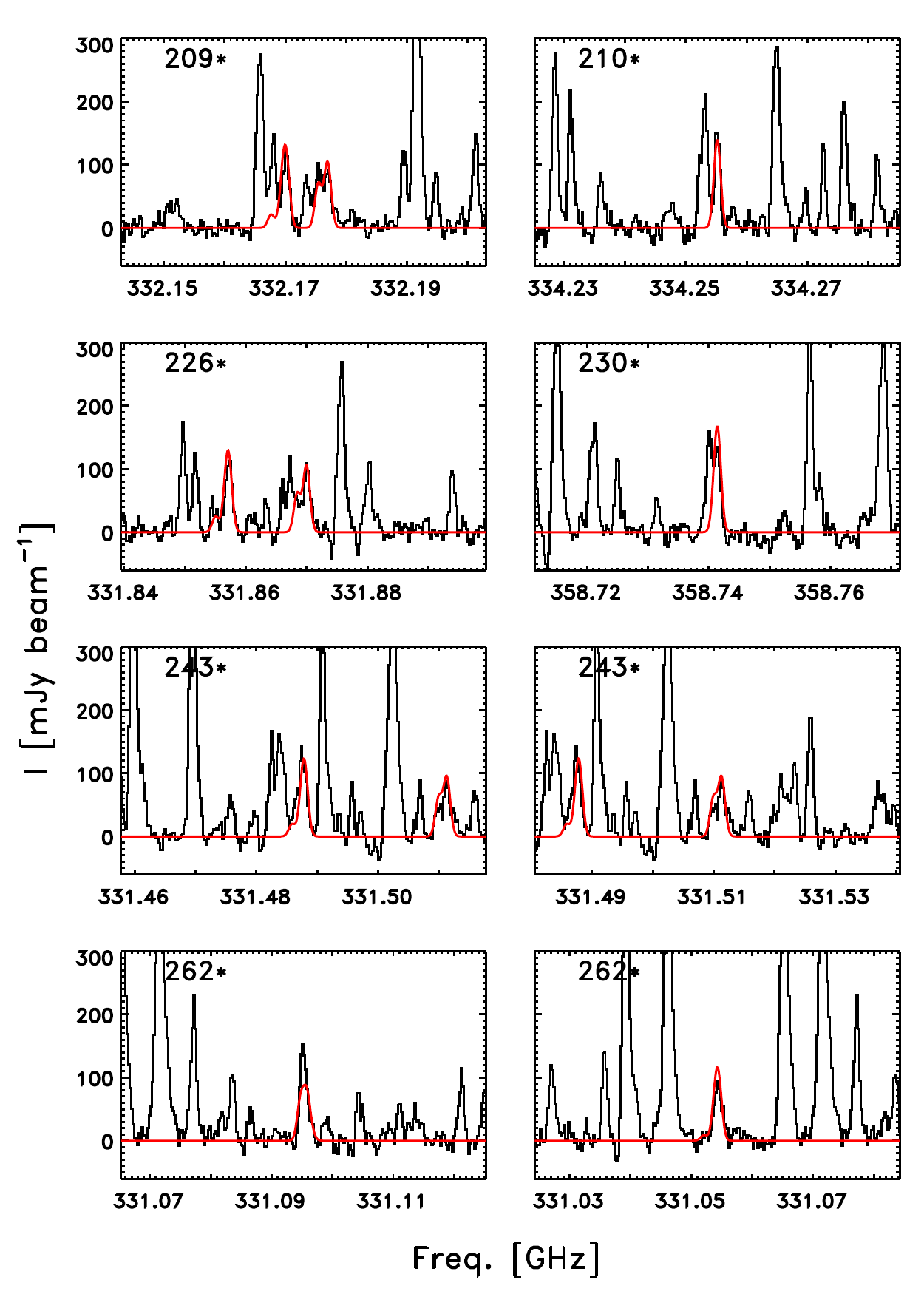}}\\
\resizebox{0.88\textwidth}{!}{\includegraphics{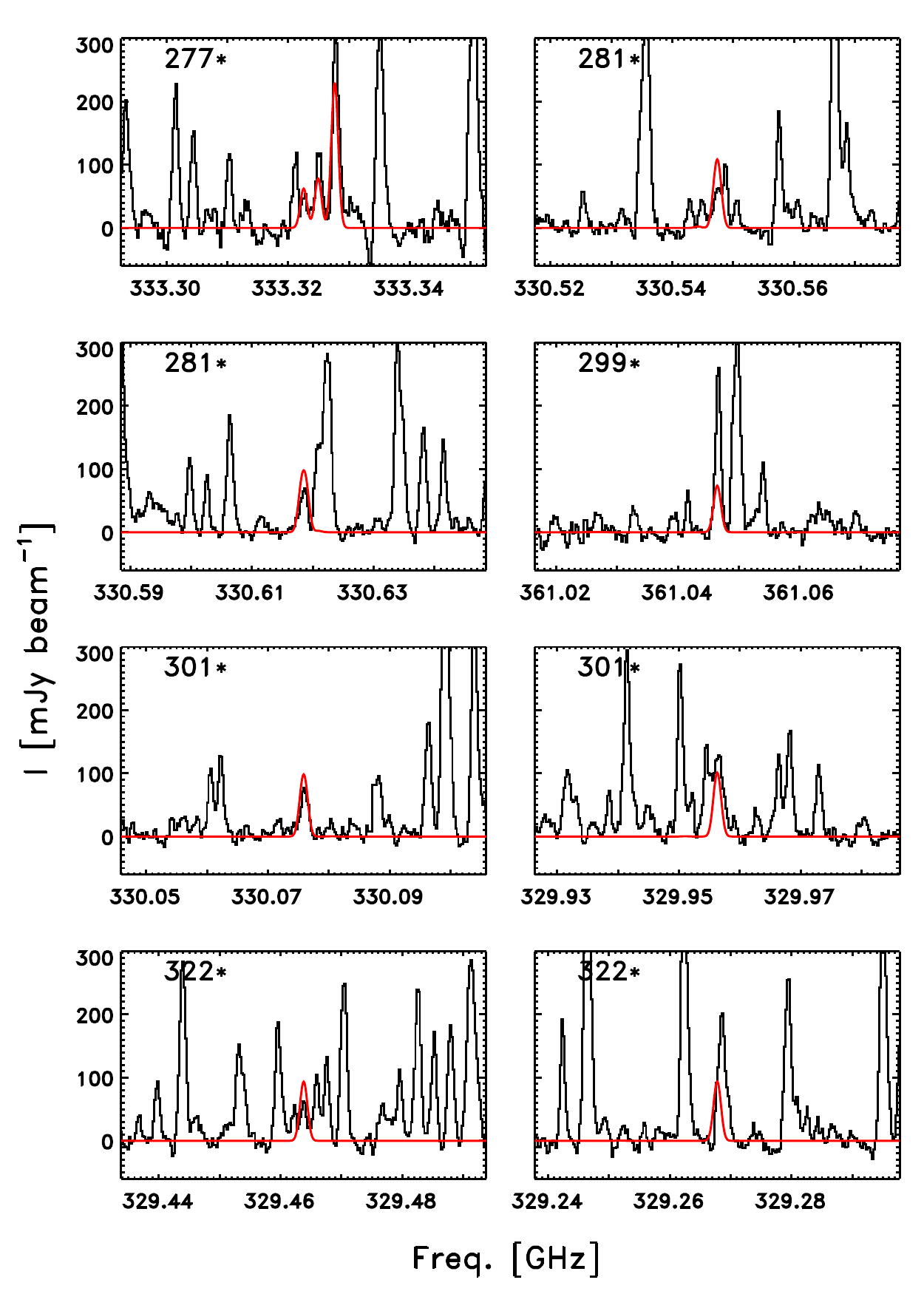}\includegraphics{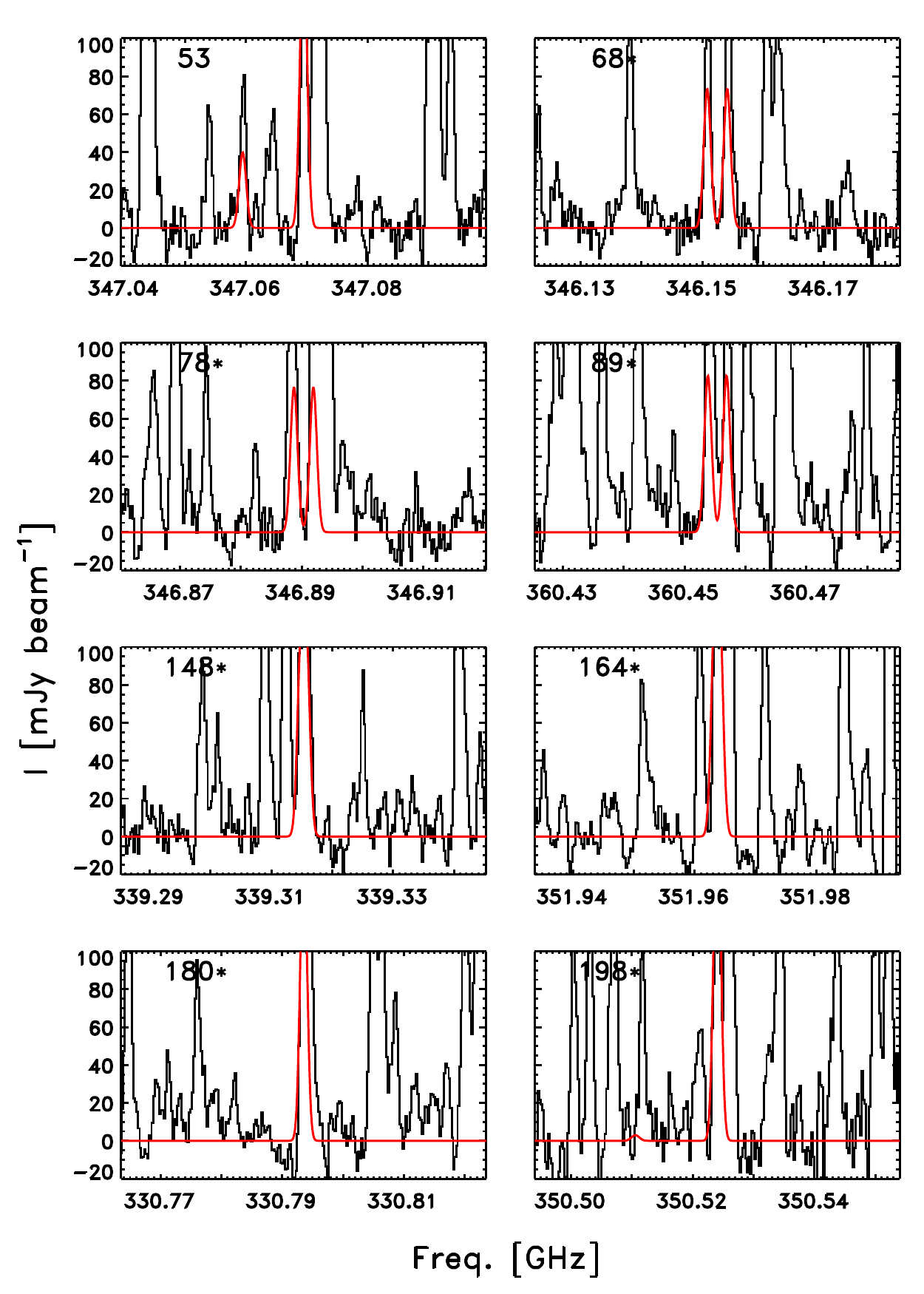}}
\captionof{figure}{As in Fig.~\ref{first_spectra} for the 24+8 
  brightest lines of the deuterated isotopologues of dimethyl ether:
  the antisymmetric form in the upper panels and lower left and the
  (rarer) symmetric form in the lower right panel  (see also Fig.~\ref{dme_spectra}).}\label{deutdme_spectra}
\end{minipage}
\clearpage

\subsection{Acetaldehyde}
\begin{minipage}{\textwidth}
\resizebox{0.88\textwidth}{!}{\includegraphics{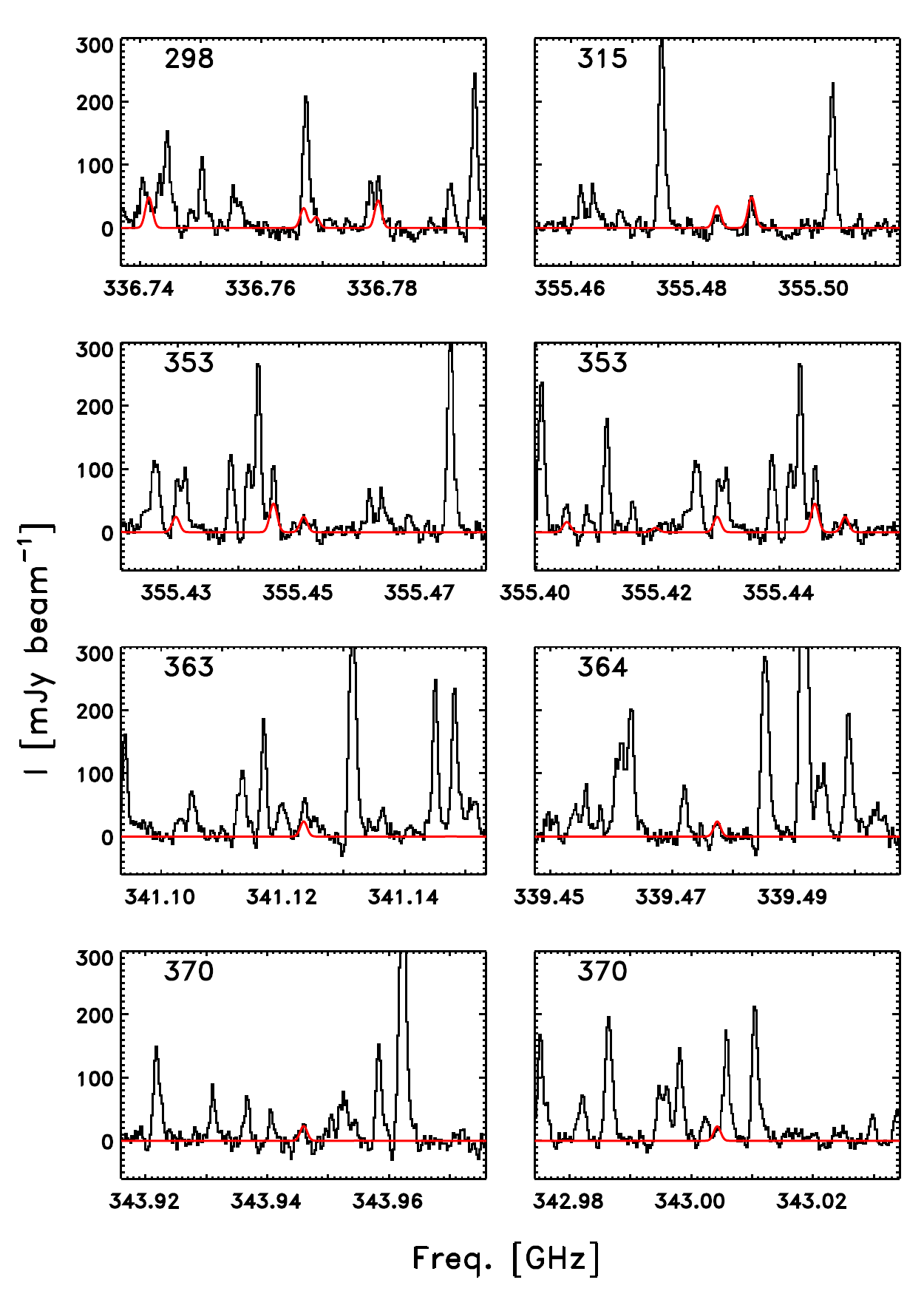}\includegraphics{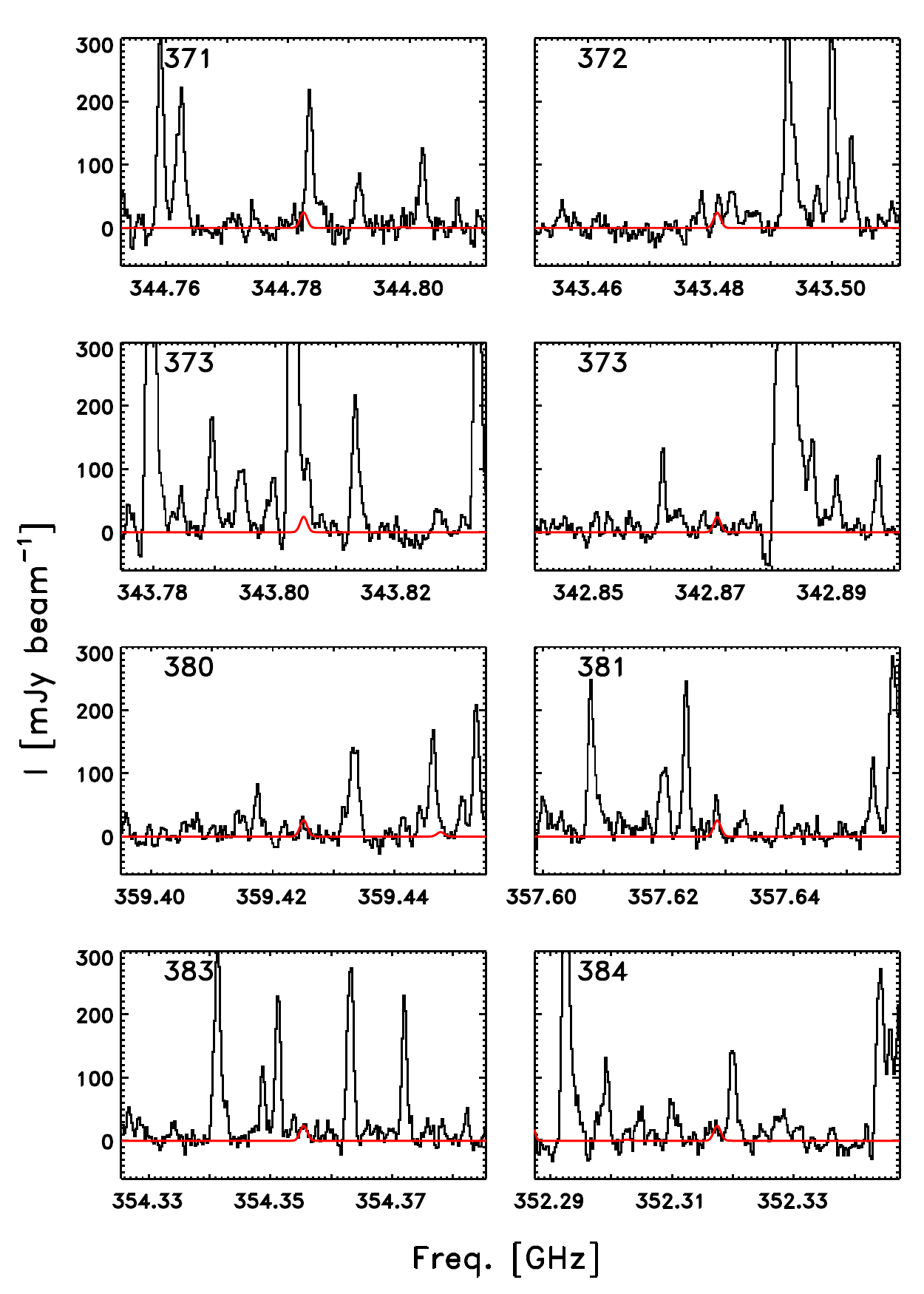}}
\resizebox{0.44\textwidth}{!}{\includegraphics{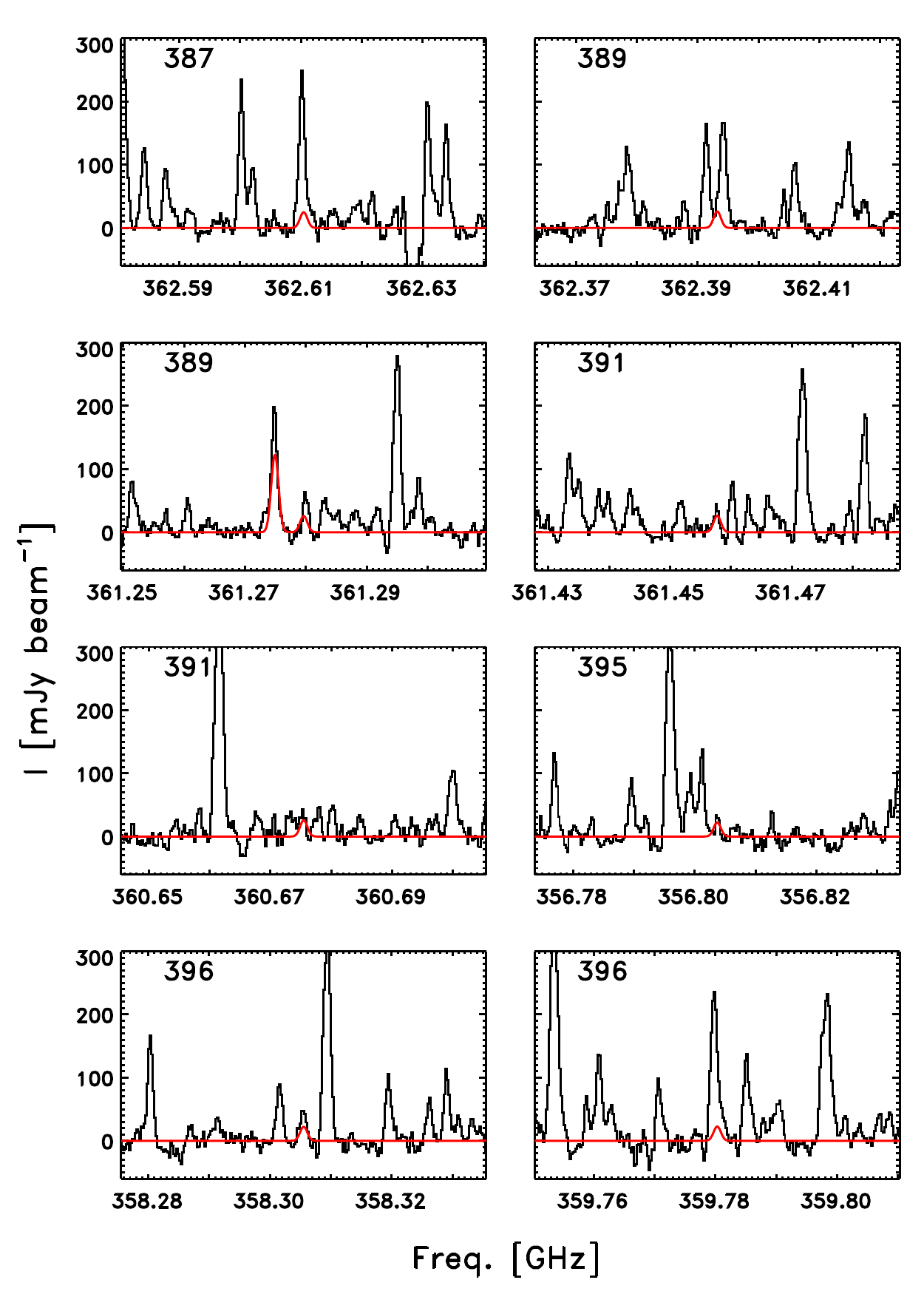}}
\captionof{figure}{As in Fig.~\ref{first_spectra} for the 24 brightest
  lines of one of the $^{13}$C isotopologues of acetaldehyde, $^{13}$CH$_3$CHO, as expected from the synthetic spectrum.}\label{acetaldehyde13_spectra0}
\end{minipage}

\clearpage

\begin{minipage}{\textwidth}
\resizebox{0.88\textwidth}{!}{\includegraphics{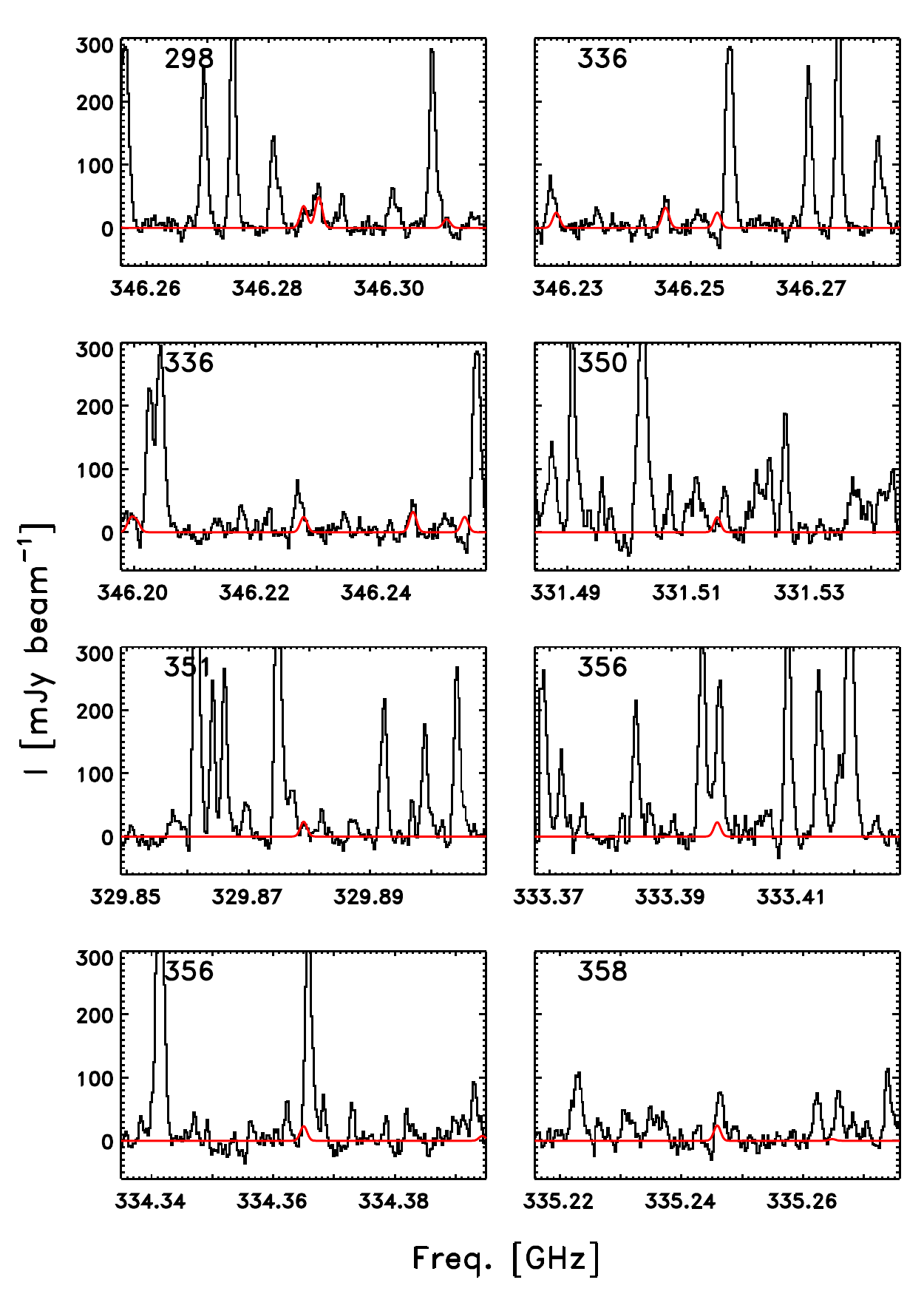}\includegraphics{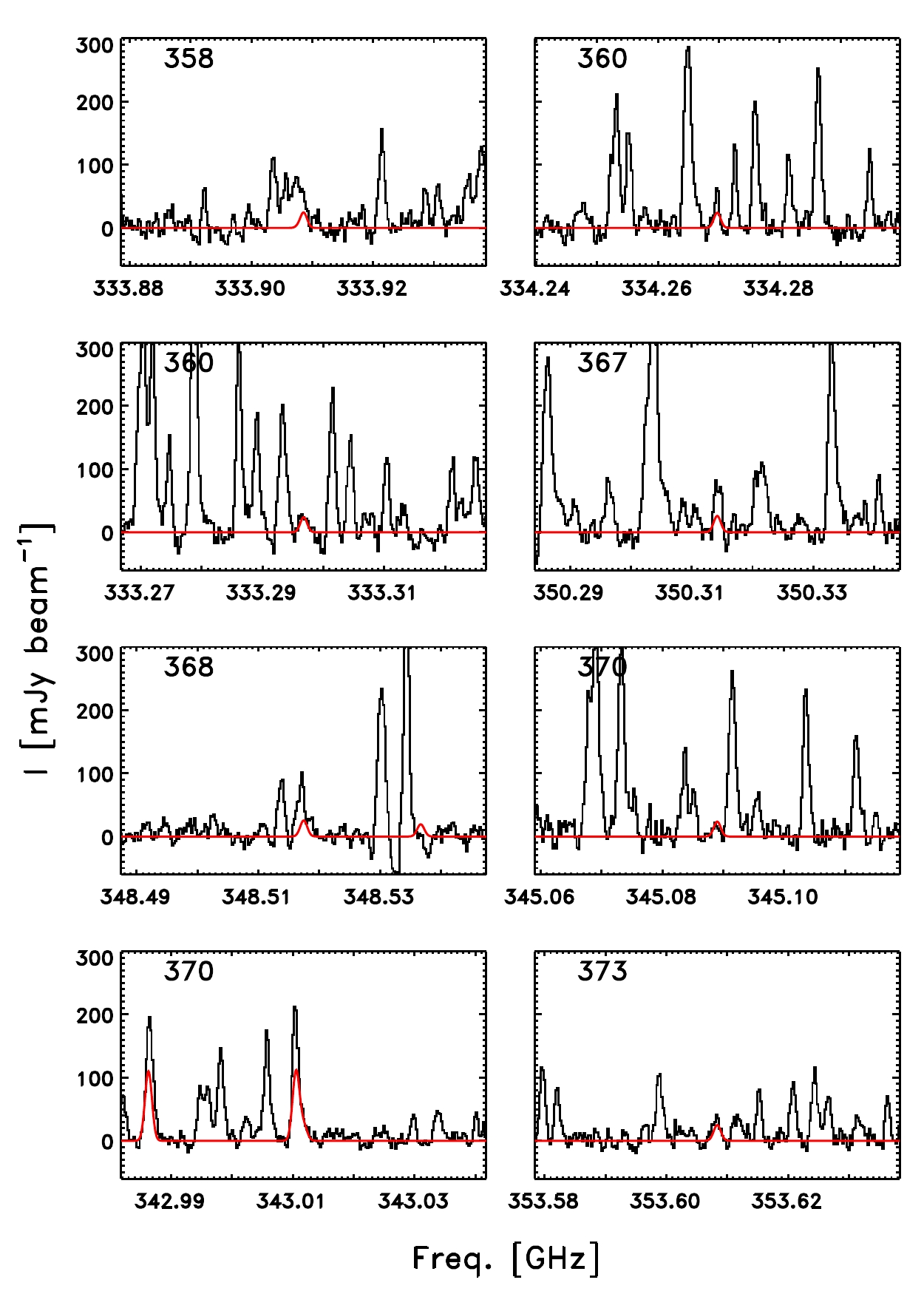}}
\resizebox{0.44\textwidth}{!}{\includegraphics{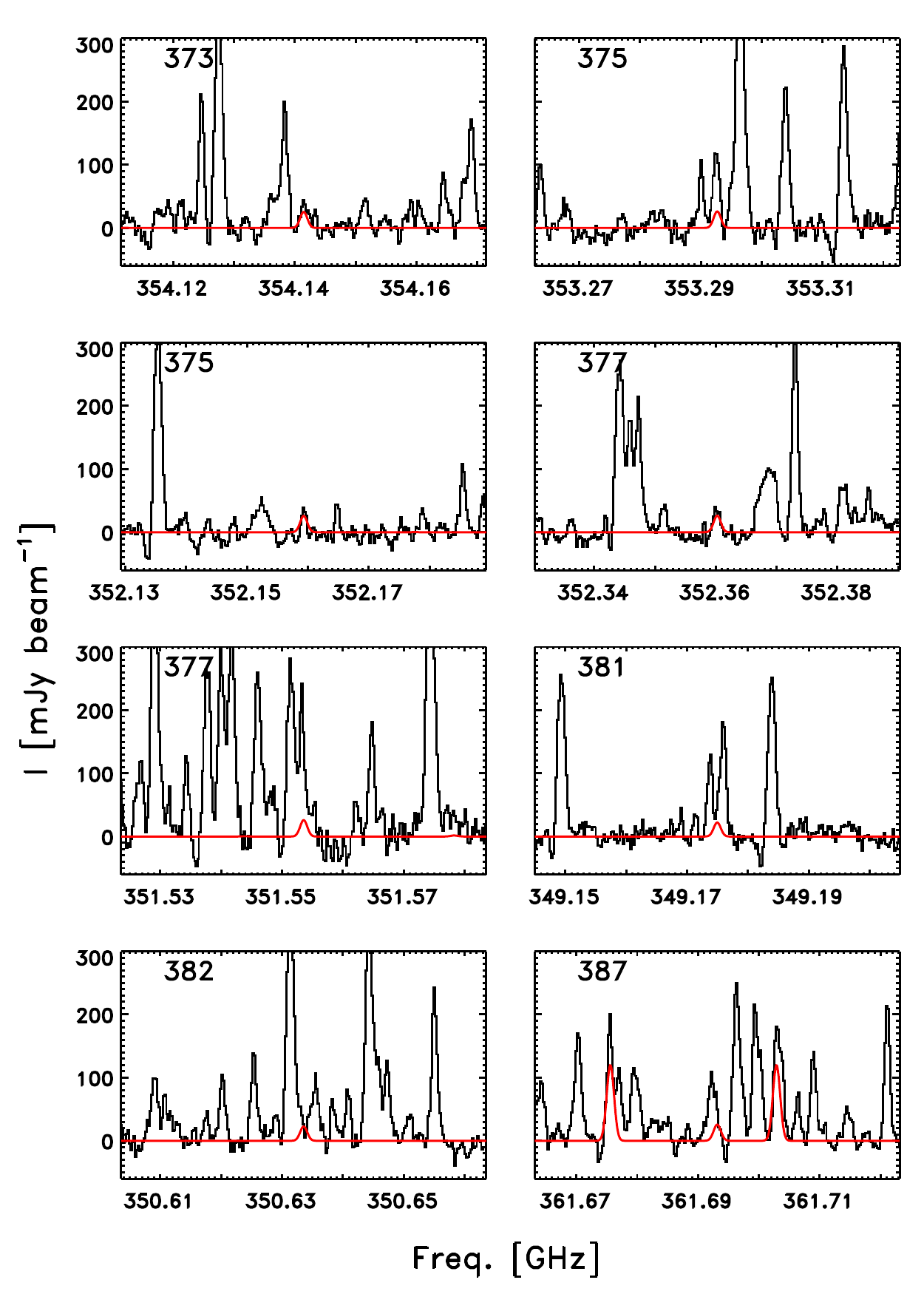}}
\captionof{figure}{As in Fig.~\ref{first_spectra} for the 24 brightest
  lines of the second $^{13}$C isotopologue of acetaldehyde, CH$_3$$^{13}$CHO, as expected from the synthetic spectrum.}\label{acetaldehyde13_spectra1}
\end{minipage}

\clearpage

\begin{minipage}{\textwidth}
\resizebox{0.88\textwidth}{!}{\includegraphics{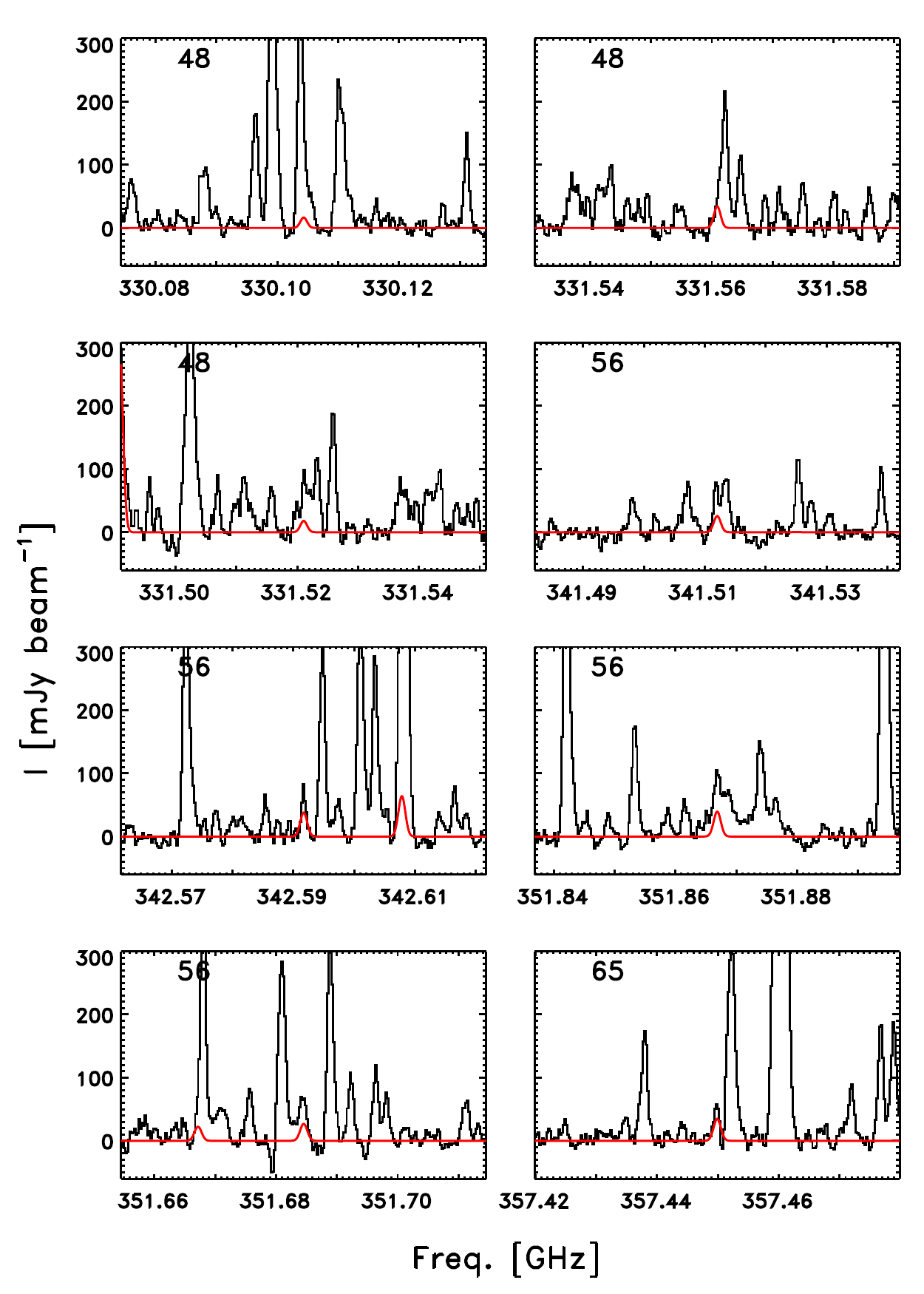}\includegraphics{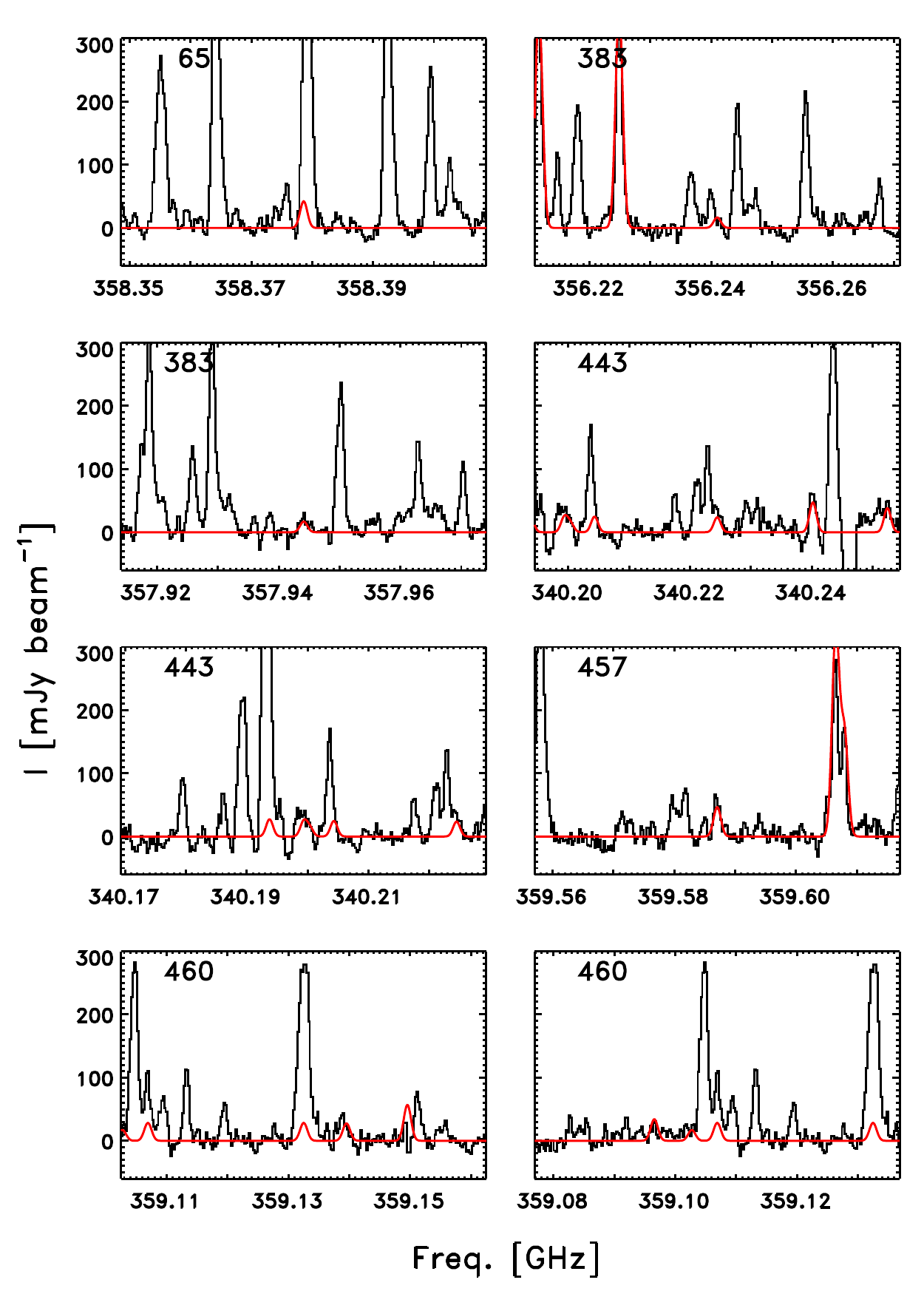}}
\resizebox{0.44\textwidth}{!}{\includegraphics{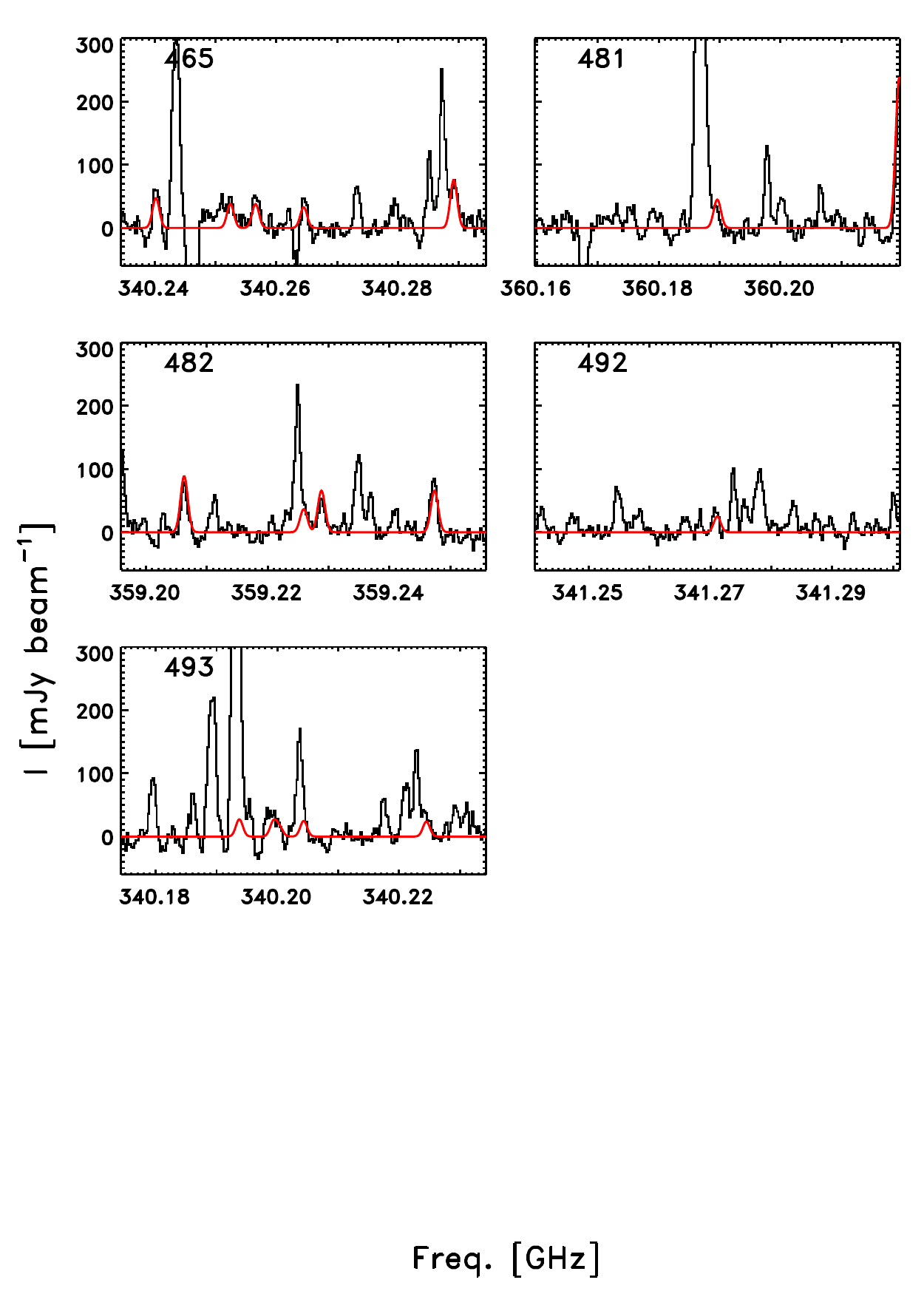}}
\captionof{figure}{As in Fig.~\ref{first_spectra} for the 21 brightest
  lines of the deuterated isotopologue of acetaldehyde, CH$_3$CDO, as expected from the synthetic spectrum.}\label{d-acetaldehyde_spectra1}
\end{minipage}
\clearpage

\subsection{Formic acid}
\begin{minipage}{\textwidth}
\resizebox{0.88\textwidth}{!}{\includegraphics{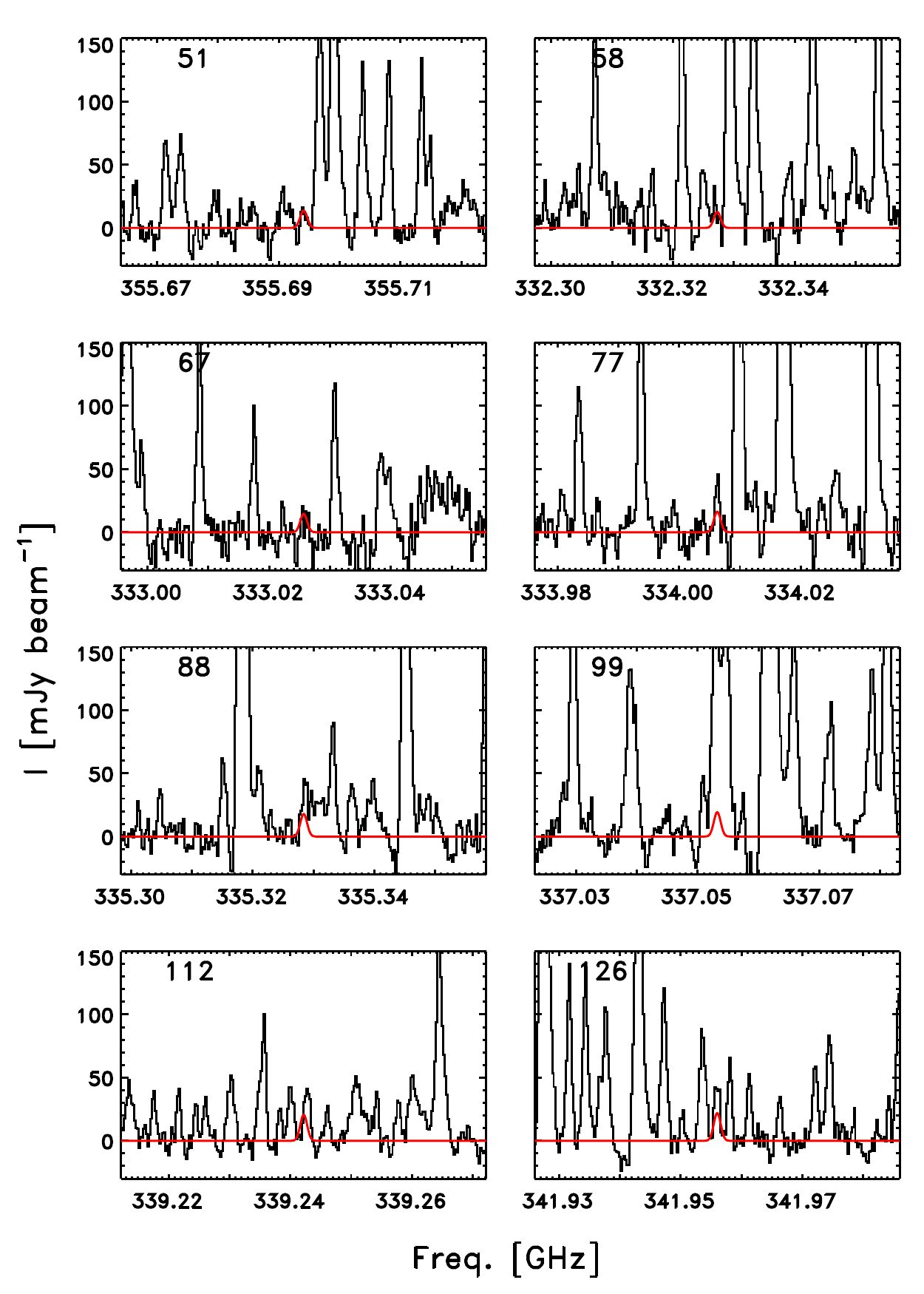}\includegraphics{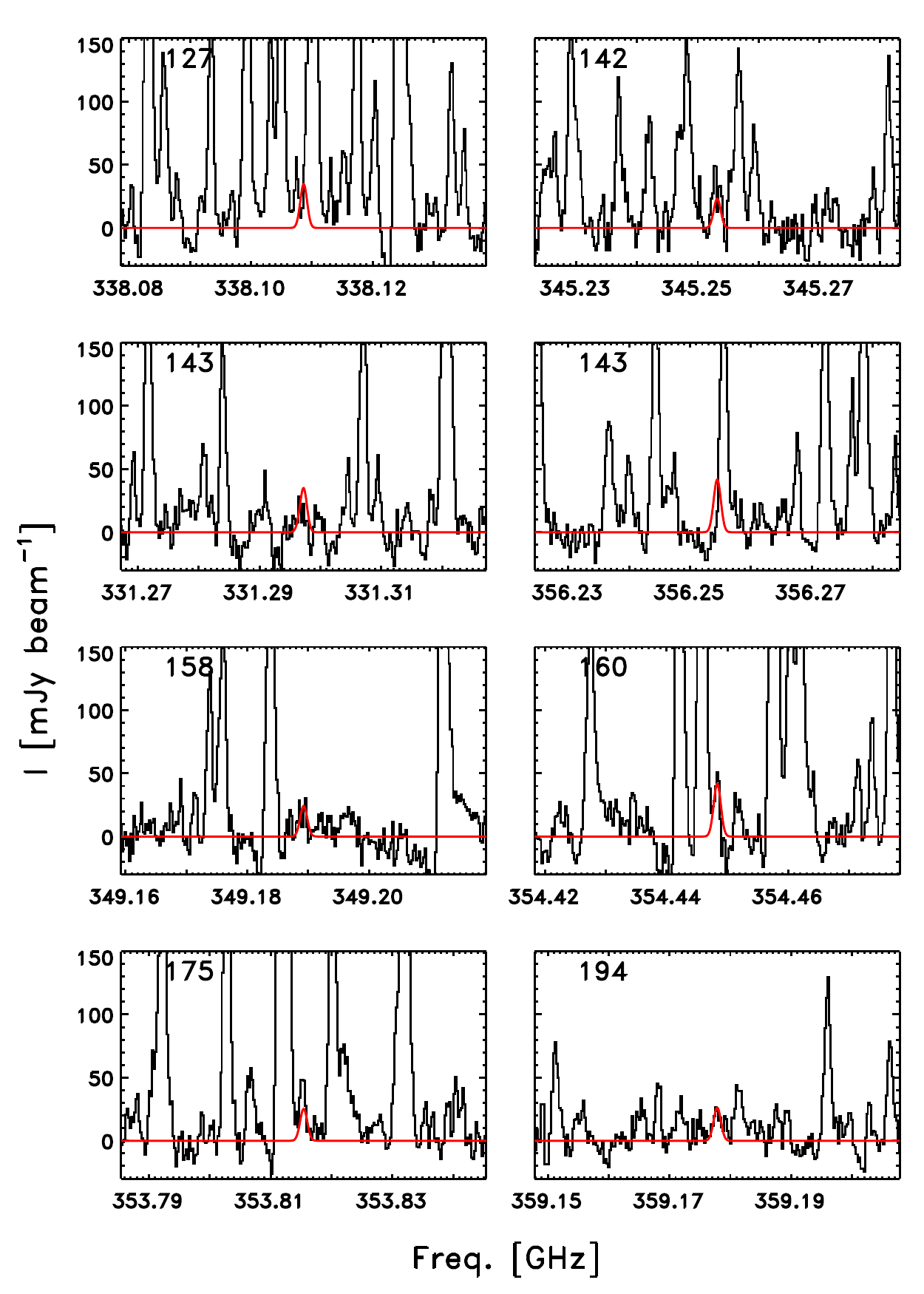}}
\resizebox{0.44\textwidth}{!}{\includegraphics{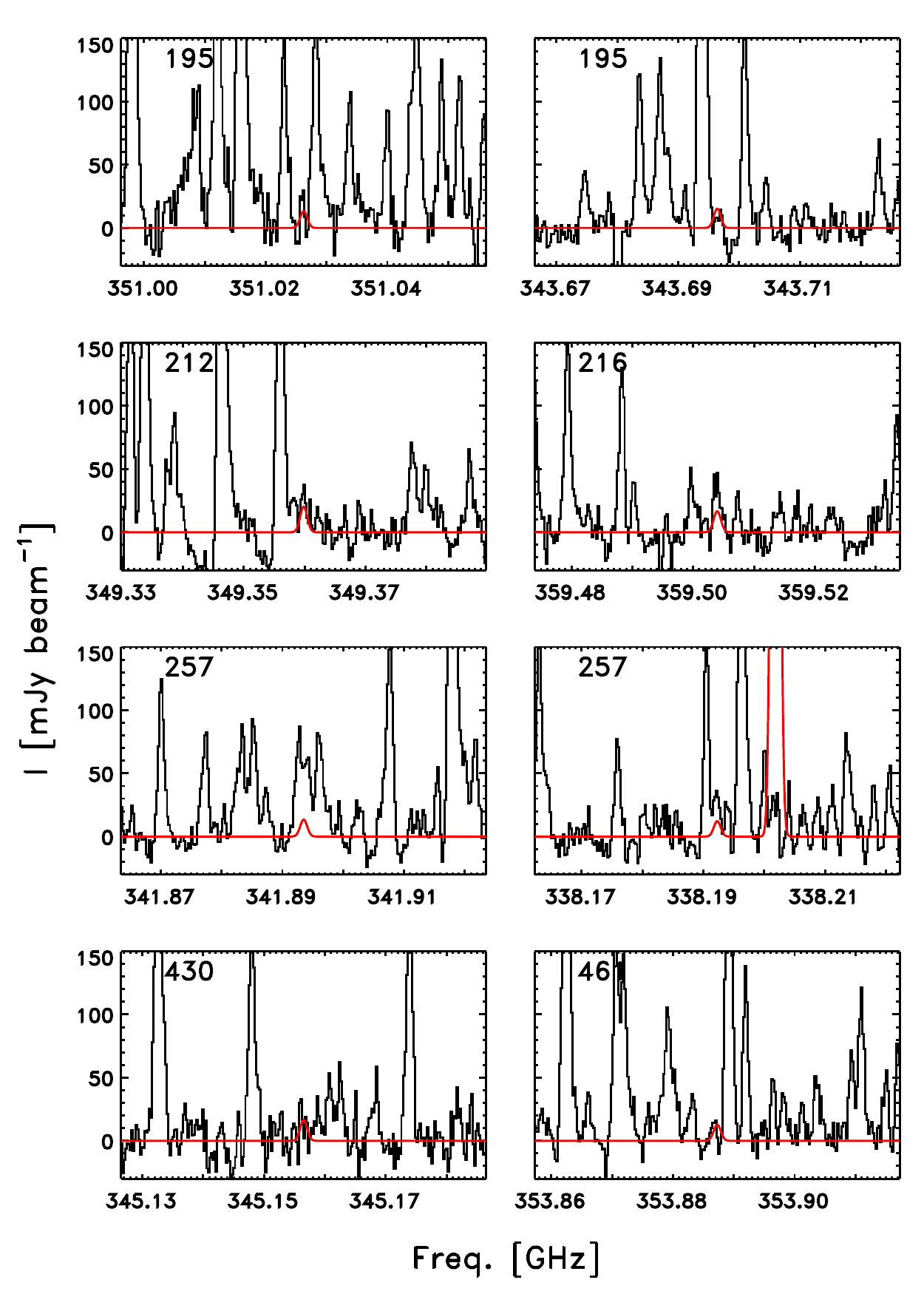}}
\captionof{figure}{As in Fig.~\ref{first_spectra} for the 24 brightest
  lines of formid acid, t-HCOOH, as expected from the synthetic
  spectrum. See comment about over-predicted line at 338.20~GHz
    in 257~K panel in Sect.~\ref{formicacid_discussion}.}\label{thcooh_spectra0}
\end{minipage}

\clearpage

\begin{minipage}{\textwidth}
\resizebox{0.88\textwidth}{!}{\includegraphics{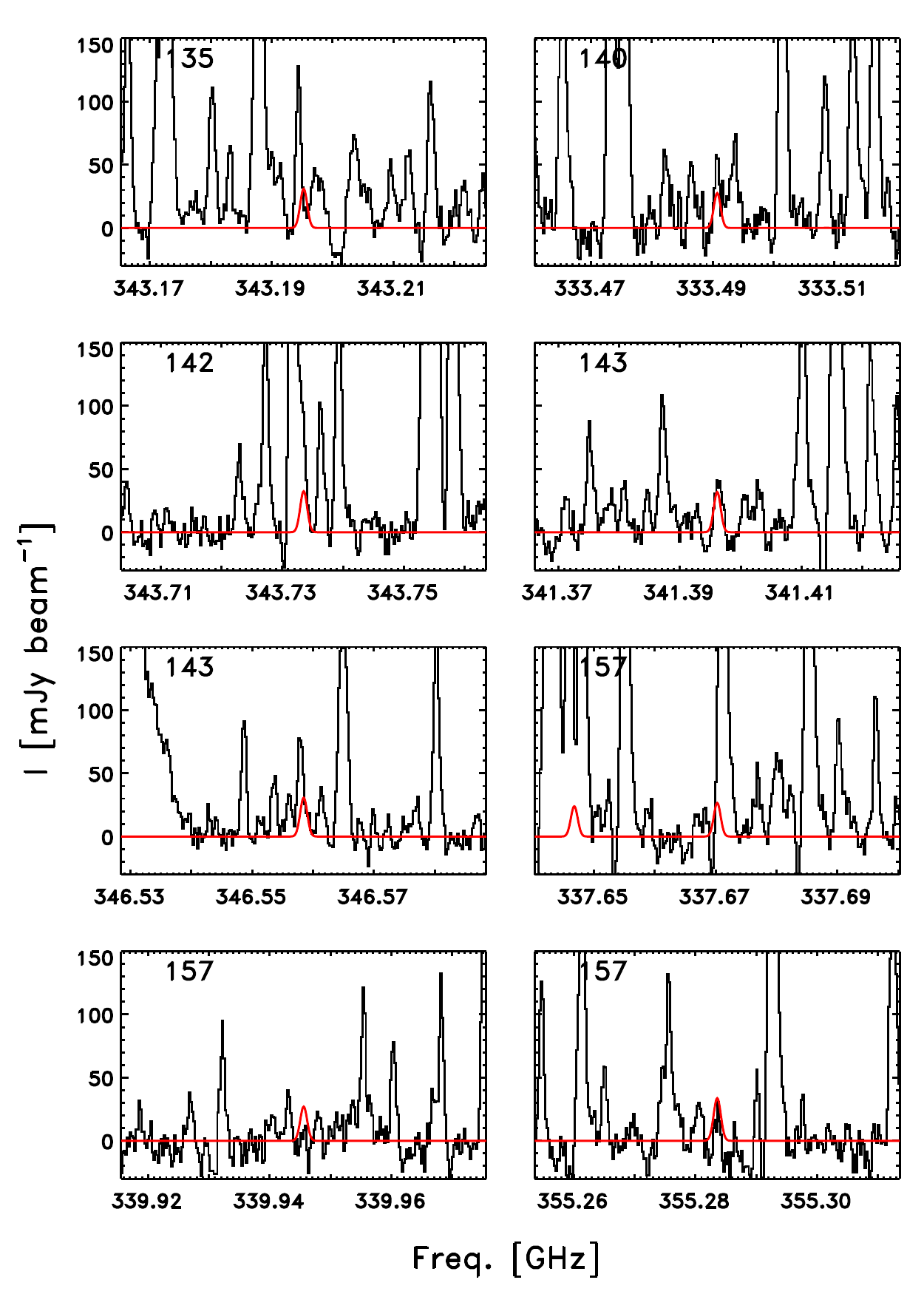}\includegraphics{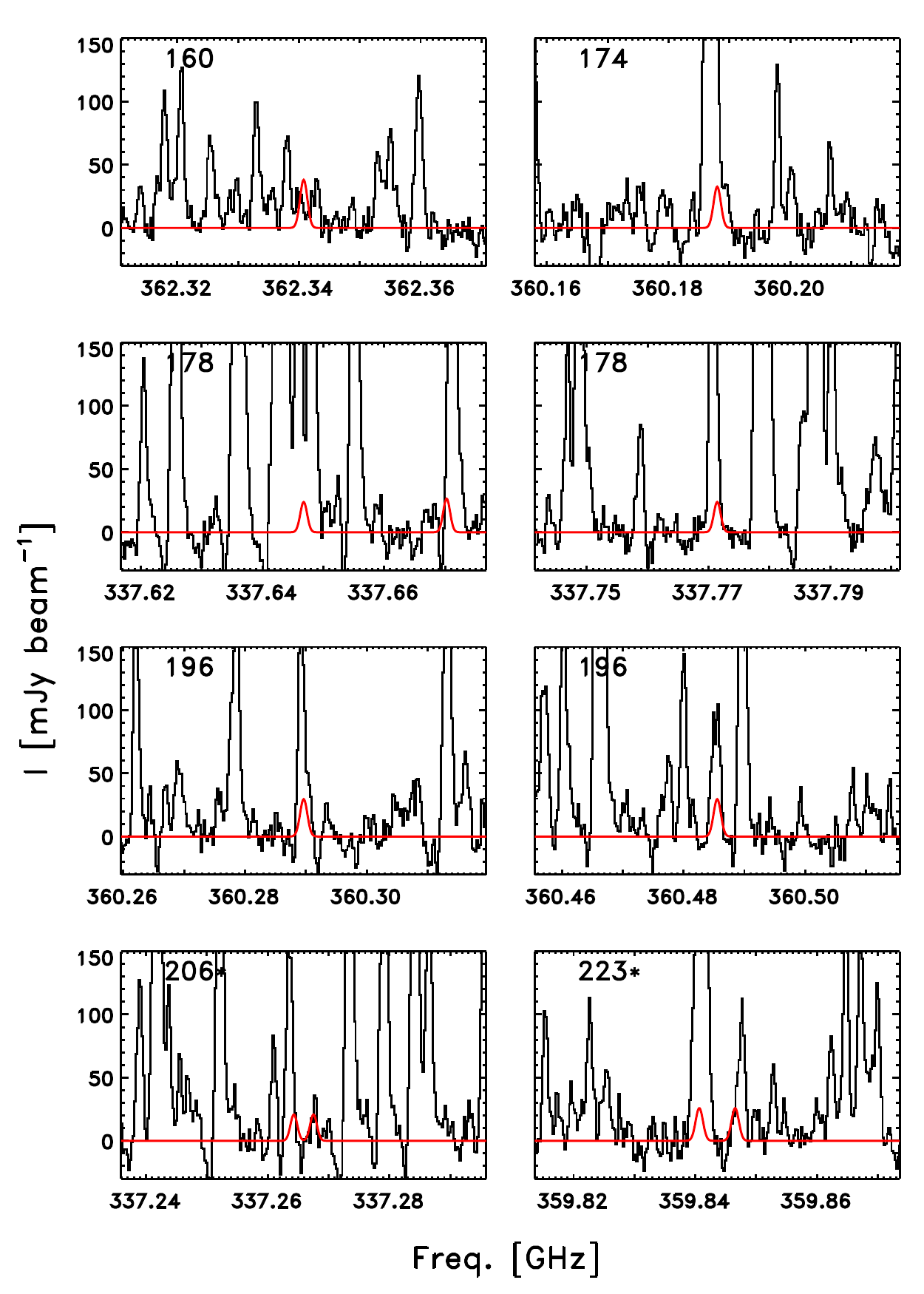}}
\resizebox{0.44\textwidth}{!}{\includegraphics{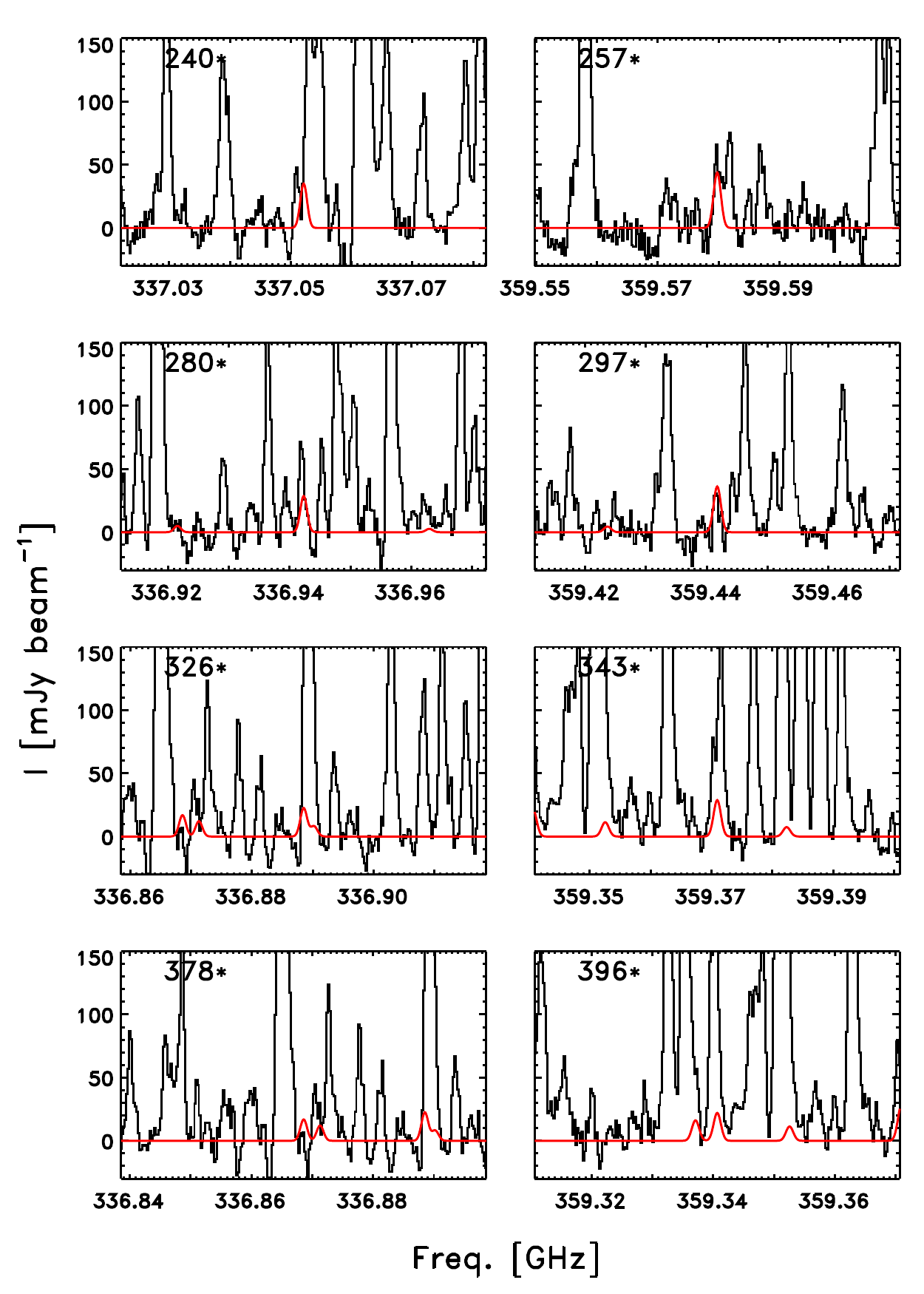}}
\captionof{figure}{As in Fig.~\ref{first_spectra} for the 24 brightest
  lines of the second $^{13}$C isotopologue of formic acid, t-H$^{13}$COOH, as expected from the synthetic spectrum.}\label{th13cooh_spectra1}
\end{minipage}

\clearpage

\begin{minipage}{\textwidth}
\resizebox{0.88\textwidth}{!}{\includegraphics{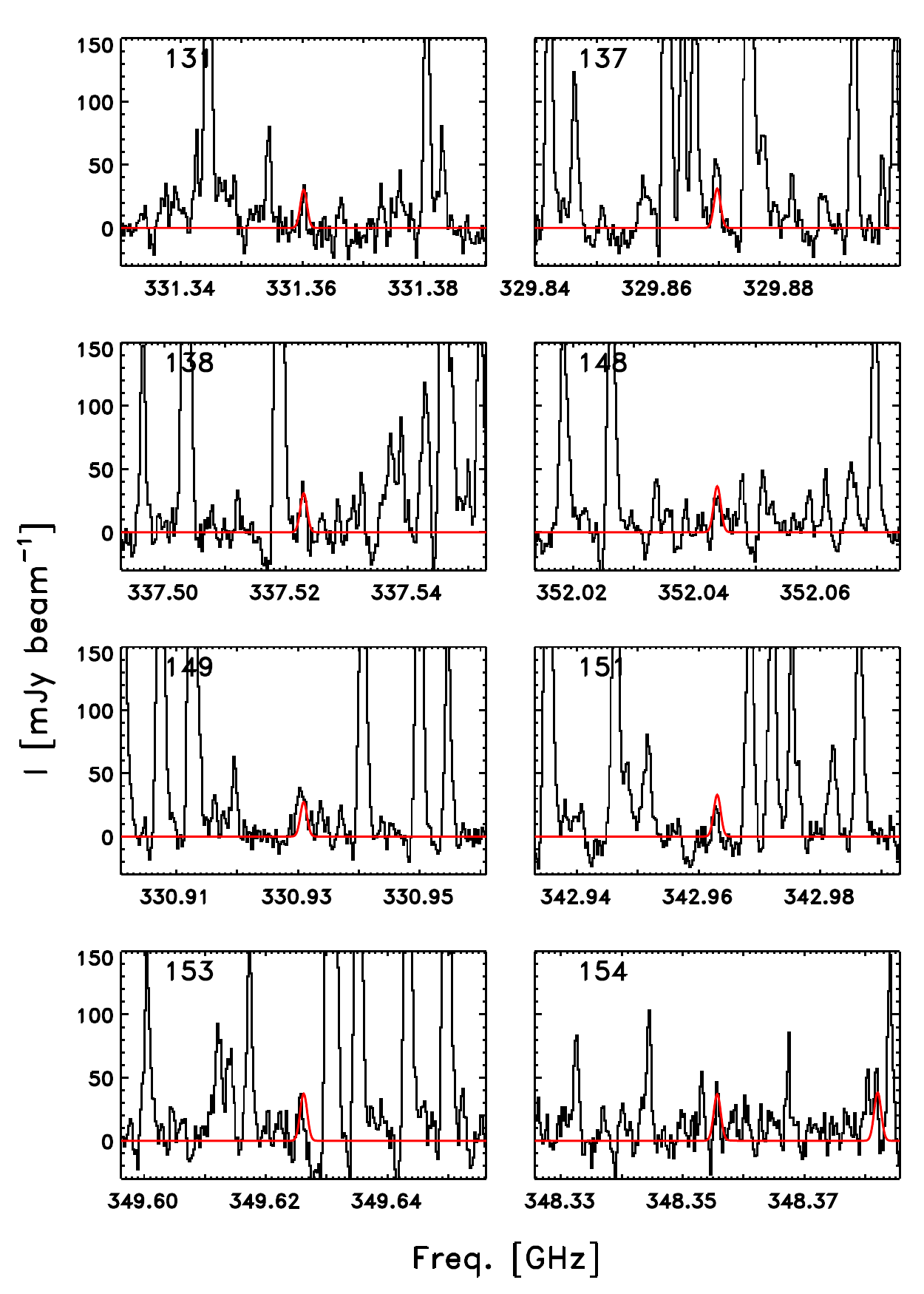}\includegraphics{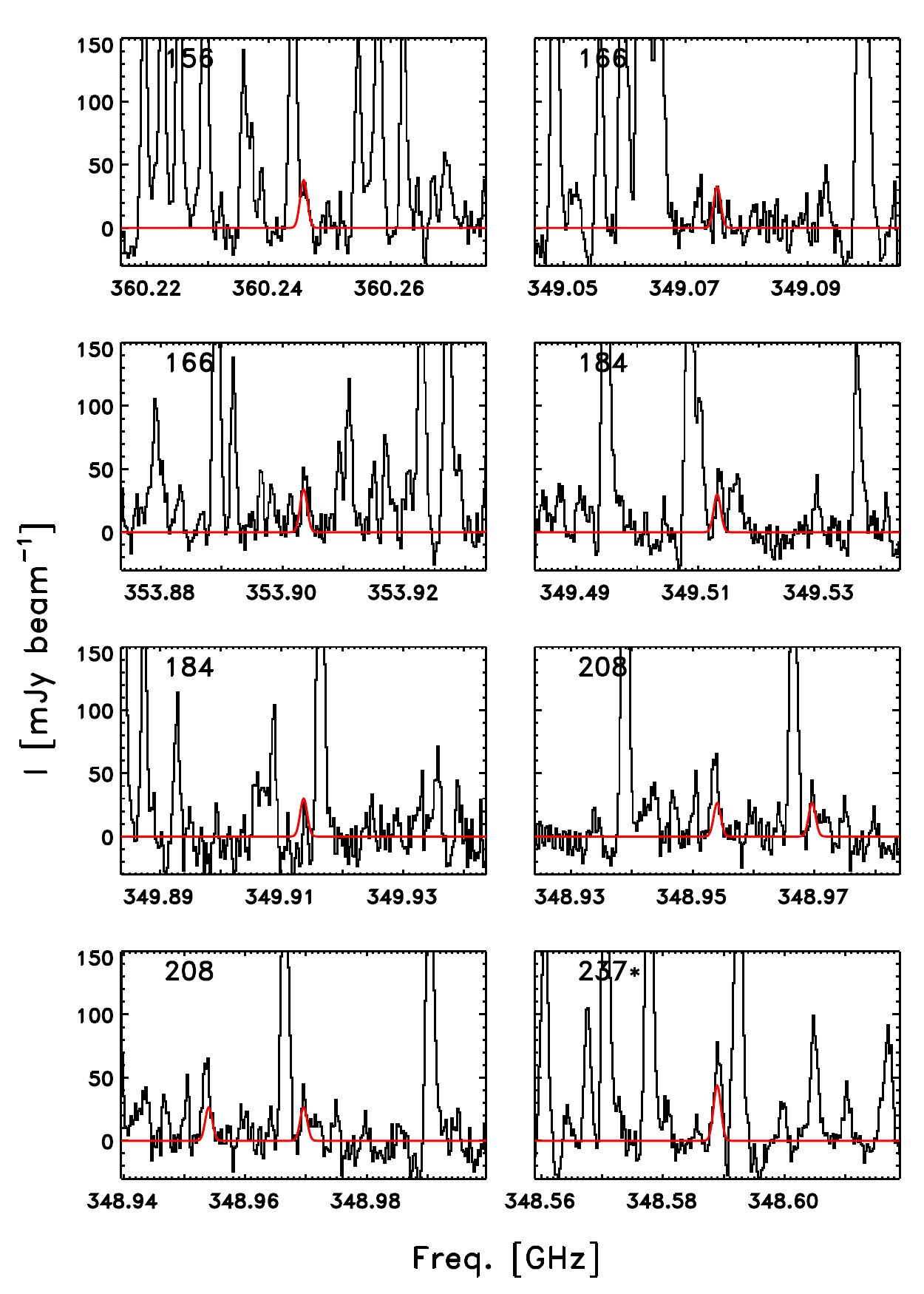}}
\resizebox{0.44\textwidth}{!}{\includegraphics{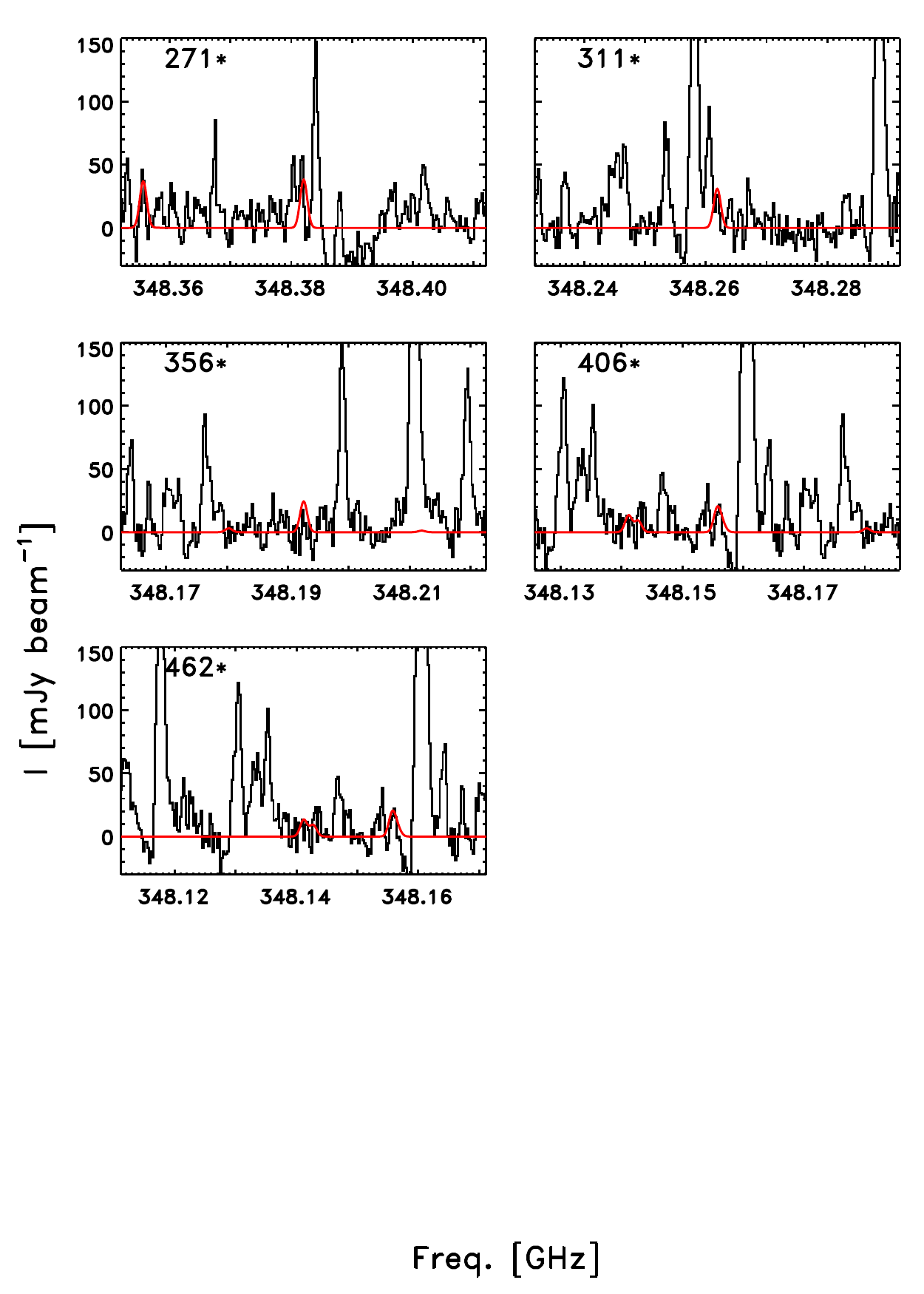}}
\captionof{figure}{As in Fig.~\ref{first_spectra} for the deuterated isotopologue of formic acid, t-HCOOD, as expected from the synthetic spectrum.}\label{t-hcood_spectra1}
\end{minipage}

\clearpage

\begin{minipage}{\textwidth}
\resizebox{0.88\textwidth}{!}{\includegraphics{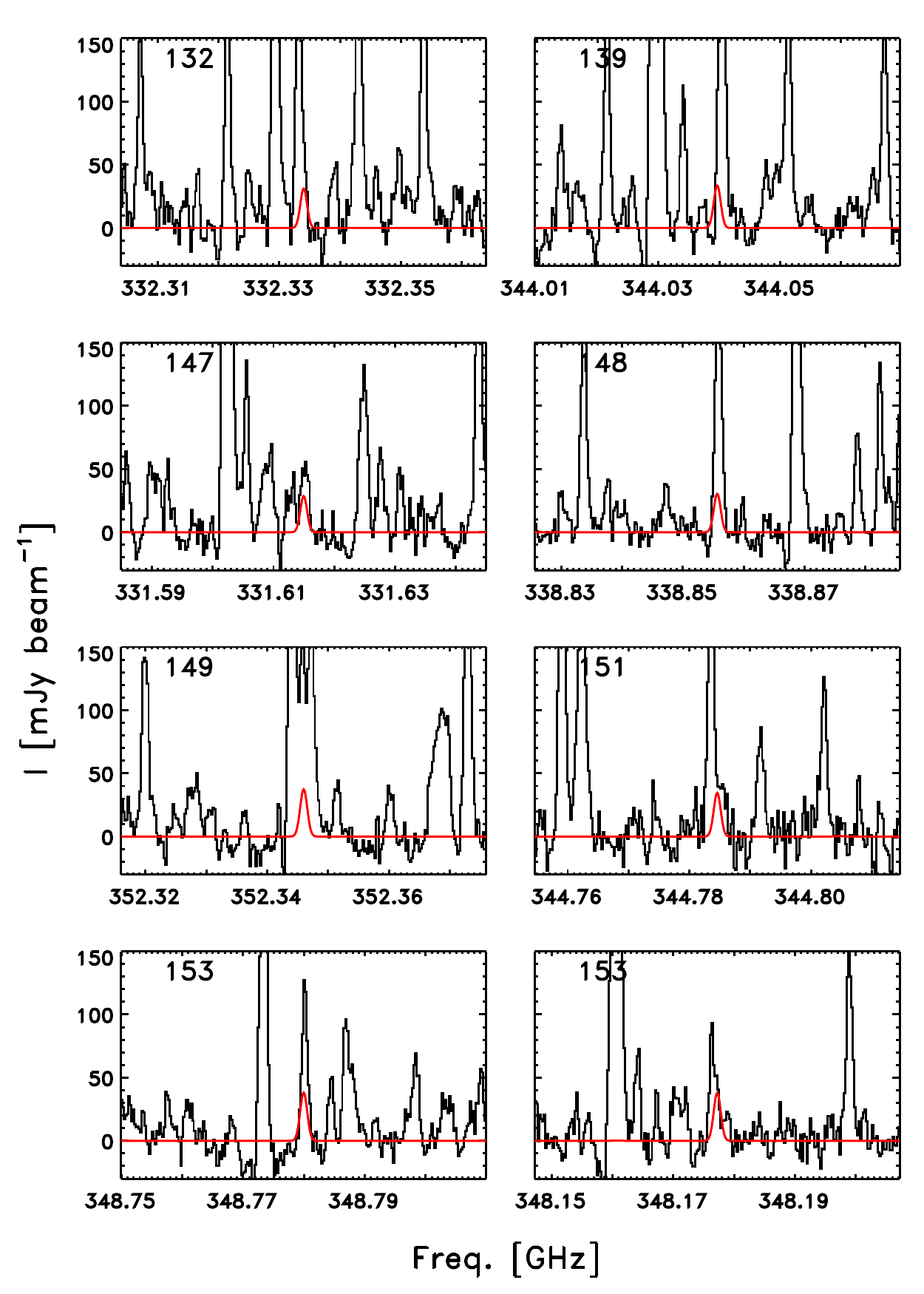}\includegraphics{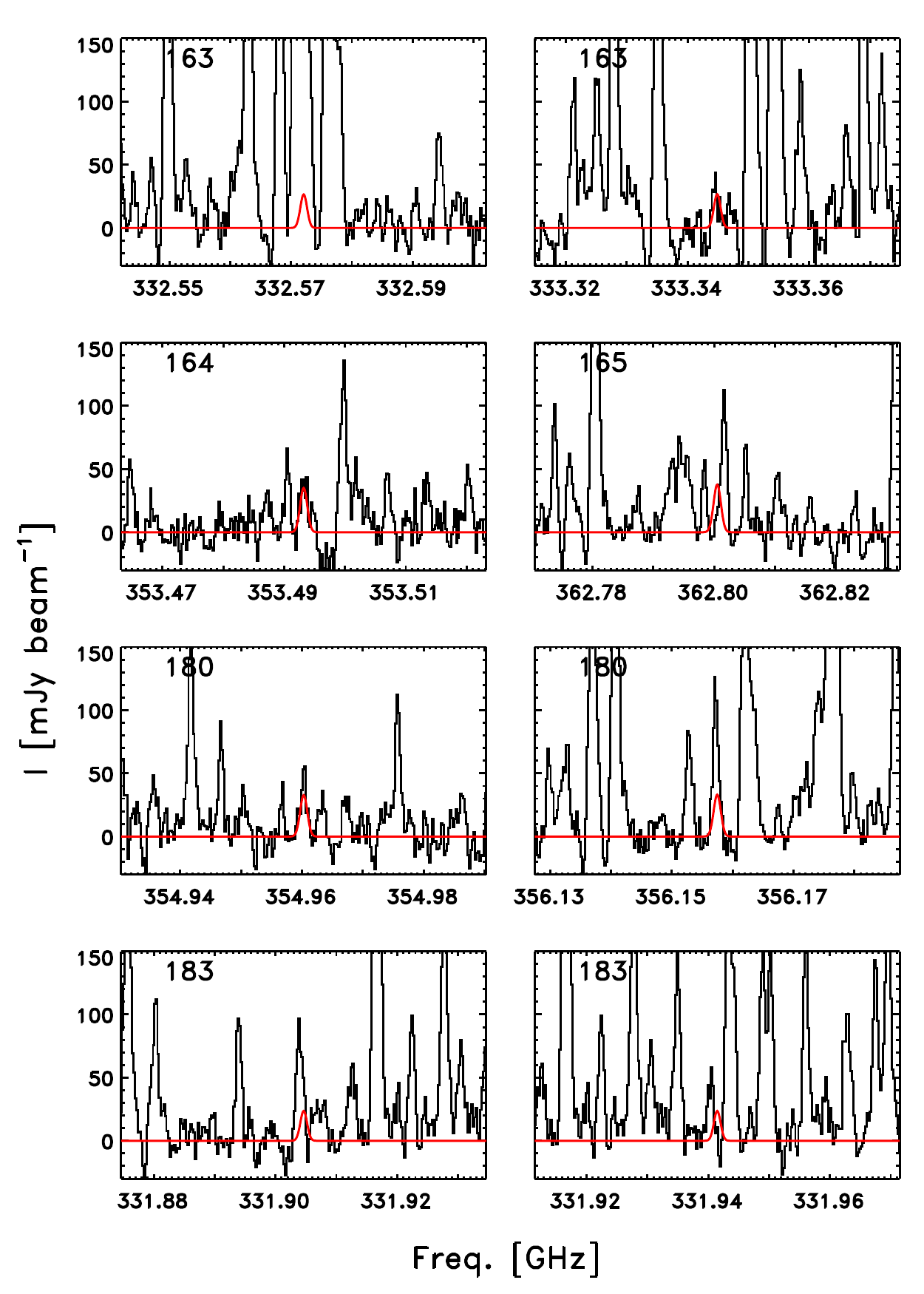}}
\resizebox{0.44\textwidth}{!}{\includegraphics{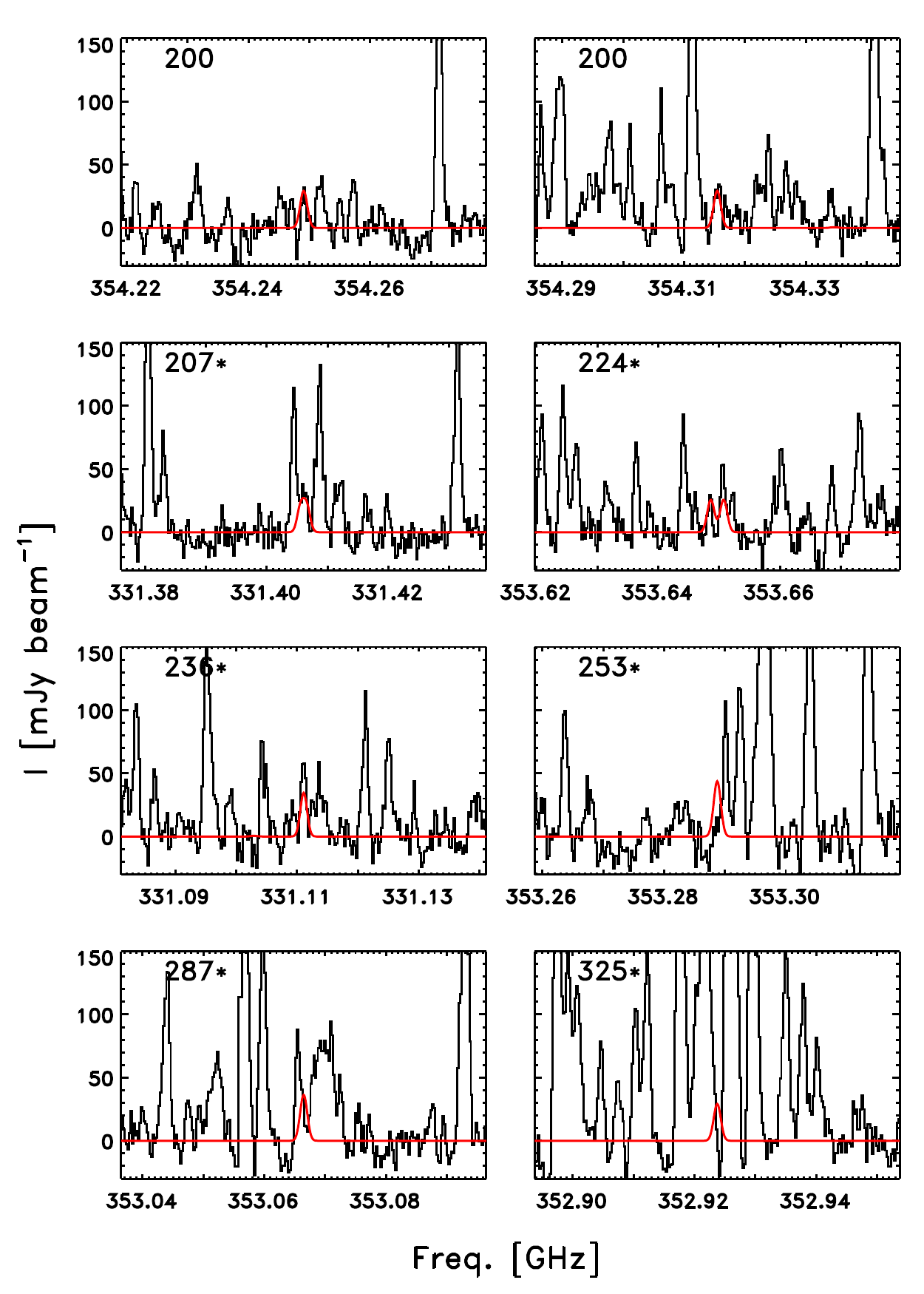}}
\captionof{figure}{As in Fig.~\ref{first_spectra} for the 24 brightest
  lines of the  deuterated isotopologue of formic acid, t-DCOOH, as expected from the synthetic spectrum.}\label{t-dcooh_spectra1}\label{last_spectra}
\end{minipage}

\end{appendix}

\end{document}